\documentclass[manuscript]{aastex}
\usepackage{longtable,lscape}
\usepackage{rotating}
\usepackage{subfigure}

\begin{document}

\newcommand{\kms}{\mbox{km~s$^{-1}$}}
\newcommand{\s}{\mbox{$''$}}
\newcommand{\mloss}{\mbox{$\dot{M}$}}
\newcommand{\my}{\mbox{$M_{\odot}$~yr$^{-1}$}}
\newcommand{\ls}{\mbox{$L_{\odot}$}}
\newcommand{\ms}{\mbox{$M_{\odot}$}}
\newcommand\mdot{$\dot{M}  $}
\newcommand{\sect}[1]{\S\ref{#1}\xspace}
\title{YOUNG PLANETARY NEBULAE: HUBBLE SPACE TELESCOPE IMAGING AND A NEW MORPHOLOGICAL CLASSIFICATION SYSTEM}
\author{Raghvendra Sahai\altaffilmark{1}, Mark Morris\altaffilmark{2}, Gregory Villar\altaffilmark{3}
}
\altaffiltext{1}{Jet Propulsion Laboratory, MS\,183-900, California
Institute of Technology, Pasadena, CA 91109}
\altaffiltext{2}{Division of Astronomy, Department of Physics and Astrophysics, UCLA, Los Angeles, CA 90095}
\altaffiltext{3}{Cal Poly, Pomona}

\begin{abstract}
Using Hubble Space Telescope images of 119 young planetary nebulae, most of which have not previously been published, we have
devised a comprehensive morphological classification system for these objects. This system generalizes a recently devised
system for pre-planetary nebulae, which are the immediate progenitors of planetary nebulae (PNs). Unlike previous
classification studies, we have focussed primarily on young PNs rather than all PNs, because the former best show the
influences or symmetries imposed on them by the dominant physical processes operating at the first and primary stage of the
shaping process. Older PNs develop instabilities, interact with the ambient interstellar medium, and are subject to the
passage of photoionization fronts, all of which obscure the underlying symmetries and geometries imposed early on. Our
classification system is designed to suffer minimal prejudice regarding the underlying physical causes of the different
shapes and structures seen in our PN sample, however, in many cases, physical causes are readily suggested by the geometry,
along with the kinematics that have been measured in some systems. Secondary characteristics in our system such as ansae
indicate the impact of a jet upon a slower-moving, prior wind; a waist is the signature of a strong equatorial concentration
of matter, whether it be outflowing or in a bound Keplerian disk, and point symmetry indicates a secular trend, presumably
precession, in the orientation of the central driver of a rapid, collimated outflow. 
{\bf [The quality of the figures as it appears in the arXiv pdf output is not up-to-par; the full ms with
high-quality figures is available by anonymous FTP at ftp:\//\//ftp.astro.ucla.edu\//pub\//morris\//AJ-360163-sahai.pdf]}.
\end{abstract}


\section{Introduction}\label{intro}
Although preplanetary nebulae (PPNs) \& planetary nebulae (PNs) evolve from initially spherically--symmetric mass-loss
envelopes around AGB stars, modern ground-based imaging surveys have shown that the vast majority of the former deviate
strongly from spherical symmetry (e.g., Schwarz, Corradi \& Melnick 1992, Manchado et al. 1996a). In a morphologically
unbiased survey of young PNs with the Hubble Space Telescope (HST), Sahai \& Trauger (1998, ST98) found no round objects, but
a variety of bipolar and multipolar morphologies. The significant changes in the circumstellar envelope (CSE) morphology
during the evolutionary transition from the AGB to the post-AGB (pAGB) phase require a primary physical agent or agents which
can break the spherical symmetry of the radiatively-driven, dusty mass-loss phase. In the ``generalised
interacting-stellar-winds" (GISW) model, the AGB CSE is assumed to be equatorially dense, and the expansion of a fast
($>1000$\,\kms) isotropic wind from the PN central star produces an aspherical PN (Balick 1987). Hydrodynamic simulations
based on this model could reproduce a variety of axisymmetric shapes (e.g., review by Balick \& Frank 2002). However, 
Soker (1997, 1990) pointed out that the GISW model could not explain the presence of point symmetry or collimated flows and ansae in
PNs. 
Jet-like outflows were first used to explain bipolar morphology in a small sample of nebulae around evolved stars (Morris 1987, 1990),
and the presence of ansae in PNs (Soker 1990). Faced with the
complexity, organization and frequent presence of point-symmetry in the morphologies of their survey PNs, ST98 proposed that
the primary agent for breaking spherical symmetry is a jet or collimated, fast wind (CFW) operating during the early post-AGB
or late AGB evolutionary phase. The CFWs are likely to be episodic, and either change their directionality (i.e., wobbling of
axis, or precession) or have multiple components operating in different directions (quasi)simultaneously (Sahai 2004). In the
ST98 model, primary shaping begins {\it prior} to the PN phase, and the variety of PN shapes and structure depends in detail
on the CFW characteristics (direction, strength, opening angle, temporal history).

Direct evidence for CFWs during the pre-PN phase has come from sensitive molecular line observations which reveal the
presence
of very fast (few$\times$100 \kms) molecular outflows in PPNs and a few very late AGB stars, with huge momentum-excesses
showing that these winds cannot be radiatively driven (e.g., Bujarrabal et al. 2001, Sahai et al. 2006). Using STIS/HST, a
carbon star, V Hya, has been ``caught in the act" of ejecting a very fast (250\,\kms), highly collimated blobby outflow
(Sahai et al. 2003a). Strong support for the ST98 model was recently provided by a (morphologically) unbiased HST imaging
survey of young PPNs which shows very close similarities in morphology between these objects and young PNs (Sahai et al.
2007a: SMSC07). If the ST98 model is correct, then the question arises: what is the engine for producing CFW's? If
point-symmetric shapes result from the flow collimator precessing or becoming unstable, then what causes the destabilization?
Can CFW's be produced by single stars or is a binary companion essential? Single-star models have invoked stellar rotation,
strong magnetic fields, or both (e.g., Garcia-Segura et al 1999, Blackman et al. 2001), and binary models have invoked the
angular momentum and/or the gravitational influence of a companion (e.g., Morris 1981,\,1987, Soker \& Livio 1994, Livio \&
Pringle 1997). Yet, in spite of vigorous debate (e.g., Bujarrabal et al. 2000), no consensus has yet emerged even as to which
of the above two broad classes of models is correct (Balick \& Frank 2002)!

Morphological classification schemes can play an important role in constraining the physical mechanism or
mechanisms that influence the mass loss process. Several PN classification schemes have been presented previously.
The major themes of the earliest classifications were based on whether objects were round or elliptical (Zuckerman
\& Aller 1986). These were refined with the inclusion of bipolar objects and recognition of objects with point
symmetry. The most detailed of such schemes has been presented in papers by Schwarz et al. (1992), Schwarz,
Corradi \& Stanghellini (1993: SCS93), and Corradi \& Schwarz (1995: CS95), based on a sample of 400 objects, with
four morphological classes: elliptical (includes round objects), bipolar, point-symmetrical, and irregular.
Manchado et al. (1996a: Metal96) have published an atlas of 243 PNs (non-refereed), and presented a morphological
classification which is similar to the one described by CS95.

However, all of these earlier schemes are based on ground-based imaging, which gives a typical angular resolution
of $\gtrsim 1{''}$. This resolution precludes recognition of the important morphological traits of
most PNs, especially if they are young and therefore physically small. But even for larger and well-resolved
objects, a major difficulty arises due to the typical brightness distribution of a bipolar or multipolar PN --
such PNs have waists\footnote{an equatorially-flattened central region separating extended lobes oriented
near/along a  polar axis} that are very bright, compared to their extended lobes, and the convolution of such a
distribution with a seeing Gaussian function whose size is comparable to the size of the waist, can result in a
shape which looks roughly elliptical. But profiting from the capabilities of HST, several PN surveys have been
carried out, revealing the structures of PNs with unprecedented detail -- these cannot be adequately described by
the previous classification schemes. Now is an appropriate time for undertaking a new classification scheme that
is more detailed, inclusive, and precise than the previous ones, and that can best elucidate the
predominant physical processes that contribute to the observed morphologies.


In this paper, we propose a comprehensive morphological classification system for these objects, based on such
a system for PPNs devised using their unbiased HST imaging survey of the latter (SMSC07). SMSC07
found a wide variety of morphologies in PPNs, qualitatively similar to those found for young PNs, which is
physically intuitive, since young PNs represent the immediate evolutionary phase after the PPN phase. We have
therefore extended the SMSC07 PPN classification system to {\it young} PNs (for an operational definition of
``young PNs", see \S\,\ref{obs}). Unlike previous classification studies, we have focussed primarily on young PNs
rather than all PNs, because the former best show the influences or symmetries imposed on them by the dominant
physical processes operating at an earlier stage of the shaping process. Older PNs lose these characteristics due
to the continued operation of the very fast central star wind and photoionization, and associated dynamical
instabilities. Further, interaction with the ISM becomes important for old PNs, and can introduce a new set of
morphological features in these objects (Dgani \& Soker 1998). We show that the morphological system for PPNs can
be adapted for young PNs directly, but with modifications and extensions.

The plan of our paper is as follows. In \S\,\ref{obs}, we summarise the selection criteria of our sample, and the various HST
surveys and other GO programs from which the images have been taken; in \S\,\ref{result}, we describe the primary classes
(\S\,\ref{primary}), the secondary characteristics (\S\,\ref{secondary}), the determination of nebular expansion ages
(\S\,\ref{age}), classification statistics (\S\,\ref{stats}), the limitations imposed by imaging resolution, sensitivity and nebular
orientation effects (\S\,\ref{effects}, application of our classification scheme to a new PN sample(\S\,\ref{newsamp}), and in \S\
\ref{discuss}, we conclude with a discussion on how
our classification scheme is relevant for understanding the formation and shaping of planetary nebulae.


\section{Observations}\label{obs}
The objects included in this work mostly come from several surveys with HST/WFPC2, most of them fitting the
selection criterion of ST98, namely that the [OIII]$\lambda$5007/H$\alpha$ flux ratio, R$_{exc}$, be less
than about unity\footnote{ST98 use R$_{exc}<1/1.5$; we have relaxed this a bit, R$_{exc}\le1$, in order to
include a larger PN sample}, used to select young PNs. ST98 argue that R$_{exc}$ is expected to be low in
young PNs because (1) the central stars have low effective temperatures (25,000-40,000\,K), resulting in a low
state of nebular excitation and therefore a low [OIII] flux for the bulk of the nebular gas, and (2) young PNs are
compact, with large dust optical depths towards their central regions, resulting in a large selective extinction
of the shorter wavelength [OIII] line compared with H$\alpha$, since [OIII] is more centrally distributed.

Two of the major surveys used in our study, GO\,6353 and 8345, (PI: Sahai) were specifically carried out using R$_{exc}\le1$ as the
selection criterion, with the goal of studying young PN morphologies. The third major survey from which we
have selected young PNs meeting the ST98 criterion is GO\,9356 (PI: A. Zijlstra), which covered Galactic Bulge PNs.
Smaller numbers of objects which fit the ST98 criterion were taken from GTO\,6221 (PI: J. Trauger), GO\,8307 (PI: S. Kwok) and GO\,6347
(PI: K.
Borkowski). For five objects (PNG051.5+00.2, PNG061.3+03.6, PNG067.9-002.2, PNG110.1+01.9, PNG332.9-09.9), that were imaged as part of
SNAPshot surveys for PPNs (GO\,9463 \& GO\,10536, PI: Sahai), we used the broad-band filter images (at 0.6\,\micron~and 0.8\,\micron)
available.  We supplemented our sample further with additional objects as follows (i) 1 PN, with
R$_{exc}$ formally larger than, but within measurement errors, not significantly different from unity, and (ii) 23 PNs
with R$_{exc}>1$ (generally small-sized, and therefore likely to be young as confirmed by our age estimates, see \S\
\ref{age}).

Most of these images were obtained in HST's SNAPshot mode, with relatively modest
integration times. The total number of objects that are included in this study is 119. A log of the observations is provided
in Table\,\ref{obslog}. The objects in the table are listed in order of increasing galactic longitude (first number in the PK
or PNG designation); when the longitude is the same, then in order of decreasing galactic latitude (second number including
sign, in the PK or PNG designation). The table first lists all objects with R$_{exc}\le1$ (from PK000+17D1 to PNG359.2+04.7) followed
by all objects where this ratio is greater than unity (Table\,\ref{obslog}). The last
column in
the table lists the name of the
dataset in the HST archive.

A significant fraction of an ongoing new large SNAPshot survey of PNs with the WFC3/HST instrument (GO 11657, PI:
Stanghellini) has recently been completely and the results are in the public domain. This survey utilises one narrow-band
filter (F502N, covering the [OIII]$\lambda$5007 line), and 3 broad band filters (F200LP, F350LP, and F814W). Inspection of
the images from this survey shows that the F200LP and F350LP bandpasses, which are extremely wide, and cover all major
nebular emission lines, including [OIII], [NII] and H$\alpha$, show the morphology most sensitively. We use the images from
this survey to demonstrate that our new morphological classification scheme is comprehensive, as it can adequately describe
all the morphologies seen so far in this survey. 

All images discussed in this paper which were taken with WFPC2 were downloaded from the HST archive of pipeline-calibrated images
maintained at the Canadian Astronomy Data Centre (CADC). The images taken with ACS or WFC3 are pipeline-calibrated images downloaded
from the StScI/MAST archive.

\section{Results}\label{result}
\begin{figure}[htb]
\vskip -0.6cm
\resizebox{0.77\textwidth}{!}{\includegraphics{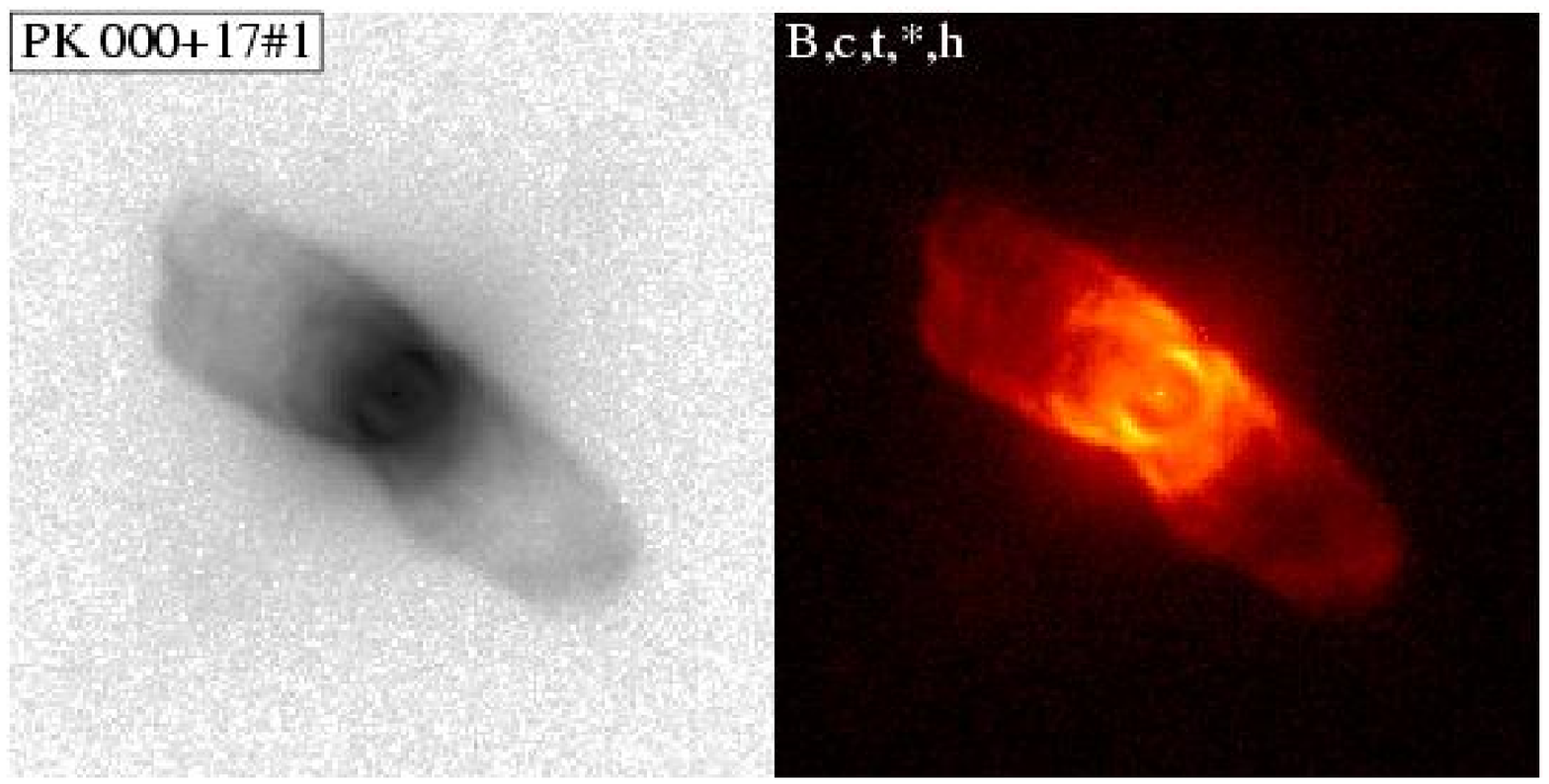}}
\vskip -0.2cm
\caption{PK\,000+17\#1 -- left panel shows HST H$\alpha$ image (log stretch, reverse grey-scale), right panel
shows the same image, processed to enhance sharp structures, in false-color. The panel length along the horizontal axis is
given in Table\,\ref{morphs} {\bf [The quality of this and following figures as it appears in the arXiv pdf output is not up-to-par;
the full ms with high-quality figures is available by anonymous FTP at 
ftp:\//\//ftp.astro.ucla.edu\//pub\//morris\//AJ-360163-sahai.pdf]}.
}
\label{000+17d1}
\end{figure}

\begin{figure}[htb]
\vskip -0.2cm
\resizebox{0.77\textwidth}{!}{\includegraphics{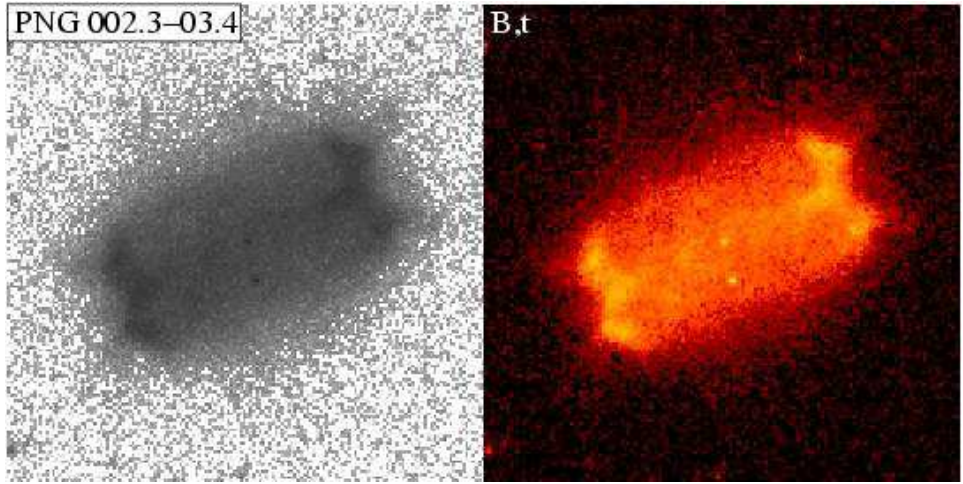}}
\vskip -0.6cm
\caption{As in Fig\,1., but for PNG002.3-03.4.
}
\label{002.3-03.4}
\end{figure}

\begin{figure}[htb]
\resizebox{0.77\textwidth}{!}{\includegraphics{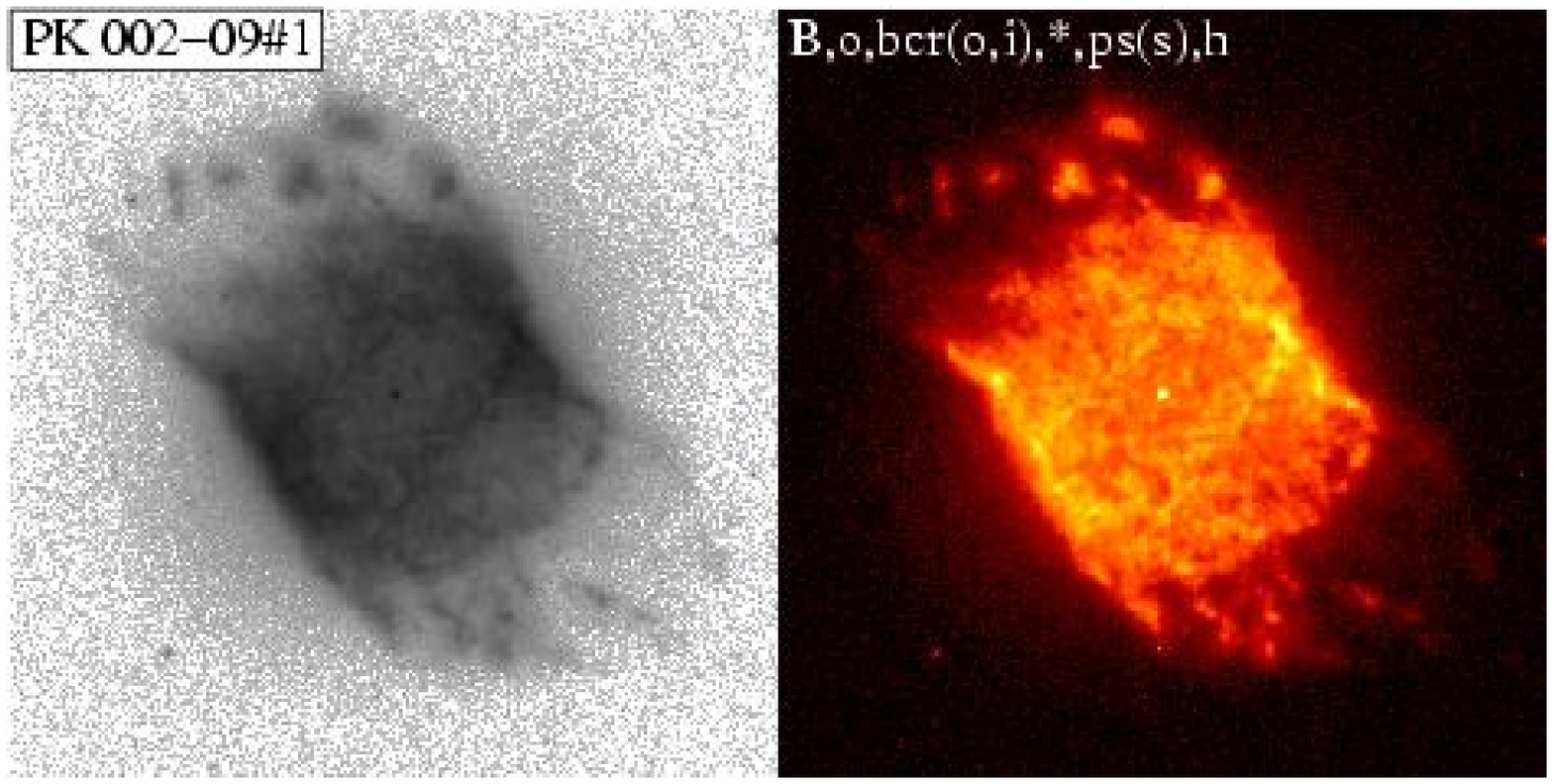}}
\caption{As in Fig\,1., but for PK\,002-09\#1.
}
\label{002-09d1}
\end{figure}

\begin{figure}[htb]
\resizebox{0.77\textwidth}{!}{\includegraphics{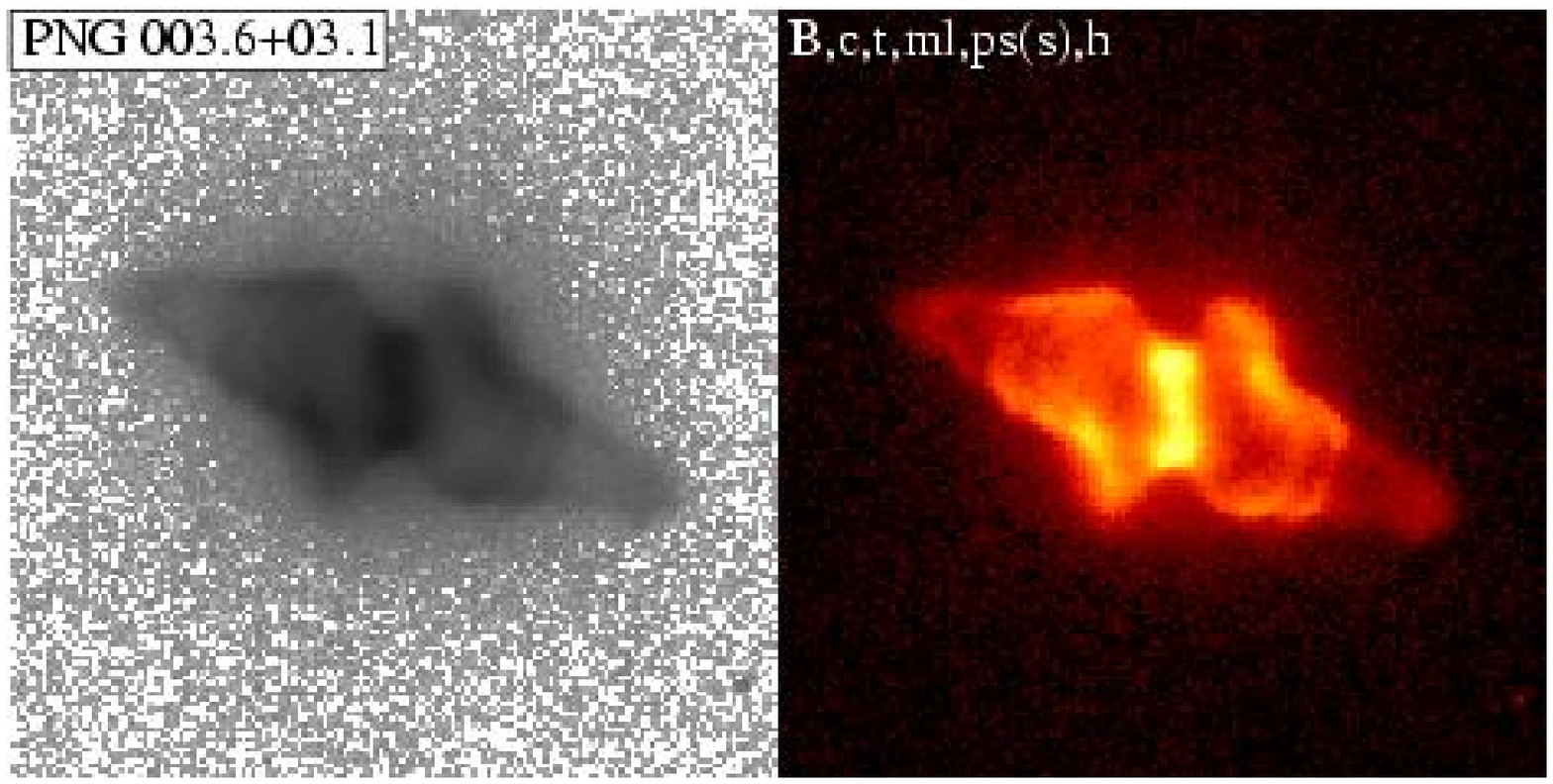}}
\caption{As in Fig\,1., but for PNG003.6+03.1.
}
\label{003.6+03.1}
\end{figure}

\begin{figure}[htb]
\vskip -0.6cm
\resizebox{0.77\textwidth}{!}{\includegraphics{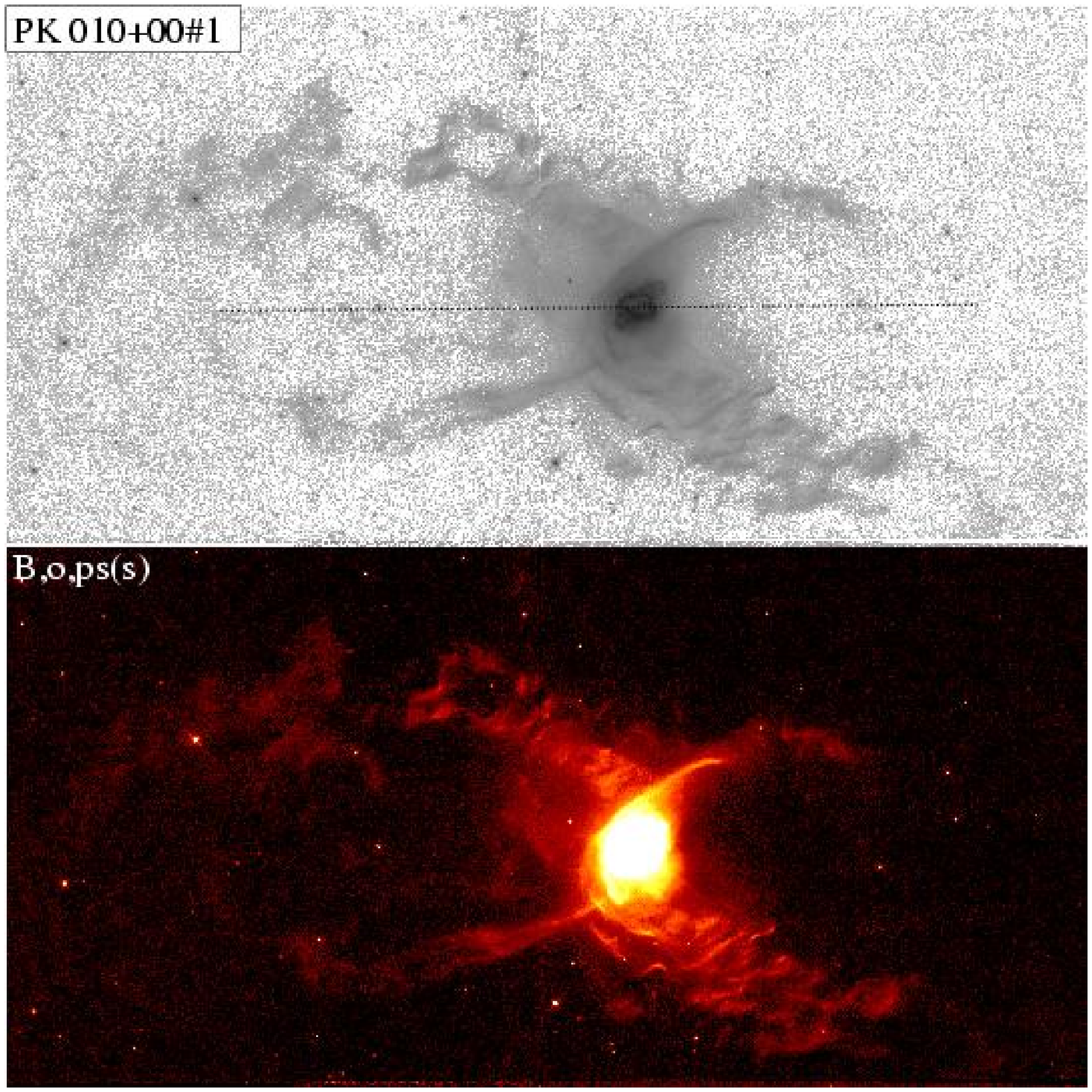}}
\caption{As in Fig\,1., but for PK\,010+00\#1
}
\label{010+00d1}
\end{figure}

\begin{figure}[htb]
\resizebox{0.77\textwidth}{!}{\includegraphics{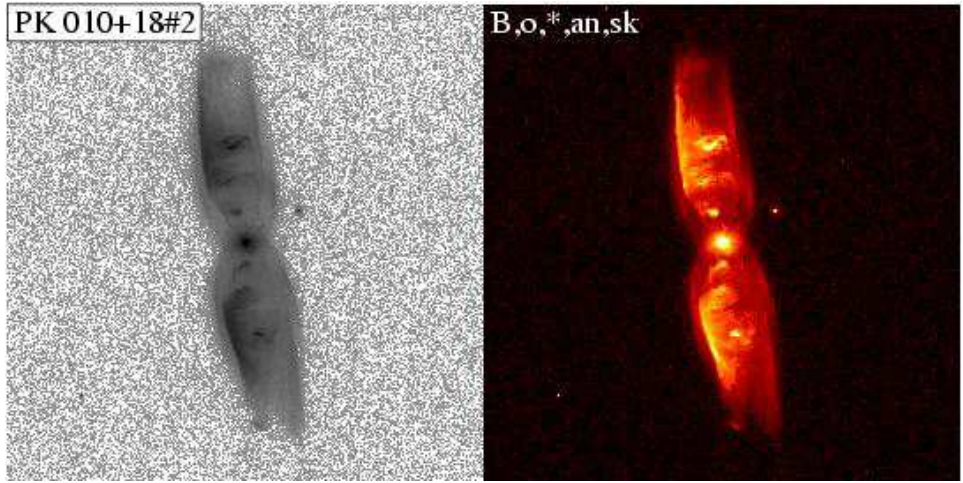}}
\caption{As in Fig\,1., but for PK\,010+18\#2.
}
\label{010+18d2}
\end{figure}

\begin{figure}[htb]
\vskip -0.6cm
\resizebox{0.77\textwidth}{!}{\includegraphics{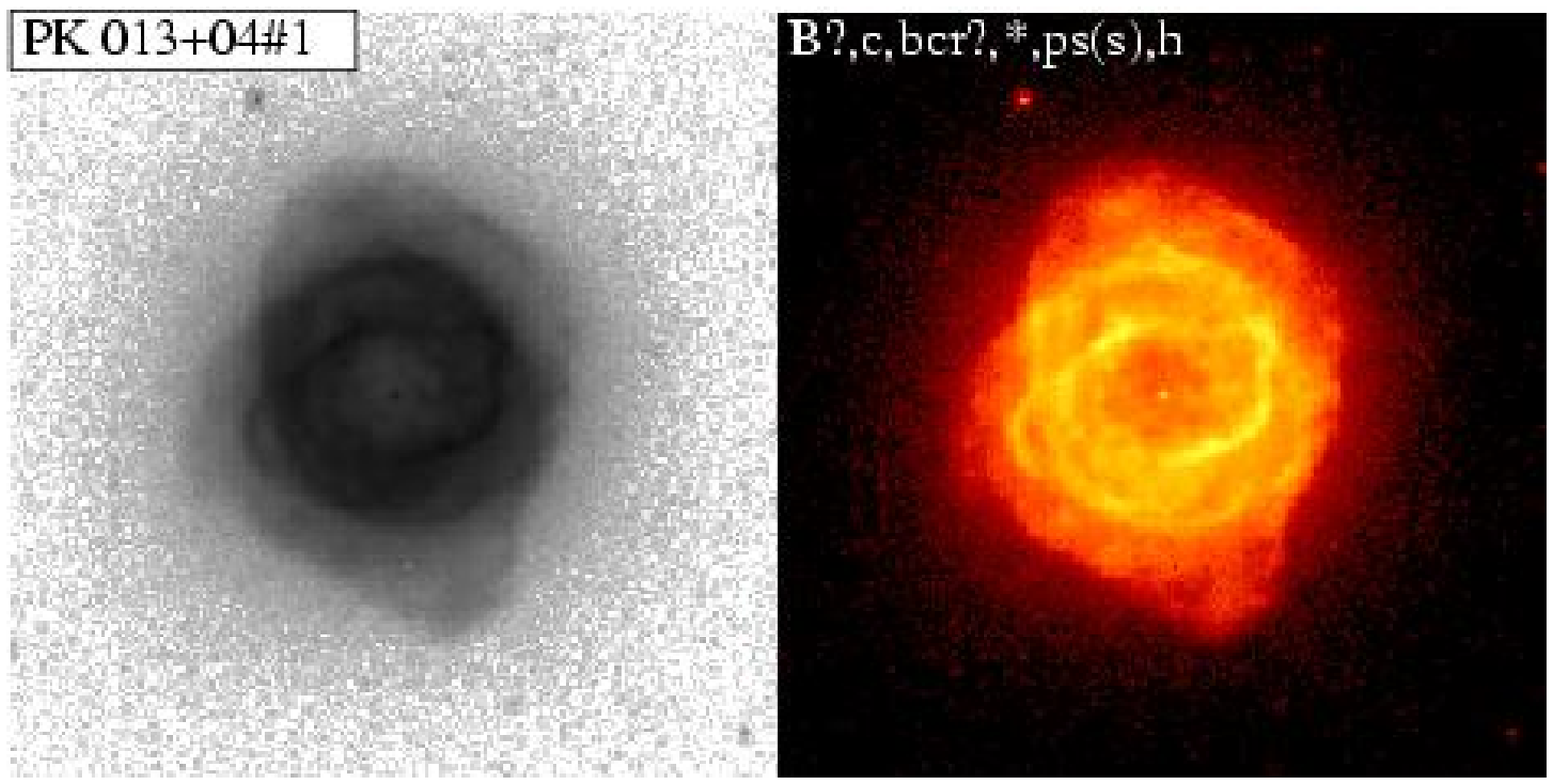}}
\caption{As in Fig\,1., but for PK\,013+04\#1 and the F658N ([NII]) filter.
}
\label{013+04d1}
\end{figure}

\begin{figure}[htb]
\vskip -0.6cm
\resizebox{0.77\textwidth}{!}{\includegraphics{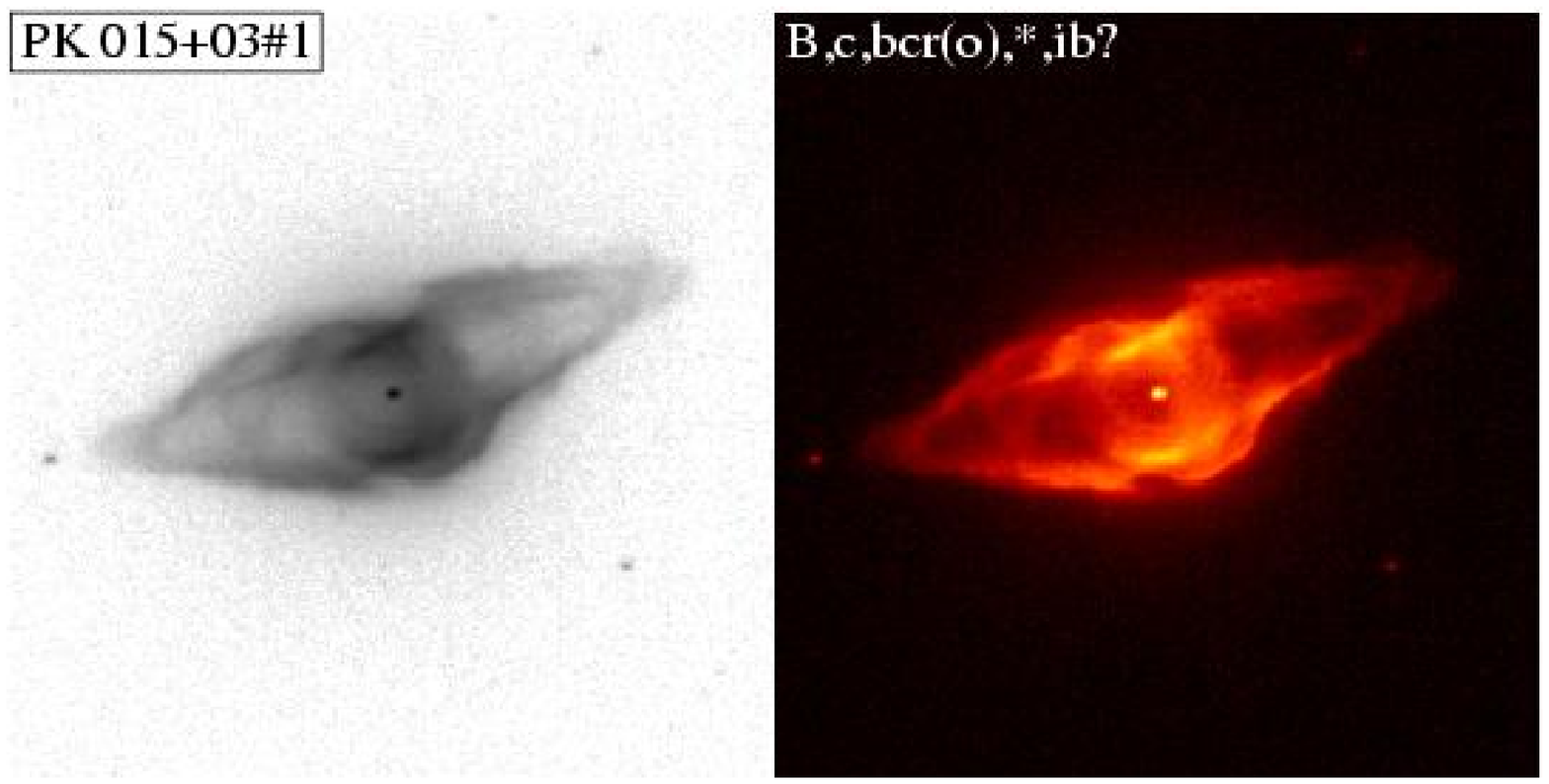}}
\caption{As in Fig\,1., but for PK\,015+03\#1.
}
\label{015+03d1}
\end{figure}

\begin{figure}[htb]
\vskip -0.6cm
\resizebox{0.77\textwidth}{!}{\includegraphics{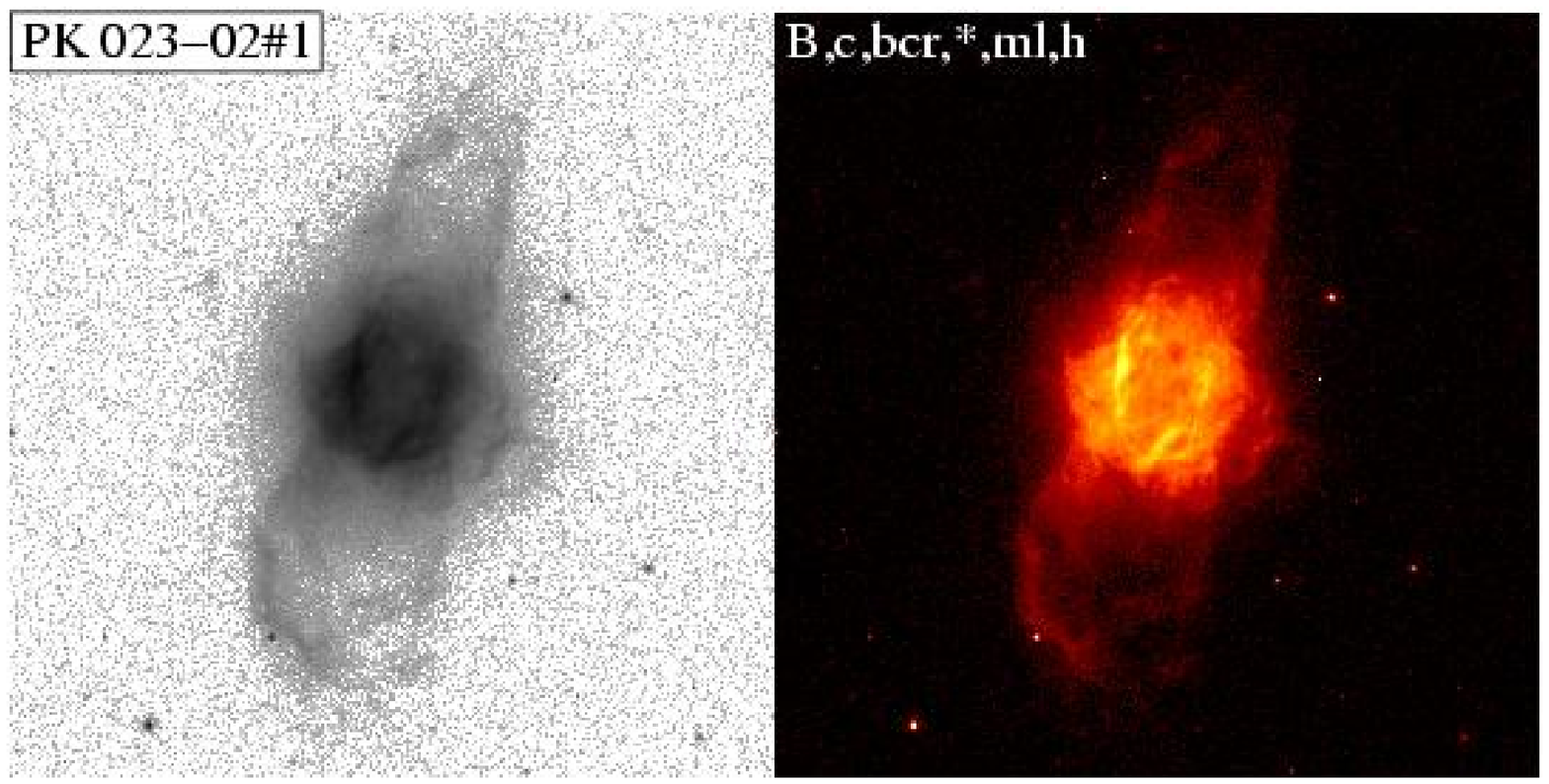}}
\caption{As in Fig\,1., but for PK\,023-02\#1.
}
\label{023-02d1}
\end{figure}

\begin{figure}[htb]
\vskip -0.6cm
\resizebox{0.77\textwidth}{!}{\includegraphics{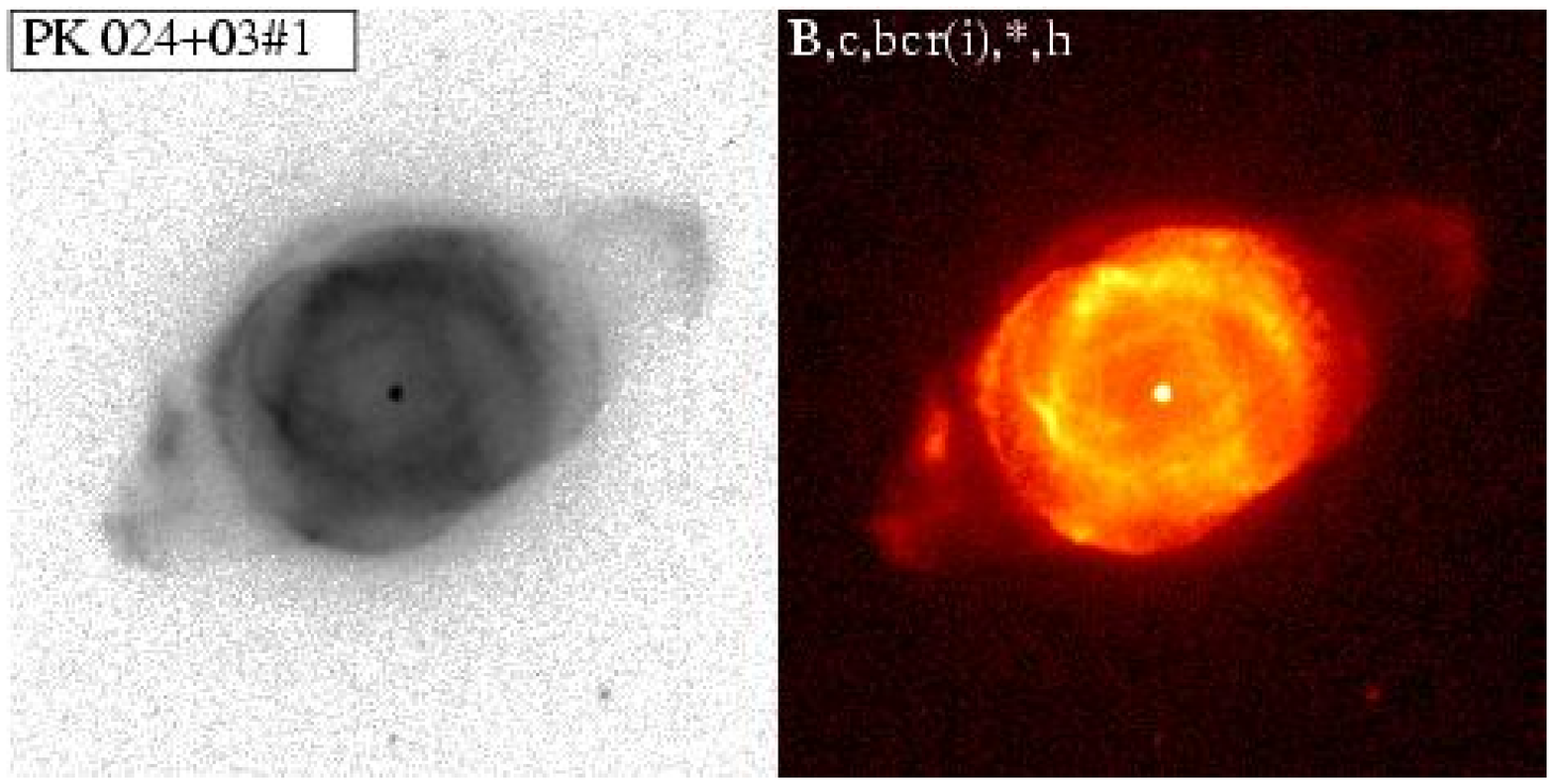}}
\caption{As in Fig\,1., but for PK\,024+03\#1.
}
\label{024+03d1}
\end{figure}

\begin{figure}[htb]
\vskip -0.6cm
\resizebox{0.77\textwidth}{!}{\includegraphics{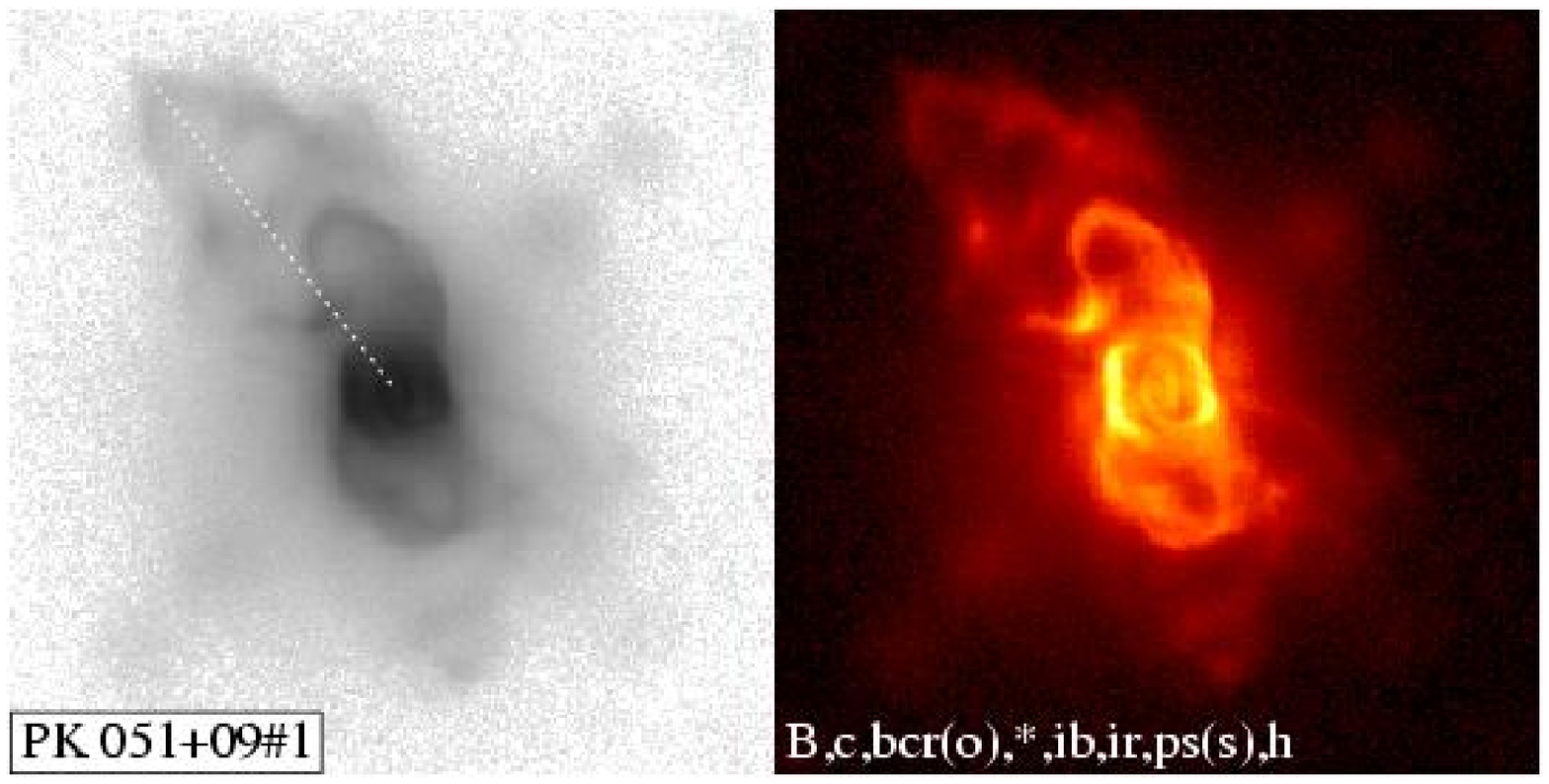}}
\caption{As in Fig\,1., but for PK\,051+09\#1.
}
\label{051+09d1}
\end{figure}

\begin{figure}[htb]
\vskip -0.6cm
\resizebox{0.77\textwidth}{!}{\includegraphics{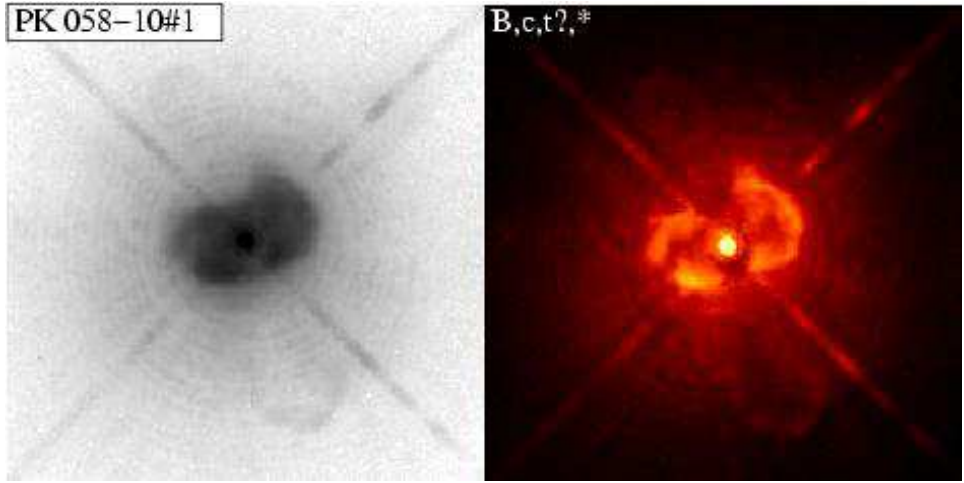}}
\caption{As in Fig\,1., but for PK\,058-10\#1.
}
\label{058-10d1}
\end{figure}

\begin{figure}[htb]
\vskip -0.2cm
\resizebox{0.77\textwidth}{!}{\includegraphics{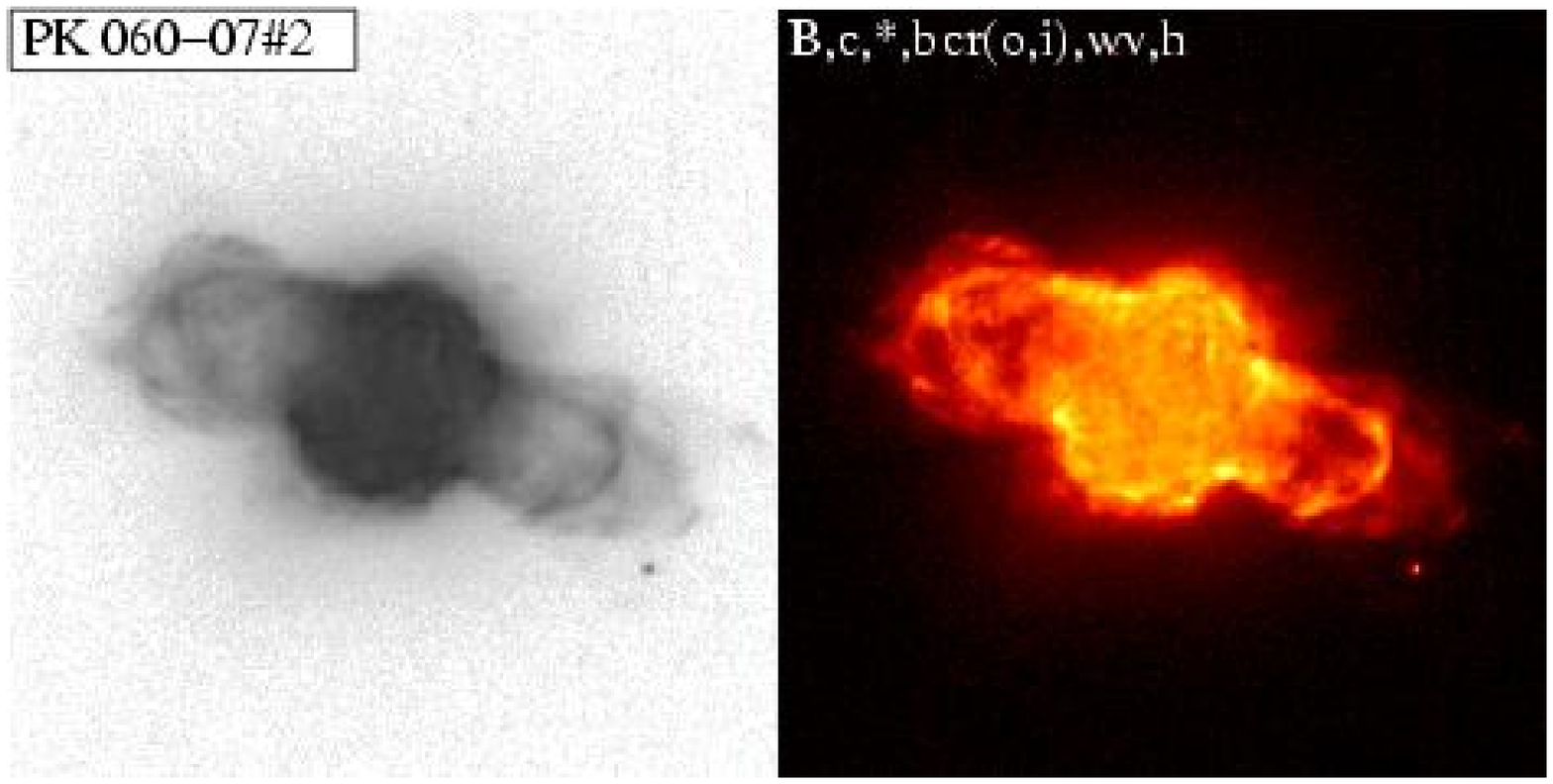}}
\caption{As in Fig\,1., but for PK\,060-07\#2.
}
\label{060-07d2}
\end{figure}

\begin{figure}[htb]
\vskip -0.2cm
\resizebox{0.77\textwidth}{!}{\includegraphics{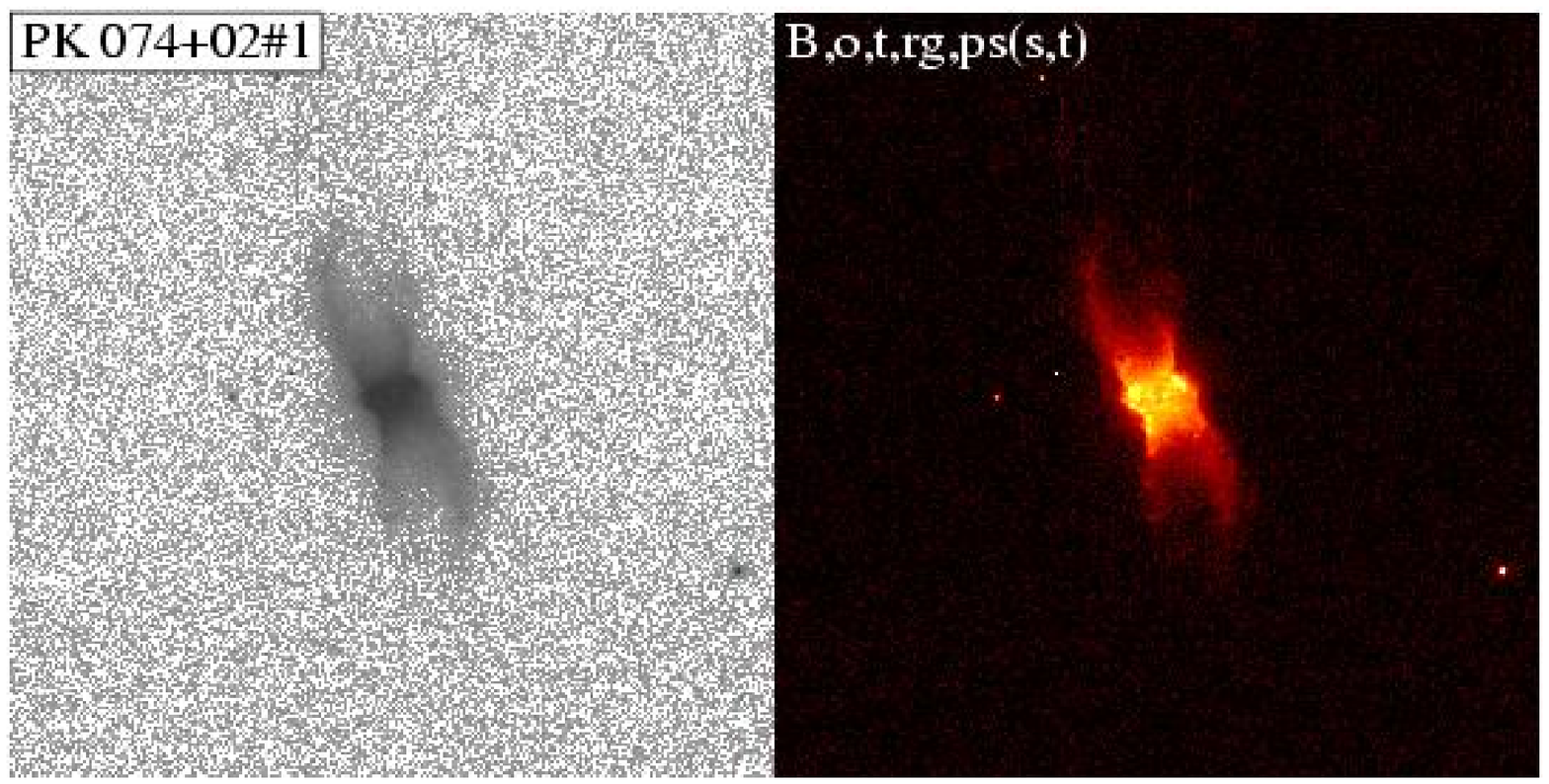}}
\caption{As in Fig\,1., but for PK\,074+02\#1.
}
\label{074+02d1}
\end{figure}

\begin{figure}[htb]
\vskip -0.6cm
\resizebox{0.77\textwidth}{!}{\includegraphics{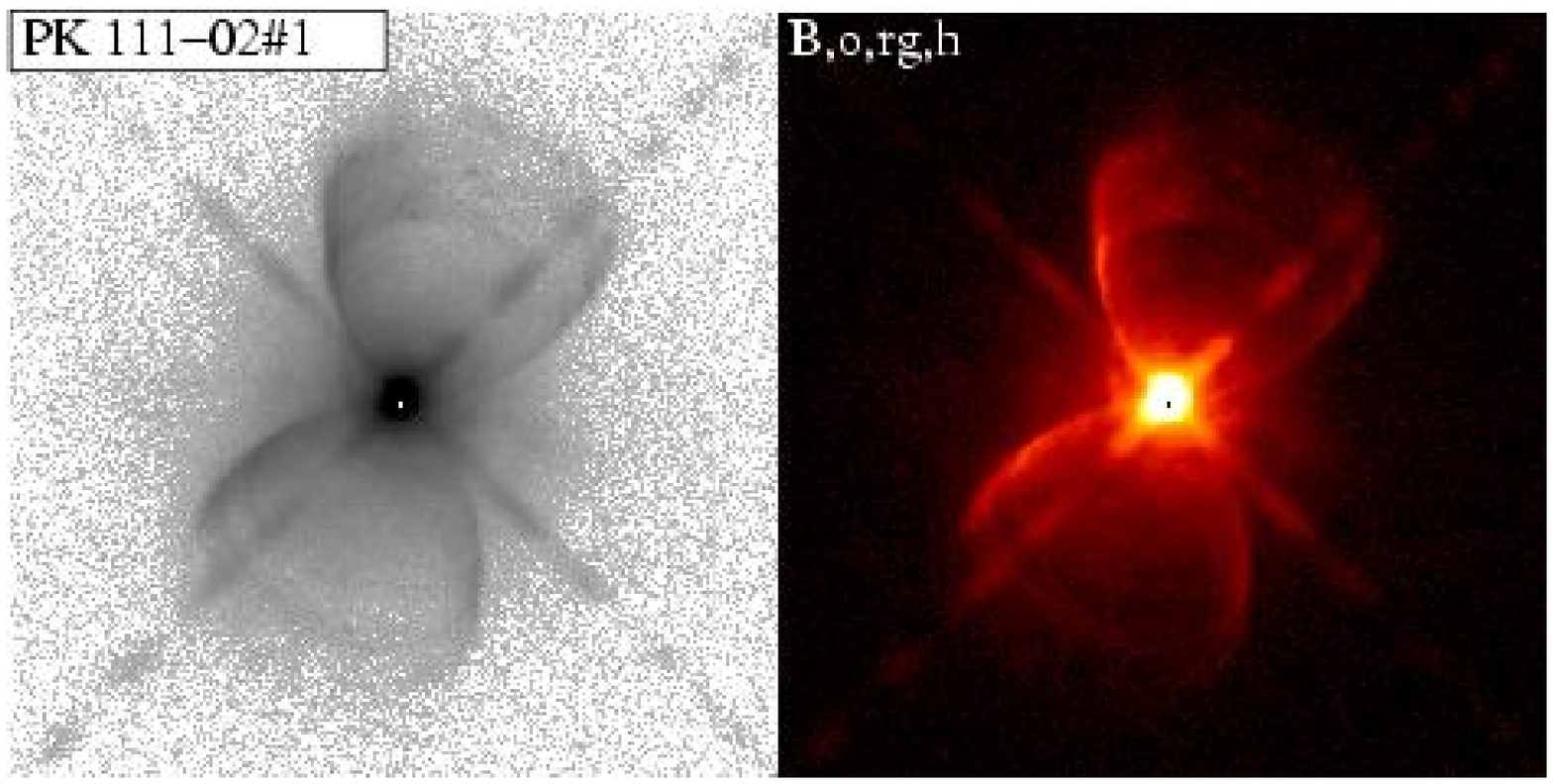}}
\caption{As in Fig\,1., but for PK\,111-02\#1 (adapted from ST98).
}
\label{111-02d1}
\end{figure}

\begin{figure}[htb]
\vskip -0.6cm
\resizebox{0.77\textwidth}{!}{\includegraphics{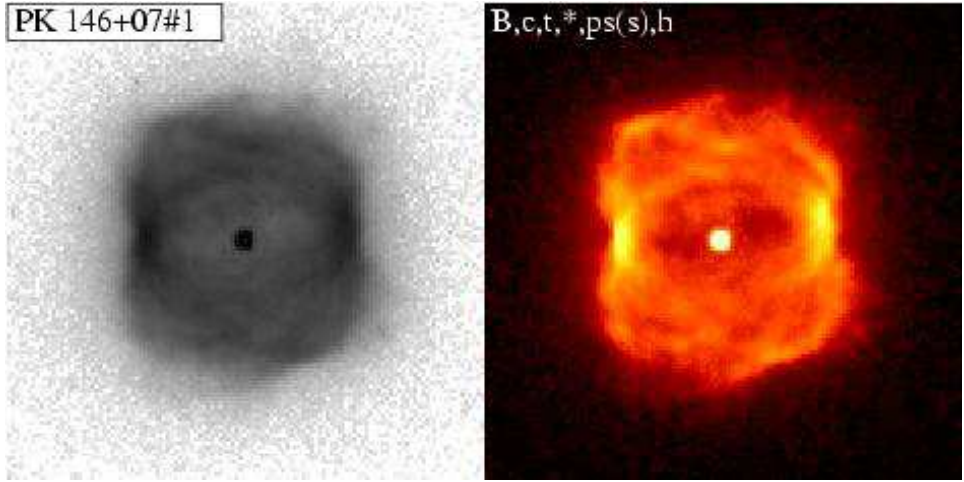}}
\caption{As in Fig\,1., but for PK\,146+07d1 (adapted from ST98).
}
\label{146+07d1}
\end{figure}

\begin{figure}[htb]
\vskip -0.6cm
\resizebox{0.77\textwidth}{!}{\includegraphics{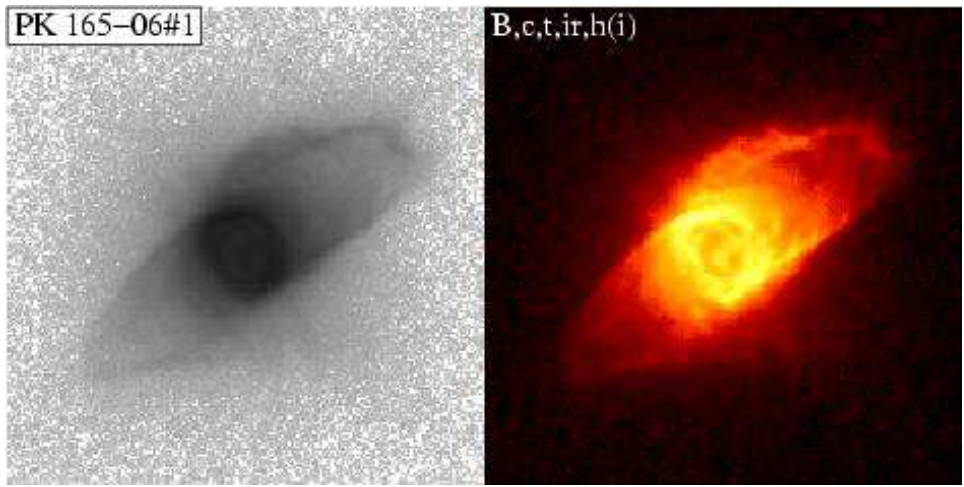}}
\caption{As in Fig\,1., but for PK\,165-06\#1.
}
\label{165-06d1}
\end{figure}

\begin{figure}[htb]
\resizebox{0.77\textwidth}{!}{\includegraphics{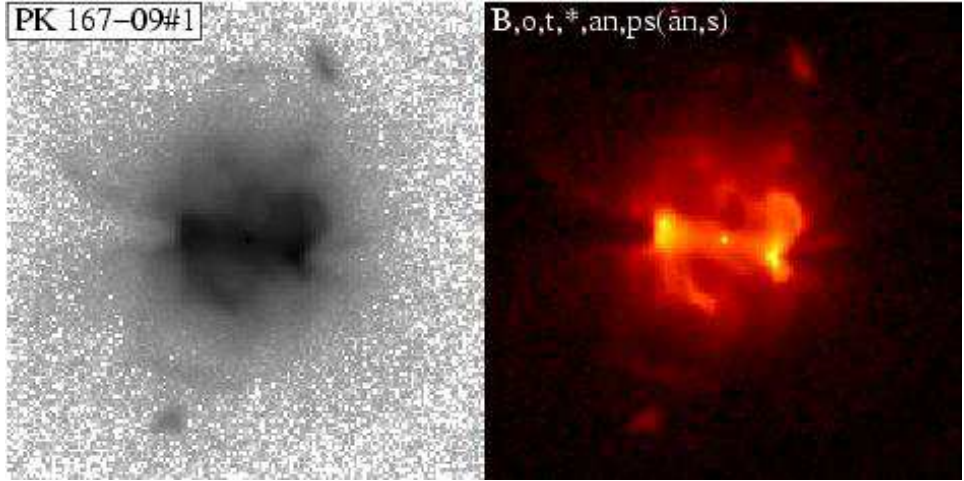}}
\caption{As in Fig\,1., but for PK\,167-09\#1 (adapted from ST98).
}
\label{167-09d1}
\end{figure}

\begin{figure}[htb]
\vskip -0.6cm
\resizebox{0.77\textwidth}{!}{\includegraphics{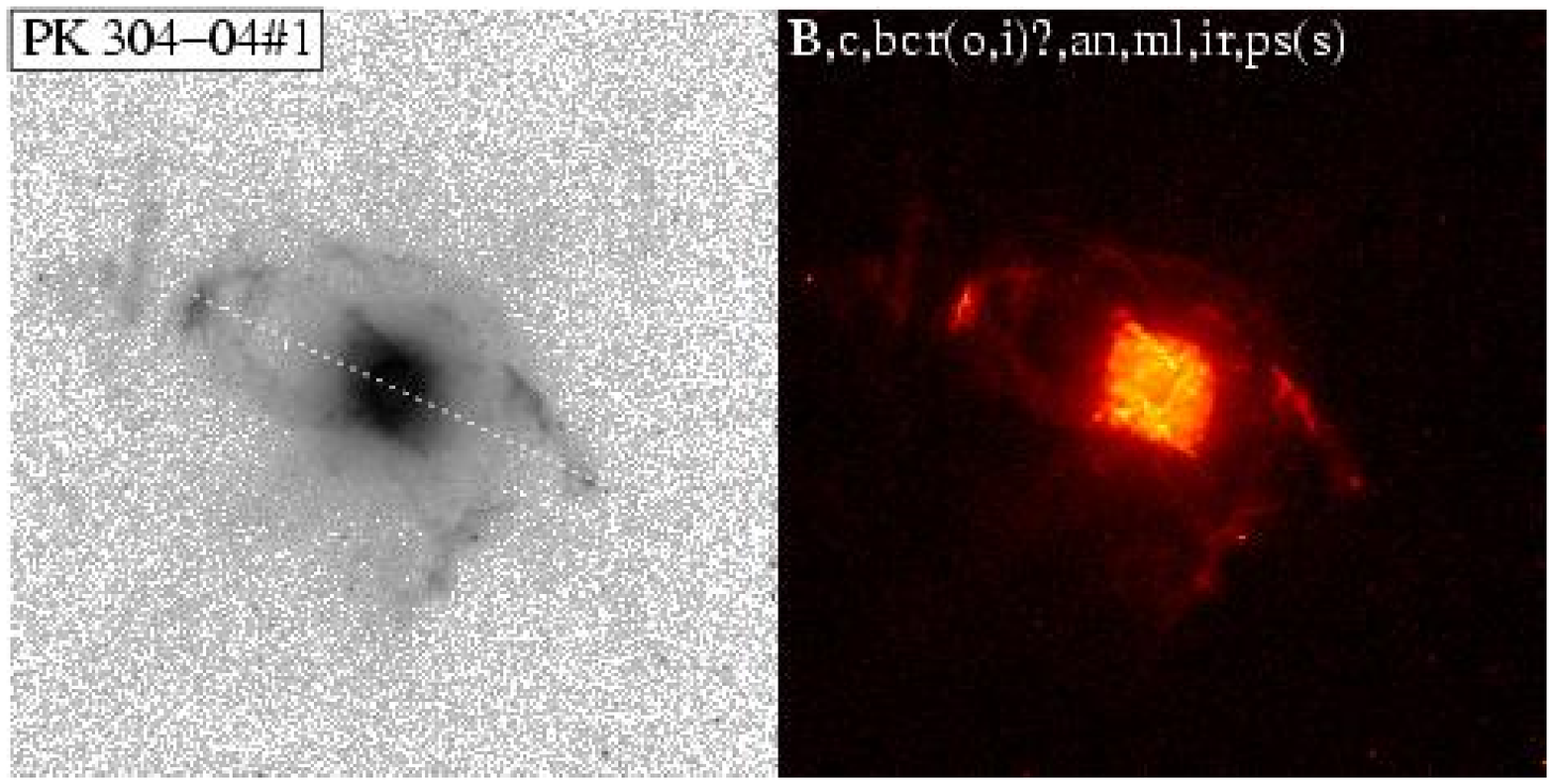}}
\caption{As in Fig\,1., but for PK\,304-04\#1.
}
\label{304-04d1}
\end{figure}

\begin{figure}[htb]
\resizebox{0.77\textwidth}{!}{\includegraphics{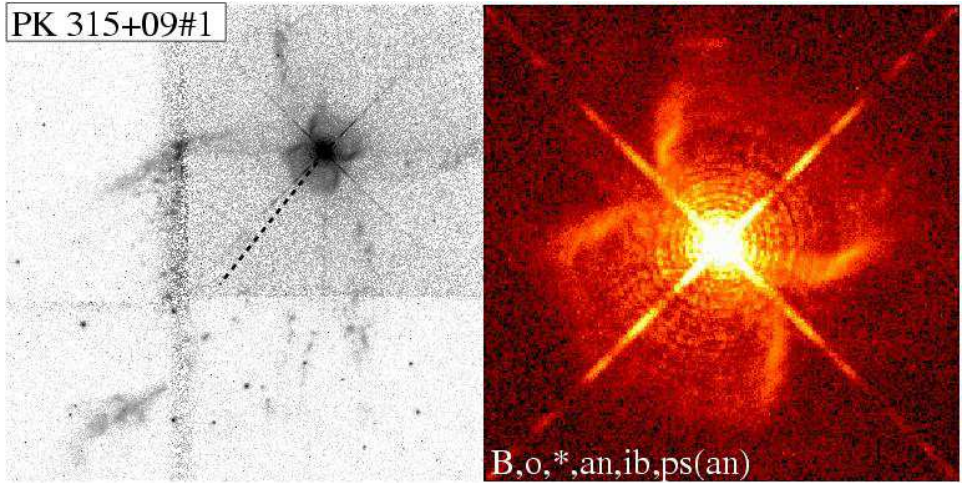}}
\caption{As in Fig\,1., but for PK\,315+09\#1  (adapted from ST98). The field-of-view in the left 
panel does not cover the ansa feature diametrically opposed to the one seen in the lower left corner (see Corradi \& Schwarz 1993 for a
full view). The right panel shows only the central region ($12.2{''}\times\,12.2{''}$).
}
\label{315+09d1}
\end{figure}

\begin{figure}[htb]
\resizebox{0.77\textwidth}{!}{\includegraphics{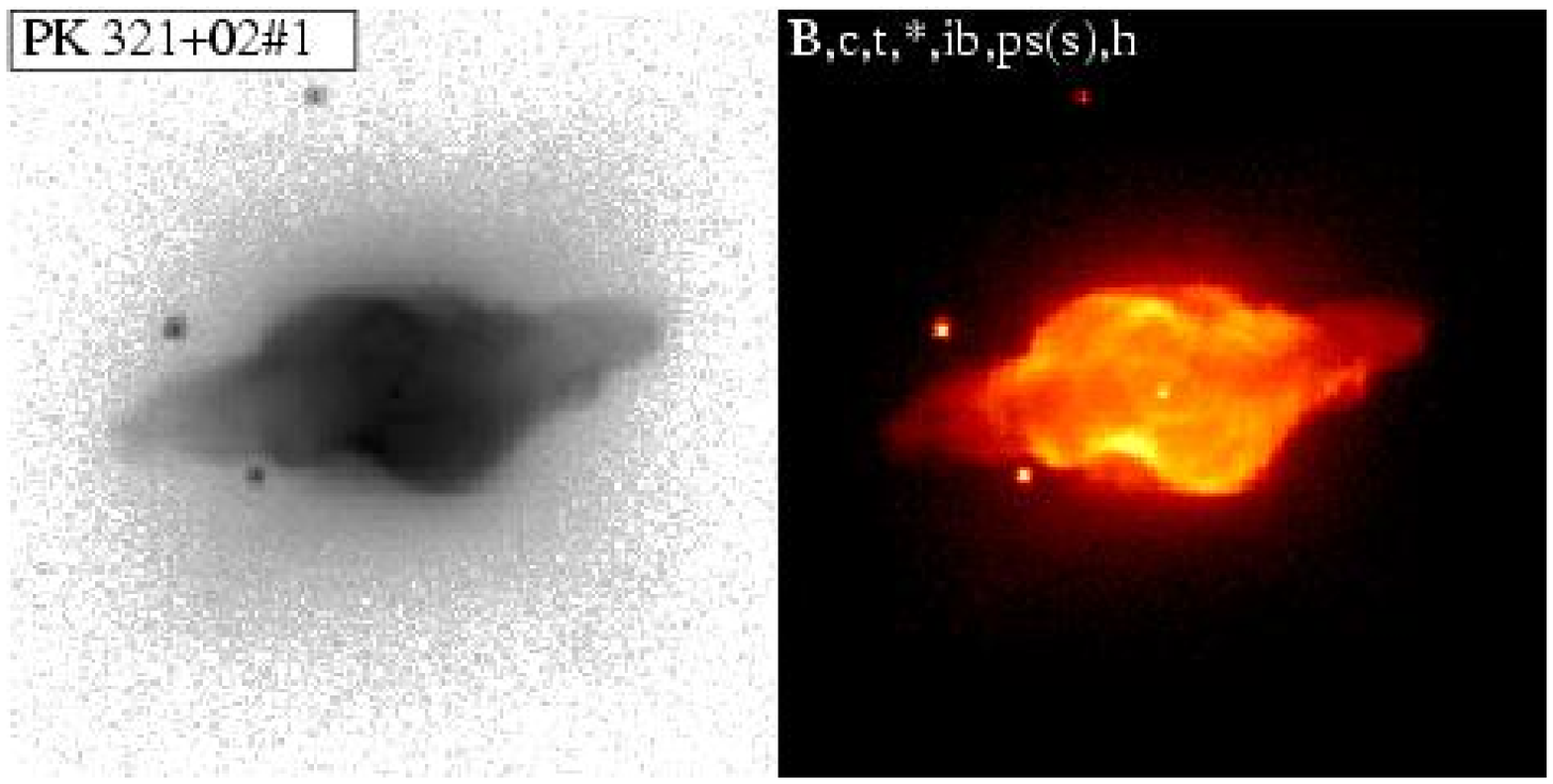}}
\vskip -0.3cm
\caption{As in Fig\,1., but for PK\,321+02\#1 (adapted from ST98).
}
\label{321+02d1}
\end{figure}

\begin{figure}[htb]
\vskip -0.2cm
\resizebox{0.77\textwidth}{!}{\includegraphics{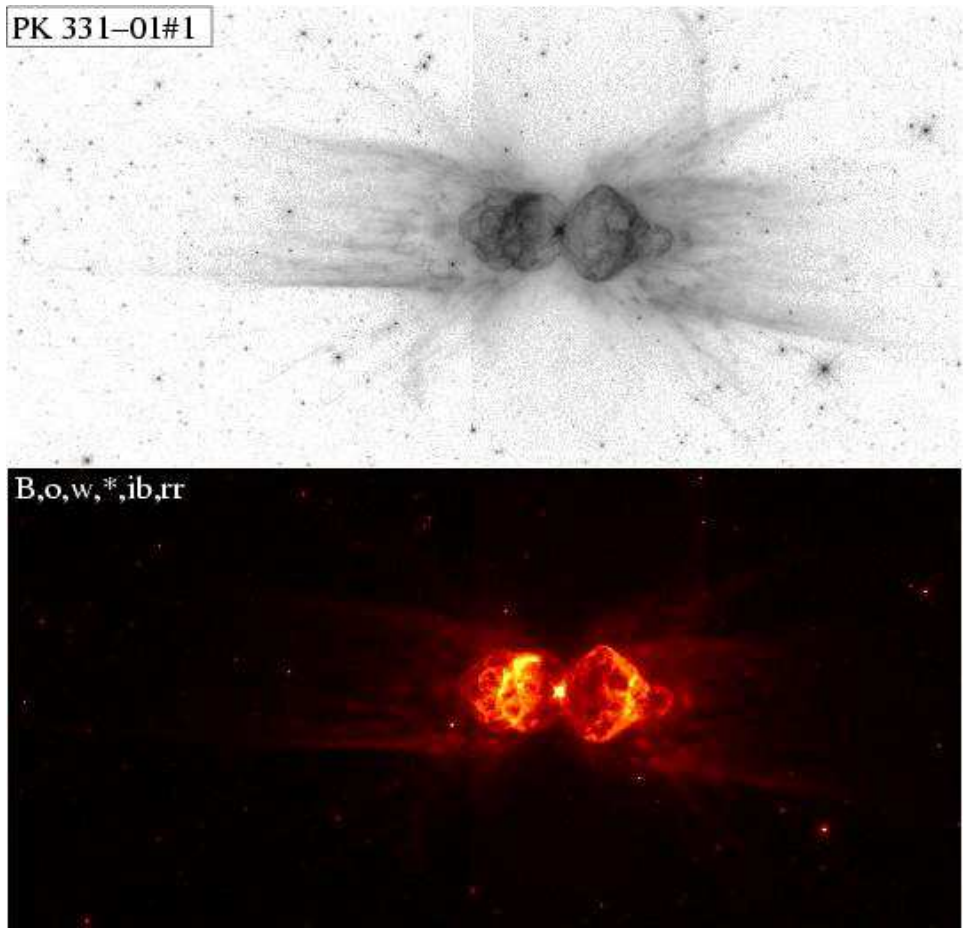}}
\vskip -0.4cm
\caption{As in Fig\,1., but for PK\,331-01\#1.
}
\label{331-01d1}
\end{figure}

\begin{figure}[htb]
\vskip -0.6cm
\resizebox{0.77\textwidth}{!}{\includegraphics{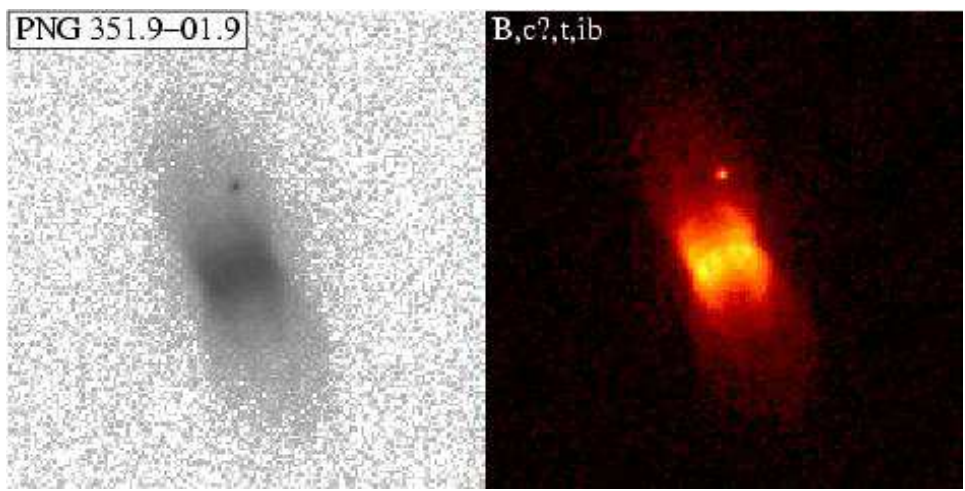}}
\caption{As in Fig\,1., but for PNG351.9-01.9.
}
\label{351.9-01.9}
\end{figure}

\begin{figure}[htb]
\resizebox{0.77\textwidth}{!}{\includegraphics{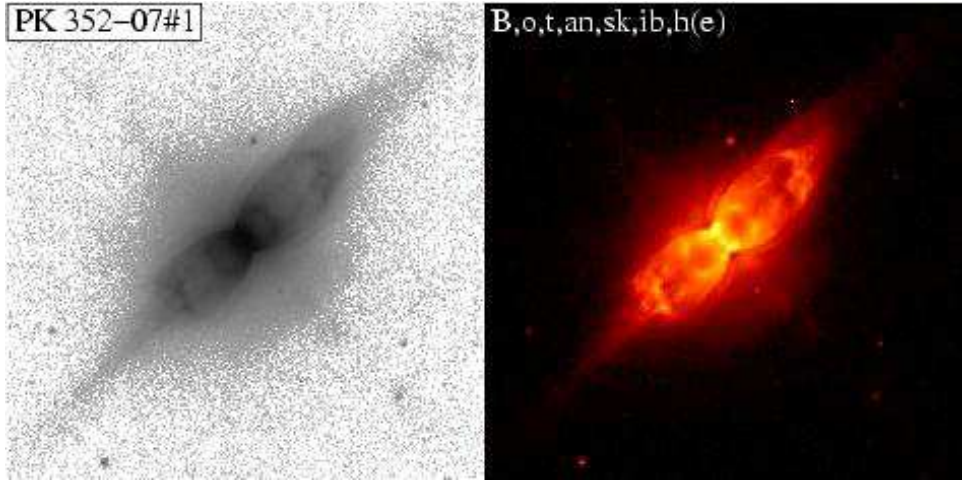}}
\caption{As in Fig\,1., but for PK\,352-07\#1.
}
\label{352-07d1}
\end{figure}

\begin{figure}[htb]
\vskip -0.6cm
\resizebox{0.77\textwidth}{!}{\includegraphics{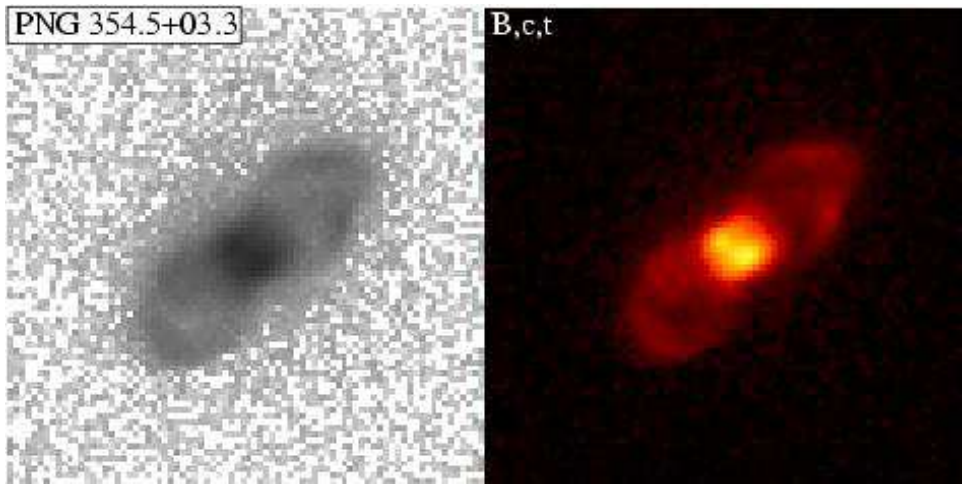}}
\caption{As in Fig\,1., but for PNG354.5+03.3.
}
\label{354.5+03.3}
\end{figure}

\begin{figure}[htb]
\vskip -0.6cm
\resizebox{0.77\textwidth}{!}{\includegraphics{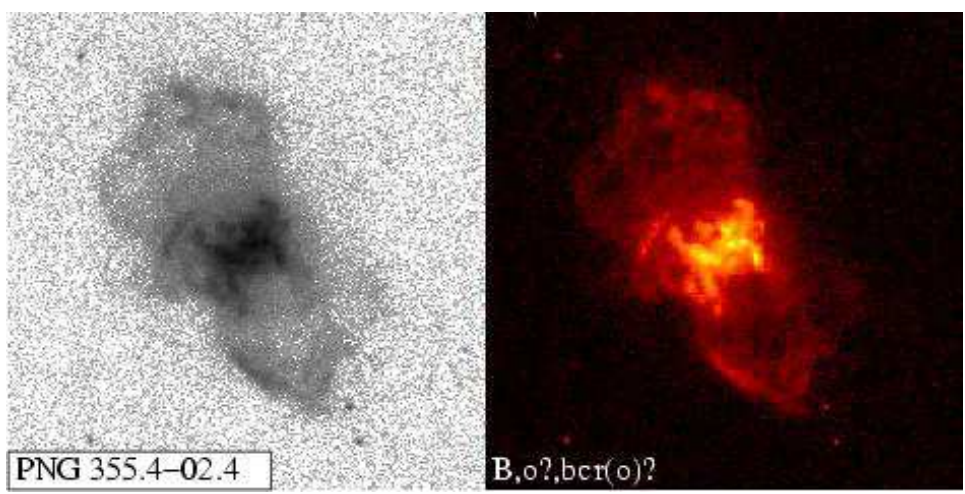}}
\vskip -0.4cm
\caption{As in Fig\,1., but for PNG355.4-02.4.
}
\label{355.4-02.4}
\end{figure}

\begin{figure}[htb]
\vskip -0.2cm
\resizebox{0.77\textwidth}{!}{\includegraphics{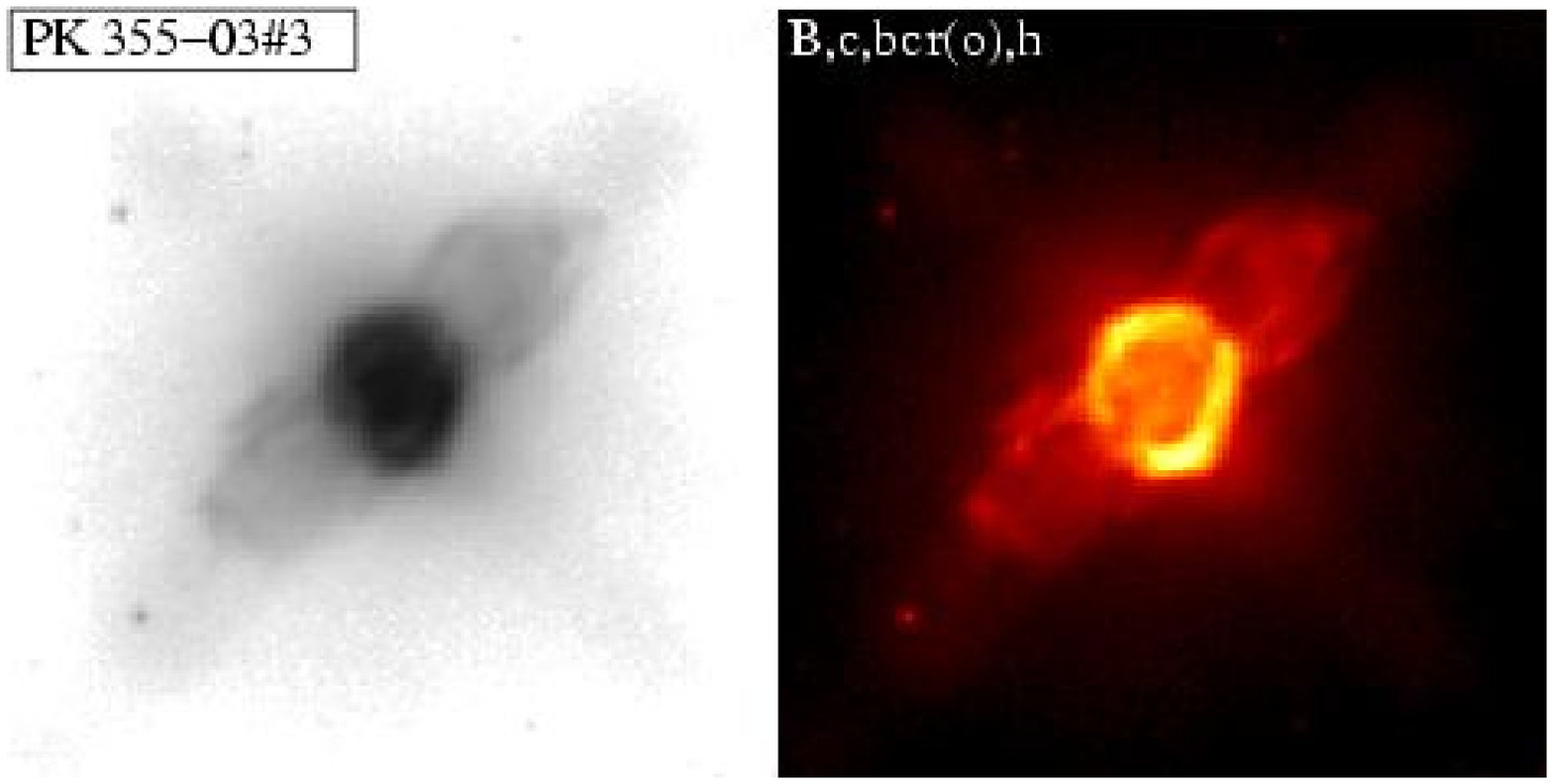}}
\caption{As in Fig\,1., but for PK\,355-03\#3.
}
\label{355-03d3}
\end{figure}

\begin{figure}[htb]
\vskip -0.6cm
\resizebox{0.77\textwidth}{!}{\includegraphics{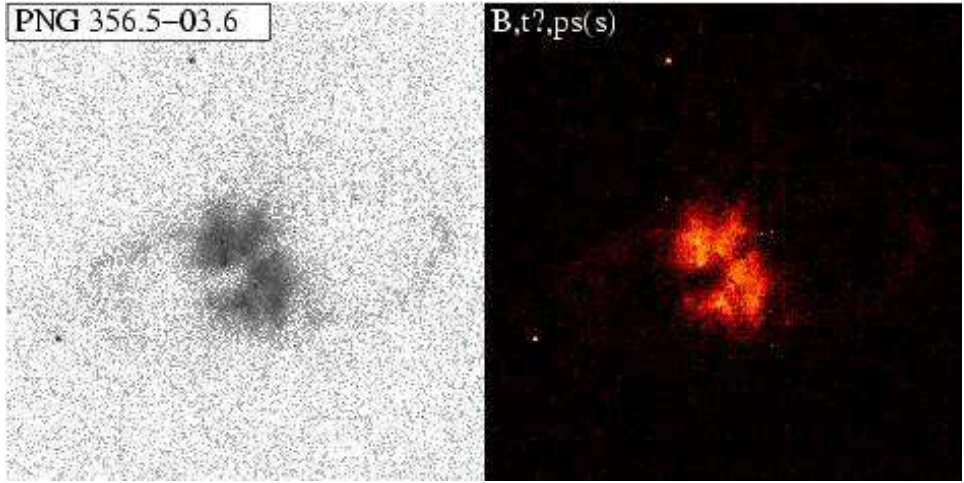}}
\caption{As in Fig\,1., but for PNG356.5-03.6.
}
\label{356.5-03.6}
\end{figure}

We present the images of all the objects in our sample, with the exception of MyCn\,18 (PNG307.5$-$04.9)\footnote{images in Sahai et
al. (1999)}, in Figs.\,\ref{000+17d1}--\ref{093.9-00.1}. For most of our objects,
we have used the H$\alpha$ (F656N) images, since these are best suited to showing the overall nebular structure. In a few
cases, H$\alpha$ images are not available so [NII] (F658N) images have been used. A comparison of the morphologies seen in
H$\alpha$ and [NII] images, when both are available, shows that there is very little difference between these in
determining their morphological classifications. For the 4 objects imaged in the GO\,9463 \& GO\,10536 programs, we only have
images in two broad-band filters (i.e., taken with either the F435W \& F606W, or the F606W \& F814 filters). With the
exception of one of these (PNG332.9-09.9, in which the F606W image is a relatively short exposure one and so the F435W image
is used), we present the F606W images, since this filter includes the H$\alpha$ line, which most likely represents the
dominant contribution to the emission seen in the images. For the PNs imaged in emission-lines, the intensity is
proportional to the square of the density, hence the dynamic range is quite large. Therefore, in order to show the nebular
structure optimally, we have used a log-stretch black-and-white image in reverse grey-scale, as well as a false-color one in
which the intensity has been processed in order to enhance sharp features\footnote{The processed image, $Im_P =
Im_O/(Im_O+0.04 Im_S)$, where $Im_O$ is the original image, 
and $Im_S$ is obtained by smoothing $Im_O$, as in ST98, Fig.\,1}. For the 4 PNs observed in broad-band filters, the
log-stretch black-and-white images show all features of the nebulae adequately, hence only these are shown. All figures and tabulated
results
form this paper will be made available to the community via the Vizier service of the Centre de Donnes astronomiques de Strasbourg
(CDS).

\subsection{Primary Classes}\label{primary}
\begin{figure}[htb]
\vskip -0.6cm
\resizebox{0.77\textwidth}{!}{\includegraphics{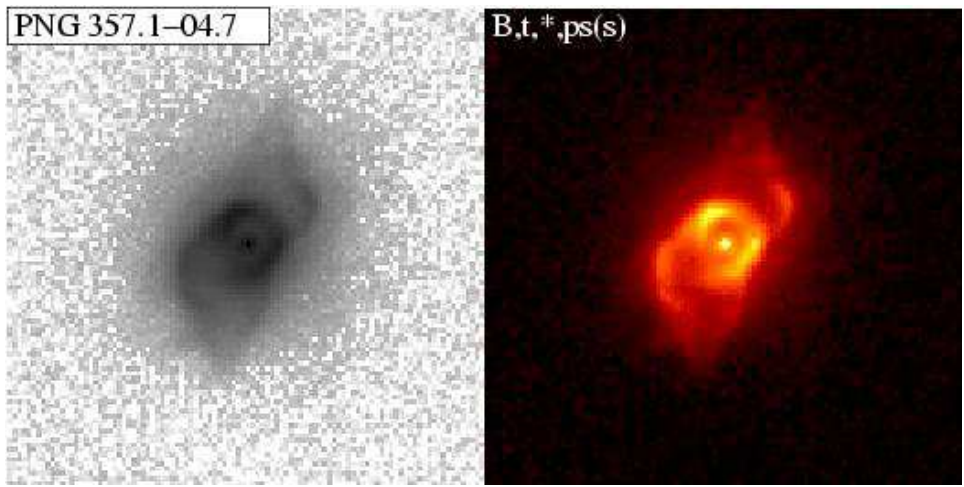}}
\caption{As in Fig\,1., but for PNG357.1-04.7.
}
\label{357.1-04.7}
\end{figure}

\begin{figure}[htb]
\vskip -0.6cm
\resizebox{0.77\textwidth}{!}{\includegraphics{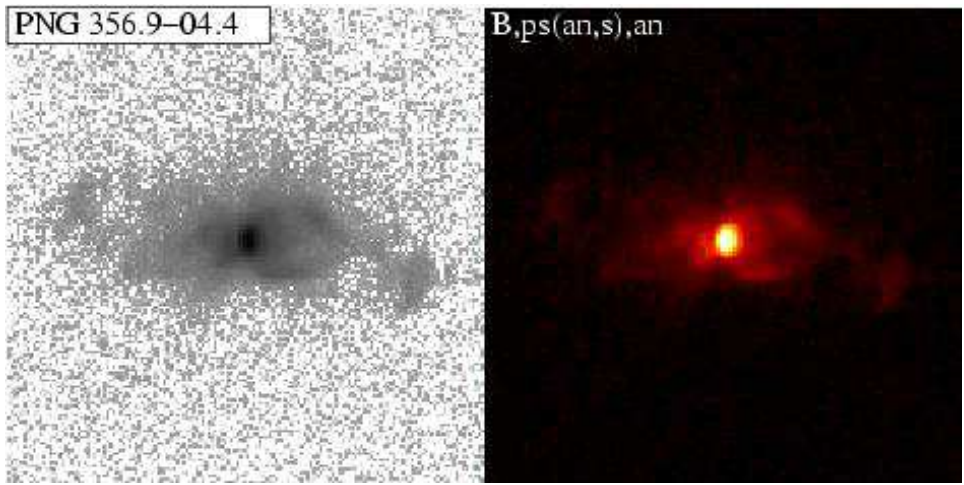}}
\caption{As in Fig\,1., but for PNG356.9-04.4.
}
\label{356.9_04.4}
\end{figure}

\begin{figure}[htb]
\vskip -0.6cm
\resizebox{0.77\textwidth}{!}{\includegraphics{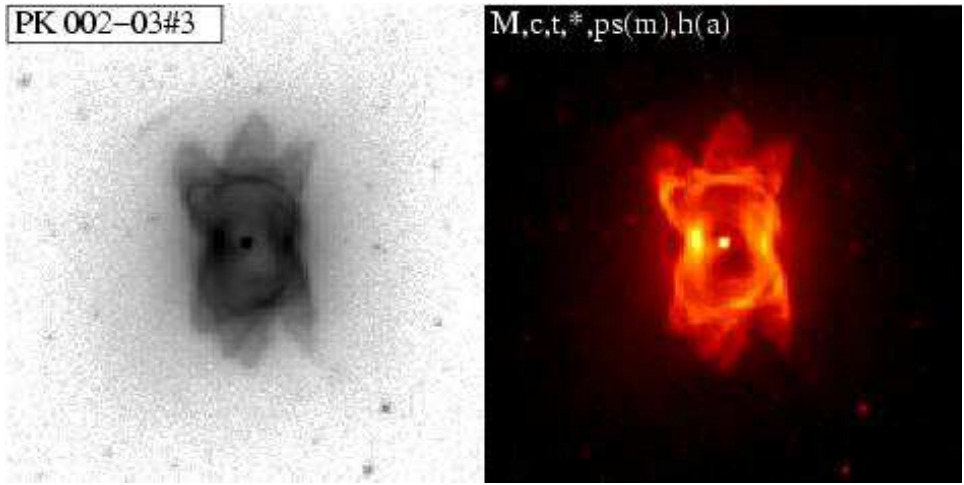}}
\caption{As in Fig\,1., but for PK\,002-03\#3 (adapted from Sahai 2000).
}
\label{002-03d3}
\end{figure}

\begin{figure}[htb]
\vskip -0.4cm
\resizebox{0.77\textwidth}{!}{\includegraphics{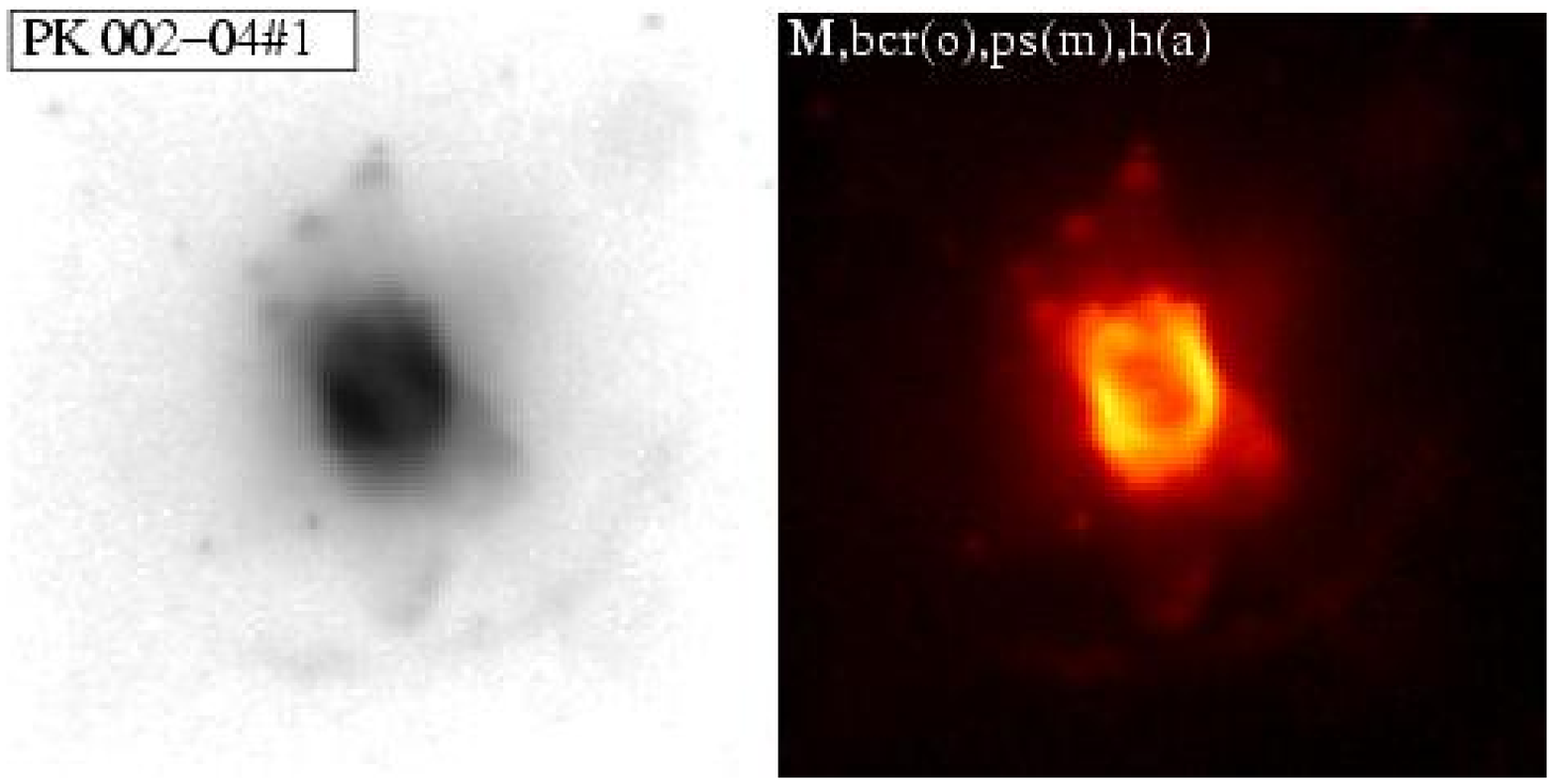}}
\caption{As in Fig\,1., but for PK\,002-04\#1.
}
\label{002-04d1}
\end{figure}

\begin{figure}[htb]
\vskip -0.6cm
\resizebox{0.77\textwidth}{!}{\includegraphics{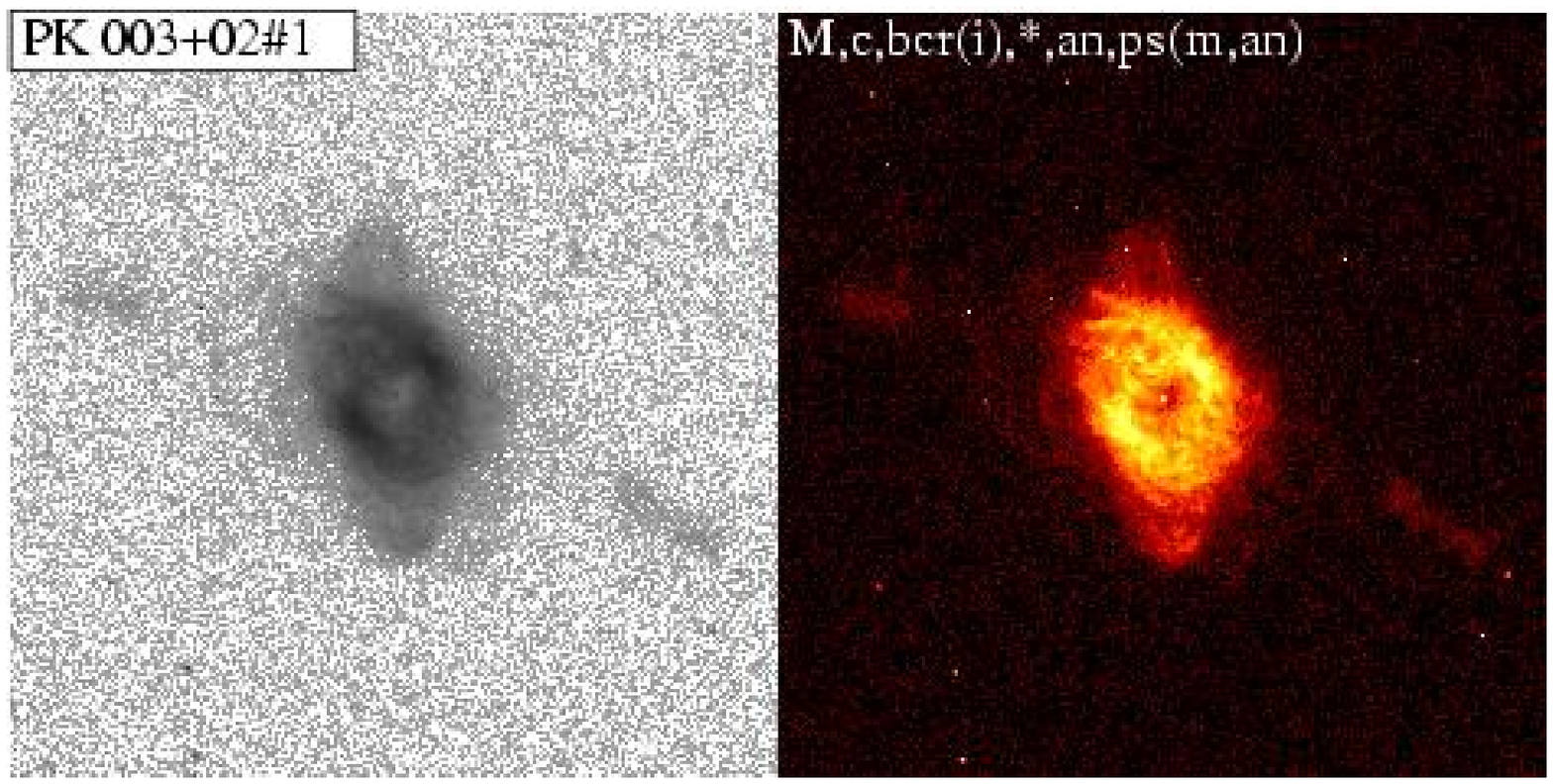}}
\caption{As in Fig\,1., but for PK\,003+02\#1.
}
\label{003+02d1}
\end{figure}

\begin{figure}[htb]
\vskip -0.4cm
\resizebox{0.77\textwidth}{!}{\includegraphics{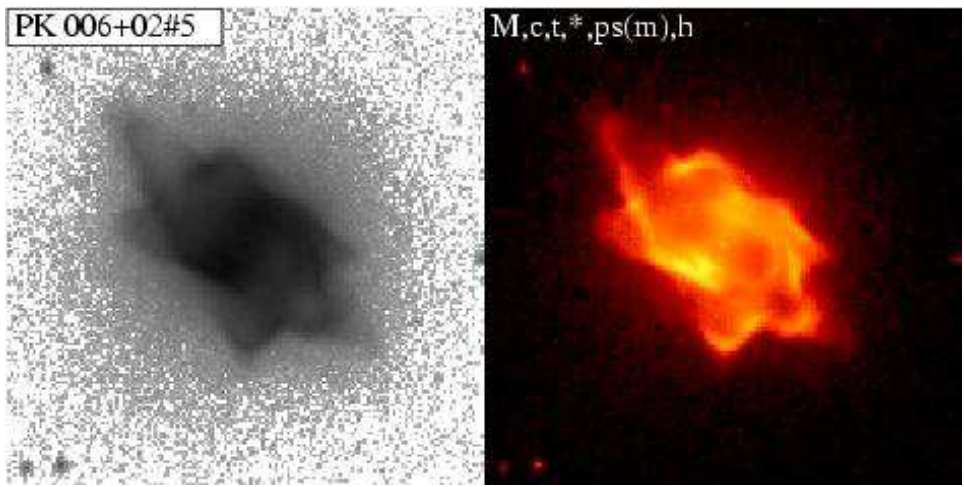}}
\caption{As in Fig\,1., but for PK\,006+02\#5.
}
\label{006+02d5}
\end{figure}

\begin{figure}[htb]
\vskip -0.6cm
\resizebox{0.77\textwidth}{!}{\includegraphics{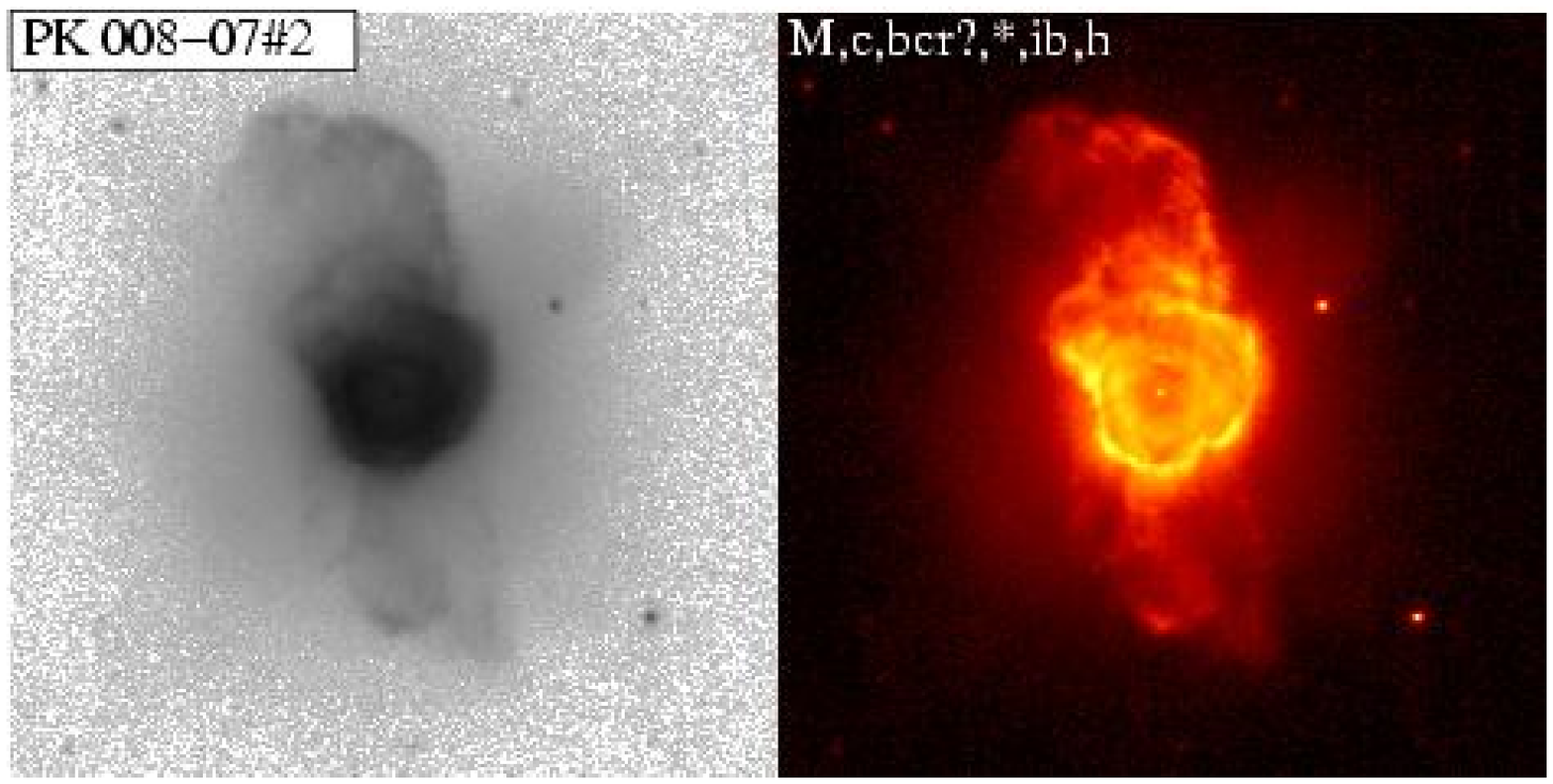}}
\caption{As in Fig\,1., but for PK\,008-07\#2.
}
\label{008-07d2}
\end{figure}

\begin{figure}[htb]
\vskip -0.4cm
\resizebox{0.77\textwidth}{!}{\includegraphics{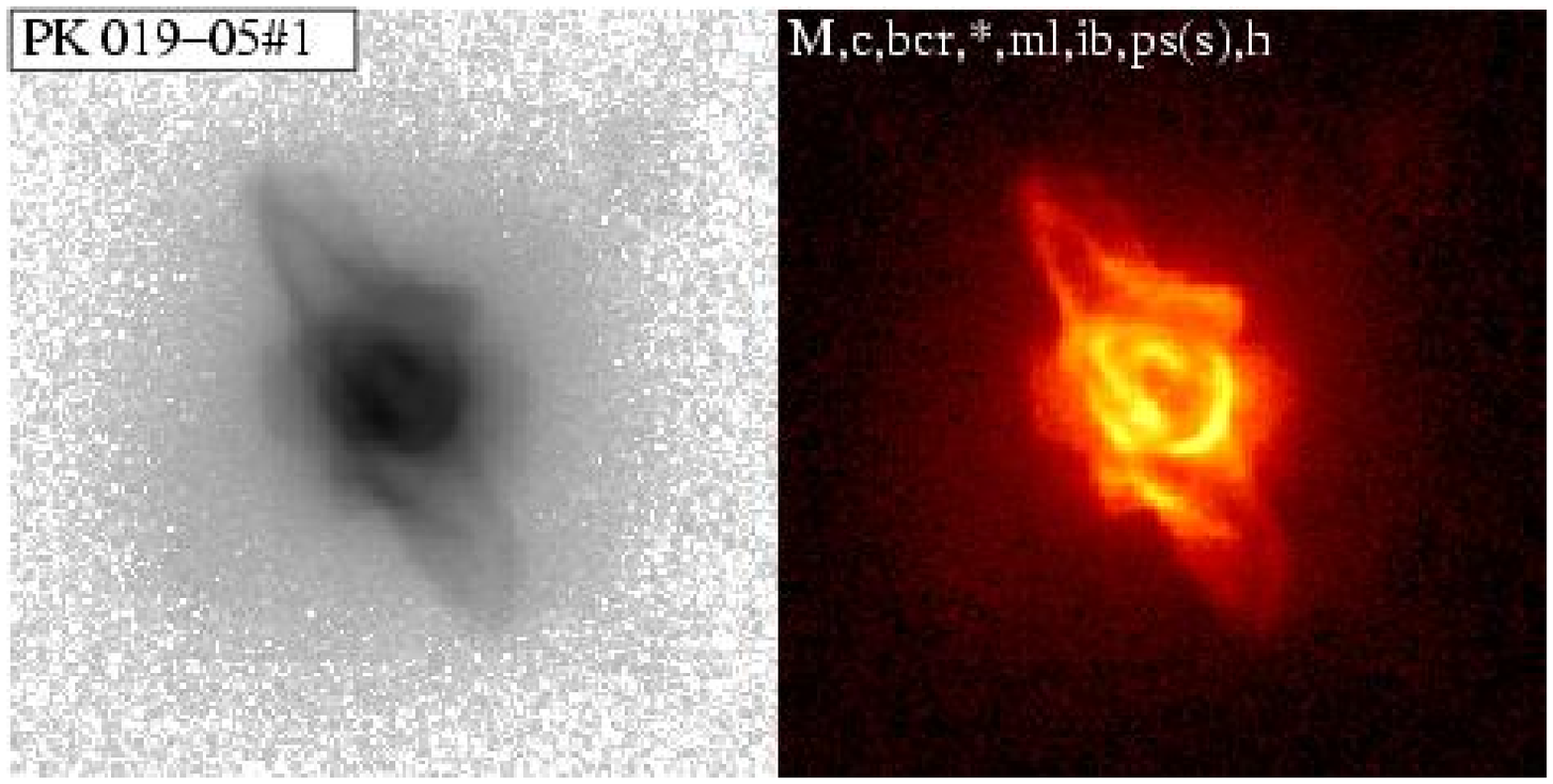}}
\caption{As in Fig\,1., but for PK\,019-05\#1.
}
\label{019-05d1}
\end{figure}

\begin{figure}[htb]
\vskip -0.6cm
\resizebox{0.77\textwidth}{!}{\includegraphics{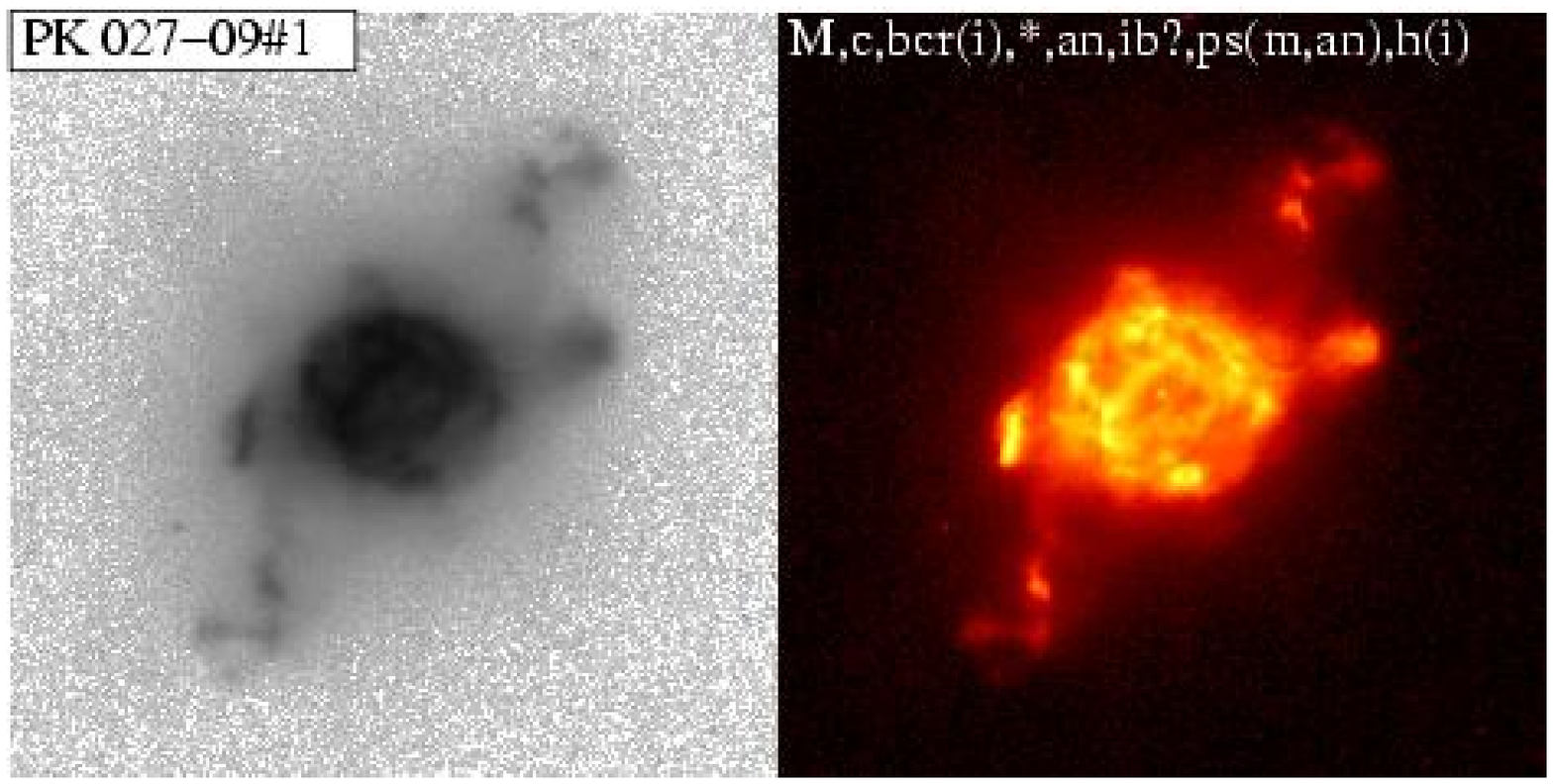}}
\caption{As in Fig\,1., but for PK\,027-09\#1.
}
\label{027-09d1}
\end{figure}

\begin{figure}[htb]
\vskip -0.6cm
\resizebox{0.77\textwidth}{!}{\includegraphics{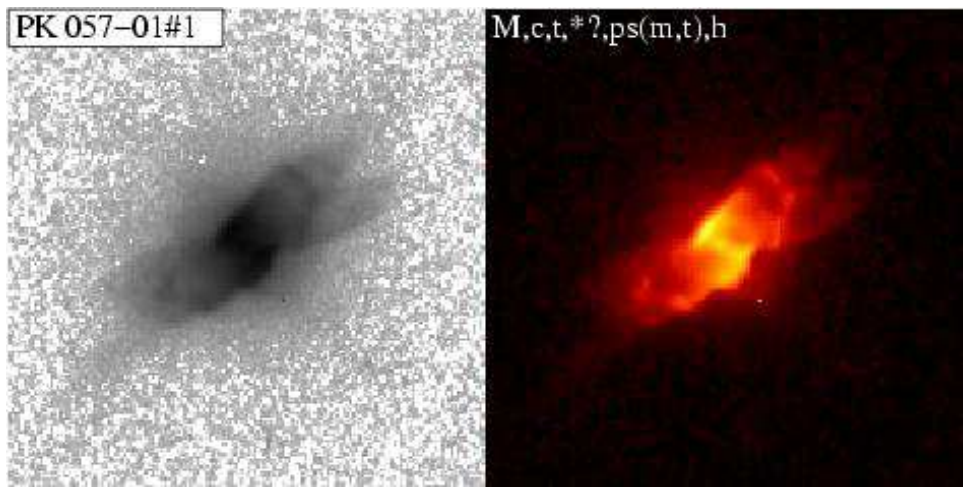}}
\caption{As in Fig\,1., but for PK\,057-01\#1.
}
\label{057-01d1}
\end{figure}

\begin{figure}[htb]
\resizebox{0.77\textwidth}{!}{\includegraphics{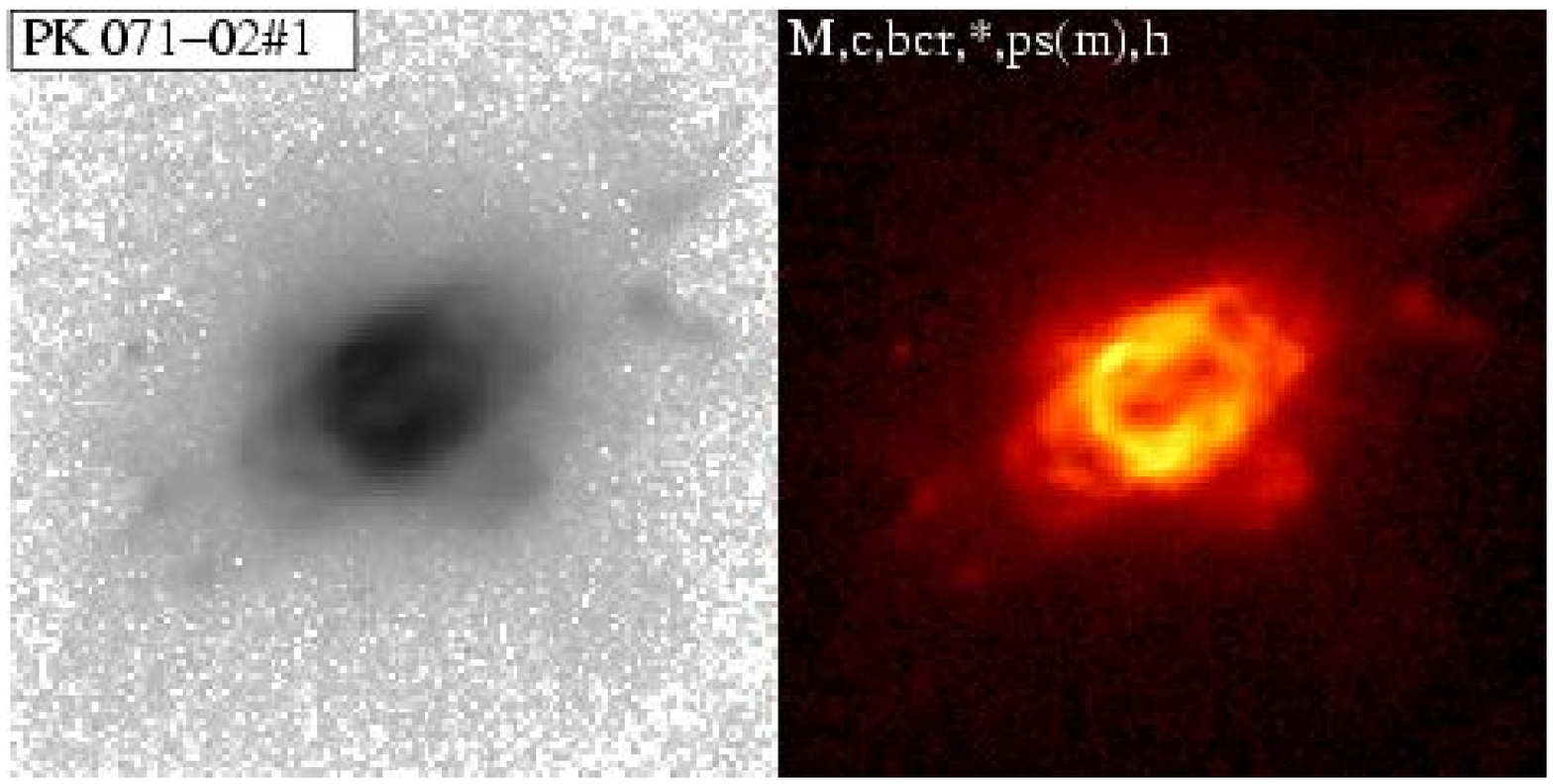}}
\caption{As in Fig\,1., but for PK\,071-02\#1.
}
\label{071-02d1}
\end{figure}

\begin{figure}[htb]
\vskip -0.6cm
\resizebox{0.77\textwidth}{!}{\includegraphics{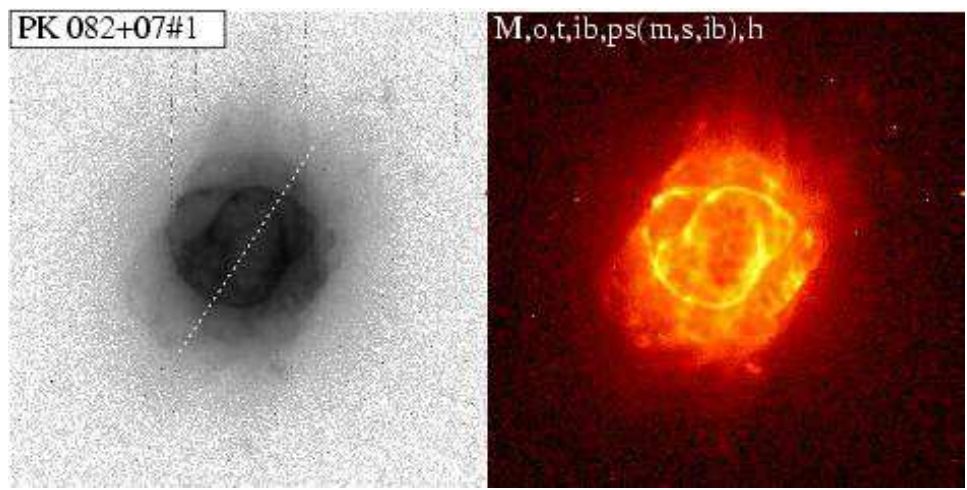}}
\caption{As in Fig\,1., but for PK\,082+07\#1.
}
\label{082+07d1}
\end{figure}

\begin{figure}[htb]
\vskip -0.6cm
\resizebox{0.77\textwidth}{!}{\includegraphics{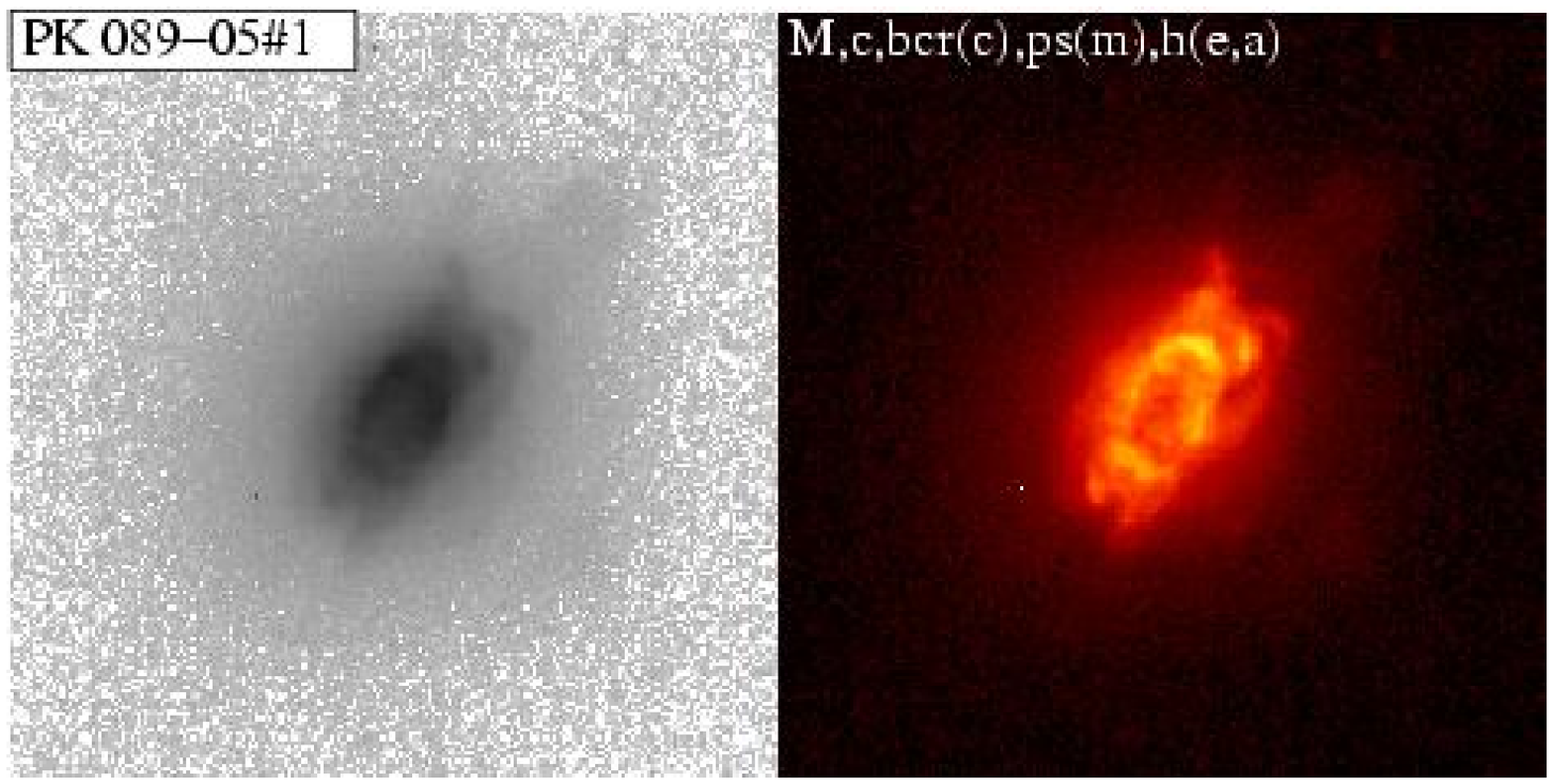}}
\caption{As in Fig\,1., but for PK\,089-05\#1.
}
\label{089-05d1}
\end{figure}

\begin{figure}[htb]
\vskip -0.6cm
\resizebox{0.77\textwidth}{!}{\includegraphics{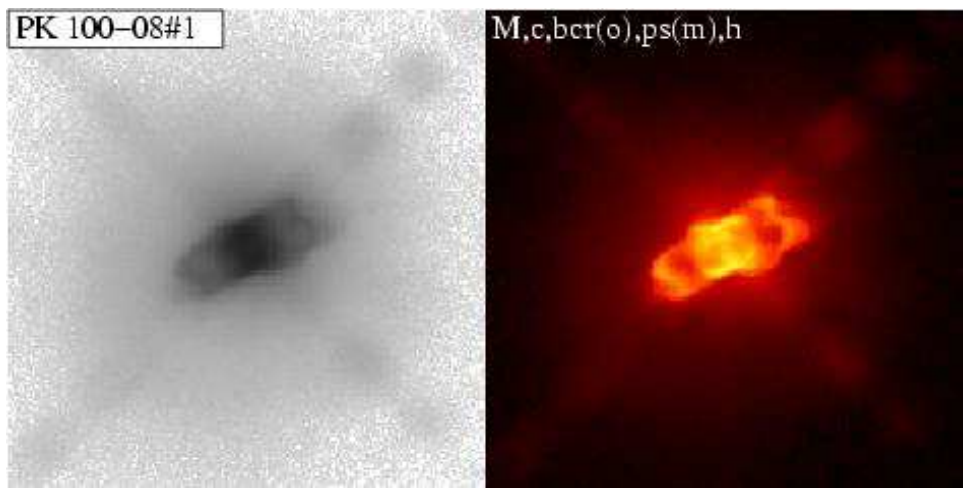}}
\caption{As in Fig\,1., but for PK\,100-08\#1 (adapted from ST98).
}
\label{100-08d1}
\end{figure}

\begin{figure}[htb]
\resizebox{0.77\textwidth}{!}{\includegraphics{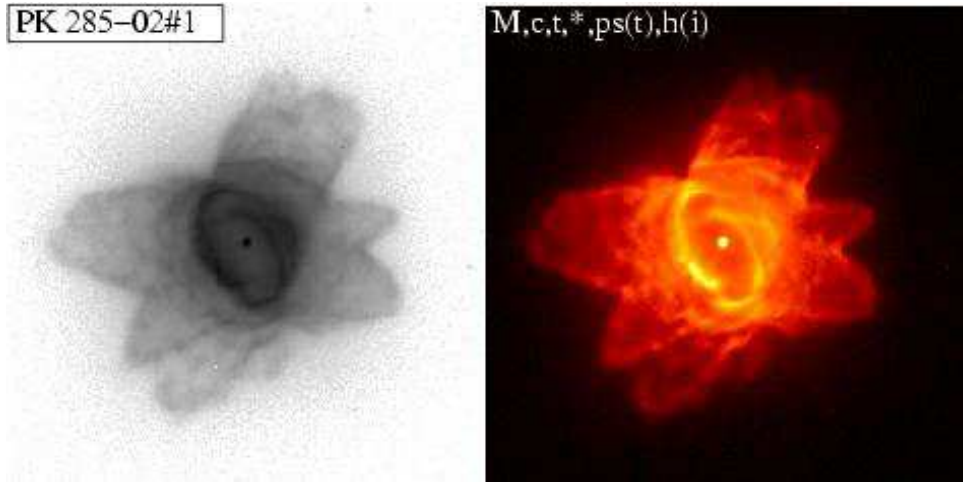}}
\caption{As in Fig\,1., but for PK\,285-02\#1 (adapted from Sahai 2000).
}
\label{285-02d1}
\end{figure}

\begin{figure}[htb]
\vskip -0.6cm
\resizebox{0.77\textwidth}{!}{\includegraphics{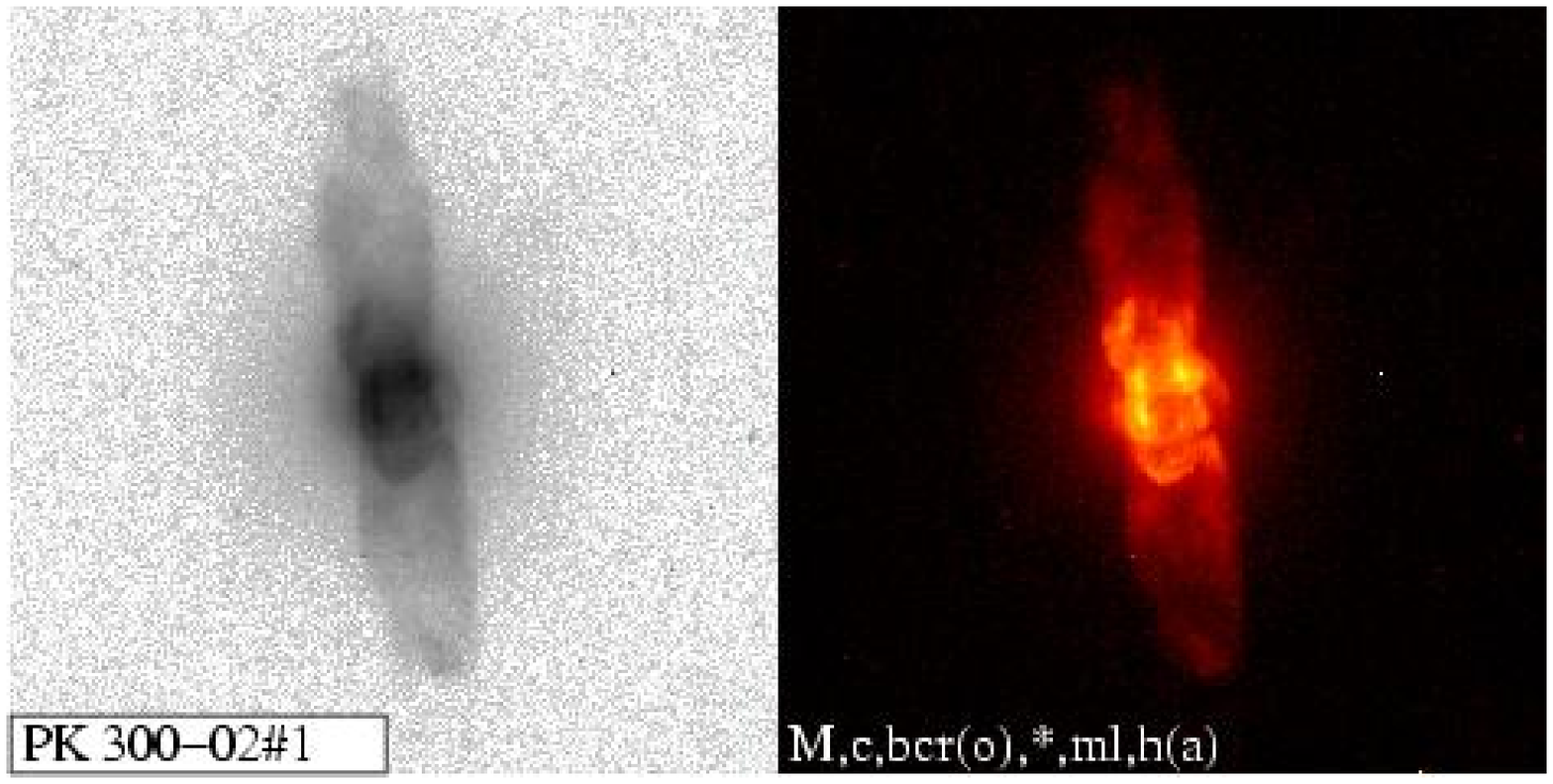}}
\caption{As in Fig\,1., but for PK\,300-02\#1.
}
\label{300-02d1}
\end{figure}

\begin{figure}[htb]
\vskip -0.6cm
\resizebox{0.77\textwidth}{!}{\includegraphics{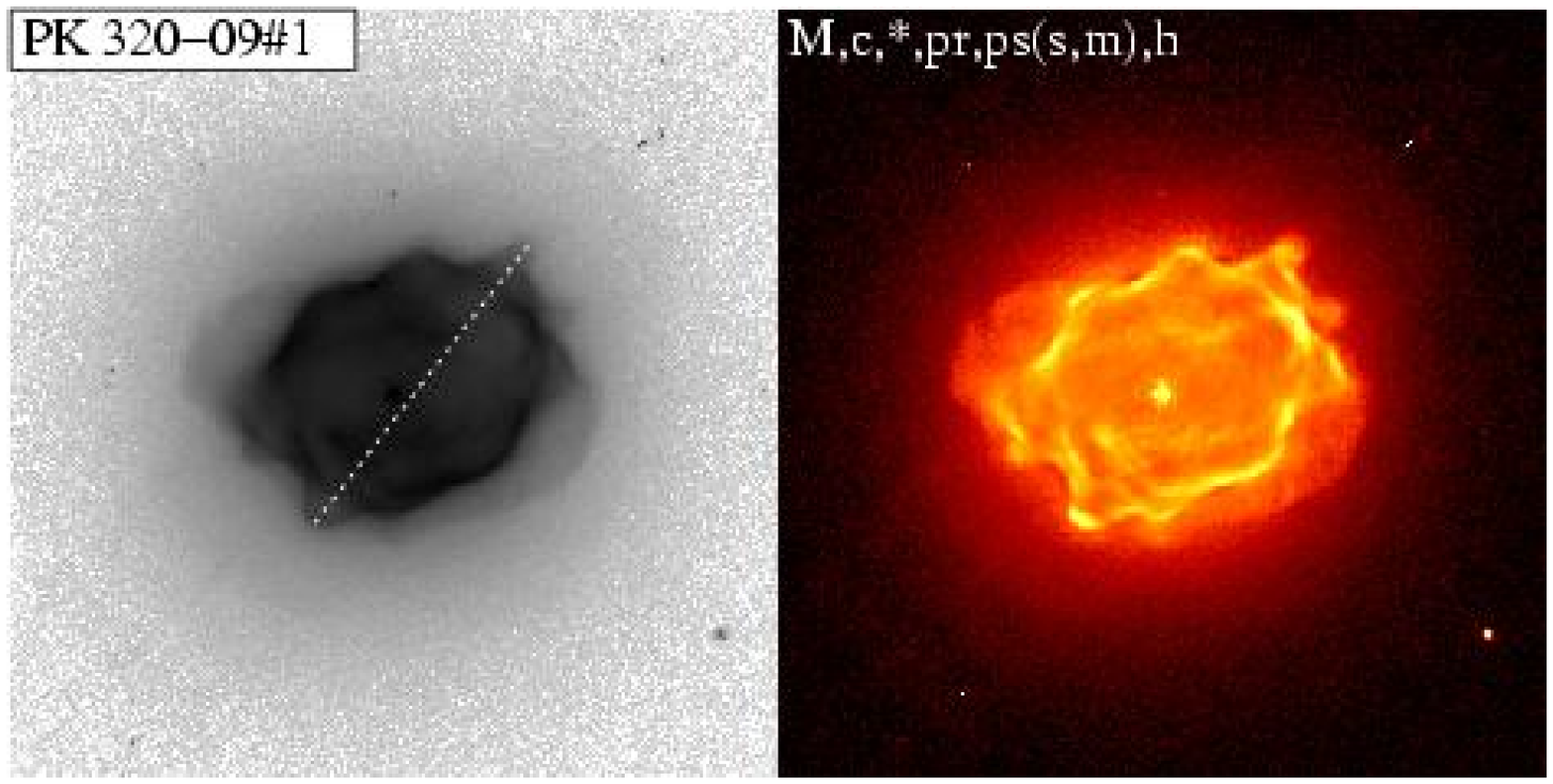}}
\caption{As in Fig\,1., but for PK\,320-09\#1 (adapted from ST98).
}
\label{320-09d1}
\end{figure}

\begin{figure}[htb]
\resizebox{0.77\textwidth}{!}{\includegraphics{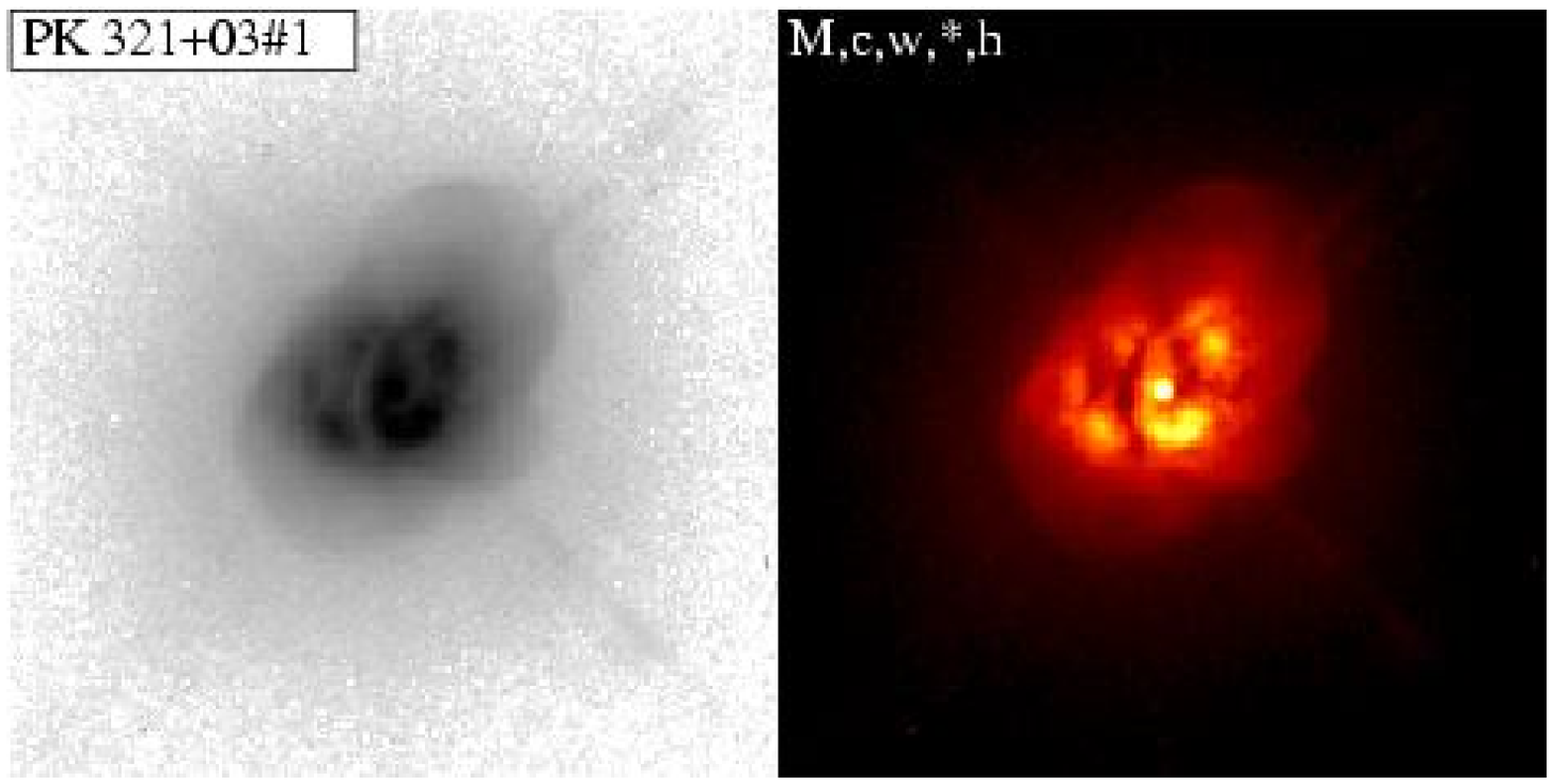}}
\caption{As in Fig\,1., but for PK\,321+03\#1.
}
\label{321+03d1}
\end{figure}

\begin{figure}[htb]
\resizebox{0.77\textwidth}{!}{\includegraphics{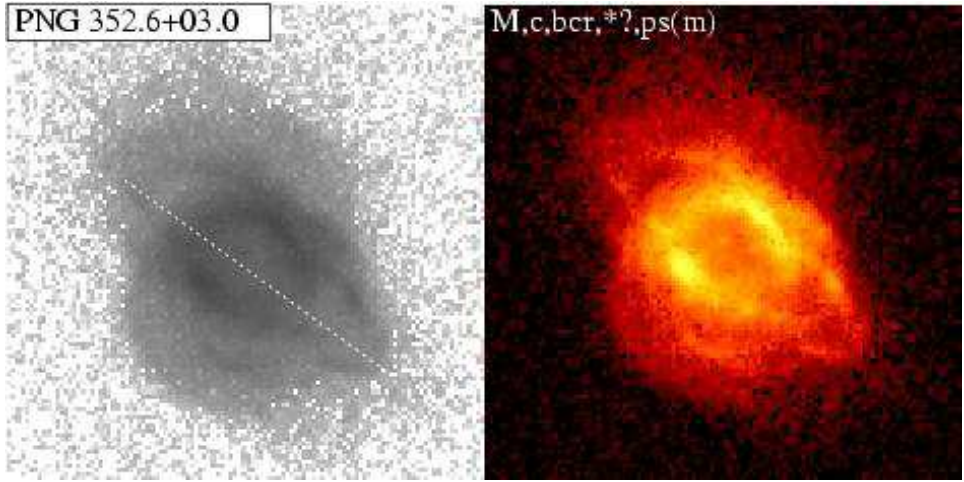}}
\caption{As in Fig\,1., but for PNG352.6+03.0.
}
\label{352.6+03.0}
\end{figure}

\begin{figure}[htb]
\resizebox{0.77\textwidth}{!}{\includegraphics{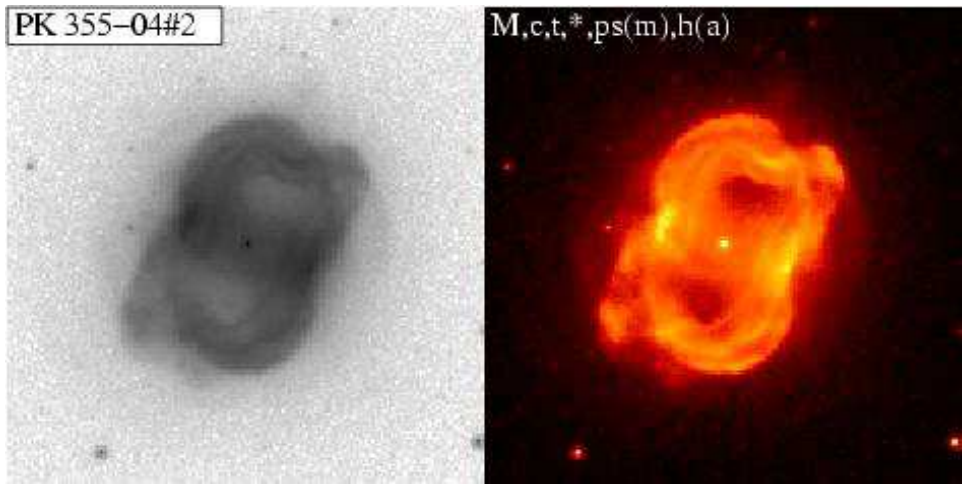}}
\caption{As in Fig\,1., but for PK\,355-04\#2.
}
\label{355-04d2}
\end{figure}

\begin{figure}[htb]
\vskip -0.6cm
\resizebox{0.77\textwidth}{!}{\includegraphics{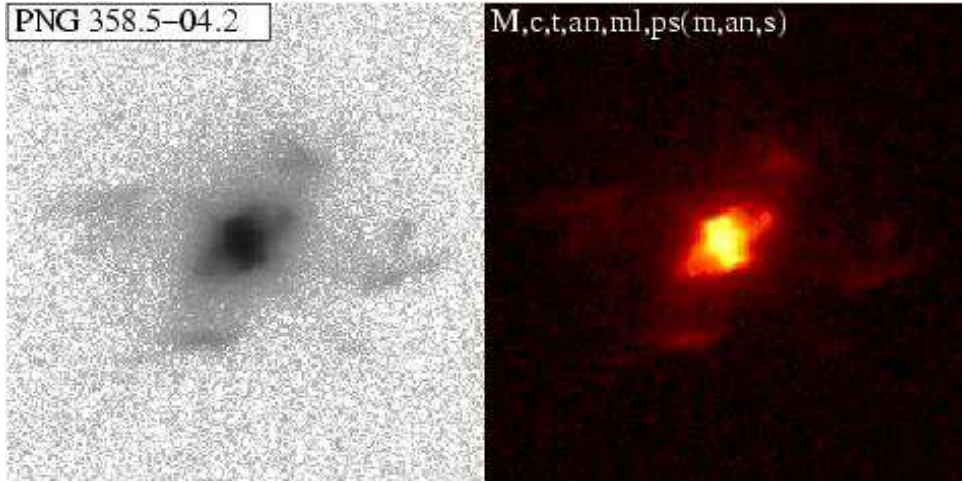}}
\caption{As in Fig\,1., but for PNG358.5-04.2.
}
\label{358.5-04.2}
\end{figure}

\begin{figure}[htb]
\vskip -0.6cm
\resizebox{0.77\textwidth}{!}{\includegraphics{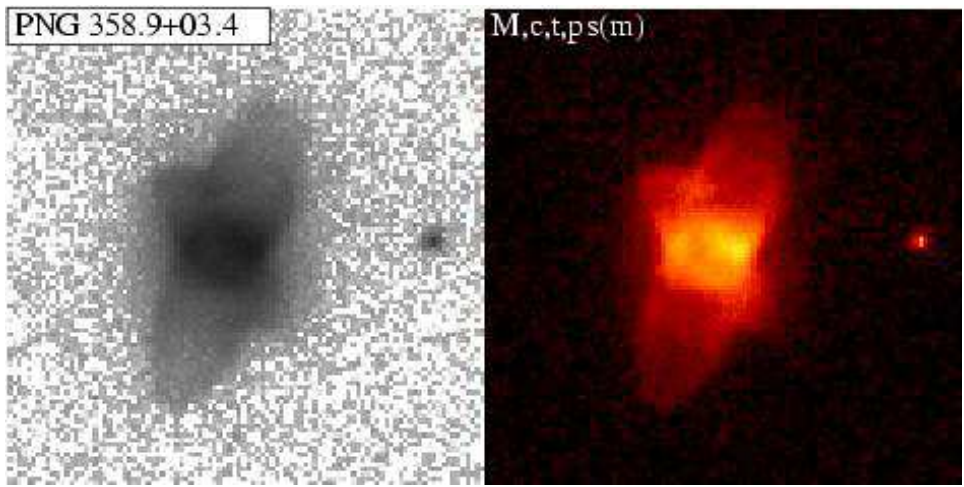}}
\caption{As in Fig\,1., but for PNG358.9+03.4.
}
\label{358.9+03.4}
\end{figure}

\begin{figure}[htb]
\resizebox{0.77\textwidth}{!}{\includegraphics{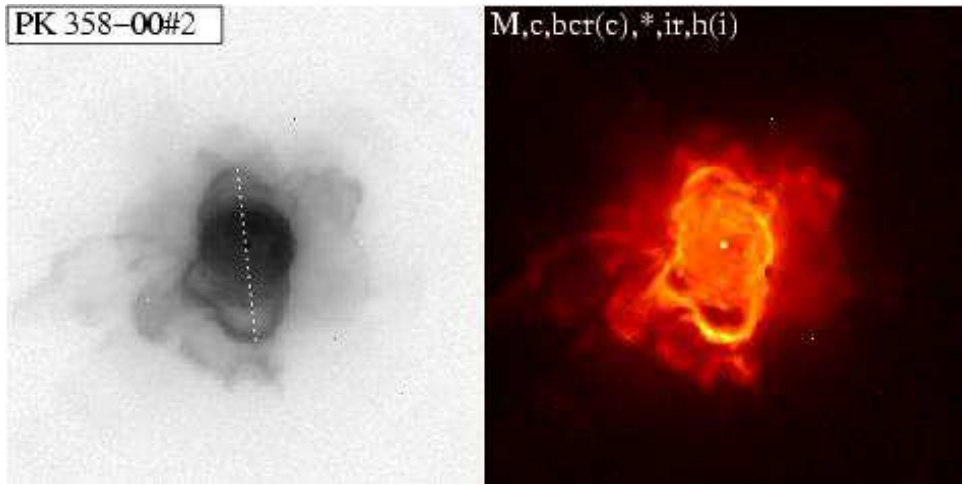}}
\caption{As in Fig\,1., but for PK\,358-00\#2 (adapted from ST98).
}
\label{358-00d2}
\end{figure}

\begin{figure}[htb]
\vskip -0.6cm
\resizebox{0.77\textwidth}{!}{\includegraphics{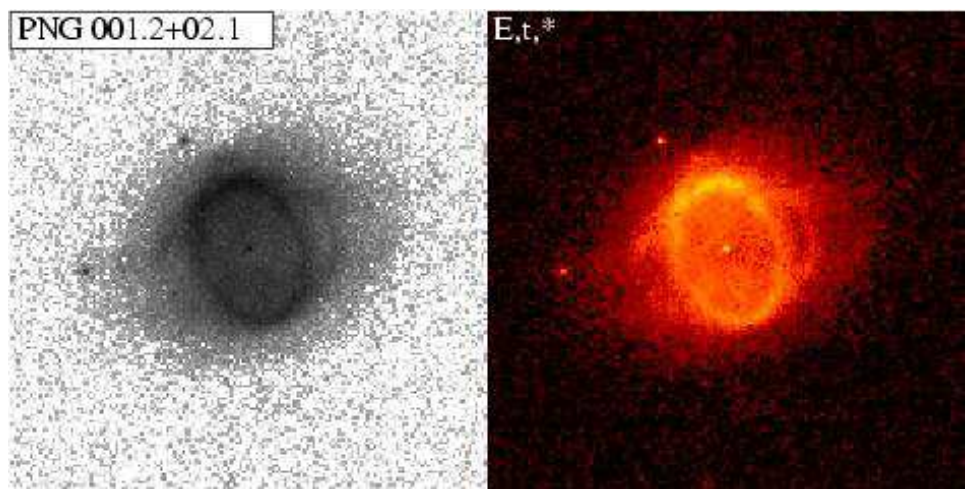}}
\caption{As in Fig\,1., but for PNG001.2+02.1.
}
\label{001.2+02.1}
\end{figure}

\begin{figure}[htb]
\vskip -0.6cm
\resizebox{0.77\textwidth}{!}{\includegraphics{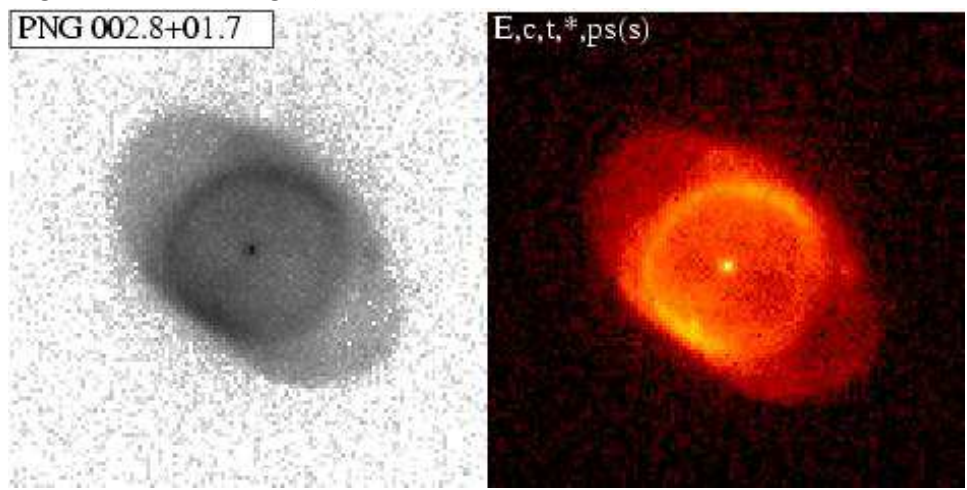}}
\caption{As in Fig\,1., but for PNG002.8+01.7.
}
\label{002.8+01.7}
\end{figure}

\begin{figure}[htb]
\vskip -0.6cm
\resizebox{0.77\textwidth}{!}{\includegraphics{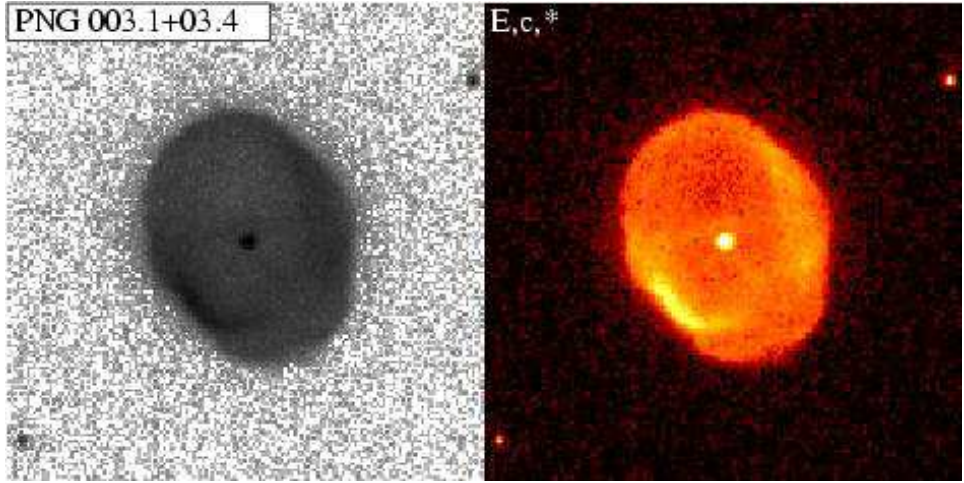}}
\caption{As in Fig\,1., but for PNG003.1+03.4.
}
\label{003.1+03.4}
\end{figure}

\begin{figure}[htb]
\vskip -0.6cm
\resizebox{0.77\textwidth}{!}{\includegraphics{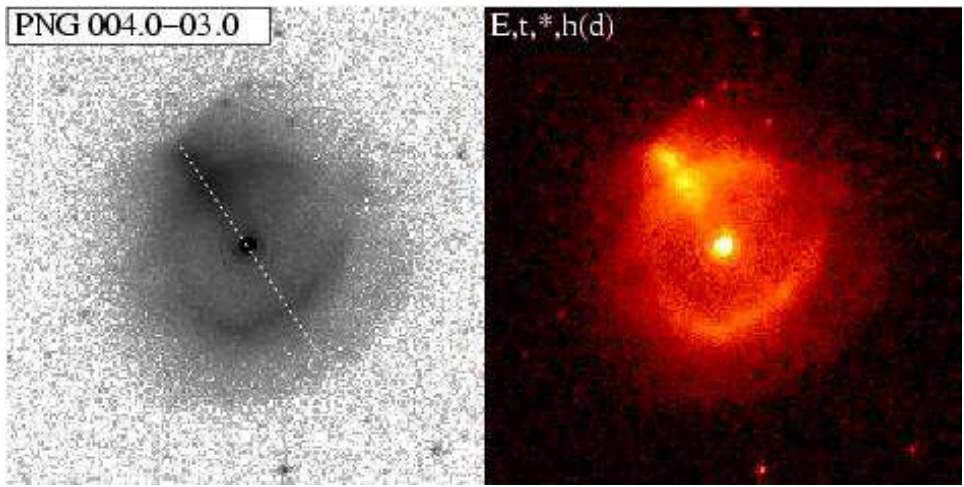}}
\caption{As in Fig\,1., but for PNG004.0-03.0.
}
\label{004.0-03.0}
\end{figure}

\begin{figure}[htb]
\vskip -0.6cm
\resizebox{0.77\textwidth}{!}{\includegraphics{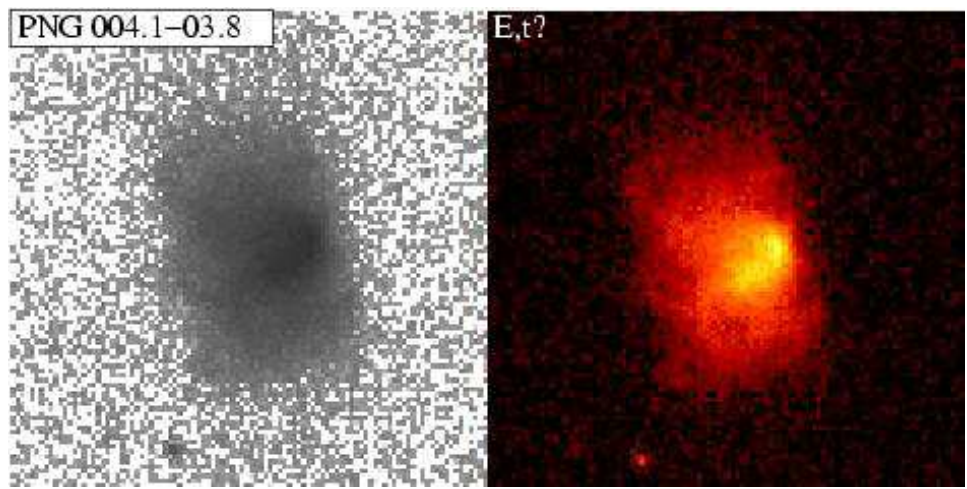}}
\caption{As in Fig\,1., but for PNG004.1-03.8.
}
\label{004.1-03.8}
\end{figure}

\begin{figure}[htb]
\vskip -0.6cm
\resizebox{0.77\textwidth}{!}{\includegraphics{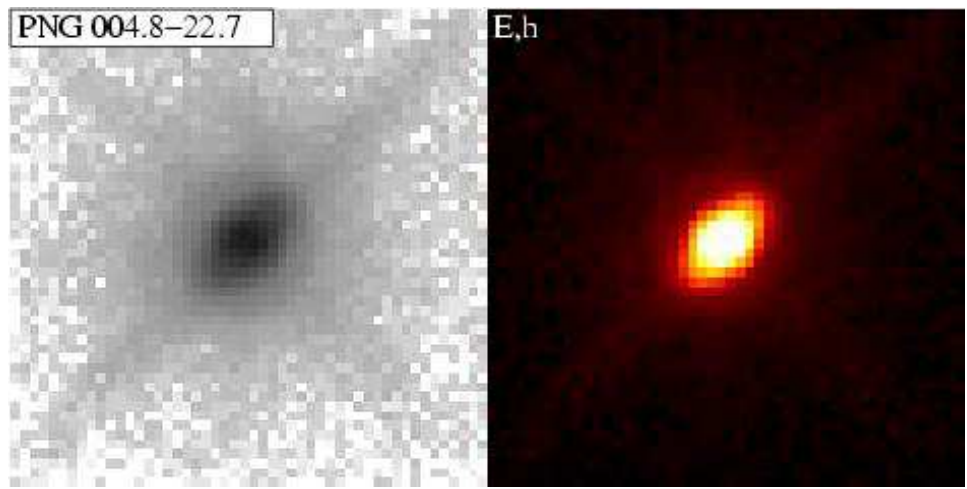}}
\caption{As in Fig\,1., but for PNG004.8-22.7.
}
\label{004.8-22.7}
\end{figure}

\begin{figure}[htb]
\vskip -0.6cm
\resizebox{0.77\textwidth}{!}{\includegraphics{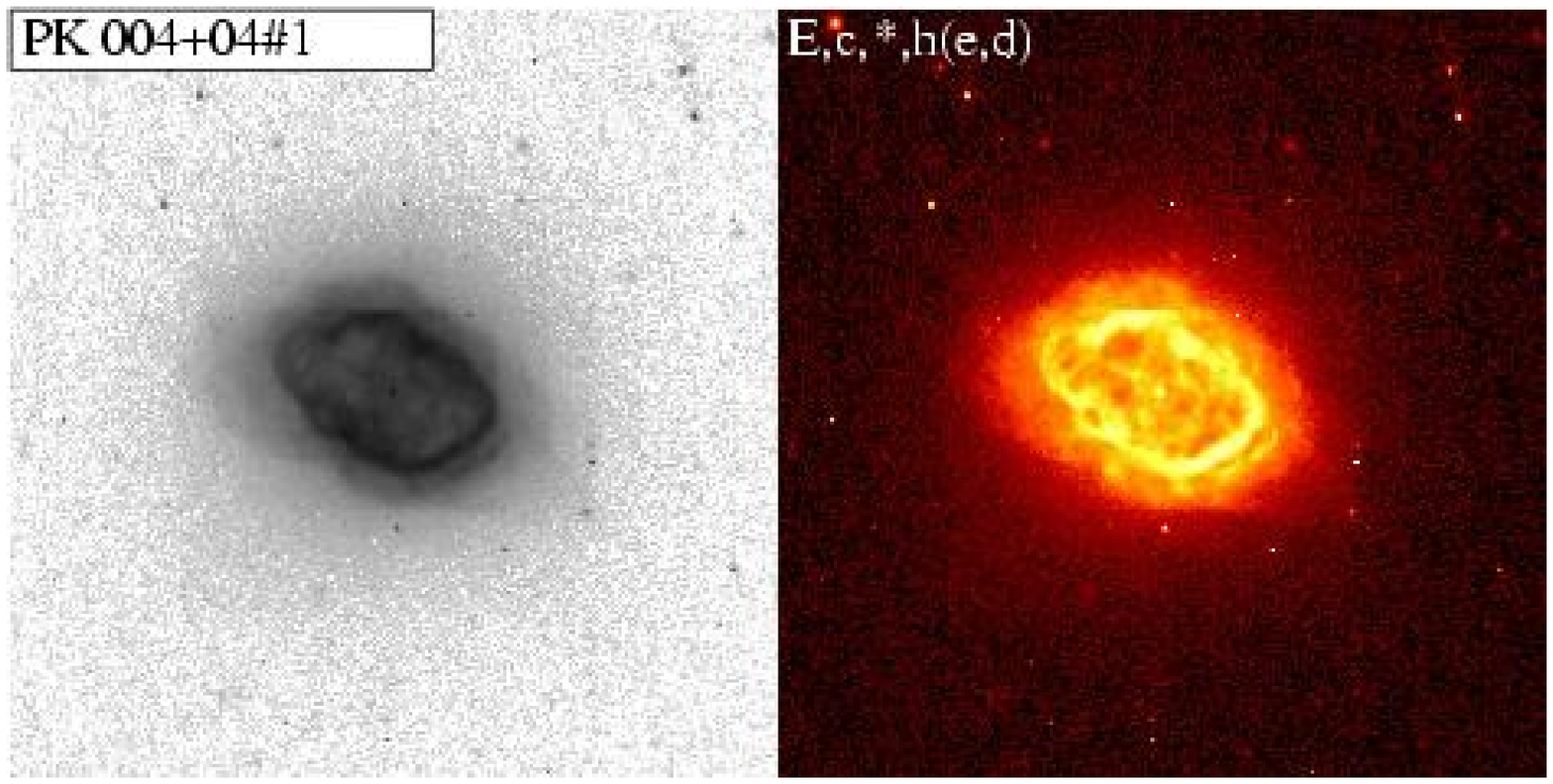}}
\caption{As in Fig\,1., but for PK\,004+04\#1.
}
\label{004+04d1}
\end{figure}

\begin{figure}[htb]
\vskip -0.6cm
\resizebox{0.77\textwidth}{!}{\includegraphics{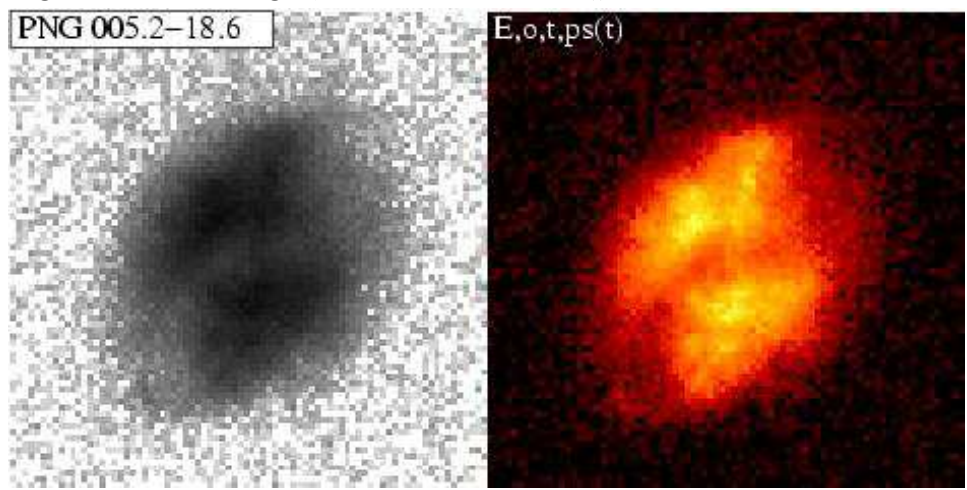}}
\caption{As in Fig\,1., but for PNG005.2-18.6.
}
\label{005.2-18.6}
\end{figure}

\begin{figure}[htb]
\vskip -0.6cm
\resizebox{0.77\textwidth}{!}{\includegraphics{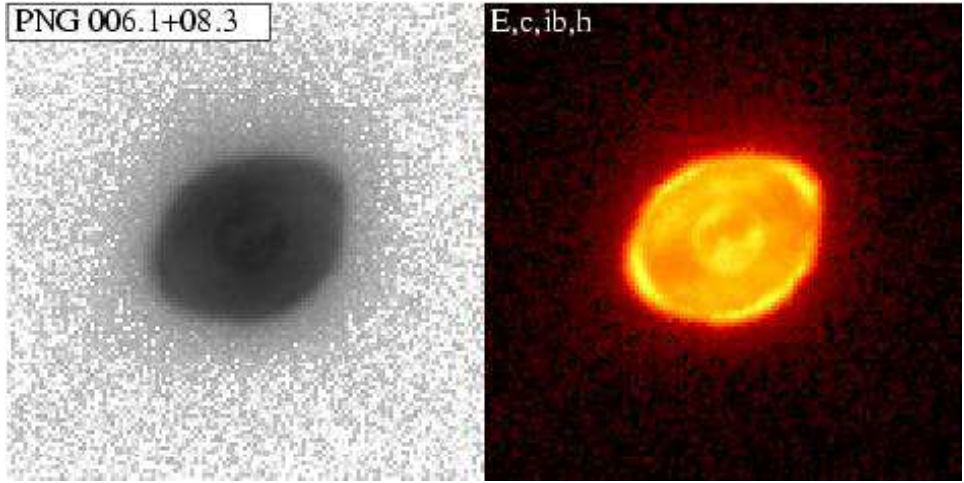}}
\caption{As in Fig\,1., but for PNG006.1+08.3.
}
\label{006.1+08.3}
\end{figure}

\begin{figure}[htb]
\vskip -0.6cm
\resizebox{0.77\textwidth}{!}{\includegraphics{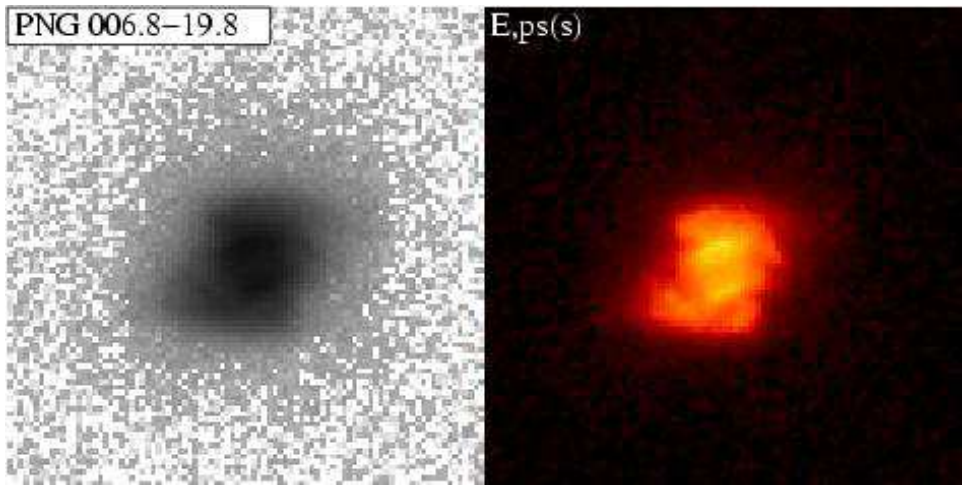}}
\caption{As in Fig\,1., but for PNG006.8-19.8.
}
\label{006.8-19.8}
\end{figure}

\begin{figure}[htb]
\vskip -0.6cm
\resizebox{0.77\textwidth}{!}{\includegraphics{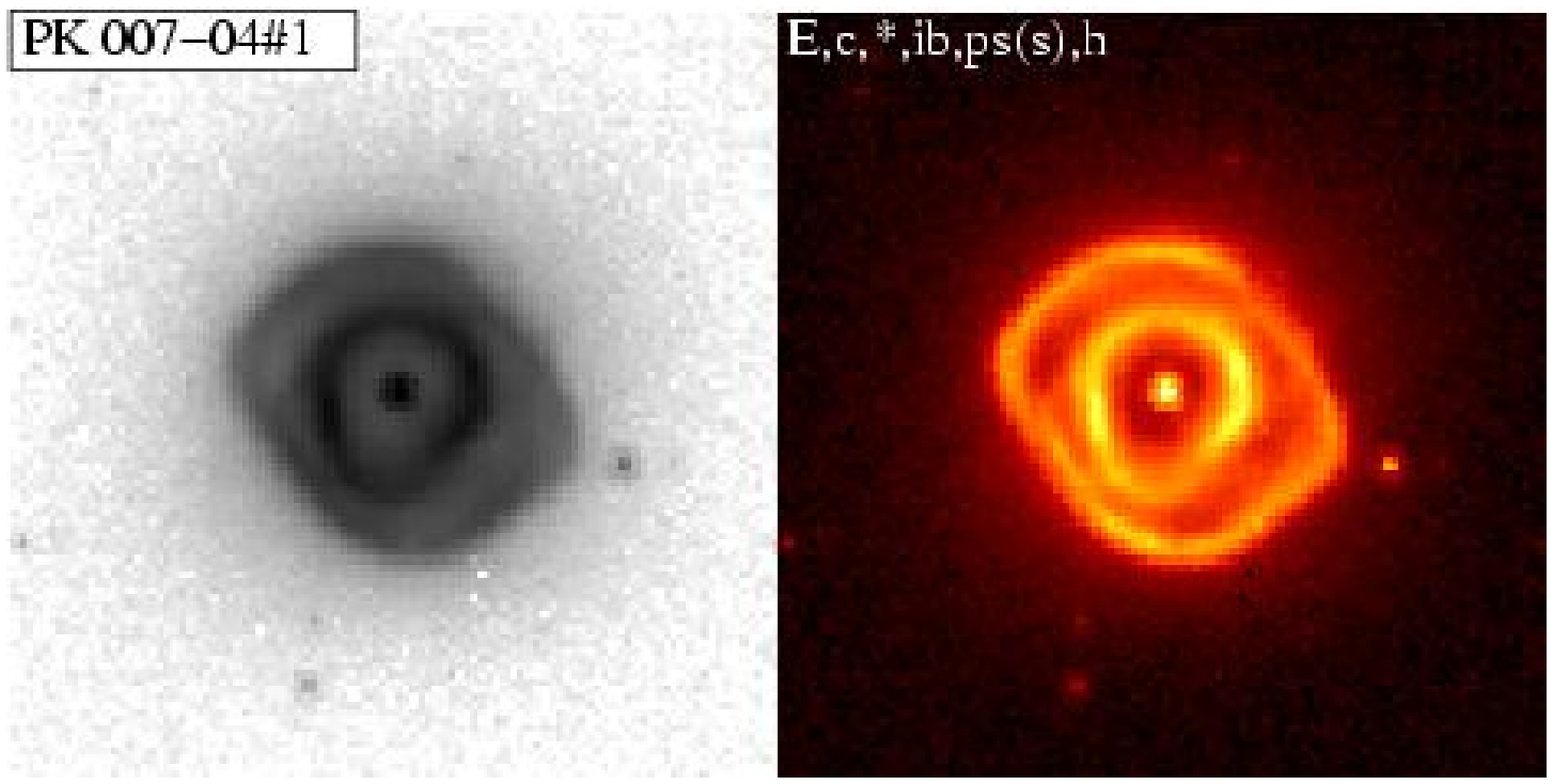}}
\caption{As in Fig\,1., but for PK\,007-04\#1.
}
\label{007-04d1}
\end{figure}

\begin{figure}[htb]
\vskip -0.6cm
\resizebox{0.77\textwidth}{!}{\includegraphics{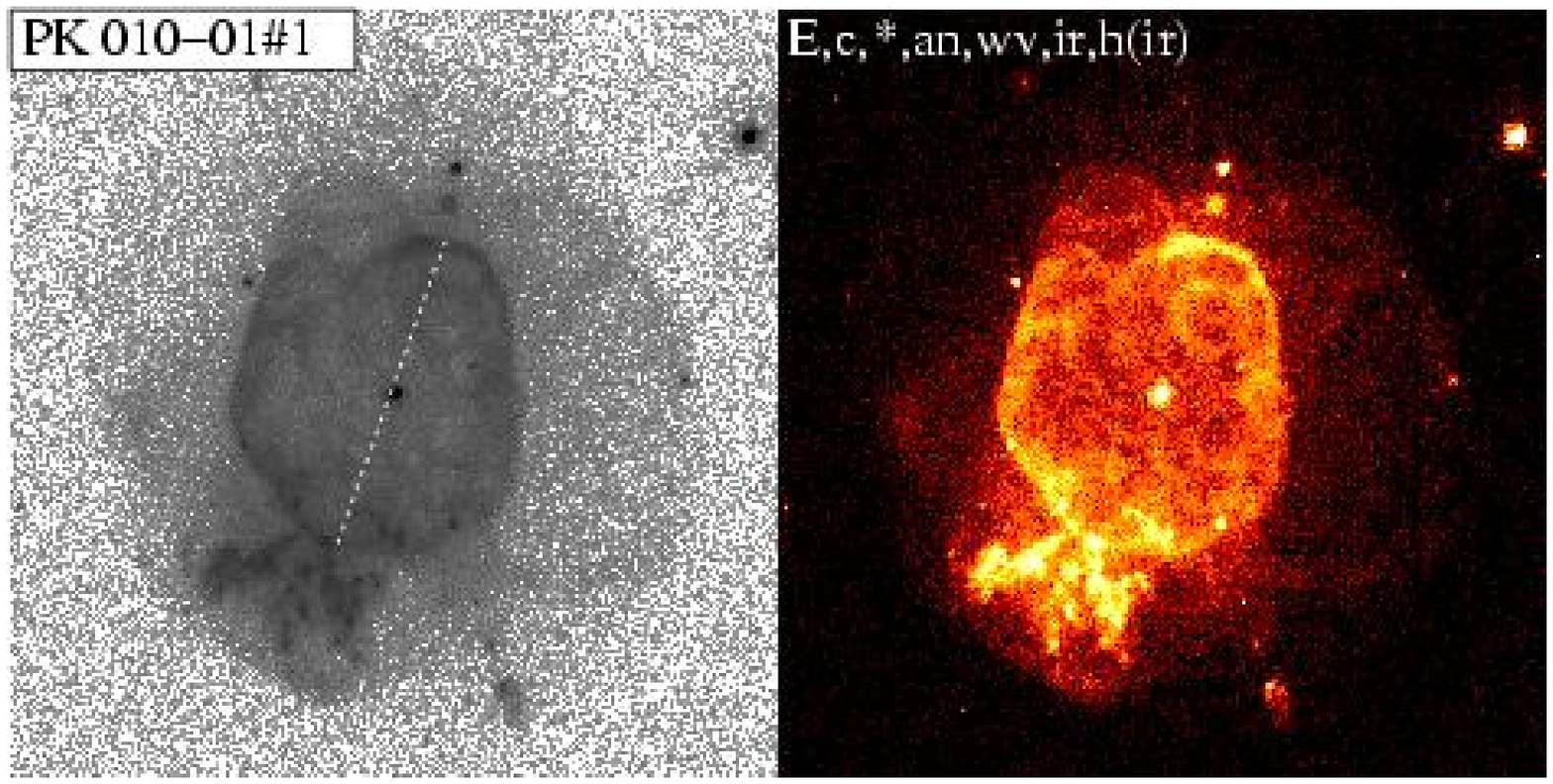}}
\caption{As in Fig\,1., but for PK\,010-01\#1 and the F658N ([NII]) filter.
}
\label{010-01d1}
\end{figure}

\begin{figure}[htb]
\vskip -0.6cm
\resizebox{0.823\textwidth}{!}{\includegraphics{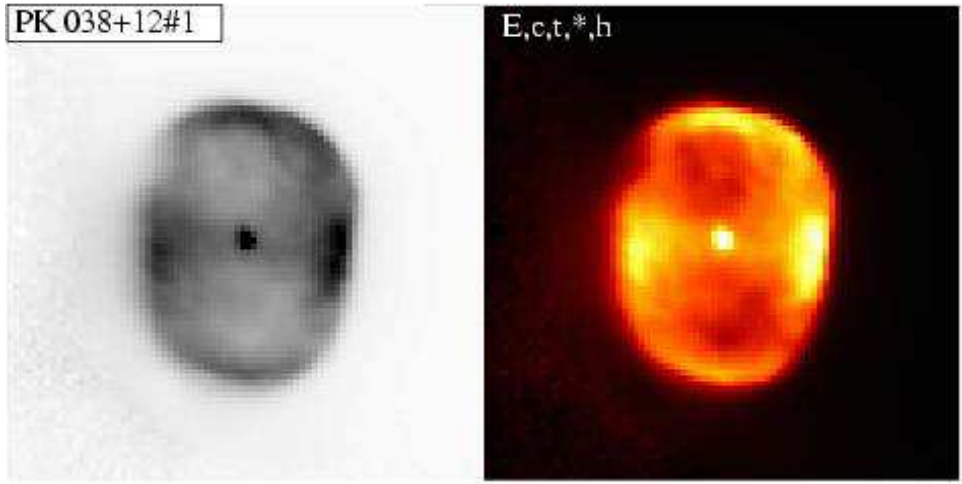}}
\caption{As in Fig\,1., but for PK\,038+12\#1.
}
\label{038+12d1}
\end{figure}

\begin{figure}[htb]
\vskip -0.6cm
\resizebox{0.77\textwidth}{!}{\includegraphics{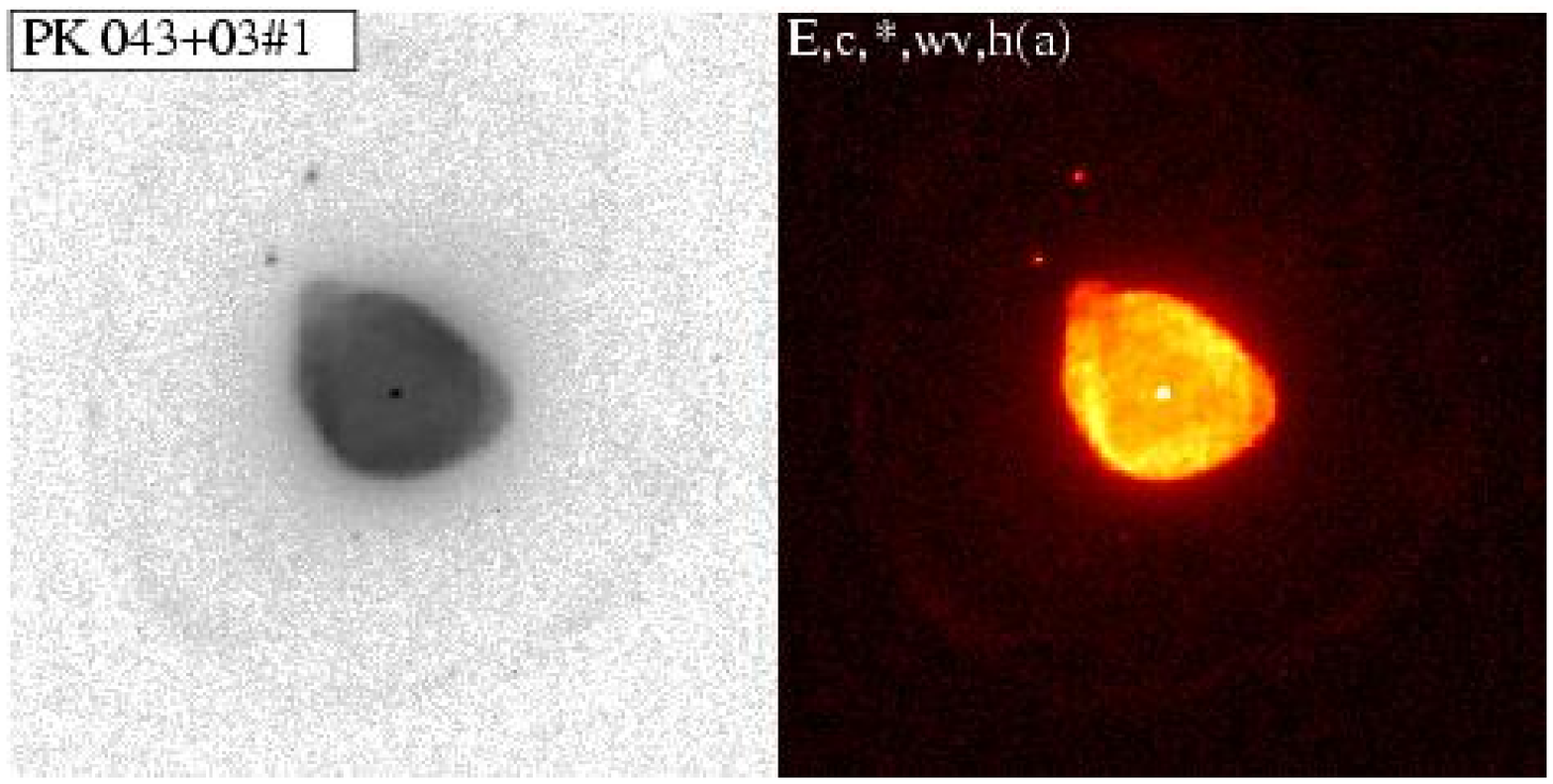}}
\caption{As in Fig\,1., but for PK\,043+03\#1.
}
\label{043+03d1}
\end{figure}

\begin{figure}[htb]
\vskip -0.6cm
\resizebox{0.77\textwidth}{!}{\includegraphics{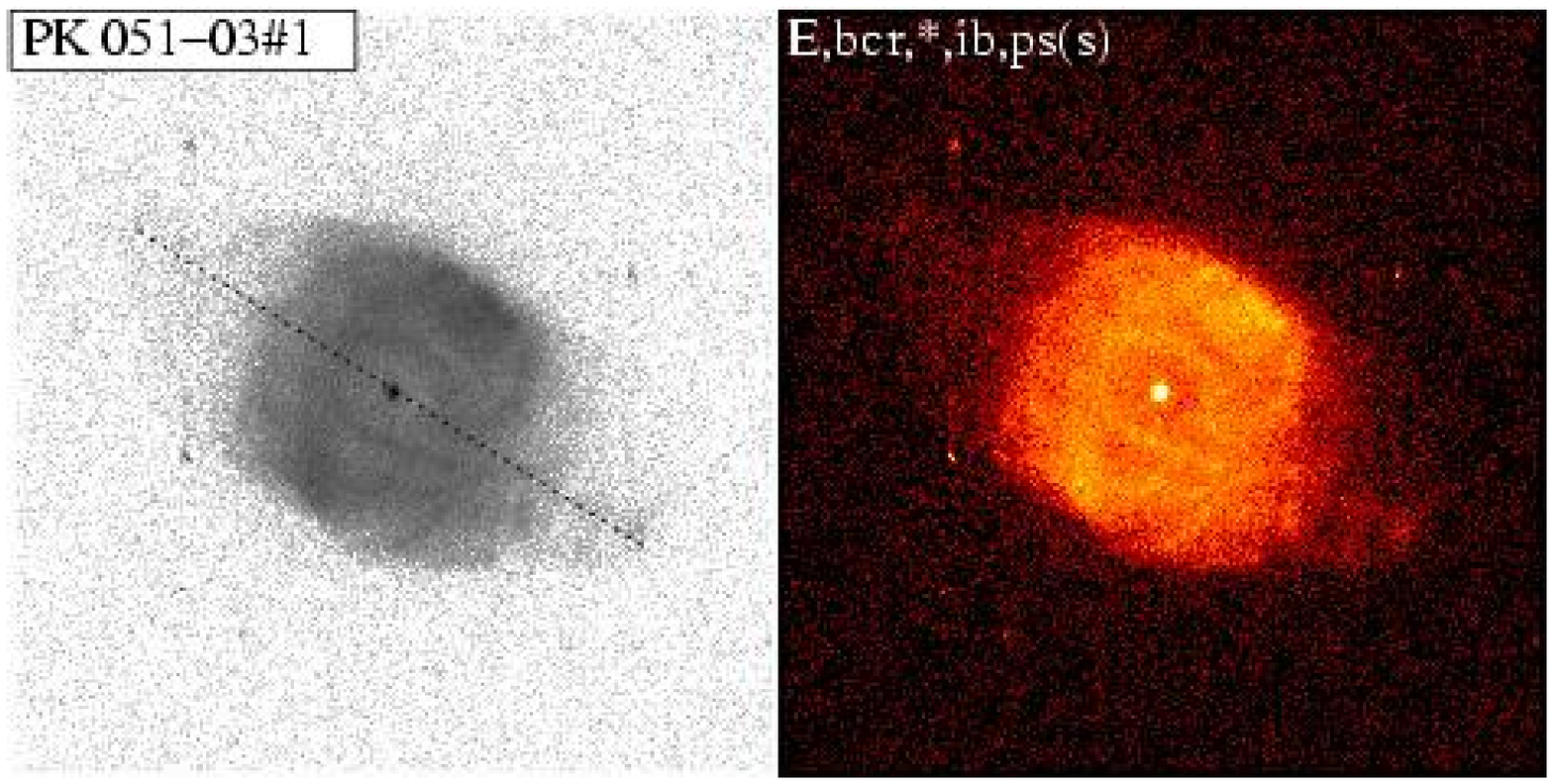}}
\caption{As in Fig\,1., but for PK\,051-03\#1.
}
\label{051-03d1}
\end{figure}

\begin{figure}[htb]
\vskip -0.6cm
\resizebox{0.77\textwidth}{!}{\includegraphics{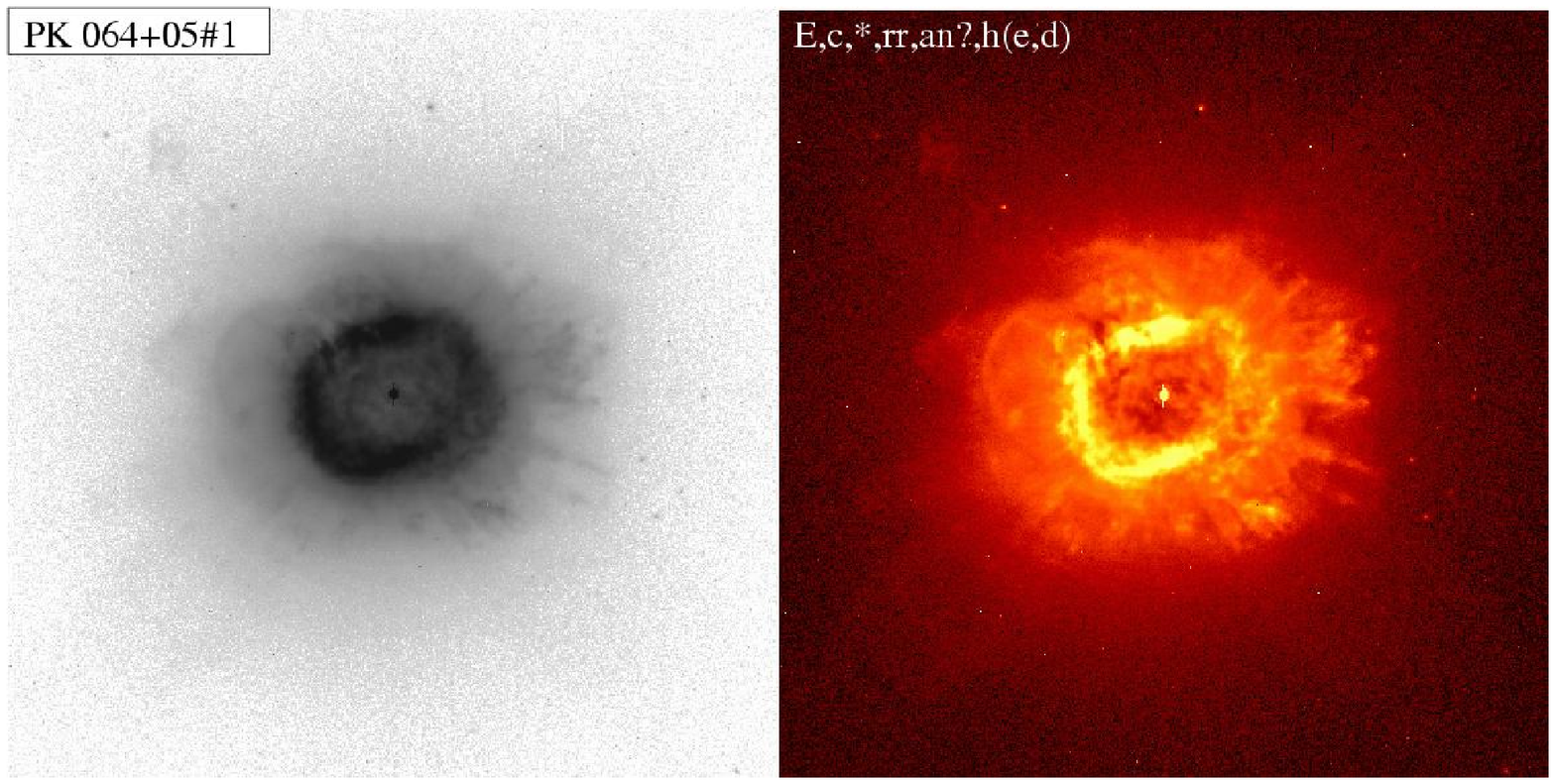}}
\caption{As in Fig\,1., but for PK\,064+05\#1 (adapted from ST98).
}
\label{064+05d1}
\end{figure}

\begin{figure}[htb]
\vskip -0.6cm
\resizebox{0.77\textwidth}{!}{\includegraphics{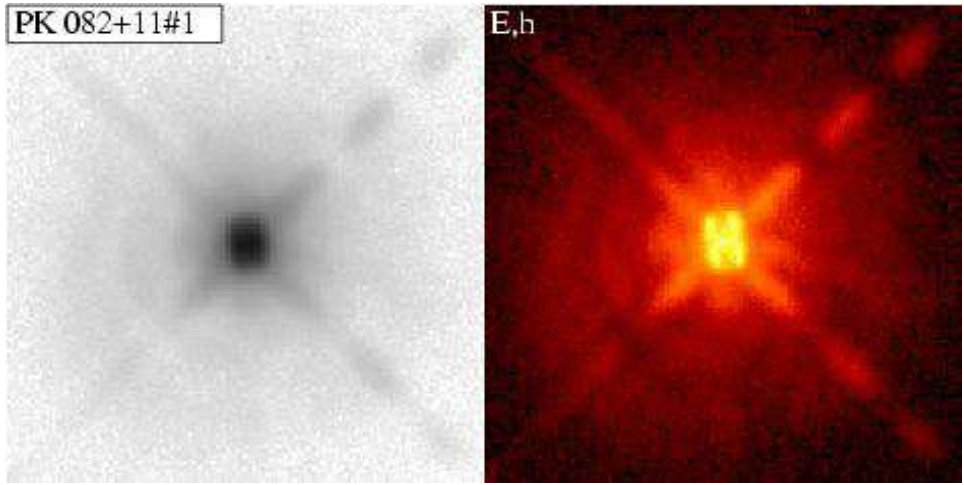}}
\caption{As in Fig\,1., but for PK\,082+11\#1.
}
\label{082+11d1}
\end{figure}

\begin{figure}[htb]
\vskip -0.6cm
\resizebox{0.77\textwidth}{!}{\includegraphics{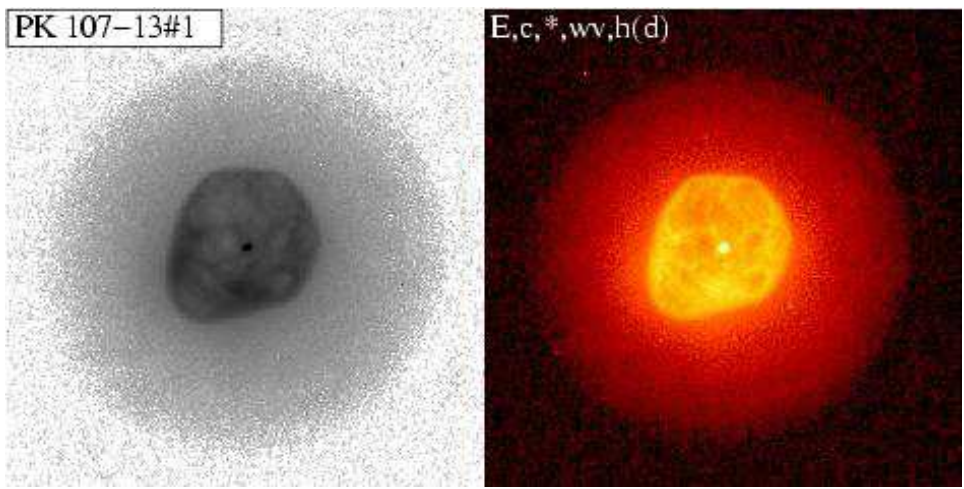}}
\caption{As in Fig\,1., but for PK\,107-13\#1.
}
\label{107-13d1}
\end{figure}

\begin{figure}[htb]
\vskip -0.6cm
\resizebox{0.77\textwidth}{!}{\includegraphics{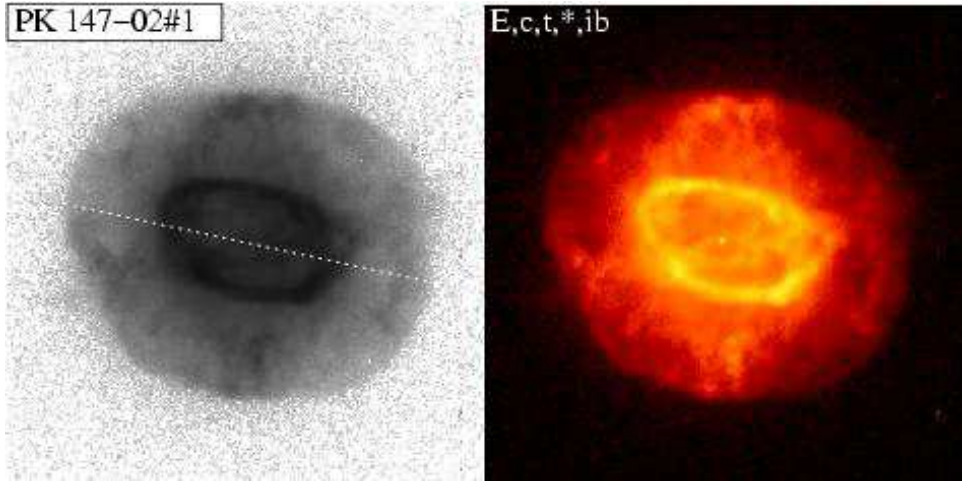}}
\caption{As in Fig\,1., but for PK\,147-02\#1.
}
\label{147-02d1}
\end{figure}

The PPN classification system of SMSC07 consisted of 4 primary classes based on the overall nebular shape, and
several categories of secondary characteristics related to specific properties of the lobes, waist, and haloes,
and the presence of point-symmetry. This scheme is summarised in Table\,\ref{codes} (non-italicised text). The four
existing primary classes, and the ones that we extend to PNs as well, are: B (bipolar), M (multipolar), E
(elongated), and I (irregular). The B class (illustrated for PNs in Figs.\,\ref{000+17d1}--\ref{356.9_04.4})
represents objects which show two primary, diametrically opposed lobes, centered on the central star or its
expected location. The pair of lobes must have
a``pinched in" shape in the region around the center from where they emanate, and/or the lobes should be visible
on both sides of a minimum-light-intensity central region (due to an obscuring dust lane). The M class  (Figs.\
\ref{002-03d3}--\ref{358-00d2}) represents
objects having two or more primary lobe pairs whose axes are not aligned. The E
class (Figs.\
\ref{001.2+02.1}--\ref{359.2+04.7}) is simply one in which objects are elongated along a specific axis, i.e., are not round.
The I class  (Fig.\ \ref{003.8+05.3}--\ref{350+04d1}) represents objects in which extended circumstellar structure can be
seen,
but where no obvious lobe or shell-like structures can be identified, and which therefore do not fit in any of the
previous categories. As the name implies, class-I objects usually do
not display any obvious geometrical symmetry such as axial or point-symmetry.

We extend the SMSC07 system to young PNs, by adding new primary classes as well as secondary characteristics
(italicised text in Table\,\ref{codes}). An additional three primary classes have been added. The first is R  
(Figs.\,\ref{004.8+02.0}--\ref{012.2+04.9}), which
describes round objects. The maximum asymmetry for an object to be classified as R (rather than E) is $<10\%$, i.e., the  widest extent
of the
object should be a factor $<1.1$ times its average extent.
Round objects are rare: of a total of 119 PNs, we only find 4. Note that in the PPN study, SMSC07
did
not find a single round object. The
second is L (Figs.\,\ref{001.7-04.4}--\ref{356-03d3}), which
describes objects having collimated lobes, but which show no constriction in the central, waist region -- i.e.,
the lobes are not pinched-in towards the waist region, which is a requirement for being classified as
B, so we do not include them in
the B class even though they may be closely related (see \S\,\ref{waistevol}). In order for an object to have an L,
rather than an E classification,
we require the collimation factor (defined as the ratio of the total tip-to-tip extent of the lobes to their 
lateral width) to be 3 or larger.

The third new primary class is designated S (Figs.\,\ref{002.9-03.9}--\ref{356.8+03.3}), which describes a small set of
objects in which the projection on the
sky of the most prominent nebular structure has a two-armed spiral shape. The
apparent spiral structure first becomes
evident at some finite radius out from the center, i.e., the spiral-shaped features do not go all the way in to
the center. No lobe or shell structures can be seen, although diffuse nebulosity may be present. PNG\,356.8+03.3
and PK032+07\#2 are the best examples (Figs.\,\ref{356.8+03.3},\,\ref{032+07d2}). Two other examples, PNG\,002.9-03.9 and
PNG\,008.6-02.6 (Figs.\,\ref{002.9-03.9},\,\ref{008.6-02.6}), show additional structures. PNG357.1-04.7 (Fig.\
\ref{357.1-04.7}), also shows a
bright, two-armed spiral feature, but each arm is part of the opposing peripheries of a pair of lobes.
This object, classified as B, appears therefore to be a connecting link between the S and the B or L classes.

The well-known, nearby PN, NGC\,7293 (the Helix Nebula), with its two spiral-arm like structures
observed in molecular-line (CO) emission (Young et al. 1999), shows considerable similarity to our class S objects. Two additional
well-resolved examples of the S shape are NGC6309
(ground-based images in SCM92, V{\'a}zquez et al. 2008), and K4-55 (ground-based image in the IAC catalog: Metal96)\footnote{also
recently imaged with HST
as the last image taken with the WFPC2 instrument before it was returned to Earth}.
In each of these two cases, one can discern that the spiral-shaped features are the highly brightened sides of very faint lobes.
So it is not entirely clear whether S represents a new primary morphological class, or whether it should be regarded as a
special case of bipolar with point-symmetric shape, B, $ps(s)$ (see \S\,\ref{secondary}). While we suspect the latter
to be plausible, we retain S as a separate class in view of the fact that some of the S objects show no signs of bipolar
lobes. Deeper observations of the S systems can in principle resolve this question. 



In the PPN classification study by SMSC07, the morphologies are those of the optical continuum, resulting from
dust-scattered light, whereas in this study, the morphologies that we are examining and classifying are
dominated by H$\alpha$ emission, and are therefore indicative of the ionized gas. This difference should be
kept in mind when comparing the two studies. However, as discussed by SMSC07, the observed morphologies in
both cases are a very good indicator of the geometric shapes of the lobe walls and other nebular structures. 

\subsection{Secondary Characteristics}\label{secondary}
The presence of secondary structural features in the nebulae is denoted with lower-case letters
following the capital letter representing the major
class. Below, we provide a brief summary of the secondary descriptors used for our PPN scheme, and introduce new
ones which have had to be added in order to accommodate new features seen in PN morphologies. In general, 
the features to which we have chosen to assign descriptors, are those that appear to be common to multiple PNs, and which 
display some geometric symmetry or order.

For PPN, we first added secondary characeristics related to the lobes in the B, M, or E classes; if the lobes are open (i.e.,
like a vase) or closed at their outer ends (i.e., have a bubble-structure) they are denoted by $o$ or $c$, respectively. We
then characterised the central region of PPNs, where the presence of a dark obscuring band along the short axis of the
nebula (i.e., a minimum in an intensity cut taken along the primary long axis of the nebula, and usually described as the
``waist" of the nebula) was denoted by $w$. Evidence for point-symmetry in the nebular structure was denoted by $ps$. This
classification was not applied to axially symmetric objects, even though axial symmetry is a special case of point-symmetry.
The point-symmetry could be of three general types, resulting from (i) the presence of 2 or more pairs of
diametrically-opposed lobes, denoted as $ps(m)$, (ii) the distribution of ansae point-symmetrically about the center,
denoted as $ps(an)$, and (iii) the overall geometrical shape of the lobes being point-symmetric, denoted as $ps(s)$.

Three additional nebular characteristics were included: (i) bright, compact knots in diametrically-opposed pairs, normally
referred to as ansae, were denoted by $an$\footnote {sometimes only one bright ansa-like knot (instead of a pair) is seen,
we designate this as $an?$}; (ii) the presence of minor lobes as, for example, seen in the Frosty Leo Nebula (Sahai et al. 2000a),
was denoted by $ml$, and (iii) a skirt-like structure around the primary lobes was denoted by $sk$. We define minor lobes as
being distinctly smaller and thinner than the main lobes, noting that lobes whose lengths are significantly smaller but
whose widths are comparable to those of the major lobes, are likely being foreshortened by projection -- in such cases the
object is classified as M (e.g., PK019-05\#1,\,Fig.\,\ref{019-05d1}, and PK300-02\#1,\,Fig.\,\ref{300-02d1}). 



The presence of a halo\footnote{a diffuse structure which lies outside the bright primary nebular structure, and that can be clearly
seen above the background sky; the halo surface brightness is typically a factor 10 or more less than that of the bright primary
structure} was denoted with $h$, with a qualifier $e$ if it had an elongated shape, i.e., as $h(e)$. If the halo
shape could not be determined reasonably, we added $i$, meaning that the shape was indeterminate, that is, as $h(i)$. The
presence of arc-like structures in the halo, as for example seen in the Egg Nebula (Sahai et al. 1998b), was denoted by $a$: $h(a)$.
Note that it is possible for a halo to have a smaller visible radial extent than the nebular lobes in the
images shown. However, this does not necessarily imply that the physical radius to which the halo extends is smaller than
the lobes, since halo sizes are likely to be brightness limited.

All of these secondary characteristic descriptors have been retained in our PN scheme; but we have had to add new
ones in order to account for the greater diversity of morphological features in PNs. These are described below.

\subsubsection{Equatorial Waist \& Central Region}\label{waist}
The most important distinction between the PPN and PN secondary classifications is related to the appearance of the
waist
region. Since most surveys of PNs have been carried out in emission-line filters covering lines such as H$\alpha$,
[NII]6583 and [OIII]5007, the waist almost always appears as a bright feature, rather than a dark feature. Indeed,
for a bipolar, multipolar or collimated lobe planetary nebula, the waist is often the brightest structural
component, which is understandable since the waist region is much more dense than the lobes. Note also that if the
waist region is in expansion, then it will continue to flow radially outward from the star as a PPN evolves into a
PN. The central regions of PNs with waists are thus, in general, expected to be more exposed and visible than
those of the PPNs from which they evolve. Consequently, the region between the central star and the periphery of
the waist is quite often well-resolved and is usually fainter than the periphery, giving the waist the appearance
of a belt or toroid. Hence, we denote the presence of a bright waist region with $t$ (for torus). Quite often,
the torus region shows significant structure -- if this is point-symmetric, as first found in PK285-02\#1 (Fig.\
\ref{285-02d1}) by Sahai (2000; see his Fig.\,1), we add the qualifier $t$, to the $ps$ descriptor, i.e., as $ps(t)$.
Additional examples of PNs which have point-symmetric torii are seen in Figures\,\ref{057-01d1}, \ref{005.2-18.6},
and \ref{008.2+06.8}.

Of course, very young PNs such as NGC\,6302 (Matsuura et al. 2005) may still have optically thick
waists fully or partially obscuring their central stars, and we retain the $w$ descriptor for these. However, we
exclude the $b$ qualifier for the waist descriptor, which, in the case of PPNs, was used to denote a waist region
with
a sharp radial boundary, because in PNs, an abrupt, outer radial boundary is naturally generated in an
ionization-bounded
medium and
therefore does not represent the physical boundary of a radial density distribution.

In many bipolar (B), multipolar (M) or collimated-lobe (L) PNs, the bright central region departs significantly from the
geometry of a torus, i.e., its extent along the long axis of the nebula is equal to or larger than its extent in the
equatorial plane, and quite often has a barrel-shaped appearance. In these cases, the $w$ or $t$ descriptor is inadequate
for describing this region; we therefore define a new descriptor called ``barrel-shaped central region" or $bcr$ -- Figs.\
\ref{bcr-c} and\,\ref{bcr-o} show a collection of twelve such objects from our sample. If the
ends of the barrel clearly appear to be closed or open, we add the qualifier $c$ or $o$, respectively, to this descriptor,
i.e., $bcr(c)$ (Fig.\,\ref{bcr-c}) or $bcr(o)$ (Fig.\,\ref{bcr-o}); 
if no judgment can be made, this qualifier is excluded. If the brightness distribution inside the central region has
significant structure that  
appears irregular, we add the qualifier $i$: $bcr(i)$ (e.g., PK\,002-09\#1,\,PK\,024+03\#1, and PK\,003+02\#1: Figs.\,\ref{002-09d1},\
\ref{024+03d1}, and \ref{003+02d1}). If the $bcr$ shows point-symmetry (e.g., PK\,235-03\#1 and PK\,356-03\#3: Figs.\
\ref{235-03d1}, and \ref{356-03d3}), we denote this by adding the qualifier $bcr$ to the point-symmetry descriptor:
$ps(bcr)$.



Athough ($i$) a torus seen with 
its axis at some large angle to the sky plane and ($ii$) a barrel-shaped region seen with its axis nearly in the sky plane will
both appear elliptical and thus superficially similar in shape, it is usually possible to distingush between these because
for case $i$ the major axis of the ellipse will be oriented orthogonally to the long axis of the nebula, whereas in
case $ii$ it will be oriented along the long axis. Thus, for example, PNG\,001.7-04.4 (Fig.\,\ref{001.7-04.4}) is classified as
$bcr(o)$,
because the long axis of its bright central region lies along the long axis of the nebula. 

The full set of descriptors for
the central region (comprised of the $bcr$, $w$ and $t$
descriptors) is collected under the title ``structured central region", replacing the phrase ``obscuring waist"  
which was used in our PPN morphological scheme.

The torus in PK 331-02\#2 (Fig.\,\ref{331-02d2}) shows a remarkable knotty appearance, giving it a necklace-like appearance, similar to
that seen in the ground-based image of the newly discovered PN IPHASXJ194359.5+170901 (Corradi et al. 2010). These authors suggest
that this torus results from a common envelope (CE) ejection episode (the central star in this PN is a known binary), and the knotty
appearance may be either due to
density fluctuations created during CE ejection and later exacerbated by the action of the expanding ionization
front or the post-AGB fast wind, or fragmentation of the ejecta due to radiative shocks. Another example of a knotty torus is found in
PNG097.6-02.4 (Fig.\,\ref{097.6-02.4}), which belongs to a new sample of compact PN recently imaged with HST (see \S\,\ref{newsamp}).
If several additional objects with such knotty torii are found, we would recommend adding a qualifier to the $t$ descriptor to capture
this feature.


\subsubsection{Other Nebular Characteristics}\label{otherneb}
\vspace{0.1in} \noindent{\it Inner Bubbles}\\
Some PNs show the presence of small inner bubble structures which appear to lie entirely within
the defining primary geometric structure characterised by the primary class. These are
addressed by adding a new descriptor, $ib$, to the set of ``other
nebular characteristics". Some examples of PNs with inner bubbles are PK215-24\#1,
PK258-00\#1, PK352-07\#1, PK051-03\#1, PK007-04\#1, PK082+07\#1 (Fig.\,\ref{ibmosaic}), the
Southern Crab (Corradi 
\& Schwarz 1993), and MyCn18 (Sahai et al. 1999). 
The walls of these inner
bubbles appear to contain much more highly-excited gas than the primary nebular shell, as
indicated by a comparison of the [OIII] and H$\alpha$ images for 
PK215-24\arcdeg1 (Fig.\,\ref{oiii-ha-ic418}): the average surface brightness of the inner
bubble, as a fraction of the surface brightness of the primary shell, is much larger in the
[OIII] image than in the H$\alpha$ image. The inner bubble often has a pronounced
point-symmetric shape (e.g., PK258-00\#1, PK082+07\#1), which is denoted by adding the qualifier $ib$ to
the point-symmetry descriptor: $ps(ib)$.


\vspace{0.1in} \noindent{\it Rings, Arcs, Radial Rays and Microstructure}\\
Some PNs show ring-like features projected onto the lobes, e.g., the Etched Hourglass Nebula, MyCn18 (Sahai et al.
1999), NGC\,6881 (Kwok \& Su 2005), and Hb\,12 (Kwok \& Hsia 2007), which we denote with the descriptor $rg$.
Both NGC\,6881 (PK074+02\#1: Fig \ \ref{074+02d1}) and Hb\,12 (PK\,111-02\#1: Fig.\,\ref{111-02d1}) are included in
our survey. However, the above studies show that the ring structures are generally less prominent in the H$\alpha$
images compared to the [NII] images; and  Hb\,12 is the only object in our sample where the H$\alpha$ image also
shows these rings very faintly. There are two PPNs, the Red Rectangle (Cohen et al. 2004), and CRL\,618 (Trammell
\& Goodrich 2002), which also show such rings. The 2-dimensional rings, which are generally co-axial with the long
axis of the lobes in which they are found, appear to be structures which girdle the walls of the nebular lobes,
like etchings on a wine glass. These are to be distinguished from the circular arcs seen in some PPNs and PNs,
which are limb-brightened projections of 3-dimensional, geometrically-thin shell structures around the nebular
center (denoted by the $a$ qualifier of the $h$ descriptor, i.e., as $h(a)$). Not many objects in our sample show
arcs or rings in their halo. Deep
imaging of the low surface brightness haloes by Corradi et al (2004) revealed several PNs having arcs; those
authors summarise the new results from their work as well as objects already known to have these from earlier
studies (NGC\,6543, NGC\,7027, NGC\,3918, and Hb\,5).

\begin{figure}[htb]
\vskip -0.6cm
\resizebox{0.77\textwidth}{!}{\includegraphics{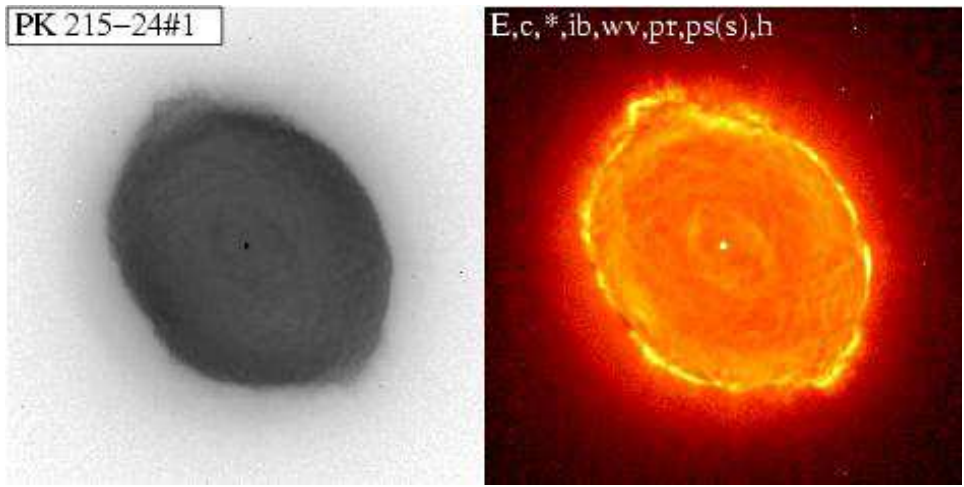}}
\caption{As in Fig\,1., but for PK\,215-24\#1.
}
\label{215-24d1}
\end{figure}
\begin{figure}[htb]
\vskip -0.6cm
\resizebox{0.77\textwidth}{!}{\includegraphics{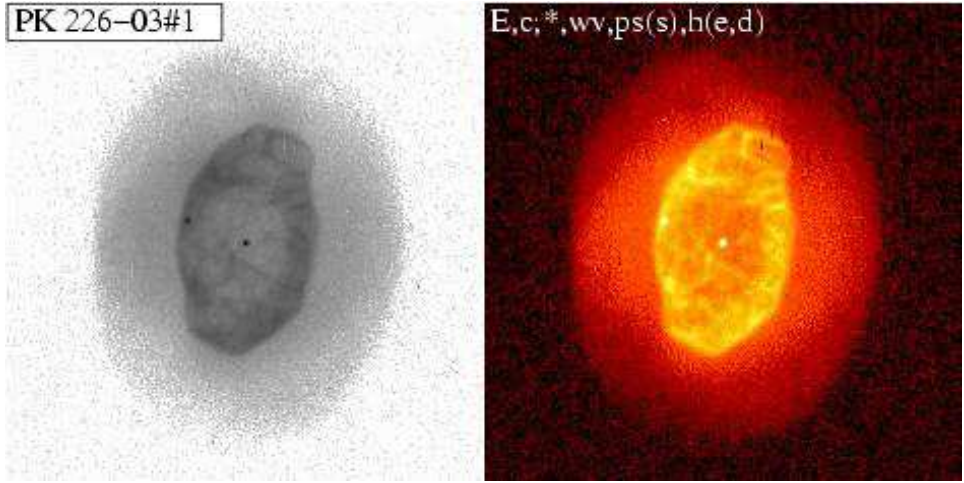}}
\caption{As in Fig\,1., but for PK\,226-03\#1.
}
\label{226-03d1}
\end{figure}
\begin{figure}[htb]
\vskip -0.6cm
\resizebox{0.77\textwidth}{!}{\includegraphics{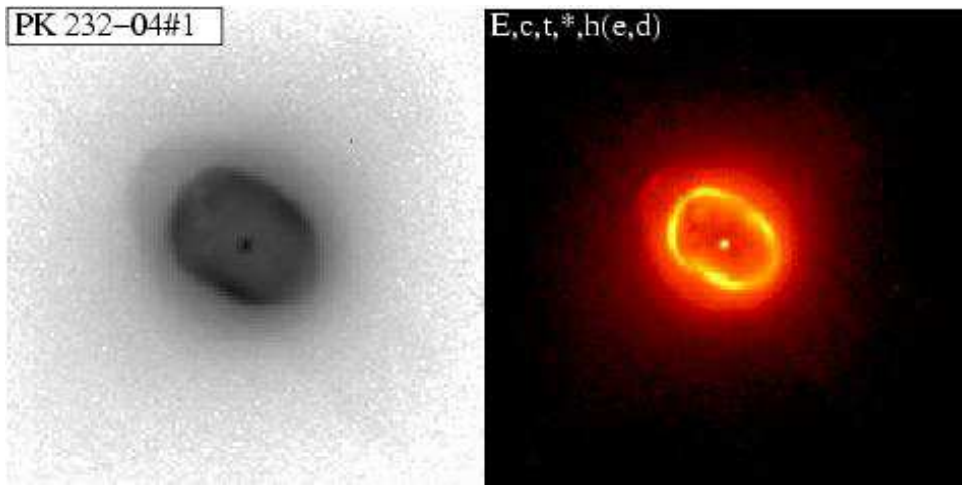}}
\caption{As in Fig\,1., but for PK\,232-04\#1.
}
\label{232-04d1}
\end{figure}
\begin{figure}[htb]
\vskip -0.6cm
\resizebox{0.77\textwidth}{!}{\includegraphics{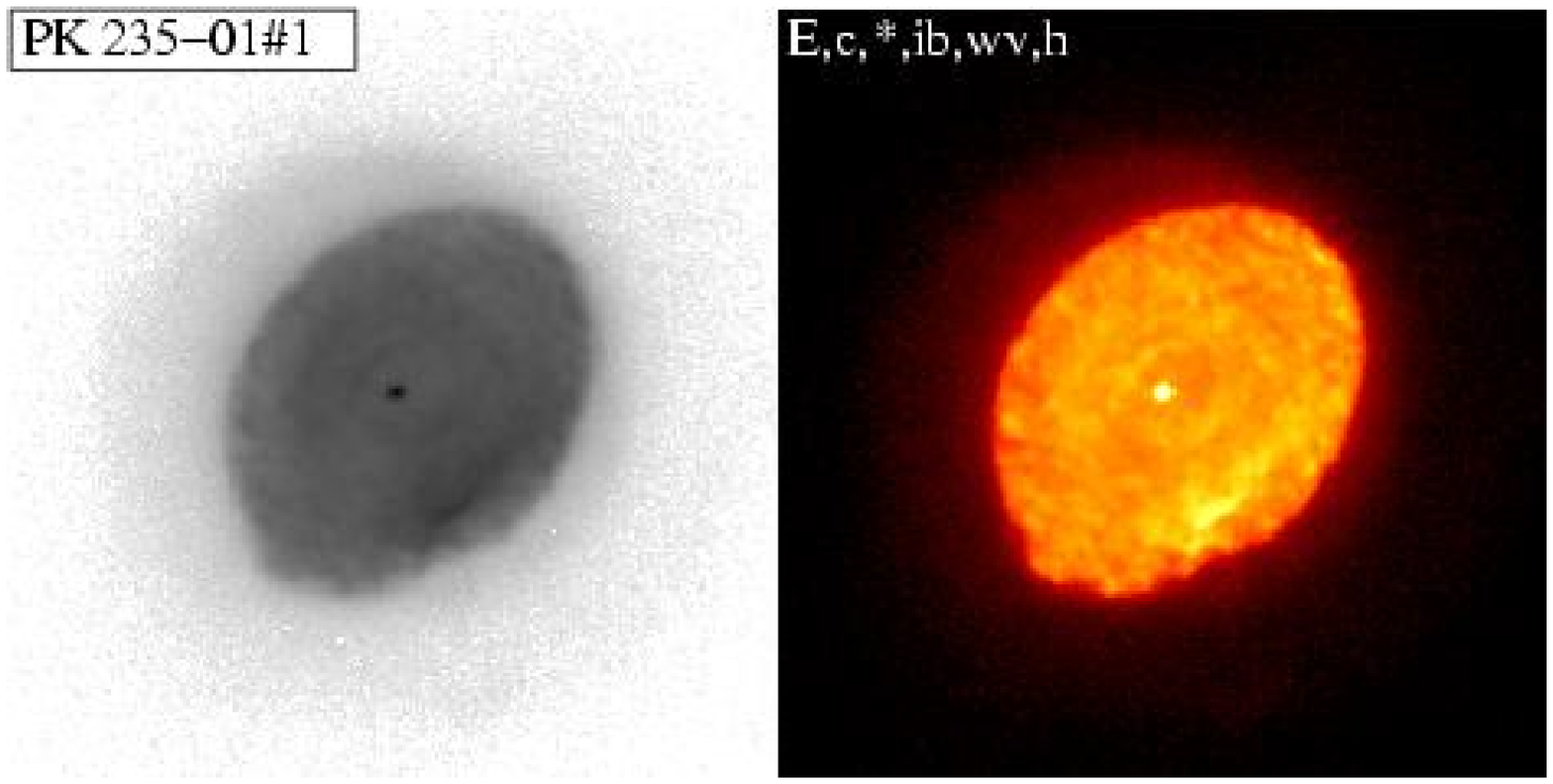}}
\caption{As in Fig\,1., but for PK\,235-01\#1.
}
\label{235-01d1}
\end{figure}
\begin{figure}[htb]
\vskip -0.6cm
\resizebox{0.77\textwidth}{!}{\includegraphics{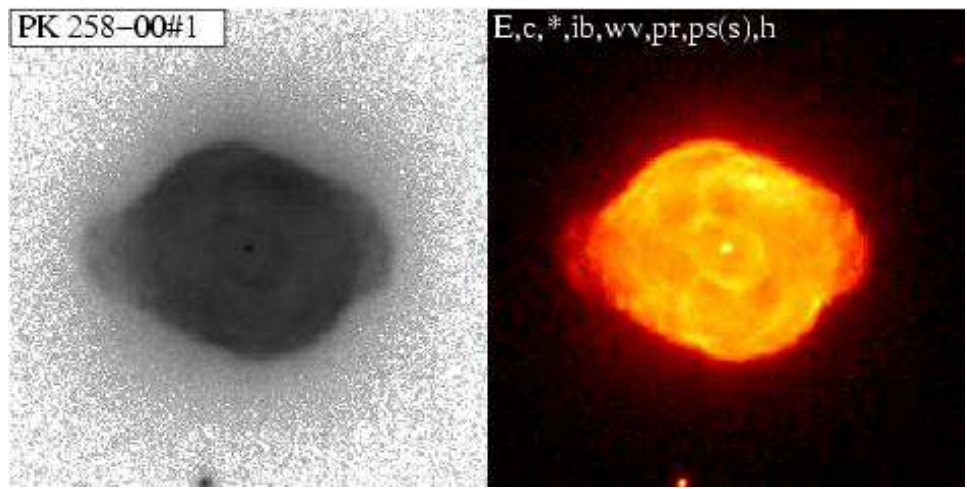}}
\caption{As in Fig\,1., but for PK\,258-00\#1.
}
\label{258-00d1}
\end{figure}
\begin{figure}[htb]
\vskip -0.6cm
\resizebox{0.77\textwidth}{!}{\includegraphics{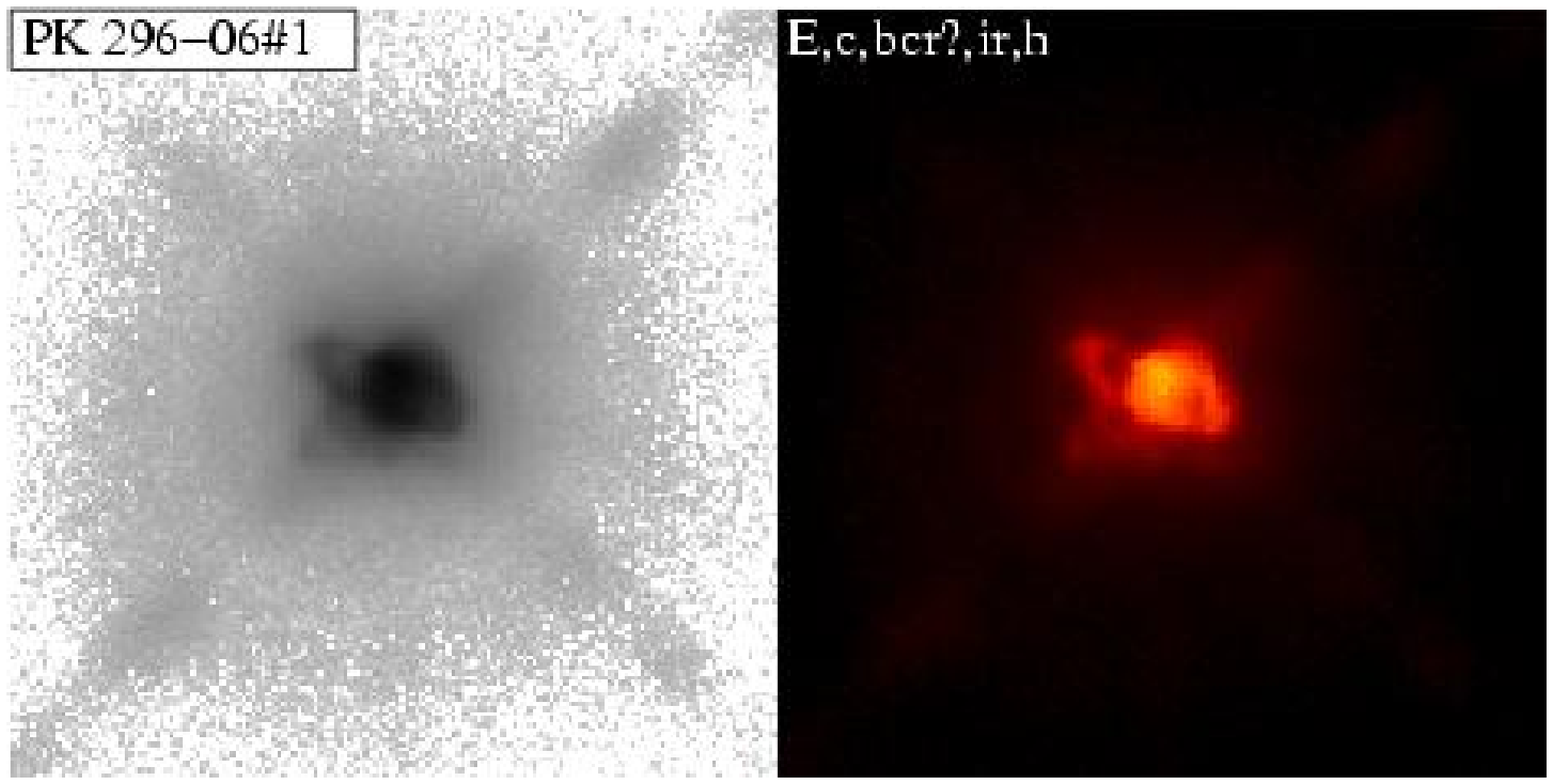}}
\caption{As in Fig\,1., but for PK\,296-06\#1.
}
\label{296-06d1}
\end{figure}
\begin{figure}[htb]
\vskip -0.6cm
\resizebox{0.77\textwidth}{!}{\includegraphics{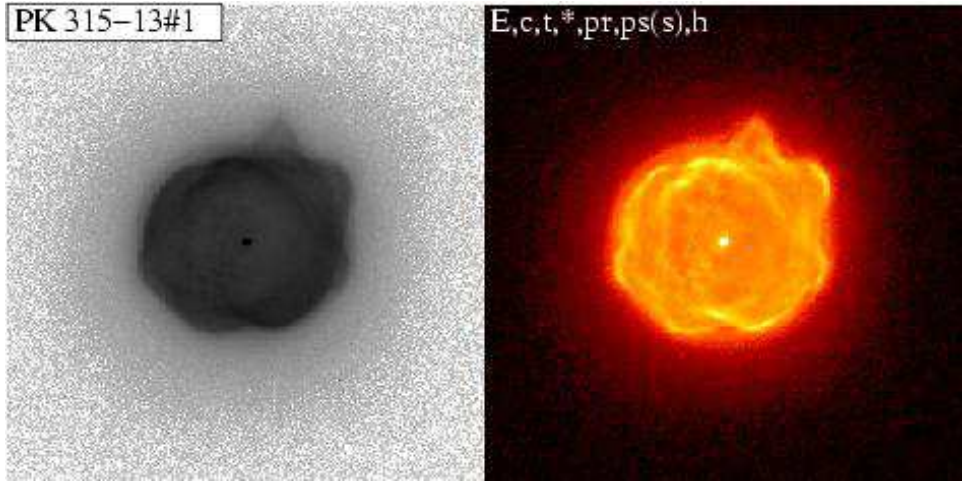}}
\caption{As in Fig\,1., but for PK\,315-13\#1 (adapted from ST98).
}
\label{315-13d1}
\end{figure}
\begin{figure}[htb]
\vskip -0.6cm
\resizebox{0.77\textwidth}{!}{\includegraphics{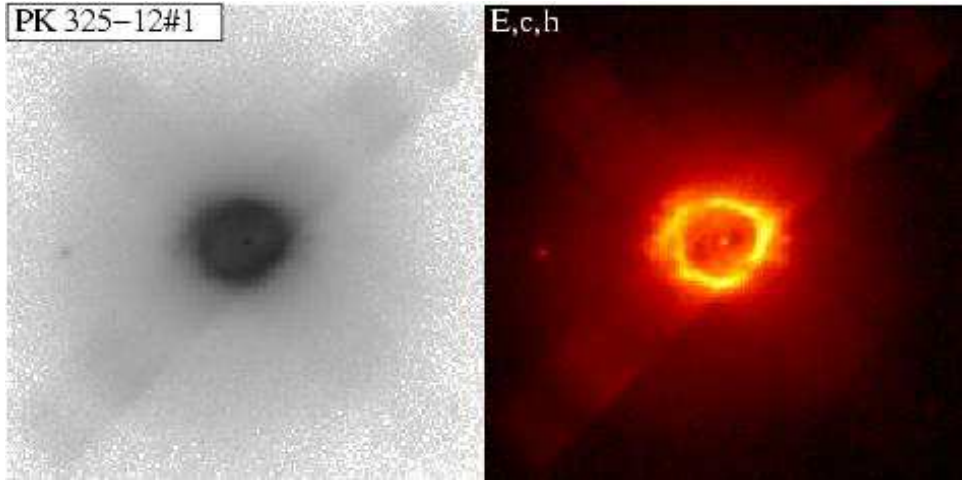}}
\caption{As in Fig\,1., but for PK\,325-12\#1.
}
\label{325-12d1}
\end{figure}
\begin{figure}[htb]
\vskip -0.6cm
\resizebox{0.77\textwidth}{!}{\includegraphics{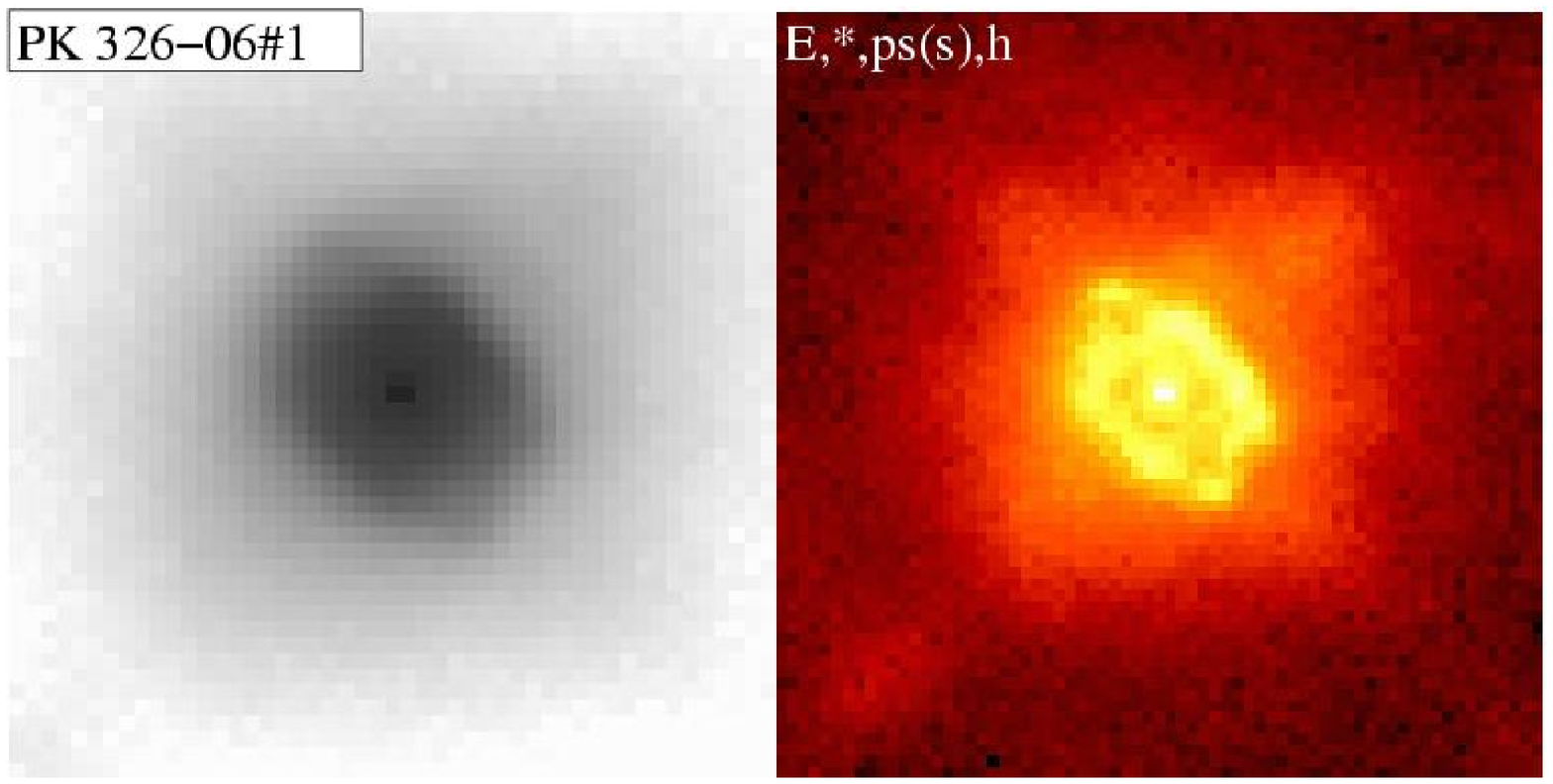}}
\caption{As in Fig\,1., but for PK\,326-06\#1.
}
\label{326-06d1}
\end{figure}
\begin{figure}[htb]
\vskip -0.6cm
\resizebox{0.77\textwidth}{!}{\includegraphics{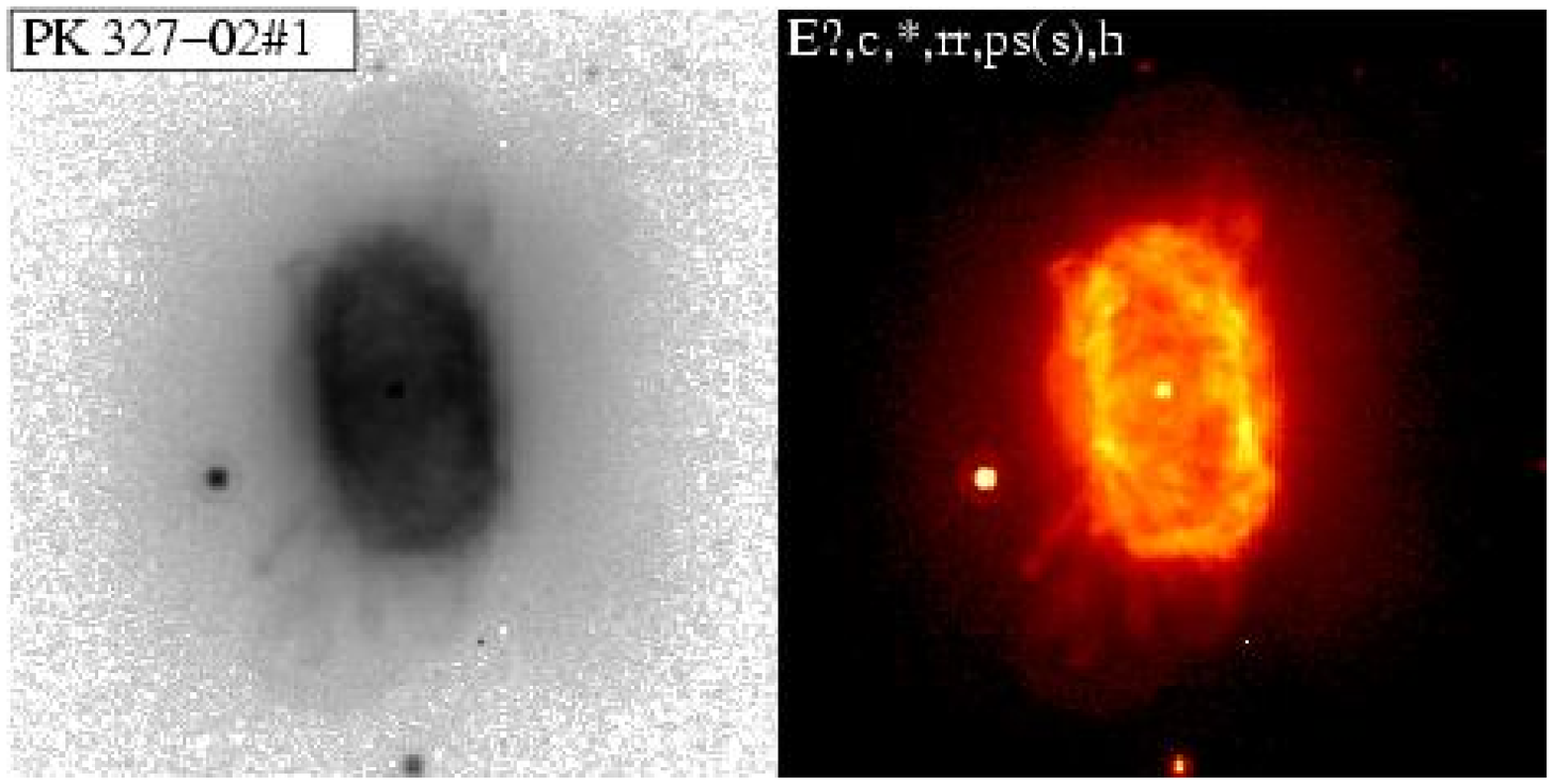}}
\caption{As in Fig\,1., but for PK\,327-02\#1 (adapted from ST98).
}
\label{327-02d1}
\end{figure}
\begin{figure}[htb]
\vskip -0.6cm
\resizebox{0.77\textwidth}{!}{\includegraphics{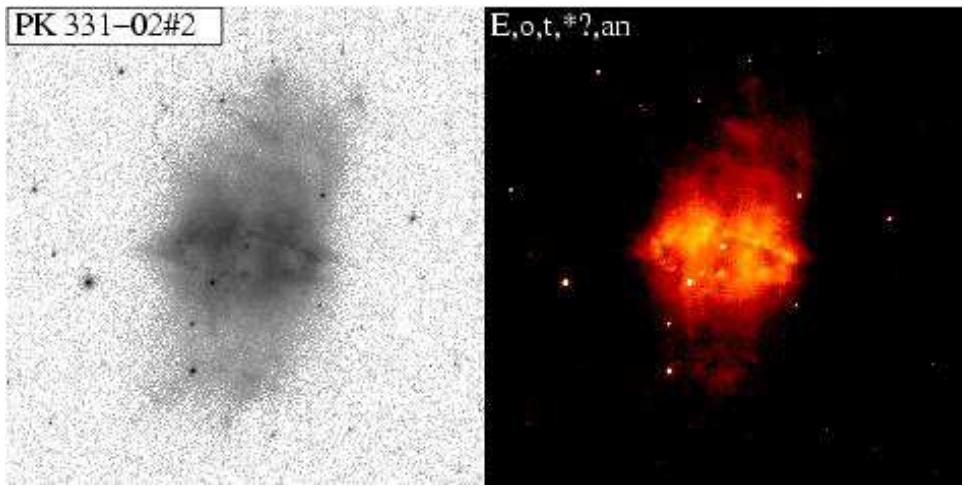}}
\caption{As in Fig\,1., but for PK\,331-02\#2.
}
\label{331-02d2}
\end{figure}
\begin{figure}[htb]
\vskip -0.6cm
\resizebox{0.77\textwidth}{!}{\includegraphics{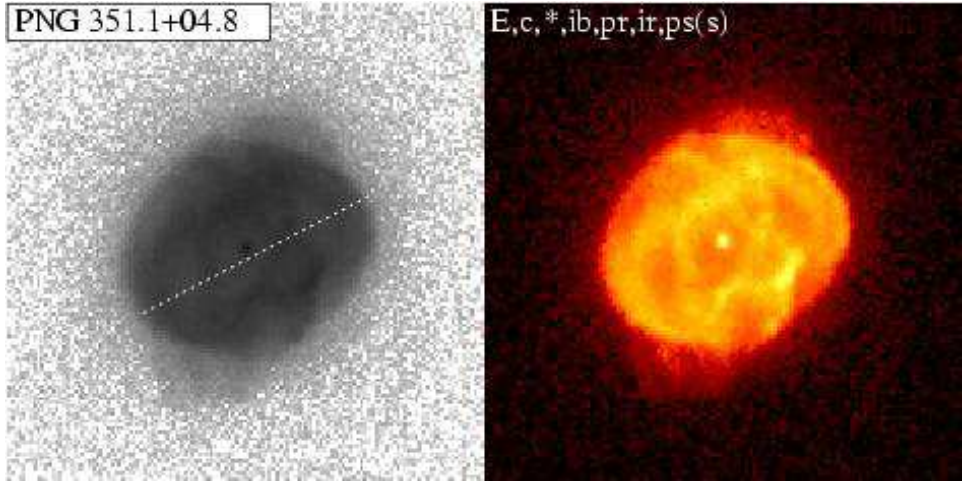}}
\caption{As in Fig\,1., but for PNG351.1+04.8.
}
\label{351.1+04.8}
\end{figure}
\begin{figure}[htb]
\vskip -0.6cm
\resizebox{0.77\textwidth}{!}{\includegraphics{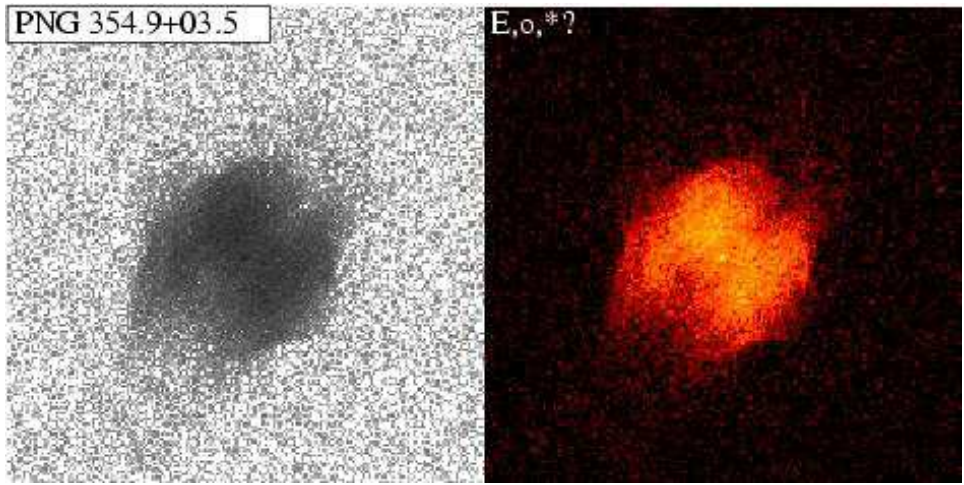}}
\caption{As in Fig\,1., but for PNG354.9+03.5.
}
\label{354.9+03.5}
\end{figure}
\begin{figure}[htb]
\vskip -0.6cm
\resizebox{0.77\textwidth}{!}{\includegraphics{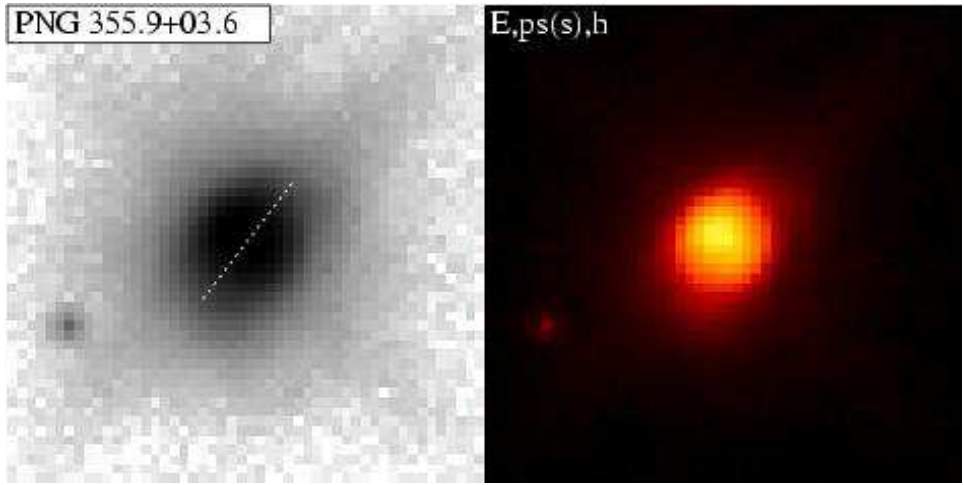}}
\caption{As in Fig\,1., but for PNG355.9+03.6.
}
\label{355.9+03.6}
\end{figure}
\begin{figure}[htb]
\vskip -0.6cm
\resizebox{0.77\textwidth}{!}{\includegraphics{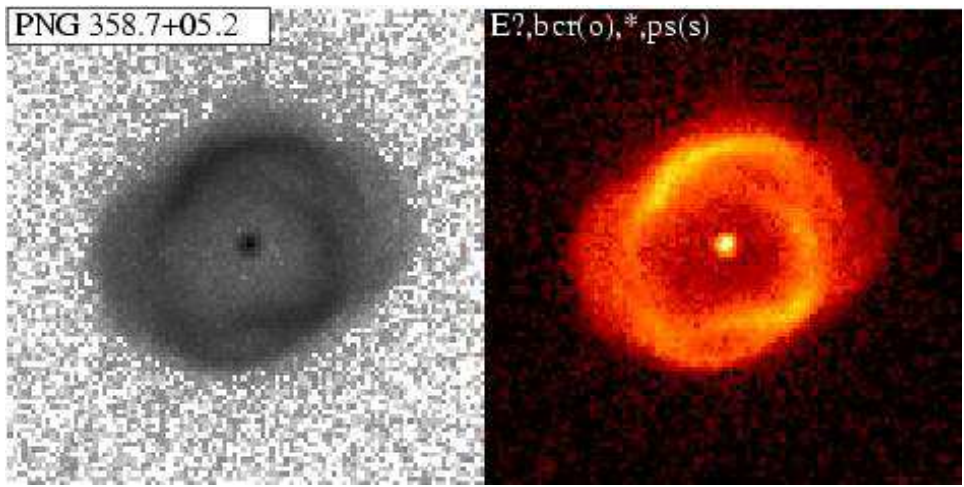}}
\caption{As in Fig\,1., but for PNG358.7+05.2.
}
\label{358.7+05.2}
\end{figure}
\begin{figure}[htb]
\vskip -0.6cm
\resizebox{0.77\textwidth}{!}{\includegraphics{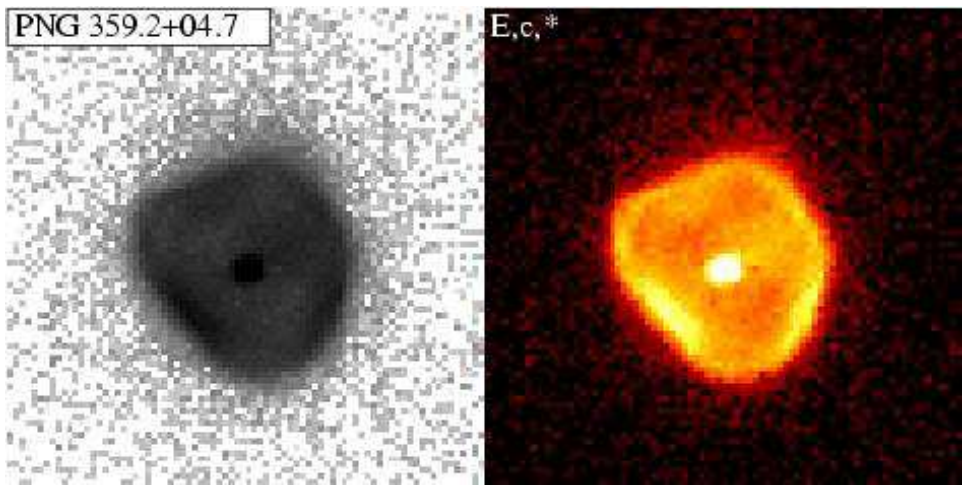}}
\caption{As in Fig\,1., but for PNG359.2+04.7.
}
\label{359.2+04.7}
\end{figure}
\begin{figure}[htb]
\vskip -0.6cm
\resizebox{0.77\textwidth}{!}{\includegraphics{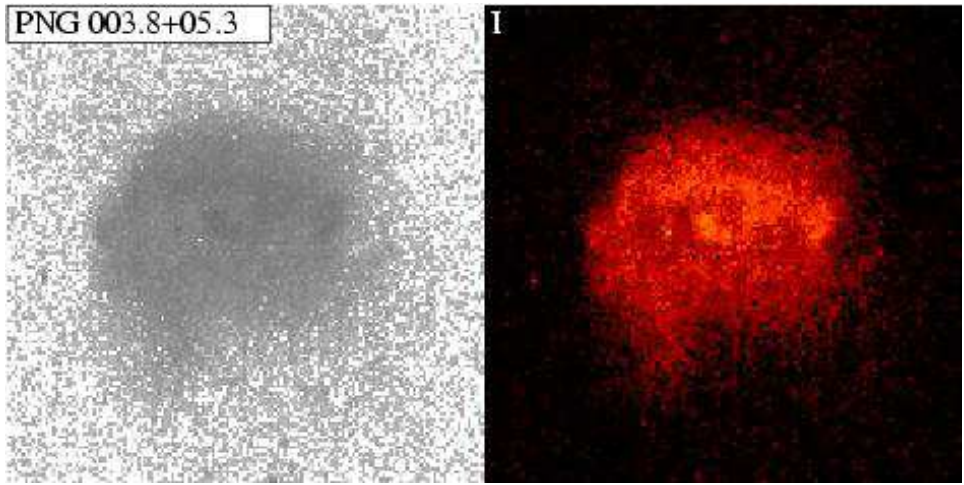}}
\caption{As in Fig\,1., but for PNG003.8+05.3.
}
\label{003.8+05.3}
\end{figure}
\begin{figure}[htb]
\vskip -0.6cm
\resizebox{0.77\textwidth}{!}{\includegraphics{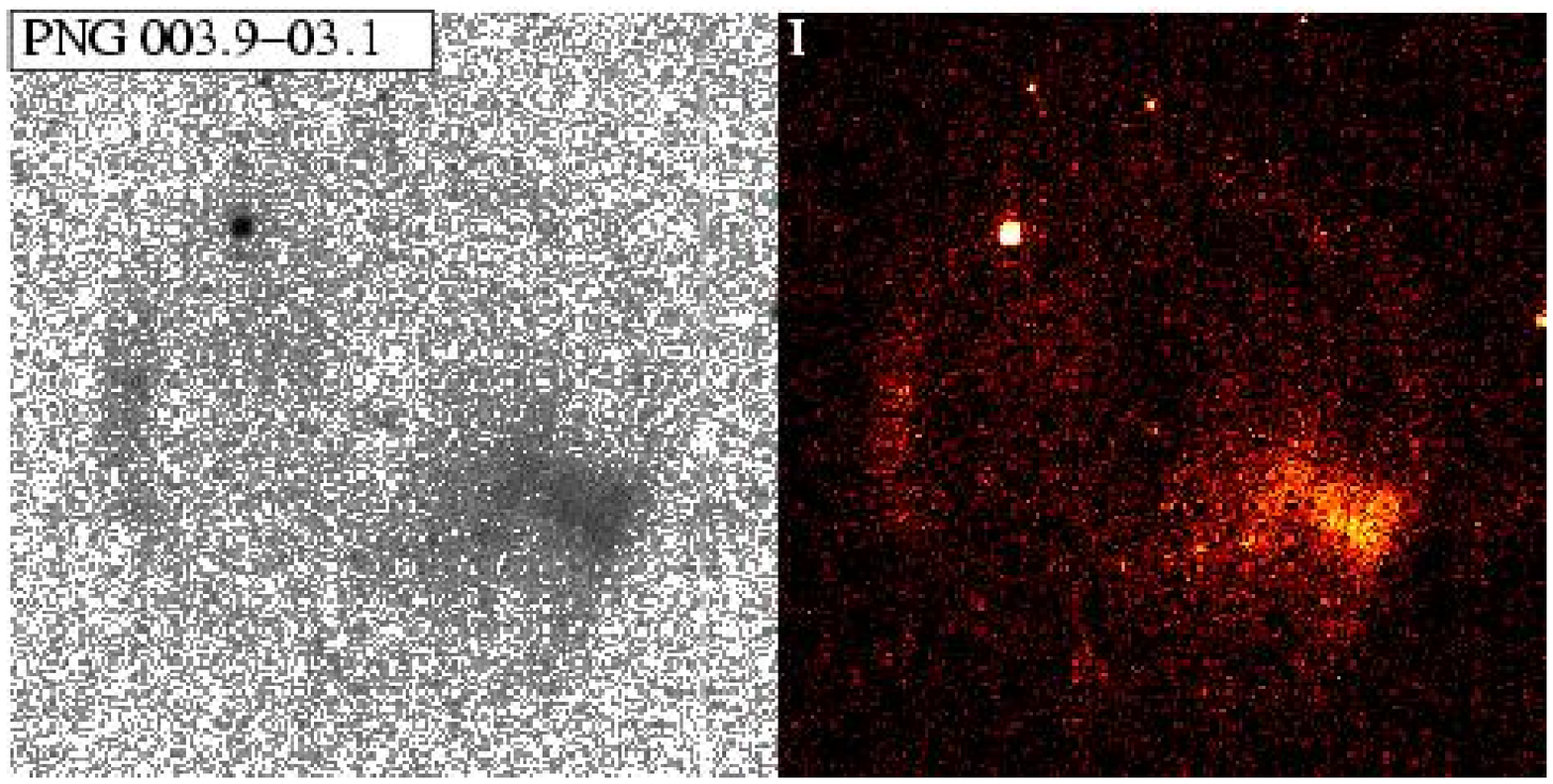}}
\caption{As in Fig\,1., but for PNG003.9-03.1.
}
\label{003.9-03.1}
\end{figure}
\begin{figure}[htb]
\vskip -0.6cm
\resizebox{0.77\textwidth}{!}{\includegraphics{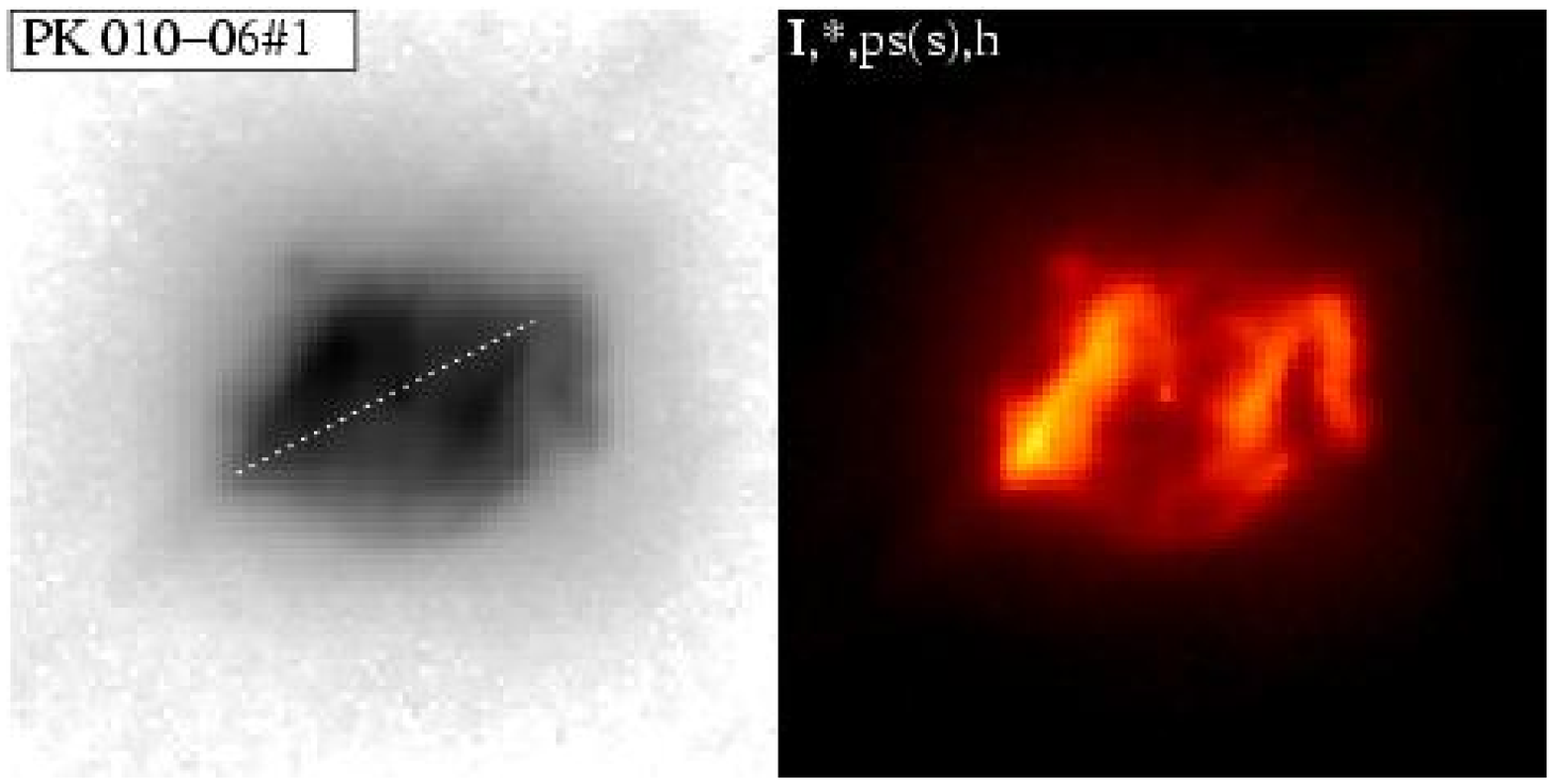}}
\caption{As in Fig\,1., but for PK\,010-06\#1.
}
\label{010-06d1}
\end{figure}
\begin{figure}[htb]
\vskip -0.6cm
\resizebox{0.77\textwidth}{!}{\includegraphics{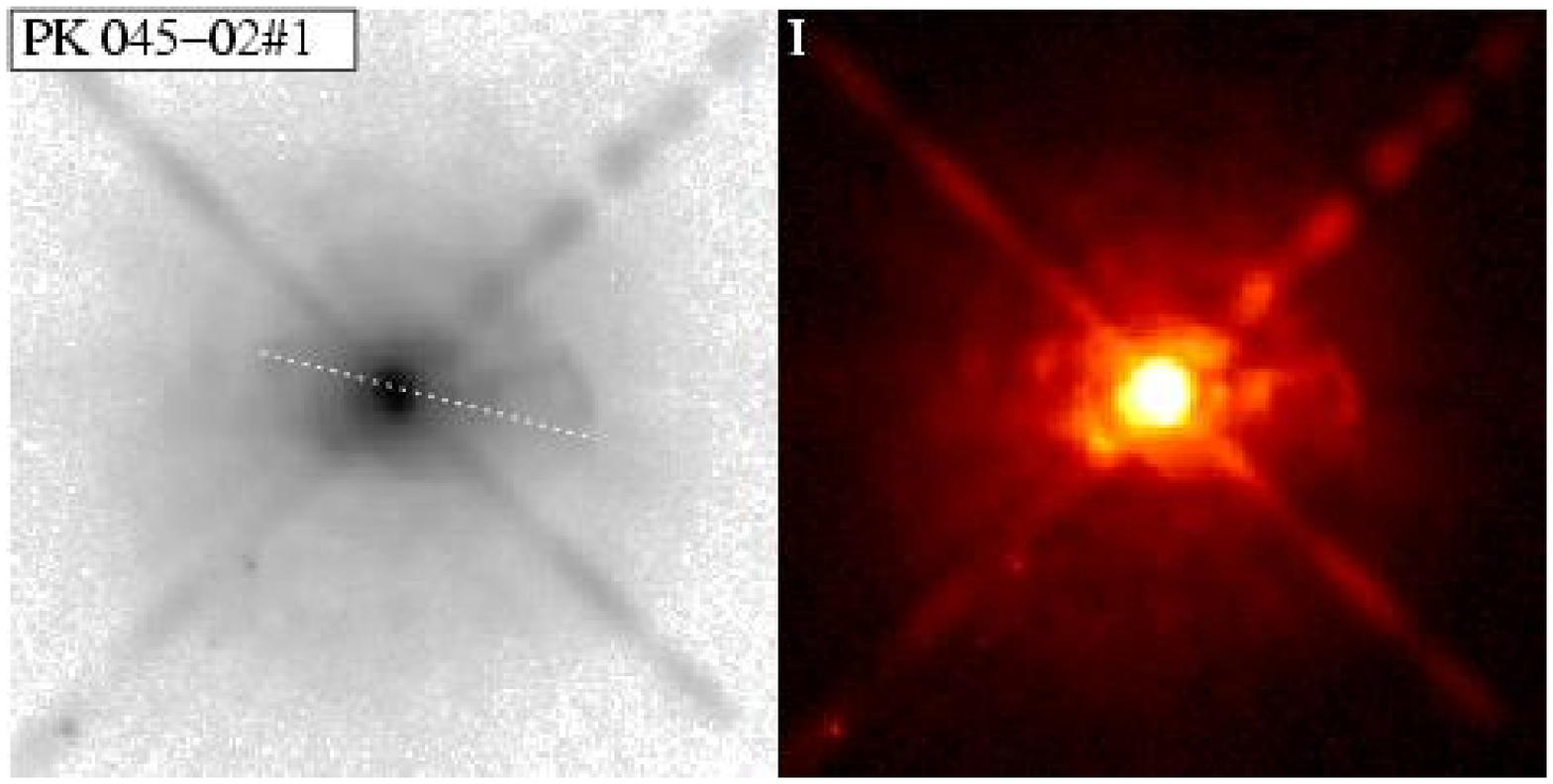}}
\caption{As in Fig\,1., but for PK\,045-02\#1.
}
\label{045-02d1}
\end{figure}
\begin{figure}[htb]
\vskip -0.6cm
\resizebox{0.77\textwidth}{!}{\includegraphics{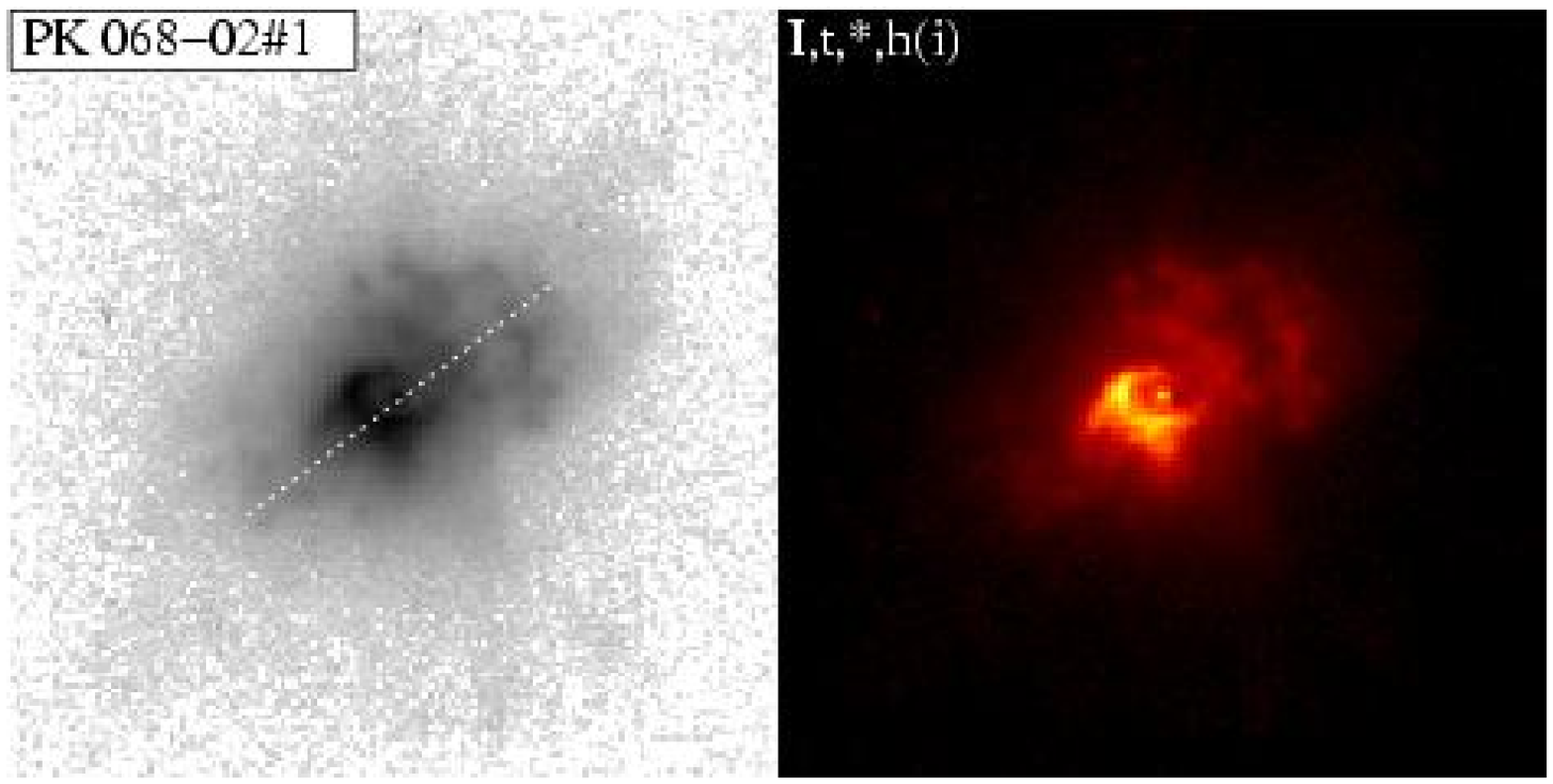}}
\caption{As in Fig\,1., but for PK\,068-02\#1 (adapted from ST98).
}
\label{068-02d1}
\end{figure}
\begin{figure}[htb]
\vskip -0.6cm
\resizebox{0.77\textwidth}{!}{\includegraphics{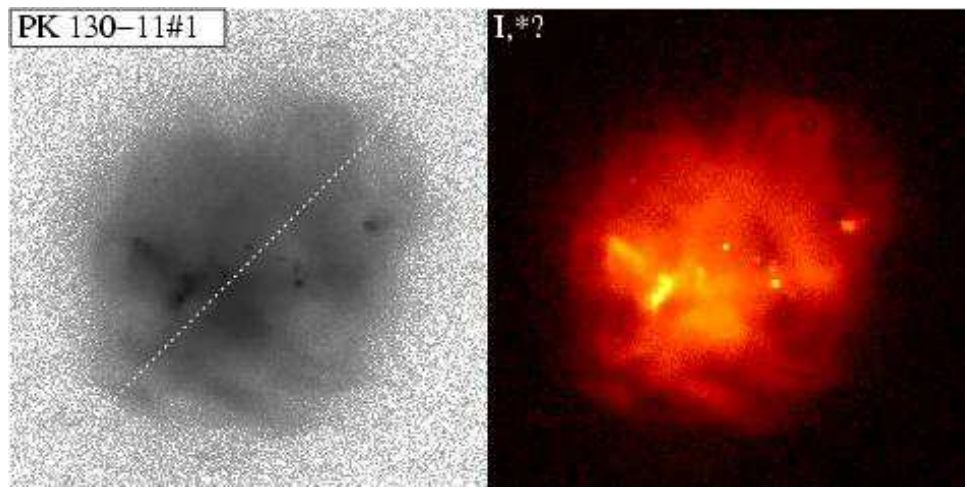}}
\caption{As in Fig\,1., but for PK\,130-11\#1.
}
\label{130-11d1}
\end{figure}
\begin{figure}[htb]
\vskip -0.6cm
\resizebox{0.77\textwidth}{!}{\includegraphics{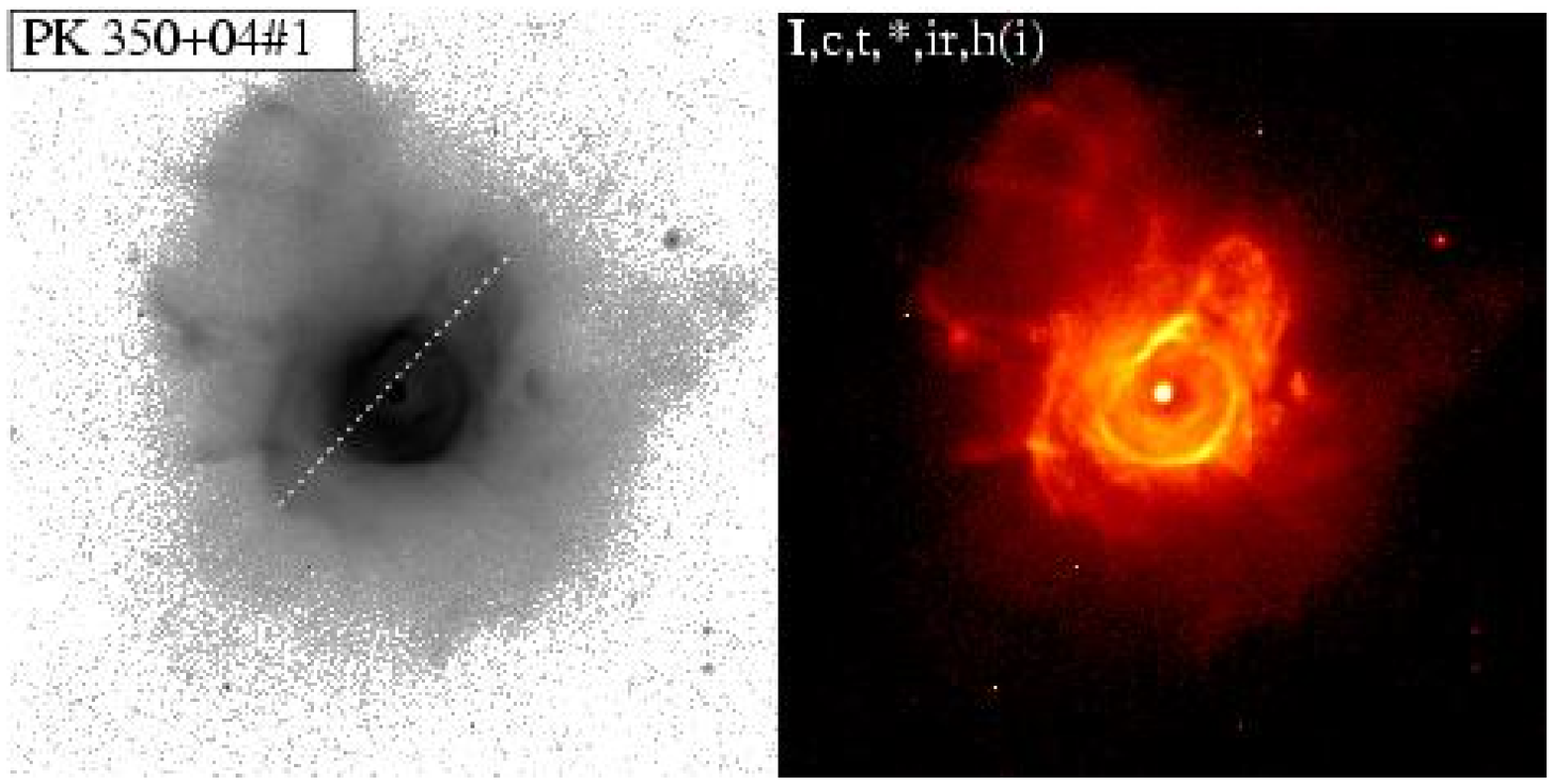}}
\caption{As in Fig\,1., but for PK\,350+04\#1.
}
\label{350+04d1}
\end{figure}
\begin{figure}[htb]
\vskip -0.6cm
\resizebox{0.77\textwidth}{!}{\includegraphics{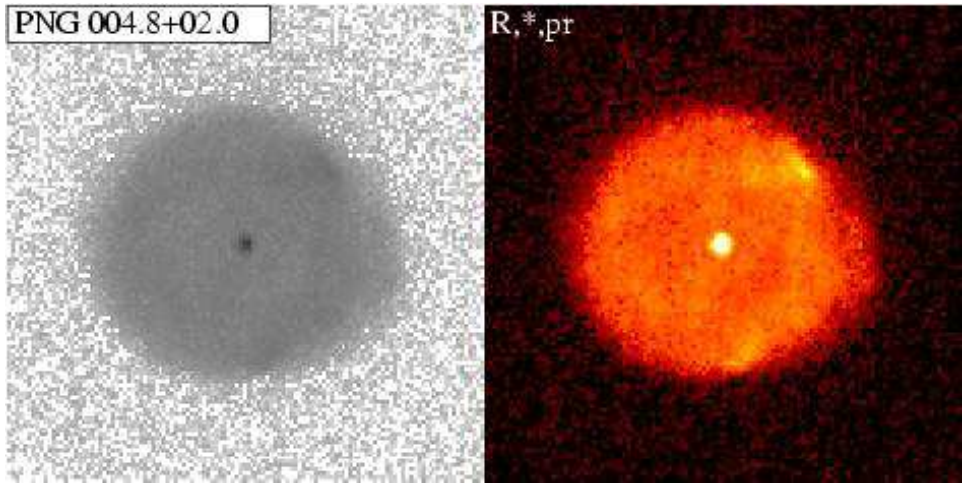}}
\caption{As in Fig\,1., but for PNG004.8+02.0. We believe that the small departures from a round shape visible in the image are caused
by a pair of small, diametrically-opposed protrusions along an axis oriented at $pa\sim-100\deg$.
}
\label{004.8+02.0}
\end{figure}
\begin{figure}[htb]
\vskip -0.6cm
\resizebox{0.77\textwidth}{!}{\includegraphics{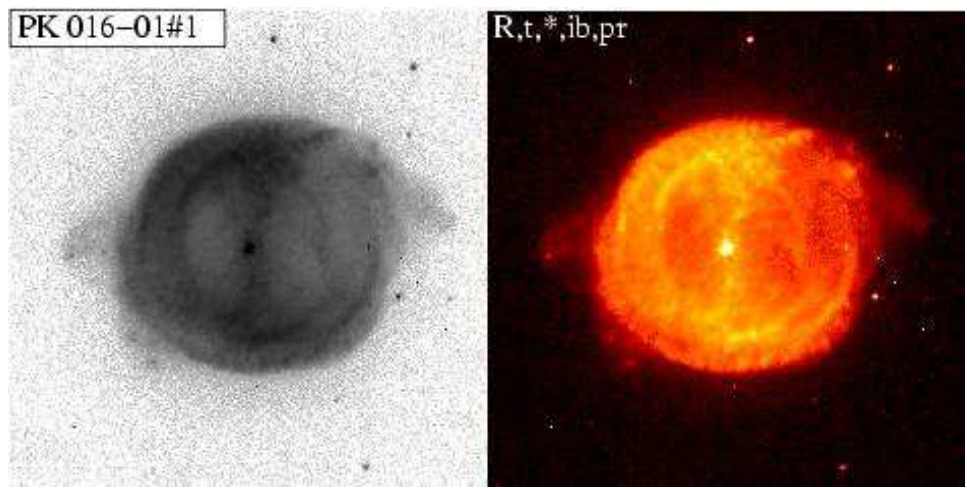}}
\caption{As in Fig\,1., but for PK\,016-01\#1.
}
\label{016-01d1}
\end{figure}
\begin{figure}[htb]
\vskip -0.6cm
\resizebox{0.77\textwidth}{!}{\includegraphics{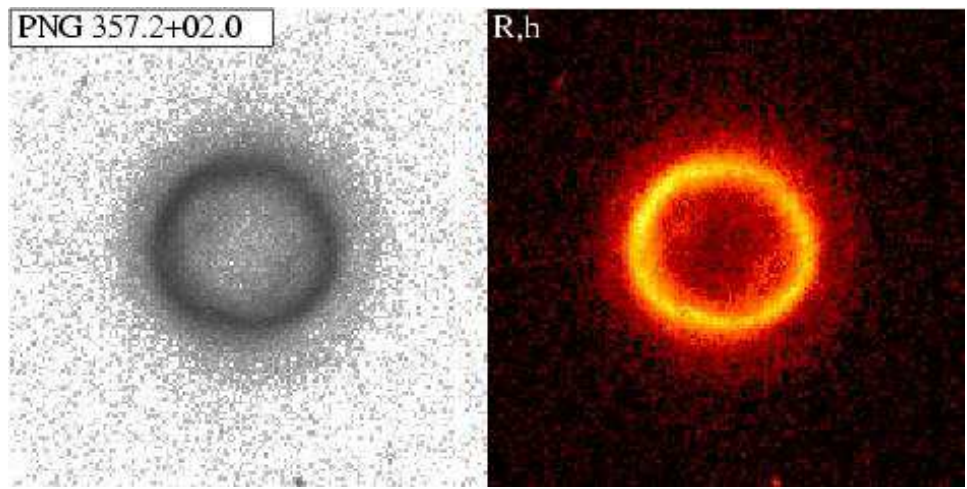}}
\caption{As in Fig\,1., but for PNG357.2+02.0.
}
\label{357.2+02.0}
\end{figure}
\begin{figure}[htb]
\vskip -0.6cm
\resizebox{0.77\textwidth}{!}{\includegraphics{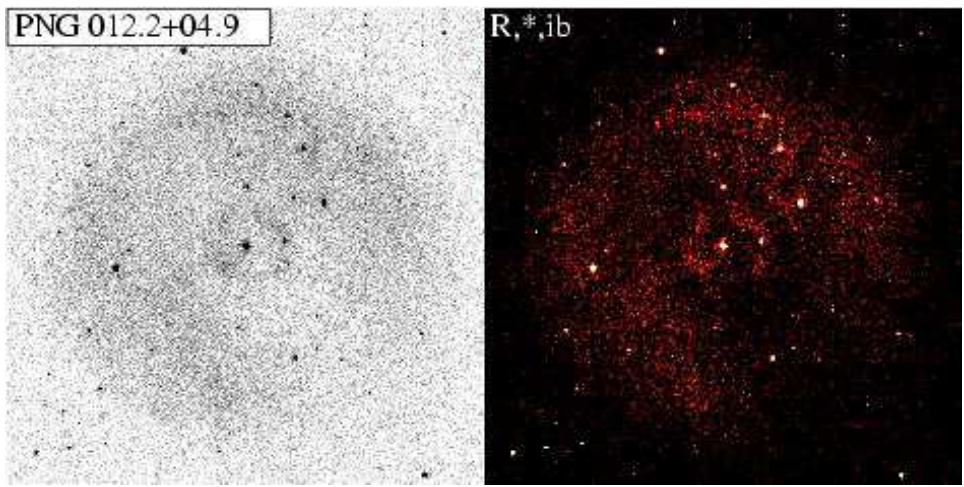}}
\caption{As in Fig\,1., but for PNG012.2+04.9.
}
\label{012.2+04.9}
\end{figure}
\begin{figure}[htb]
\vskip -0.6cm
\resizebox{0.77\textwidth}{!}{\includegraphics{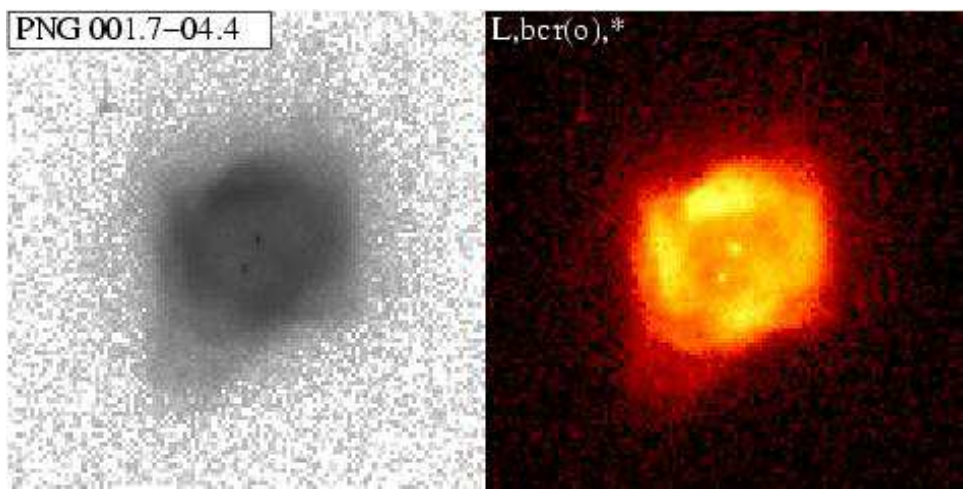}}
\caption{As in Fig\,1., but for PNG001.7-04.4.
}
\label{001.7-04.4}
\end{figure}
\begin{figure}[htb]
\vskip -0.6cm
\resizebox{0.77\textwidth}{!}{\includegraphics{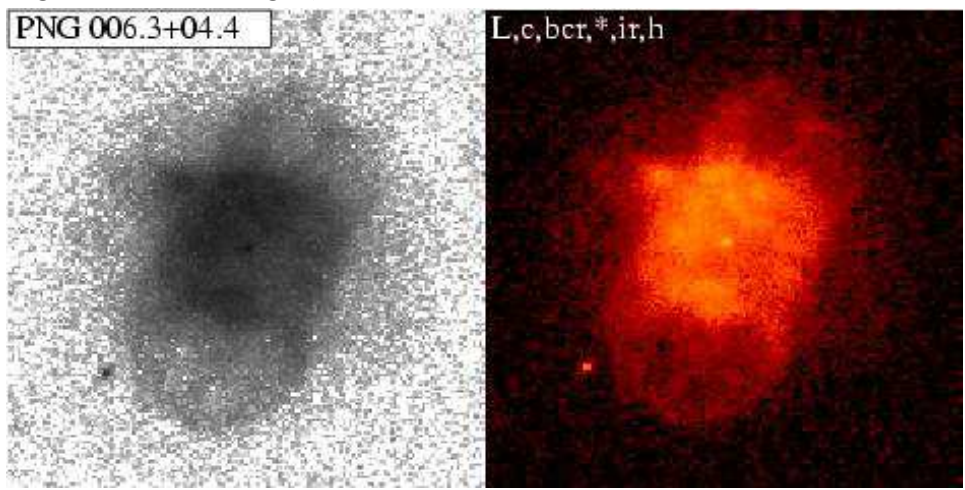}}
\caption{As in Fig\,1., but for PNG006.3+04.4.
}
\label{006.3+04.4}
\end{figure}
\begin{figure}[htb]
\vskip -0.6cm
\resizebox{0.77\textwidth}{!}{\includegraphics{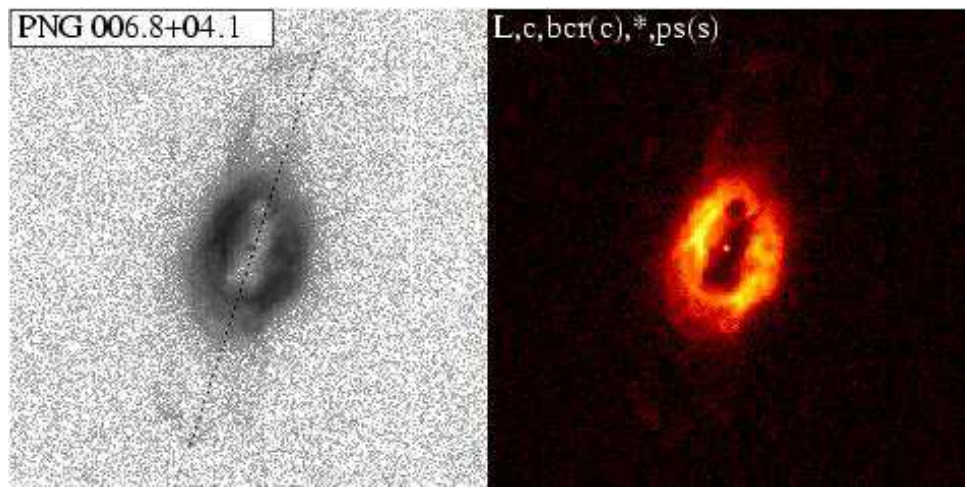}}
\caption{As in Fig\,1., but for PNG006.8+04.1.
}
\label{006.8+04.1}
\end{figure}
\begin{figure}[htb]
\vskip -0.6cm
\resizebox{0.77\textwidth}{!}{\includegraphics{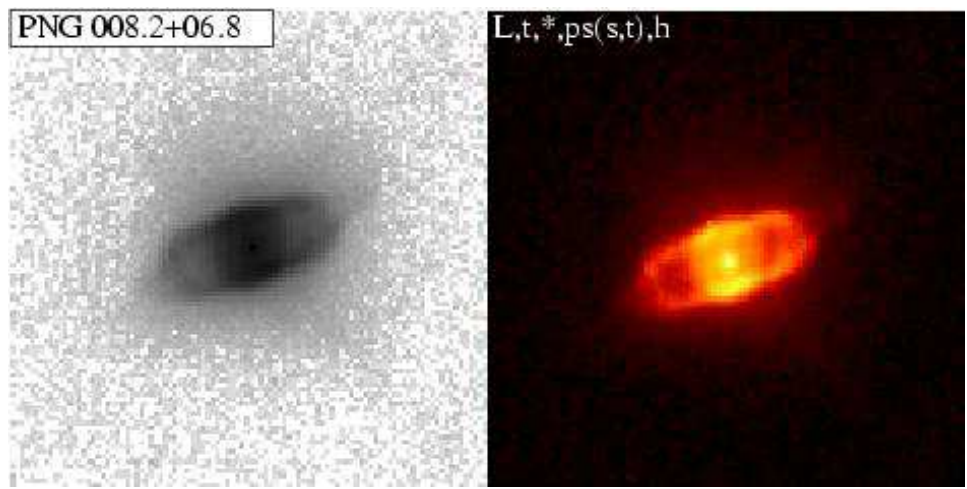}}
\caption{As in Fig\,1., but for PNG008.2+06.8.
}
\label{008.2+06.8}
\end{figure}
\begin{figure}[htb]
\vskip -0.6cm
\resizebox{0.77\textwidth}{!}{\includegraphics{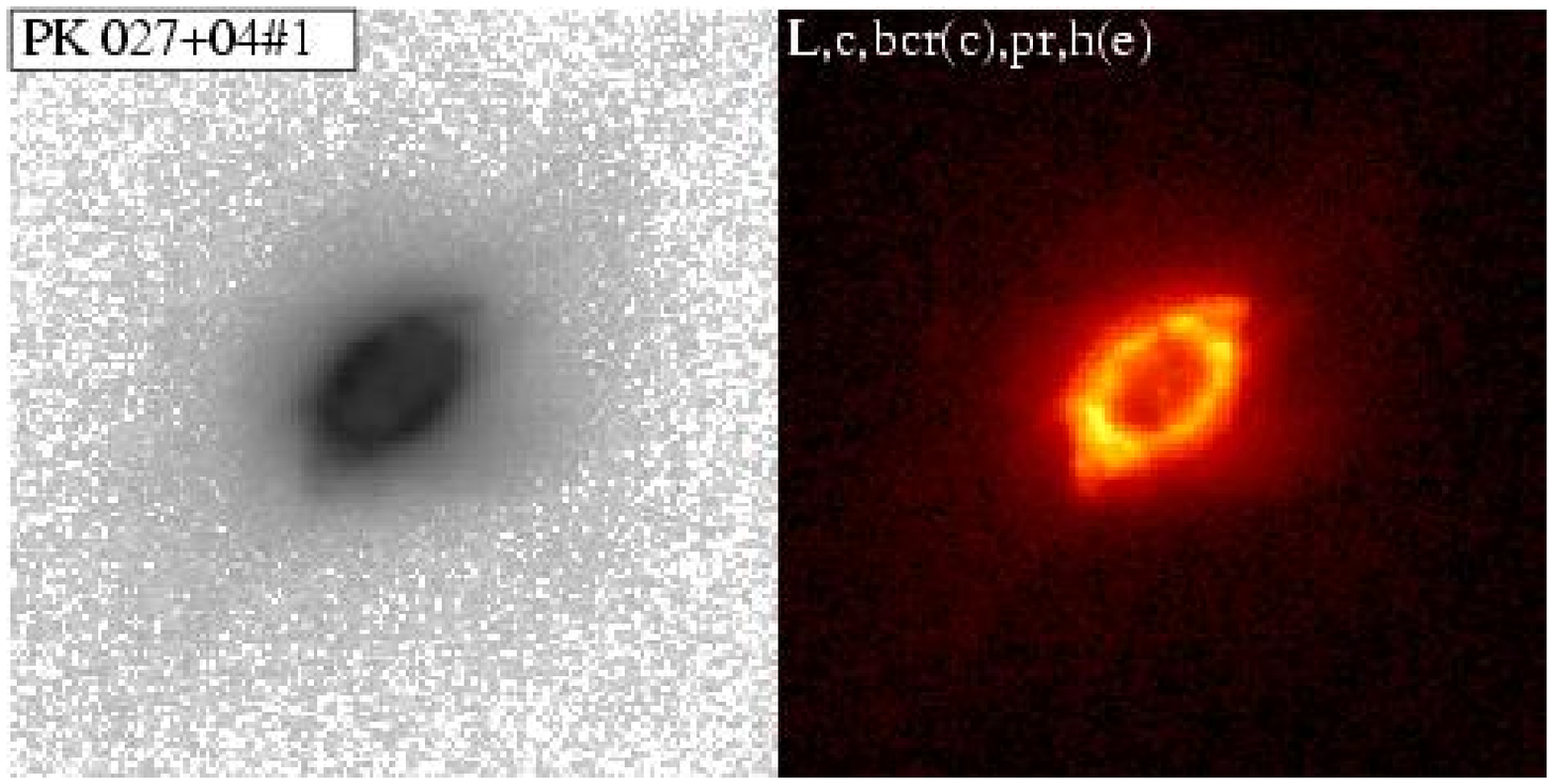}}
\caption{As in Fig\,1., but for PK\,027+04\#1.
}
\label{027+04d1}
\end{figure}
\begin{figure}[htb]
\vskip -0.6cm
\resizebox{0.77\textwidth}{!}{\includegraphics{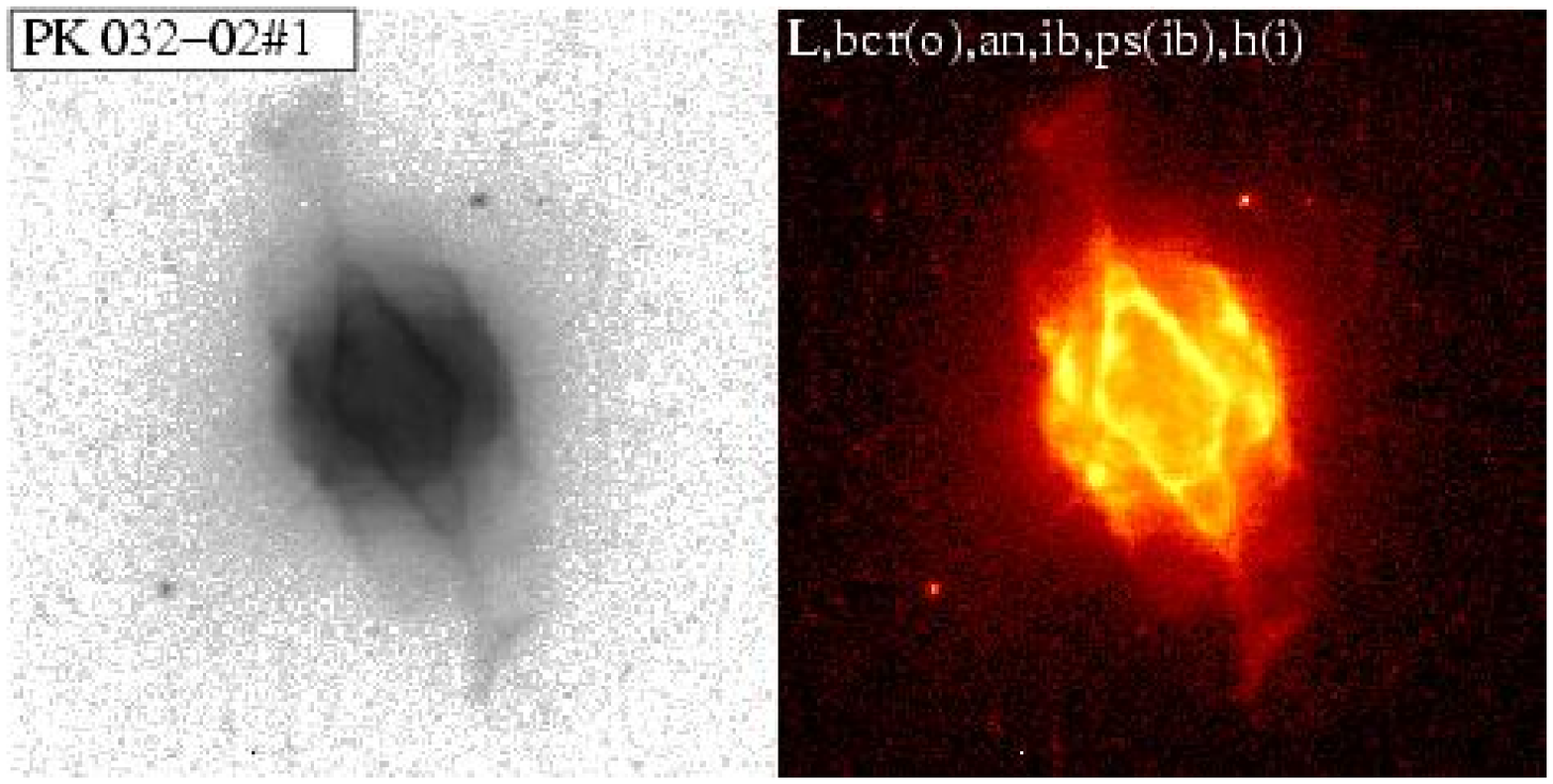}}
\caption{As in Fig\,1., but for PK\,032-02\#1.
}
\label{032-02d1}
\end{figure}
\begin{figure}[htb]
\vskip -0.6cm
\resizebox{0.77\textwidth}{!}{\includegraphics{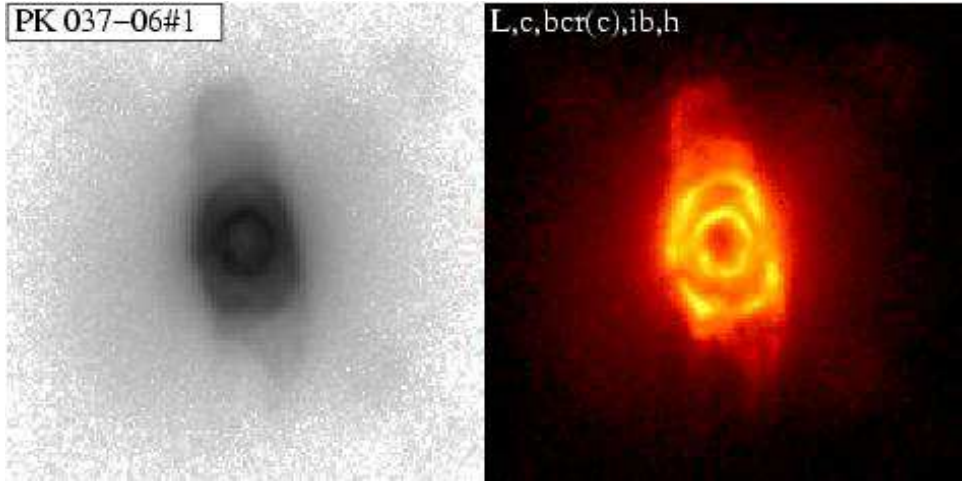}}
\caption{As in Fig\,1., but for PK\,037-06\#1.
}
\label{037-06d1}
\end{figure}
\begin{figure}[htb]
\vskip -0.6cm
\resizebox{0.77\textwidth}{!}{\includegraphics{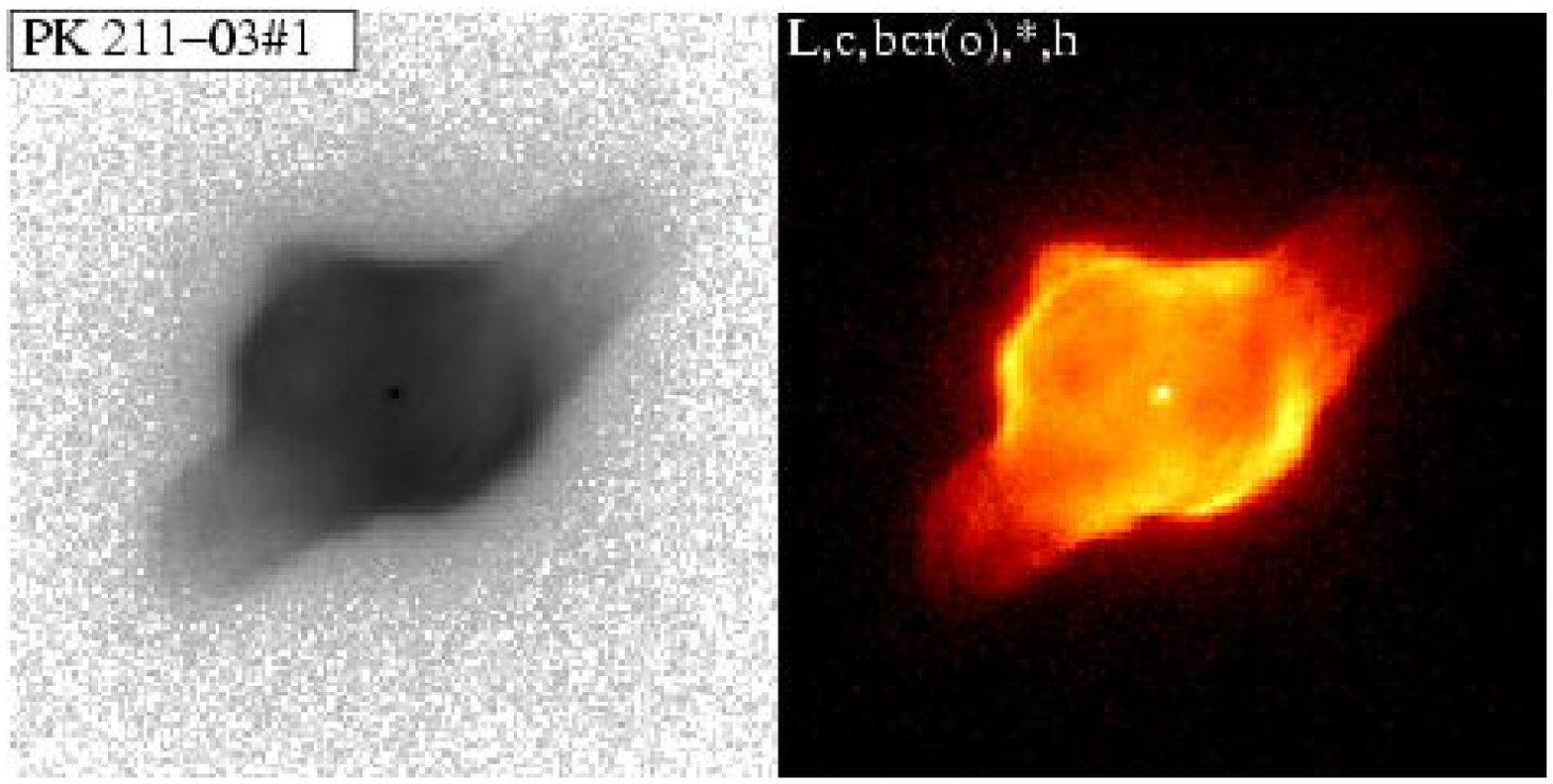}}
\caption{As in Fig\,1., but for PK\,211-03\#1.
}
\label{211-03d1}
\end{figure}
\begin{figure}[htb]
\vskip -0.6cm
\resizebox{0.77\textwidth}{!}{\includegraphics{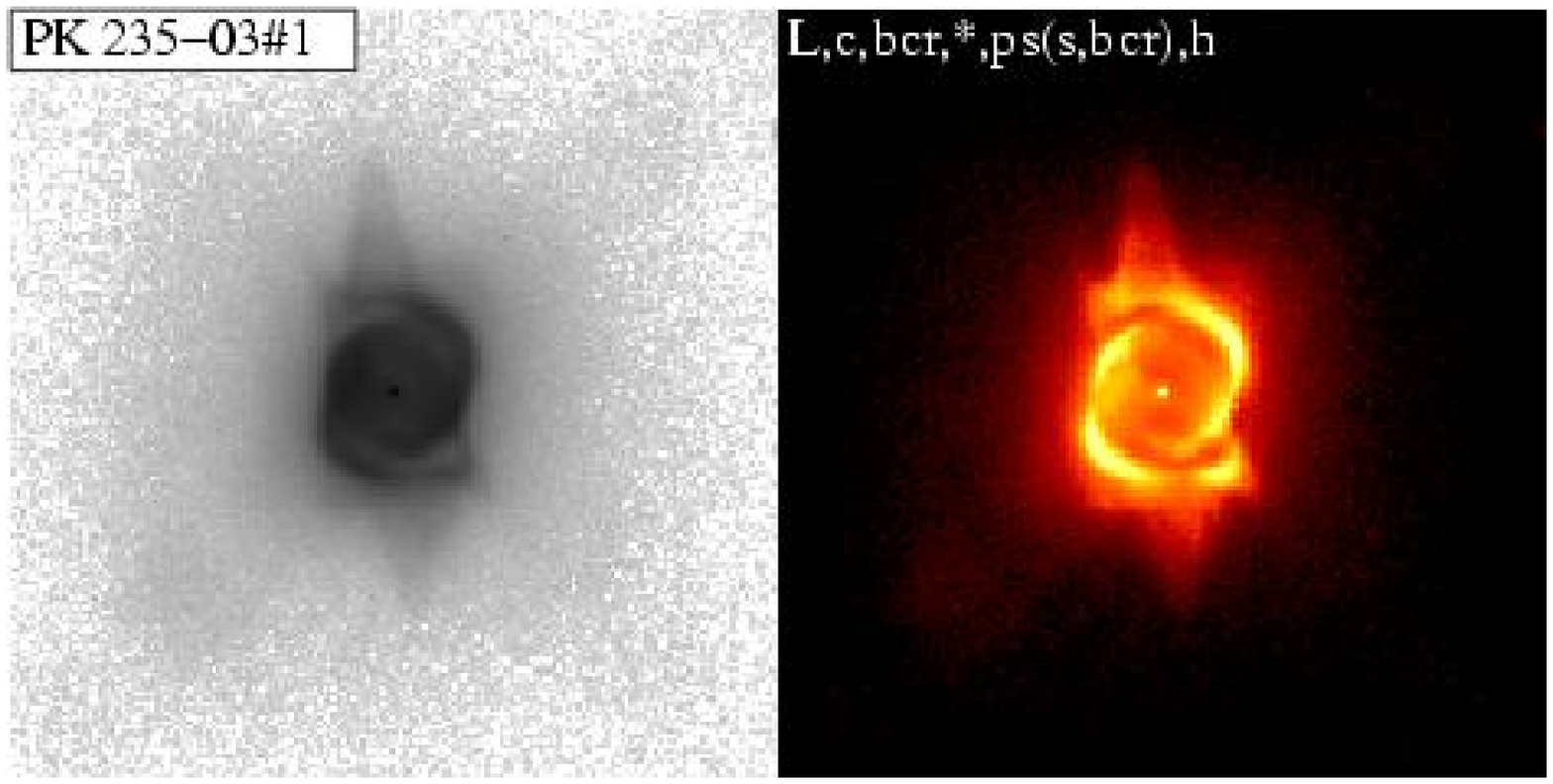}}
\caption{As in Fig\,1., but for PK\,235-03\#1.
}
\label{235-03d1}
\end{figure}
\begin{figure}[htb]
\vskip -0.6cm
\resizebox{0.77\textwidth}{!}{\includegraphics{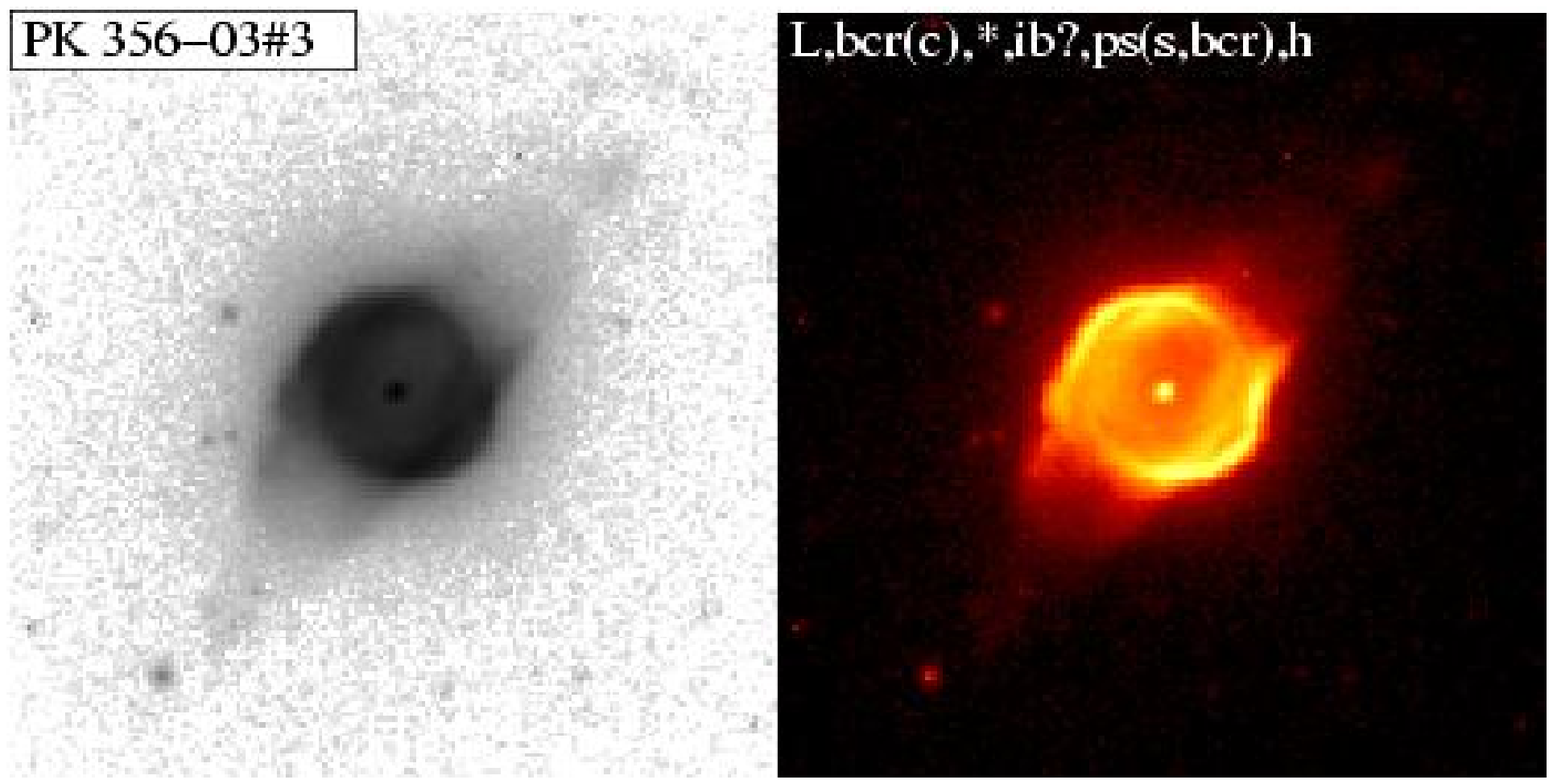}}
\caption{As in Fig\,1., but for PK\,356-03\#3.
}
\label{356-03d3}
\end{figure}
\begin{figure}[htb]
\vskip -0.6cm
\resizebox{0.77\textwidth}{!}{\includegraphics{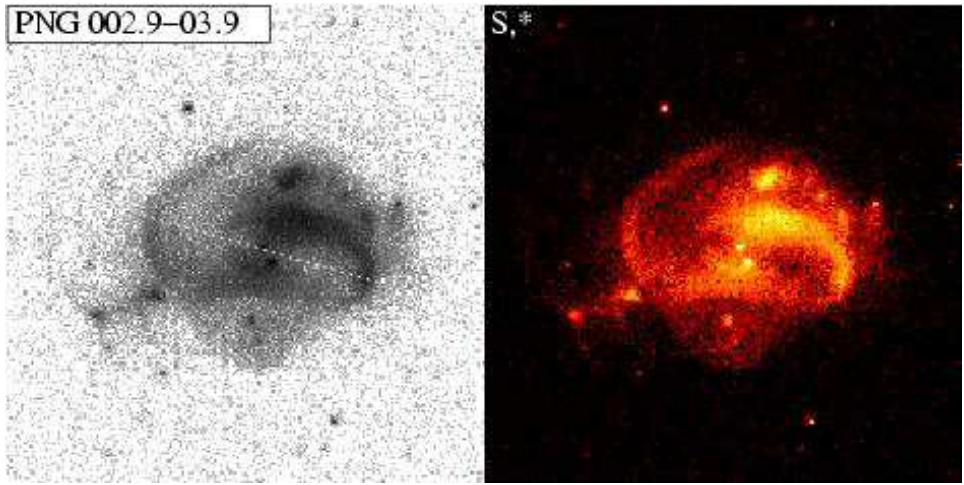}}
\caption{As in Fig\,1., but for PNG002.9-03.9.
}
\label{002.9-03.9}
\end{figure}
\begin{figure}[htb]
\vskip -0.6cm
\resizebox{0.77\textwidth}{!}{\includegraphics{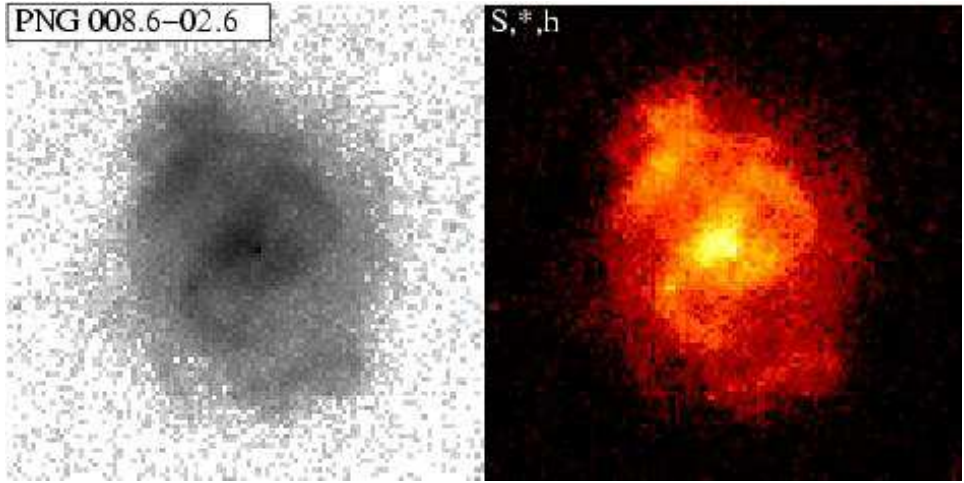}}
\caption{As in Fig\,1., but for PNG008.6-02.6.
}
\label{008.6-02.6}
\end{figure}
\begin{figure}[htb]
\vskip -0.6cm
\resizebox{0.77\textwidth}{!}{\includegraphics{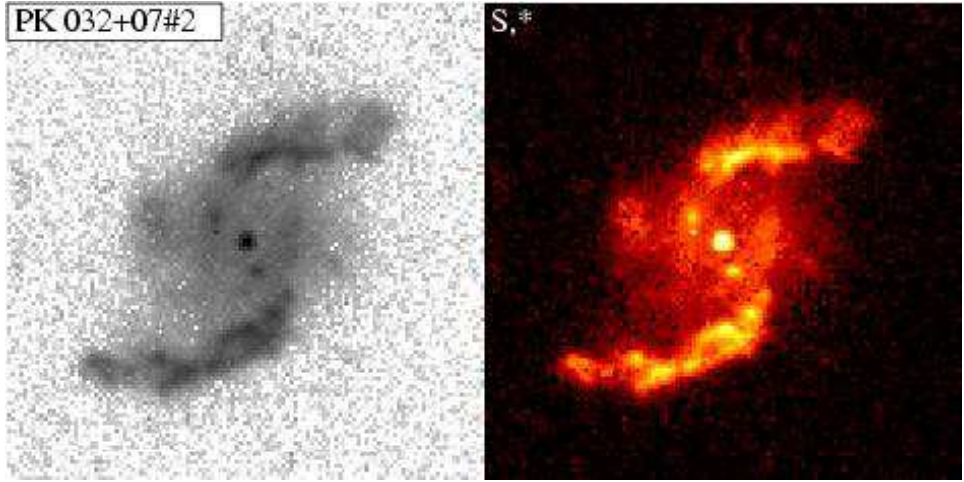}}
\caption{As in Fig\,1., but for PK\,032+07\#2 and the F658N ([NII]) filter.
}
\label{032+07d2}
\end{figure}
\begin{figure}[htb]
\vskip -0.6cm
\resizebox{0.77\textwidth}{!}{\includegraphics{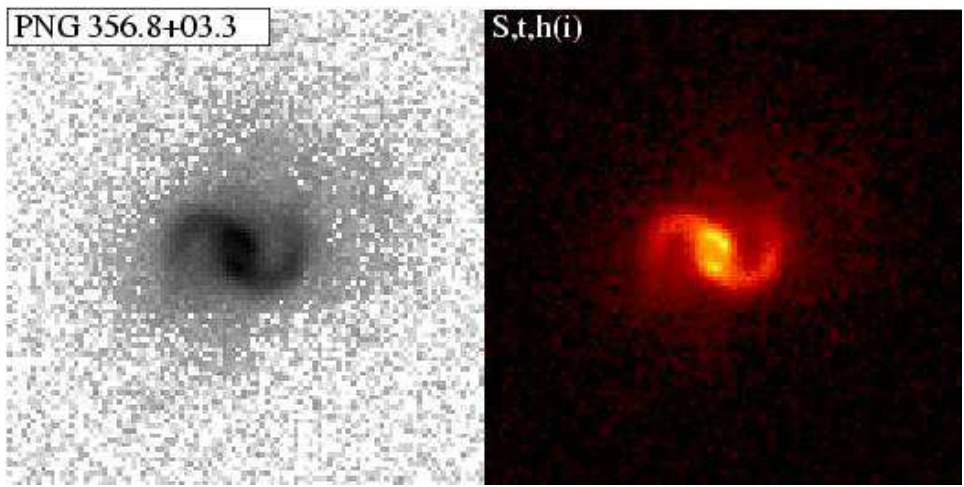}}
\caption{As in Fig\,1., but for PNG356.8+03.3.
}
\label{356.8+03.3}
\end{figure}

Other PNs show distinct microstructure, i.e. small-scale patterns of surface brightness variations, possibly
weave-like in their appearance, over the body of their primary shell structure. For example, PK215-24\#1 and
PK258-00\#1 (Figs.\,\ref{215-24d1},\ref{258-00d1}) show a weave-like pattern with large-scale order; the Etched Hourglass
nebula shows a mottled pattern in the outer parts of the hourglass (Figs.\,1 \& 4, Sahai et al. 1999) -- these are
denoted by the descriptor $wv$.

Some PNs show radial rays which, when projected inwards, appear to emanate from the central star. The best example
of such features is NGC\,6543. Balick (2004) finds that these features appear dark in an 
[OIII]/H$\alpha$ ratio image, and have bright counterparts in [NII]$\lambda$6584 and other low-ionization
lines. Balick infers that these rays are low-ionization structures, and most likely caused by
``ionization-shadows" produced by dense knots opaque to stellar ionizing photons, and their ionization is the
result of soft, diffuse UV (recombination) emission from neighboring gas. In support of this hypothesis, Balick
finds that many of the rays can be traced back to dense knots in the inner parts of the nebula. We denote these
radial ray features with the descriptor, $rr$. These radial rays are classified separately from the searchlight
beams (listed as $h(sb)$ under halo characteristics, Table\,\ref{codes}) as they are of higher multiplicity than the latter.
The searchlight beams generally occur in pairs, and lie within a narrow angular region around the polar axis of
the bipolar PPNs in which these have been seen (CRL\,2688, IRAS18276: SMSC07).

We find that a few PNs belonging to the E (and in one case, R) primary class have small,
diametrically-opposed protrusions jutting out from an otherwise smoothly curving geometrical shape describing the
primary structure of the PN; these are accounted for by adding a new descriptor, $pr$, to the set of ``other
nebular characteristics". The prime examples of PNs showing protrusions are PK215-24\#1, PK258-00\#1 and
PK016-01\#1 (Figs.\,\ref{215-24d1},\,\ref{258-00d1}, and \ref{016-01d1}), at position angles\footnote{we define the position
angle $pa$, as the angle measured anti-clockwise 
from the vertical axis in each image}, $pa\sim35\arcdeg$, $pa\sim90\arcdeg$, and $pa\sim95\arcdeg$,
respectively.
Additional possible examples are: PNG351.1+04.8, PK027+04\#1, PK315-13\#1, and PK320-09\#1 
(Figs.\,\ref{351.1+04.8},\,\ref{027+04d1},\,\ref{315-13d1}, and \ref{320-09d1}). Both
PK315-13\#1, and PK320-09\#1 (especially the latter) show not one pair, but multiple pairs of such protrusions.

Some PNs show additional nebular structures which cannot be easily described by the above
descriptors, nor possess sufficient geometrical symmetry to merit new descriptors labelling their specific
geometries. For such objects
(e.g., PK304-04\#1: Fig.\,\ref{304-04d1}), we add $ir$ to the set of descriptors for 
``other nebular characteristics".

\vspace{0.1in} \noindent{\it Halo}\\
Although halos are expected to be filled (i.e., not limb-brightened), surface-brightness limited structures and therefore not
expected to have a well-defined outer edge, we
find a few objects where the halo has a sharp outer edge, or shows the presence of a discontinuity within it -- we describe
this pheomenon by  
adding a new qualifier, $d$, for the $h$ descriptor: $h(d)$.  Prime examples
of this phenomenon are PK226-03\#1 and PK232-04\#1 (Figs.\,\ref{226-03d1},\,\ref{232-04d1}); additional examples are
PNG004.0-03.0, PK004+04\#1, and PK107-13\#1 (Figs.\,\ref{004.0-03.0},\,\ref{004+04d1},\,\ref{107-13d1}). Amongst these,
PK004+04\#1, PK064+04\#1, and PK232-04\#1 have a discontinuity within the halo; in each of these three cases, the
discontinuous structure is elongated. We note that the halos with discontinuities occur only in class-E objects.


\subsubsection{Partially Ionized Objects}\label{partion}
If we compare the H$\alpha$ image of the PN IRAS21282+5050 (PNG093.9-00.1), with its image using the broad-band
filter (F606W) (Fig.\,\ref{093.9-00.1}), we find that the latter image shows considerably more structure, based
on which we would give it a
primary classification of M, rather than its H$\alpha$-based classification as E. It is clear that for this object,
the H$\alpha$ image is only showing the ionised inner region of the multiple lobes seen in the F606W image. This
raises the possibility that for some small fraction of very young PN, where the ionization front has not reached
the boundaries of the lobes, H$\alpha$ will not reveal the full nebular morphology. In order to assess this
fraction, we need to be able to compare H$\alpha$ and deep broad-band images for a sample of very young and dusty PNs like
IRAS21282+5050.

\begin{figure}[htb]
\resizebox{0.8\textwidth}{!}{\includegraphics{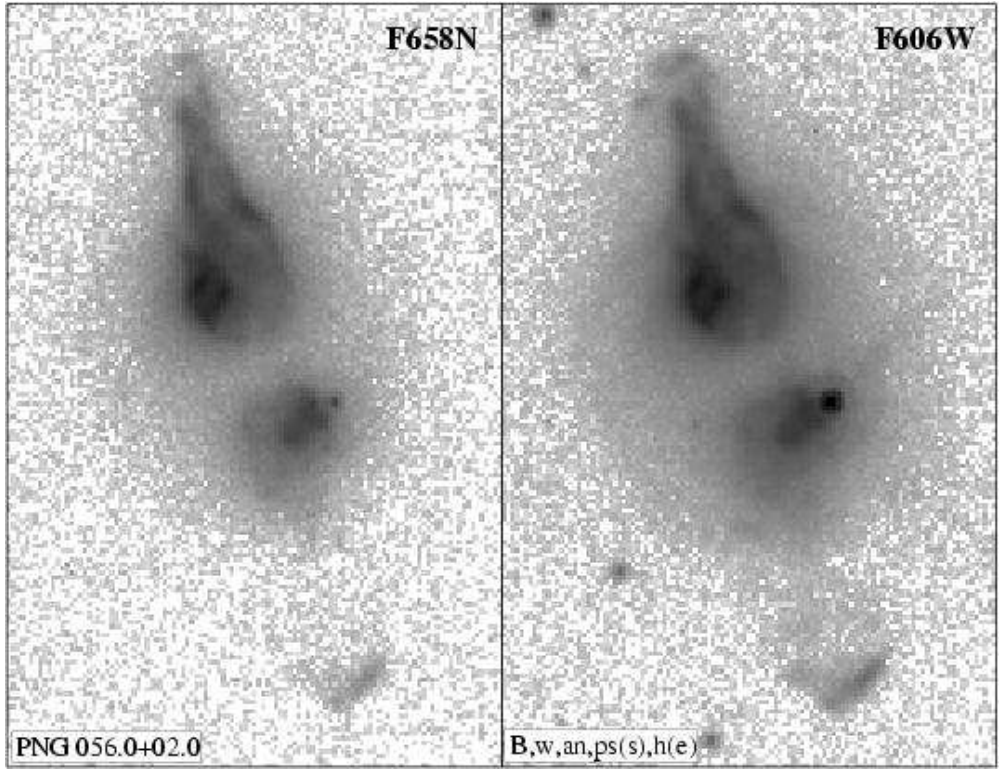}}
\vskip 0.2in
\caption{HST [NII]\,(F658N) and broad-band 0.6\micron~(filter F606W) ({\it log stretch}) images of the young
planetary nebula PN G056.0+02.0, taken with the WFPC2/PC.
}
\label{056.0+02.0-ha-cont}
\end{figure}

We have examined the HST archive for broad-band images of PNs in our sample. Unfortunately broad-band images with
adequate exposure times are not available for most of our sample (or for similar PNs). Only for a few young PNs, both
emission-line (in either H$\alpha$ or [NII]6583) and broad-band images are available; and we find that these objects have
very similar morphologies in both images, e.g.,
PN G056.0+02.0 (also known as IRAS\,19255+2123 and K\,3-35, Fig.\,\ref{056.0+02.0-ha-cont}), PK\,060-07\#2 (also known as
NGC\,6886, Fig.\,\ref{060-07d2_ha-555w}), PK\,321+03\#1 (also known as He\,2-113, Fig.\,\ref{321+03d1-ha-cont}), and
PK\,315-13\#1.  For the Etched Hourglass Nebula, MyCn18, the continuum image, taken with the medium-band filter F547M, shows
a very similar hourglass (bipolar) morphology as seen in the H$\alpha$ image, within the limitations of the lower SNR in the
F547M filter (Sahai et al. 1999).



\subsubsection{Offset Central Star}\label{staroff}
An additional feature of PNs which first became evident with the HST imaging of the PN, MyCn18 (Sahai et al. 1999)
is that the central star of the PN can be offset from the center of symmetry of the nebula. The nebular structures
in
MyCn18 have a very high degree of geometric symmetry, and the central star location has been shown to be offset
from the centers-of-symmetry of the former (such as the waist, and the hourglass lobes) (Figs. 2 \& 4 in Sahai et
al. 1999). Other examples of highly symmetric objects where the central star appears offset from the center of
symmetry are the Starfish Twins, He\,2-47 and M\,1-37 (Figs. 1 \& 2 in Sahai 2000; Figs.\,\ref{285-02d1},\,\ref{002-03d3} in this
paper). The measurements of such
offsets is not straightforward, and requires different strategies, on a case-by-case basis. We postpone the
discussion of this feature of PN morphology, and the measurements of these offsets, to a future paper. However, we
introduce a new qualifier\footnote{We defined a qualifier for the central star descriptor in our PPN survey
paper, which represented the shortest wavelength (in $\micron$) at which the central star is seen, but we do not
use that qualifier for PNs because we are using either the H$\alpha$ or [NII] filters for our classification, and
the wavelength difference between these is small} to the *
descriptor into our classification scheme (Table\,\ref{codes}), as follows. We include the offset of the central
star from the nebular center of symmetry in units of milliarcsec, with the number of significant figures giving an
indication of the accuracy or reliability of the measurement. In cases where there are several geometric
structures in the nebula relative to whose center the offset of the central star can be measured (as, e.g., in
MyCn18), we select the largest of these offsets. The main goal of this descriptor is to {\it indicate} whether the
offset is so small (closer to 0 than to $0.1{''}$) as to be not measurable, or if it is evident. If scientific
analysis is to be done based on the offset, then the actual measurements and their uncertainties in the literature
will need to be directly confronted. We use *(0) for objects in which offsets are well measured to be zero (i.e.,
closer to 0 than to $0.1{''}$), distinguishing them from those objects for which we just have * (i.e., inadequate
information on the offset). 

The morphological classifications for our sample of young PNs are given in Table\,\ref{morphs}. We classify 
MyCn\,18, a striking and demonstrative example of many of the secondary characteristics of our scheme, as
B,o,t,*(480),an,ib,wv,rg,ps(s,an).

\subsection{Nebular Ages}\label{age}

The age of each object has been calculated using its angular size, distance, and radial outflow velocity. These parameters
are
listed in Table\,\ref{morphs}. The angular size of each PN was measured along its longest axis, using the typically
well-defined edge of the nebular structure, excluding the halo (which is scattered light from the outflowing remnant of the
AGB mass-loss). For those PNs, where the edge is not clear, or does not exist because the lobes are not closed along the
long axis, we have drawn a vector showing the adopted size. In some cases, where one lobe of the nebula is much fainter than
the other (or lies outside the image field of view) so that its outer boundary cannot be determined reliably, we have used
the radial length from the central star to the tip of the bright lobe as a measure of half the nebular size. 

Our main source
for distances is the online version of the Acker et al. (1992)
catalog\footnote{http://vizier.u-strasbg.fr/viz-bin/Cat?V/84}.
There are several methods used for estimating the distances presented in that catalog. Our first choice is to take values
from the local extinction study (labelled E in the catalog). When multiple E values were found, the average was used. In the
absence of extinction values, either the kinematical distance (labelled K in the catalog) or the spectroscopic parallax
distance of binary companions (labelled S in the catalog) was used. For objects for which none of the above distances was
available, we used the median values of all the distances listed in the Acker et al. catalog. For objects without distances
in the Acker et al. catalog, we examined the published literature; for these the relevant references are listed in the
footnotes to Table,\ref{morphs}.

The [NII] expansion velocities from Acker et al. were used in calculating the sizes of each object; if the [NII] data were
missing, the [OIII] expansion velocity was used. For objects where both the [NII] and [OIII] expansion velocities were
absent,
the median of all known [NII] expansion velocities for objects
in our table was used (22\,\kms). The ages were then calculated by dividing the physical size of each object by
twice the expansion velocity. Although an accurate determination of expansion ages for our objects requires that we take into
account the inclination and a 2-dimensional model of the geometry of the expanding structures, the approximation used here
is adequate for obtaining rough estimates.
We find that most of our objects are relatively young, with a
median age of 2470\,yr, consistent with our expectation based on the ST98 selection criterion. The median age of the 23 objects
with R$_{exc}>1$ is 2880\,yr, so these are generally young objects as well.


The expansion velocities for bipolar PNs are usually found to be significantly higher in the lobes than in the
waist or central regions, as is the case for bipolar PPNs. Thus, for most bipolar and multipolar objects, our
derived ages are likely to be upper limits because the measured velocities are some global average that is less than 
the polar expansion velocities. When the expansion velocity is taken from [OIII] data, its value is
expected to be smaller than the actual expansion velocity in the lobes, since the [OIII] emission is significantly
more confined to the central region of PNs, and the expansion of the waist/central region is generally much slower than that of
the lobes: the derived age is thus again an upper limit.


\subsection{Classification Statistics}\label{stats}
We have calculated the fractions of objects in different primary classes (Table\,\ref{t5-stats}), both using the sub-sample of PNs in
which R$_{exc}\le1$ (96 objects) and the full sample (119 objects): the differences
are not statistically significant. We find that the class-B and Class-E objects represent about $30-35$\% of the total population --
although the fraction of class-E is slightly more than class-B, the difference is only at the 1$\sigma$ level. The class-M objects
represent 20\% of the population, somewhat smaller than the class-B objects. The three classes with collimated lobe structures (B, M
\& L) as a whole, represent slightly more than half the population: this must be considered a lower limit because objects in any of
these 3 classes can appear to belong to class-E if they are distant and not well-resolved, or if they are projected in such a way that
their polar axis makes a relatively small angle with respect to the line of sight. Objects in each of the remaining classes (I, R, L,
and S) represent less than 10\% each of the total. The total numbers of objects in each of these classes is rather small, hence the
differences in their populations are not significant or only marginally so. We note that point-symmetry is widespread, occurring (in
one or more of its different types) in 45\% of our sample. Although our sample is not complete, it is representative of young PNs, and
the statistics presented here are significant because the sample is drawn from observing programs which are not biased towards any
specific morphological class.

\subsection{Testing the Morphological Classification with a New PN Sample}\label{newsamp}
We apply our new morphological classification scheme to a new sample of PNe recently imaged as part of an ongoing SNAPshot
survey of PNs (Table\,\ref{t5}). Although this survey did not include narrow-band filters covering either the H$\alpha$ or
the [NII] lines, inspection of the survey images shows that the F200LP and F350LP bandpasses, which are extremely
wide, and cover all major
nebular emission lines, including [OIII], [NII] and H$\alpha$, show the morphology most sensitively (e.g., Fig.\
\ref{356.5+01.5}). A representative fraction of the imaged objects is shown in Figs.\,\ref{356.5+01.5}--\ref{344.8+03.4}. We
find that our scheme can adequately describe
all the morphologies seen in this sample, showing that it is quite comprehensive. 

\subsection{Resolution, Sensitivity and Nebular Orientation}\label{effects}

Our classification of any specific object may be affected by angular resolution, sensitivity and the object's orientation. 
Our scheme is aimed at minimizing the importance of orientation, but projection effects unavoidably
affect the classification for extreme inclinations.  A class-B or L nebula, with its long axis oriented at a small angle to
the line of sight, will  appear to be merely elongated (i.e., class E). A barrel-shaped configuration viewed from near its
axis will appear toroidal, or simply round, and an elongated halo can be projected to appear circularly symmetric. In those
cases where we suspect that the geometrical projection of the nebula plays a role in determining its classification, we have
so indicated in the notes on individual objects in Table\,\ref{morphs}. 

Sensitivity, which depends on the exposure times, has the biggest affect on the detection and classification of the halo region, which
generally has a much lower surface brightness than the bright nebular shell. So it is possible, and perhaps likely, that many of our
objects possess faint haloes which are below the limit of our detection. Given HST's resolution of $0.05{''}$, and pixel sizes
typically $0.046{''}$, we find that when the angular size is smaller than about $1.6{''}$ (e.g., Figs.\,\ref{004.8-22.7},
\ref{355.9+03.6}), but greater than about $1{''}$, most secondary descriptors listed under {\it Central Region} in Table\,\ref{codes},
some
of those listed under {\it Other Nebular Characteristics} (such as $ib$ and $wv$), and the qualifiers for point symmetry in the Central
Region, namely $ps(t,bcr,ib)$, cannot be reliably assigned. When the angular size is smaller than about $1{''}$, in addition to the
above limitations, objects that are intrinsically B, M, or L may appear to be have a primary classification of E, and most secondary
characteristics listed
under {\it Lobes}, {\it Central Region}, and {\it Other Nebular Characteristics} in Table,\,\ref{codes} cannot be reliably assigned
(although it may still
be possible to distinguish an overall point-symmetry in the shape (i.e., $ps(s)$, see e.g., Fig.\,\ref{326-06d1})).
However, only a small fraction of objects in our sample is affected by insufficient angular resolution: 8 objects in our sample have
angular sizes $\lesssim\,1.6{''}$, and 4 of these 8 have sizes $\lesssim\,1.1{''}$.

The errors in the determination of the angular extent affect the derivation of parameters such as age and diameter only in the
smallest objects (size $\lesssim\,1${''}), but are still small compared to the systematic uncertainties in these estimates due
to the poorly known distances, and/or expansion velocities. 

The morphological assignment may be ambiguous if the image is not of sufficiently good quality, and we have added question marks to
descriptors for which we judge the assignment to be somewhat ambiguous. Furthermore, if the image is too poor to assign a
classification
for any particular characteristic, we simply did not include it in the classification.  We expect the classifications to be refined,
and
possibly altered in some respects, as deeper and/or higher-resolution images become available.

\section{Discussion: The Formation and Shaping of Planetary Nebulae}\label{discuss}
The images presented in this paper forcefully demonstrate why high-resolution imaging available from a space-based
observatory like
HST, as opposed to seeing-limited ground-based facilities, is critical, even for objects
with an overall extent an order of magnitude larger than the seeing-limited resolution. In the
ground-based images, not only are important structural features not discernible, even the
qualitative assesment of the basic morphological class in a given classification scheme (e.g.
whether the PN is elliptical or bipolar) can be incorrect (\S\,\ref{intro}). This is because the waist regions of
PNs are much brighter than the polar lobes - when convolved with a large seeing disk, the spatial
spread of the waist region to a given intensity level is much larger than that of the much
fainter polar regions - so a bipolar nebula can look artificially elliptical. Two striking examples of this
effect are provided by the objects PK 300-02\#1 and PK\,000+17\#1. PK 300-02\#1, a
bipolar nebula with a very large aspect ratio (7:1) (Fig.\,\ref{300-02d1}), looks like an unremarkable
elliptical nebula with a much smaller aspect ratio (1.6:1) from the ground (G{\'o}rny et al. 1999). Similarly, PK\,000+17\#1,
whose ground-based image by SCM92 shows a featureless
elliptical blob of size about 7\farcs$\times$5\farcs8, is actually an extreme bipolar nebula
with a polar extent of $6{''}$, and a waist of width 1\farcs3 (Fig.\,\ref{000+17d1}). Thus the fraction of PNs which
are bipolar is likely to be severely underestimated from ground-based imaging studies, with
important implications for any hypothesis for their formation.

In this paper we have applied our morphological classification scheme developed for PPNs to young PNs, with modifications and
extensions. This ready applicability of our previous scheme substantiates its validity and results from the strong
morphological similarities between these two classes of objects evident in current imaging surveys with HST.
One view, which has often been expressed at a series of international conferences devoted to understanding the
formation of aspherical PNs (APN I-IV: e.g., Corradi, Manchado \& Soker 2009), is that understanding PPN shapes is not a 
prerequisite to understanding PN shapes because ionization will destroy much or all the detailed geometric
structure we see in highly-structured PPNs (e.g., multipolar objects like IRAS19024+0044). But given the striking
similarities which we report between PPN and PN shapes (e.g., between IRAS19024+0044 and the Starfish Twin PNs),
this view does not appear to be valid. The overall preservation of PPN shapes as they transform to PNs is likely a
consequence of the following two causes: (1) the ionization fronts get trapped inside the dense walls of the
nebular lobes, and (2) even if all of the lobe material in a PPN gets ionised, it maintains its shape, since its
expansion velocity significantly exceeds the sound speed in ionised gas (10\,\kms). Current observations of
several bipolar and multipolar PPNs provide evidence that their lobes are expanding at very high velocities
($\gtrsim$100\,\kms).

\subsection{The Morphological Evolution of Waists}\label{waistevol}
The comparison between the PPN and PN samples, and the resulting new morphological descriptors, highlight the task ahead of
us in trying to understand the formation of aspherical planetary nebulae. 
The emergence of new morphological features in PNs signifies the operation of new physical processes affecting the nebular
shapes, as PPNs evolve into PNs. We must first consider the shaping which occurs during the PPN phase, which is the primary
shaping stage (SMSC07). During this phase, (1) the dense mass-loss of the AGB phase has already ceased and the inner edge of
the dense circumstellar shell is advancing outward, and (2) a collimated, fast wind (i.e., the CFW) is sculpting the dense
shell from the inside out. 
As the central star evolves towards higher temperatures, the primary shaping is followed by the action of the Spherical,
Radiatively-driven, Fast Wind from the PN central star (hereafter SRFW), on the pre-shaped nebula. The fact that we see the
biggest changes between PPNs and PNs in the central waist region clearly supports this scenario, since the waist material is
the first to be encountered by the wind from the PN central star. The central regions of waists are expected to be more
exposed and visible during the PN stage compared to the PPN stage, because the waists seen in PPNs are likely expanding
structures, and further clearing of the central region occurs as a result of the SRFW. 

The closed ($c$) and open ($o$) qualifiers on the $bcr$ descriptor of the central region are diagnostic of differences in its
structure during the preceding evolutionary phases. Thus, one may imagine that the $bcr(c)$ central regions result from the
inflation of an originally compact structure that surrounded the central star in all directions, when the object was a PPN.
Such a dusty structure has been proposed to surround the central star in the well-studied PPN, CRL\,2688, by Sahai et al.
(1998b). In contrast, $bcr(o)$ central regions may be produced in objects by the inflation of a highly flared disk or
a toroidal central region. The inflating agent in both cases is the SRFW. A possible example of an object showing such inflation of a
torus into a $bcr(o)$-type central region is PK002-09\#1 (Fig.\,\ref{002-09d1}). The presence of irregular
structure in the central region, captured by the $i$ qualifier, may result from hydrodynamical instabilties produced during
its inflation; the presence of geometric structure, captured by the $g$ qualifier, is less easily understood. One
possibility is that the collimated jet-like outflows which we know operate during the PPN phase remain active as the central
star evolves to higher temperatures and starts driving the SRFW; the combined interaction of these two fast winds with
themselves and the central region then results in either irregular or geometric structure of the latter, depending on the
relative speeds and momentum fluxes of these winds.

The continued evolution of
a PN having a barrel-shaped central region ($bcr$) and a B-, M- or L- primary classification, may result in an object whose
primary structure appears to be E, because the lobe regions have become too tenuous to be seen and only the bright $bcr$
region is visible. A possible example of
this phenomenon is PK327-02\#1 (Fig.\,\ref{327-02d1}), where a pair of very faint lobes may be present, emanating from the
top and bottom parts of the periphery of the elongated central region. This possibility could be tested out using deep
imaging in low-excitation nebular lines such as [NII] and [OI], which may reveal such lobes.

There is a natural morphological continuity in primary classes from B, to L, to those which are E and have the $t$ secondary
characteristic. It is likely that class-L objects, or class-E objects having the $t$ descriptor, were B earlier on in their
evolution, but continued expansion of their waist regions as these objects evolved resulted in the loss of the pinched-in
waist as the latter expanded out to become the toroidal feature.

\subsection{The Rarity of Round PNs}\label{fewround}
Our survey shows that round PNs are rare: only 3.4\% of our sample belongs to the R class.
Soker (2002) discussed the rarity of round
PNs; he proposed that most PNs evolve from binaries, which both enhances the AGB mass-loss rates (as a ``final intensive wind") and
makes the mass outflows non-spherically symmetric, whereas round PNs are objects which evolve from single stars (i.e., have no close
companions, stellar or substellar); they also have low metallicity, so the AGB mass-loss rate is low, thus resulting in a relatively
faint PN. Hence such objects are difficult to find. In contrast to the occurrence of R objects in our sample of PNs, there is a
complete lack of round PPN (SMSC07). We hypothesize that this difference indicates that (i) a fast wind is needed in order to carve
out the aspherical cavities inside the AGB CSE for a post-AGB object to appear as a visible PPN, (ii) the fast wind is always
collimated, to a smaller or larger degree. Hence, post-AGB objects which do not develop such collimated fast winds - i.e., most likely
single stars with no close companions, stellar or substellar -- show a shell-like structure only at the PN-phase, once the central
star becomes hot enough to produce a SRFW, and ionize the swept-up circumstellar shell. This hypothesis could be tested with an
extensive search for round PPNs with HST; if it is correct, we would not find any round PPNs.

\subsection{New PN-specific Secondary Characteristics}\label{newsec}
Young PNs show more complex morphological features than PPNs. We have already discussed some of these above in connection
with the central region. The addition of new descriptors and qualifiers to the list of secondary characteristics is
indicative of new processes
that apparently do not occur during the PPN phase. For example, inner bubbles (i.e., $ib$) are likely to represent emission
from very hot
gas in the reverse shock generated by the SRFW. This interpretation is
supported most directly by the X-ray imaging observations of NGC\,6543 (Chu et al. 2001), which show that the X-ray emission
comes from an inner structure which would be classified as an inner bubble. In the case of PK215-24D1 (IC418), the [OIII]
image clearly shows enhanced emission in the inner bubble region (Fig.\,\ref{oiii-ha-ic418}). 

The $wv$ descriptor, which denotes the presence of small-scale patterns of surface brightness variations over the body of the
primary shell structure, may be the result of specific hydrodynamical instabilities. It is noteworthy that, with one
exception, all objects with $wv$ belong to the E primary class. 


The radial rays (labelled by the $rr$ descriptor) and searchlight beams (labelled by the $sb$ qualifier of the halo
descriptor) may be related phenomena. In the case of the PPN,
CRL\,2688, where these were first discovered, Setal98 presented a model in which the beams were the result of an obscuring
dust cloud aroud the central star having a specific geometry: they proposed that the obscuring cloud has annular holes
around
the symmetry axis of the nebula, allowing the preferential leakage of starlight which illuminates the extended spherical
circumstellar envelope to give the searchlight beam features. It appears plausible then, that when the central star evolves
into a hot post-AGB star with a SRFW, the latter results in partial disintegration of this cloud,
destroying its annular hole geometry and resulting in a less spatially-organized distribution of dust around the central
star. The
ionising radiation from the star can then be extincted along directions with optically thick dust clumps; it is along these
directions that the radial rays are seen, as ionization shadows.

\subsection{Comparison with Other Classification Schemes}\label{otherclass} We have compared our classification scheme with
the detailed ones by SCS93+CS95 and Metal96, and we find obvious similarities as well as important distinctions (Appendix).
We
find that our classification system is broader and more comprehensive than the
SCS93+CS95 and Metal96 systems, encompassing a
more diverse array of morphologies (by employing a larger number of secondary descriptors), and is also more precise. These
differences have mostly been motivated by the
availability of high-resolution images of PNs obtained with HST. An important difference between our scheme and
those of SCS93+CS95 and Metal96, at the primary classification level, is their inclusion of a separate
point-symmetric class. The results of our study here show that point-symmetry can be present in objects in a
variety of ways for all primary clases except those that belong to the irregular (I) class. Further, neither
Metal96 nor SCS93+CS95 have multipolar or spiral-arm classes. Our scheme includes Metal96's class Q objects in our
M (multipolar) class; it is not yet clear if the quadrupolar objects are distinct from those with more than 2
pairs of lobes. The SCS93+CS95 and Metal96 schemes do not include any descriptors for the presence of halos.

An intriguing classification of PNs, devised by Soker \& Hadar (2002: SH02), is based on the departure from an axisymmetic
shape (considering only departures along and near the equatorial plane between structures on opposite sides of the nebular
symmetry axis): SH02 discuss the connection between departure types and the physical mechanisms that may cause them, mainly
resulting from the influence of a stellar binary companion. The PN images and the classification scheme here are relevant to
the SH02 study in several ways, the most important of which is the widespread presence of point-symmetry in our PNs sample,
rather than axial symmetry, which at the very least, severely restricts the sample of PNs subject to the SH02
classification. An possible example of an object in our sample that shows a ``bent" departure from axisymmetry (as defined by SH02) is
PK\,000+17\#2 (Fig.\,\ref{000+17d1}). One of the departures discussed by SH02 relates to the offset of the central star from the
``center of the
nebula": although we have included such an offset in our scheme, we have not implemented it in this paper, because in order
to do so, one needs to define objective methods of finding the geometrical center of a nebula, and a quantitative estimate
of the resulting uncertainty (\S\,\ref{staroff}).

\subsection{Physical Mechanisms Underlying Morphology}\label{physics}
The morphological classes constituting the classification scheme presented in this paper were devised with minimal prejudice
regarding their underlying physical cause.  However, in many cases, physical causes are readily suggested by the geometry, 
supplemented by the kinematics that have been measured in some systems. Kinematic studies, mostly using high-resolution long-slit
spectroscopy (e.g., review in L{\'o}pez et al. 2004), are time-consuming, but will eventually be necessary to fully disentangle
the 3-D morphology of PNs where projection effects appear to be important. Several of the physical causes have been discussed
above
where the primary and secondary characteristics are defined. 

Collimated lobe structures seen in the B or L classes imply the presence of collimated outflows but cannot directly tell us whether the
collimation takes place near the central star(s) or is rather due to deflection by an equatorial concentration of matter, unless
point-symmetry is also present. Although many hydrodynamical studies of interacting winds, with different assumptions about their
geometries, have been carried out over the years (e.g., Mellema \& Frank 1995, Lee \& Sahai 2003, Garc{\'{\i}}a-Arredondo \& Frank
2004, Dennis et al. 2008), a new and focussed effort of hydrodynamical modelling is needed to address this issue, since simulations of
hydrodynamical collimation undertaken so far suggest that the fast outflows are intrinsically collimated. For example, comparisons of
simulations of a spherical wind interacting with an equatorially dense AGB envelope -- the so-called Generalized Interacting Stellar
Winds or GISW model (Kwok et al. 1978, Balick 1987) -- does not produce the pinched-in shape of the lobes at their base characterizing
class-B objects (see, e.g., Mellema \& Frank 1995), even with very high equatorial to polar density contrast ratios. This discrepancy
between data and models of this type highlights the importance of the pinched-in shape of the lobes in class-B PNs as a criterion for
testing models.


In all cases, {\it point symmetry} indicates a secular trend -- presumably precession -- in the orientation of the central
driver of a rapid, collimated outflow.  Point-symmetry due to shape, $ps(s)$, or the presence of point-symmetric ansae\footnote{which
most likely result from the impact of a jet
upon a slower-moving, prior wind}, $ps(an)$, implies that the outflows
are not collimated by hydrodynamical processes, but are intrinsically collimated, likely driven by a central accretion disk undergoing
precession or wobbling of its axis. Numerical simulations are of course needed to verify whether precession can produce
the several types of point-symmetry covered in our classification scheme. A recent study by Raga et al. (2009) is a good
first step in this direction. In this study, the authors show that models of accretion disks around a star in a binary
system predict that the disk will have a retrograde precession with a period a factor of $\sim$2-20 times the orbital
period, and they present an analytic, ballistic model and a three-dimensional gasdynamical simulation of a bipolar outflow from a
source in a circular orbit, and with a precessing outflow axis. They find that this combination results in a jet/counterjet
system with a small-spatial-scale spiral which is reflection-symmetric across the equatorial plane (resulting from the
orbital motion) and a larger-scale, point-symmetric spiral (resulting from the longer period precession).

The multiple lobe pairs in class M objects also require intrinsically collimated flows or ejections. Hypotheses for producing such
lobes have been discussed by Sahai et al. (2005) in connection with the starfish PPN, IRAS\,19024+0044, and include a
direction-changing bipolar jet driven by a wobbling accretion disk, or ``explosive" ejections of matter along different
directions driven by a correspondingly fast release of magnetic energy from
the central star.

A {\it waist} is the signature of a strong equatorial concentration of matter, whether it
be outflowing or in a bound Keplerian disk.  If the equatorial concentration has expanded following a diminution or a
cessation of mass loss, then an evacuated {\it toroidal} structure results, ionized and/or illuminated on its inside edge by
the radiation from the central star.  If the distribution of outflowing matter is less concentrated toward the system's
equatorial plane, then, following the cessation of mass loss, the toroidal configuration will have a large vertical extent, and the
ionization and illumination of
its inside boundary will present a {\it barrel}-shaped appearance. 

Ideally, one would like to relate nebular characteristics to fundamental, irreducible, physical variables innate to the system. The
nebular morphology, when coupled with velocity measurements, often provides access to variables that relate to the timing, such as the
temporal history of the mass loss rate, and in particular the time since the cessation of rapid mass loss. Another innate variable is
the presence of a binary companion to the mass-losing star, operationally expressed as the stellar mass ratio, separation and orbital
eccentricity. This variable can affect the degree of concentration of the mass outflow towards the system's equatorial plane (e.g.,
Mastrodemos \& Morris 1999), but in a rather complex way that requires elaborate dynamical modeling of the morphology and the velocity
field in order to access those variables. De Marco (2009), following Soker (1997), summarises five main types of PN-shaping binary
interactions as a function of just the binary separation. Amongst these types, close binaries which avoid a CE interaction (separation
few to $\lesssim$100\,AU) or result in one (separation$\lesssim$few\,AU), are the ones most likely to produce dramatic departures from
spherical symmetry in the central regions. During CE interactions in which the binary survives, the stellar envelope of the primary
star can become unbound as a result of transfer of energy and angular momentum from the secondary (e.g., Iben \& Livio 1993), and the
ejected mass can be strongly concentrated in the equatorial plane (Sandquist et al. 1998). Thus CE ejection is a promising mechanism
for producing the waist structure identified in our classification scheme for PNs and PPNs. CE interaction may also produce bipolar
nebulae more often than single progenitors, as suggested by Miszalski et al. (2009)\footnote{these authors caution that further
morphological studies are needed for a definitive conclusion on this issue} who find a ``penchant for bipolarity" in a sample of 30
post-CE PNs.

Another variable related to binarity is the rate of stellar rotation, which is likely to be significant enough to affect the geometry
of the
mass outflow only if a stellar merger has taken place during a CE interaction, or if tidal interaction in a close binary has
synchronized the primary rotation with the orbit of the secondary. 
Other fundamental variables enter the picture if a stellar magnetic field plays a role in shaping the stellar mass outflow
(e.g.,
Pascoli 1997, Chevalier \& Luo 1994; Garc{\'{\i}}a-Segura 1997, Garc{\'{\i}}a-Segura et al. 1999, Blackman et al. 2001): the
magnetic geometry, strength, and orientation.  However, Soker (2006) argues that in order to sustain a sufficiently strong global
magnetic field
for an adequate period of time during which mass can be ejected in a collimated manner, angular momentum needs to be
continuously supplied to the star, and this can only come from a companion, hence magnetic fields may play a role, but are
perhaps not the fundamental underlying agent shaping PNs. 

Additional innate variables are needed to account for the point symmetry displayed by a large fraction (almost 50\%) of the
nebulae presented here (e.g., variables relating to the coupling of orbital and rotational angular momenta might be
important in this context, but the cause of the precession of the central driver has not been identified yet in any system). 
Finally, we note that yet unidentified variables may be needed to account for the multipolarity of many systems and for the
arc
features and other discontinuities observed in the halos of many well-observed PNs and PPNs.  It is clear that this will
remain a dynamic area of research for some time to come. 

\acknowledgments
We would like to thank Bruce Balick and his students for producing their web catalog of HST 
PN images\footnote{http:\//\//www.astro.washington.edu\//users\//balick\//PNeHST}, which was helpful in the compilation of part of
the
PN sample used in this paper. We thank Noam Soker and Brent Miszalski for their reading of, and providing helpful 
comments on, an earlier version of this paper.
RS's contribution to the
research described in this publication was carried out at the Jet Propulsion Laboratory, California Institute of Technology, under a
contract with NASA. Financial support for this 
work was provided by NASA through awards from the Space Telescope Science 
Institute, operated by the Association of Universities for Research in Astronomy, 
Inc., under NASA contract NAS5-26555, as well as through a Long term Space Astrophysics award.

\vskip 0.2in
\centerline{\bf APPENDIX}\label{append}
SCS93+CS95's main
morphological classes are elliptical (E), bipolar (B), point-symmetrical (P), and irregular (I). We ignore the
trivial class of unresolved (poorly) resolved objects which SCS93+CS95 label ``stellar (gaussian)" . The {\it elliptical}
class includes both
round and elliptical shaped objects. {\it Bipolar} objects are defined as axially symmetric PNs with an
equatorial waist from which two faint extended lobes emanate, {\it point-symmetric} are objects whose
morphological components show point-symmetry around the center, {\it irregular} objects have
shapes lacking geometric symmetry and which therefore do not fall in the previous 3 classes. Metal96 have 5
primary classes: round (R), elliptical (E), bipolar (B), quadrupolar (Q), and point-symmetric (P). The
definitions of the classes in common with CS95 (i.e., R, E, B, and P) are the same. SCS93+CS95 and Metal96
add descriptors (lower-case letters) to their primary class descriptors to denote the presence of additional
structures. Thus, for E and R, Metal96, add 3 descriptors: ``s" for those with inner structures, ``a" for
those with ansae, and ``m" is when any of these exhibit multiple shells. SCS93+CS95 also use ``s" and ``m"
descriptors with similar meanings for their E class, but don't have an ``a" descriptor. For the B class,
Metal96 add a descriptor ``r" if the objects show a ``marked bright ring", whereas SCS93+CS95 add ``m" for
those which show ``multiple events". Metal96's class Q objects have two pairs of lobes, and objects which are too
irregular to fit onto the above are called non-classified or ``NC" (hence the same as SCS93+CS95's I
objects). Our round and elongated classes correspond to Metal96's R and E classes; our bipolar
and ``collimated lobe pair" classes correspond to Metal96's B class, and our irregular class is similar to
Metal96's NC class.


\clearpage

%
%

%
%
\vskip -0.7cm
%

\clearpage

%


\clearpage

\clearpage

\clearpage
\clearpage

%

%

%
%
\clearpage
%
%

\clearpage


%
%
%
%



\clearpage

%

%
%
\clearpage

%
%
\clearpage

%
%
%
\clearpage
%

\clearpage

%

%
%
\clearpage

%
%

\begin{figure}[htb]
\resizebox{1\textwidth}{!}{\includegraphics{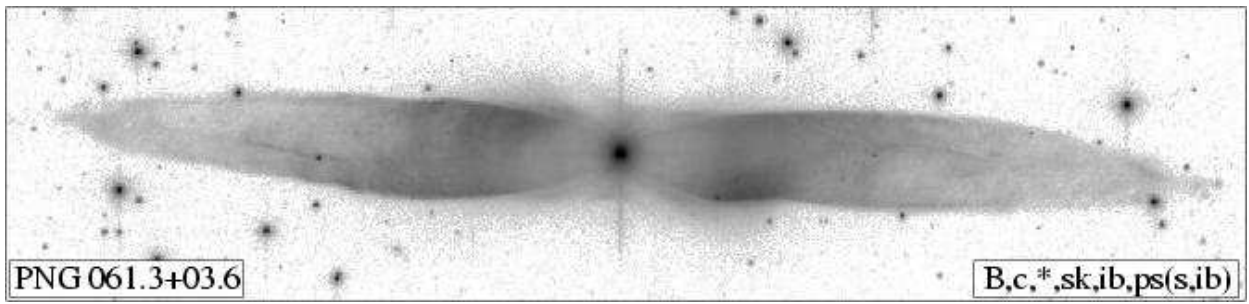}}
\caption{HST broad-band 0.6\micron~(filter F606W) ({\it log stretch}) image of the young
planetary nebula PN G061.3+03.6, taken with the ACS/WFC.
}
\label{061.3+03.6-f606w}
\end{figure}

\begin{figure}[htb]
\vskip -0.5in
\resizebox{0.8\textwidth}{!}{\includegraphics{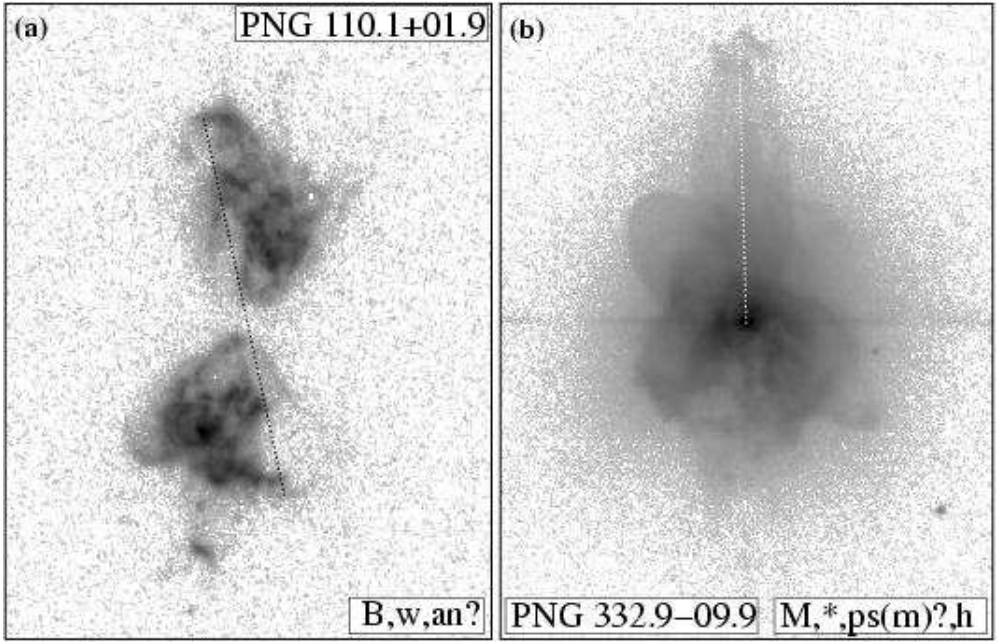}}
\vskip -0.2in
\caption{HST broad-band ({\it log stretch}) images, taken with the ACS/HRC, of two young
planetary nebulae, (a) 0.6\micron~(filter F606W) image of PNG110.1+01.9 (also IRAS\,22568+6141), and
(b) 0.43\micron~(filter F435W) image of PNG332.9-09.9 (also IRAS\,17047-5650). 
}
\label{cpd-22568}
\end{figure}

\begin{figure}[htb]
\vskip -0.2in
\resizebox{0.8\textwidth}{!}{\includegraphics{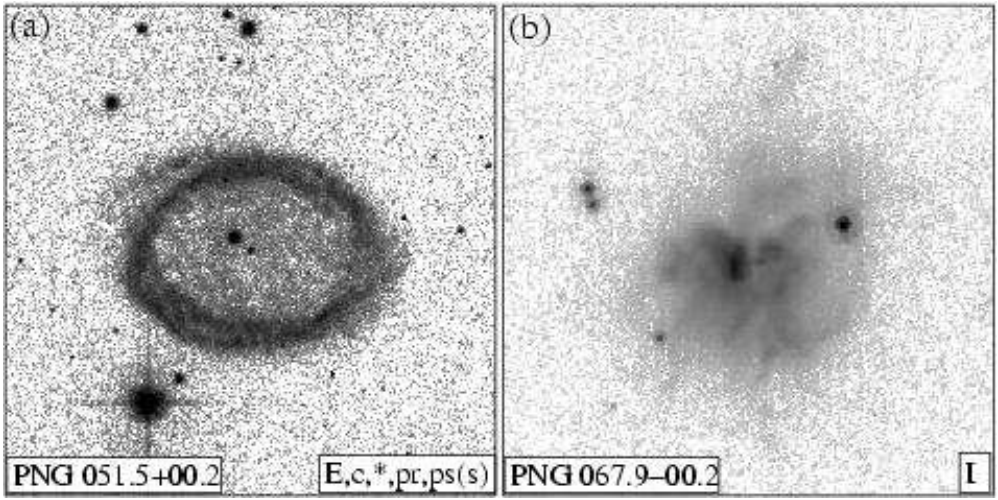}}
\vskip -0.2in
\caption{HST broad-band 0.6\micron~(filter F606W) ({\it log stretch}) images, of two young
planetary nebulae, (a) PNG 051.5+00.2 (taken with ACS/WFC), and (b) PN G067.9-00.2 (taken with ACS/HRC). 
}
\label{067.9-00.2-f606w}
\end{figure}

\begin{figure}[htb]
\vskip -0.6in
\resizebox{0.8\textwidth}{!}{\includegraphics{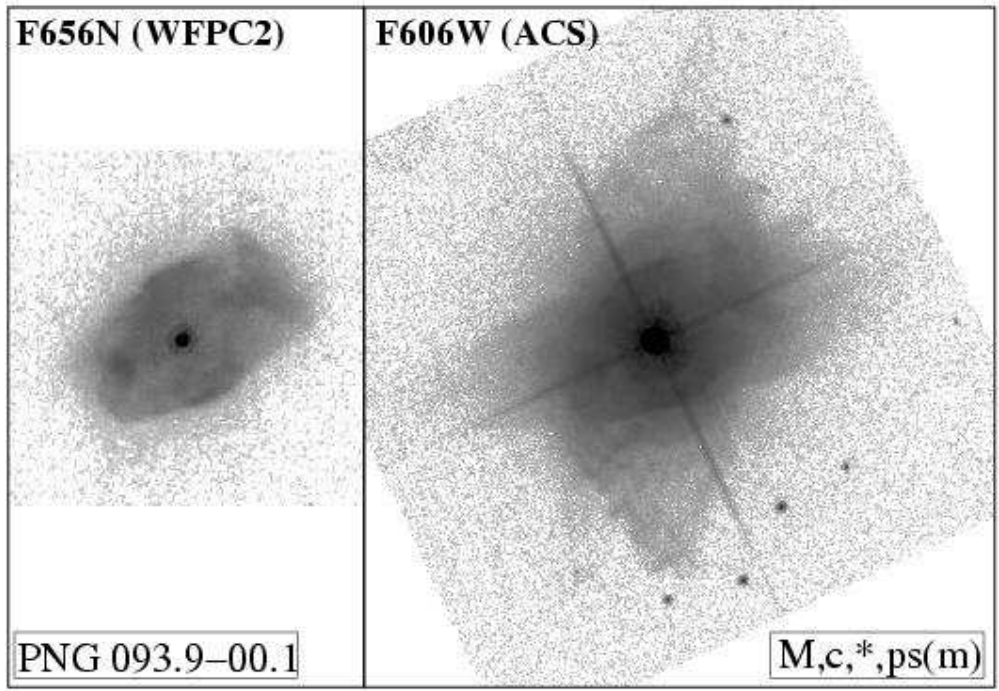}}
\vskip 0.2in
\caption{HST H$\alpha$ and broad-band 0.6\micron~(filter F606W) ({\it log stretch}) images of the young planetary
nebula PNG093.9-00.1 (IRAS\,21282+5050). The morphological classification is based on the 0.6\micron~morphology. 
}
\label{093.9-00.1}
\end{figure}

\begin{figure}[htb]
\vskip -0.6in
\resizebox{0.8\textwidth}{!}{\includegraphics{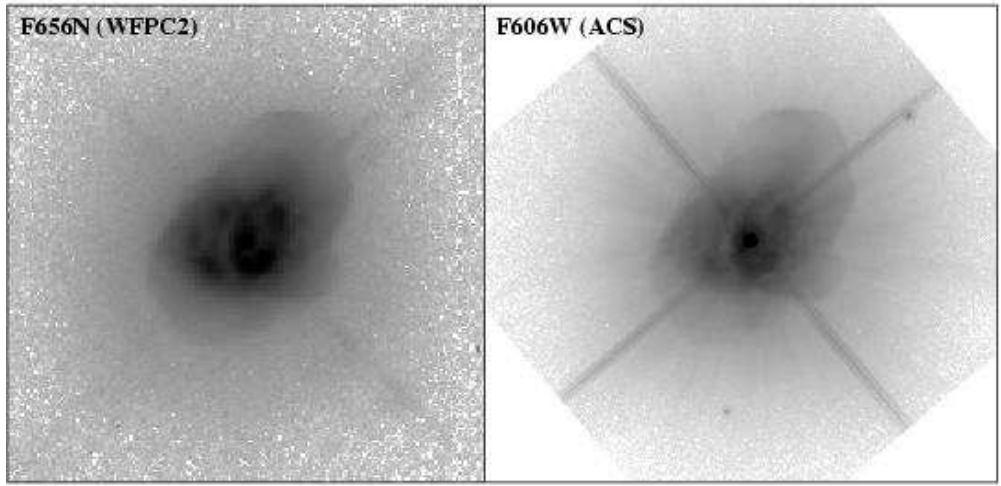}}
\vskip 0.2in
\caption{HST H$\alpha$ and broad-band 0.6\micron~(filter F606W) ({\it log stretch}) images of the young planetary
nebula PK321+03\#1 (He\,2-113) (adapted from Sahai et al. 2000)}
\label{321+03d1-ha-cont}
\end{figure}

\begin{figure}[htb]
\resizebox{0.8\textwidth}{!}{\includegraphics{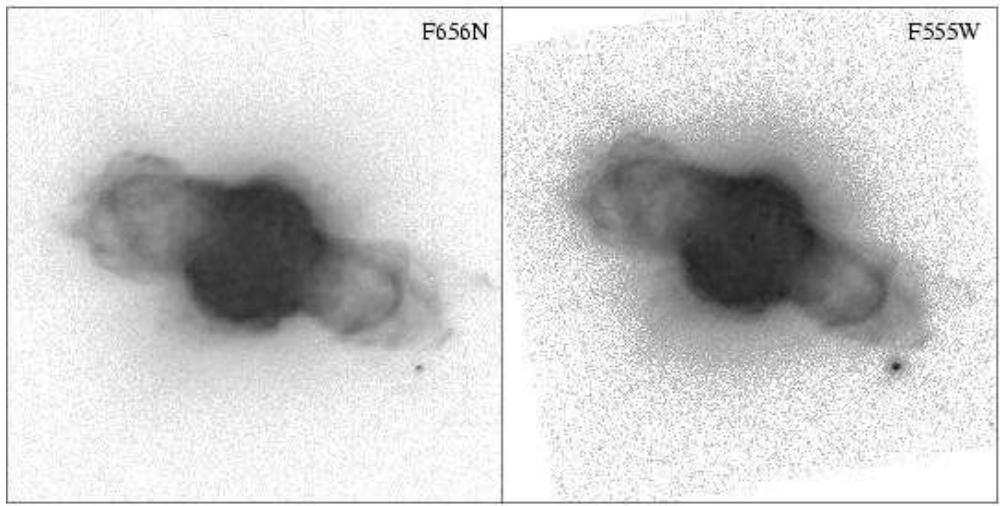}}
\vskip 0.2in
\caption{HST H$\alpha$\ and broad-band 0.55\micron~(filter F555W) ({\it log stretch}) images of the young
planetary nebula PK\,060-07\#2, taken with the WFPC2/PC.}
\label{060-07d2_ha-555w}
\end{figure}

\begin{figure}[htb]
\resizebox{0.9\textwidth}{!}{\includegraphics{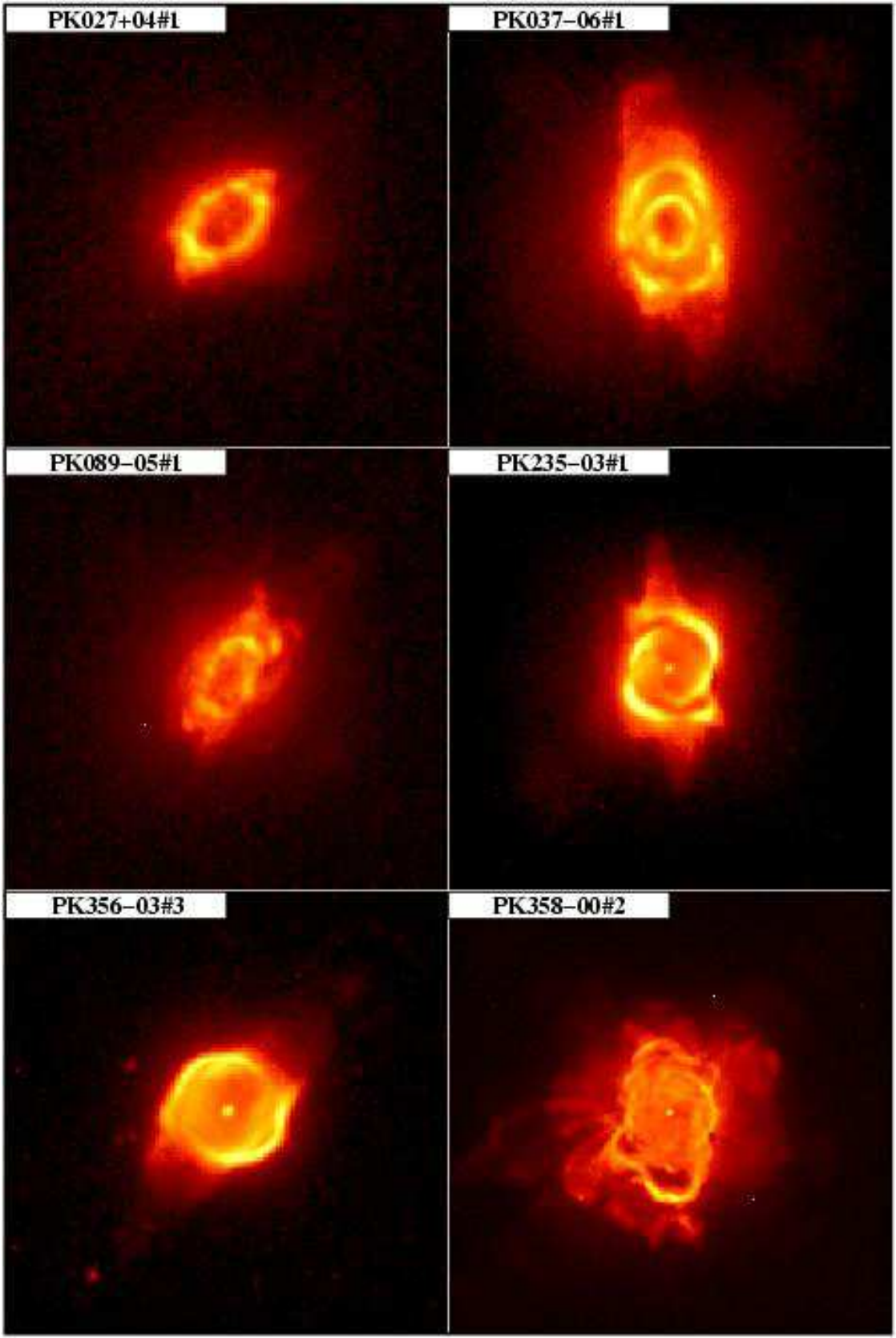}}
\caption{Young Planetary Nebulae with ``barrel-shaped central regions" which appear to be closed at the barrel ends
($bcr(c)$)
}
\label{bcr-c}
\end{figure}

\begin{figure}[htb]
\resizebox{0.9\textwidth}{!}{\includegraphics{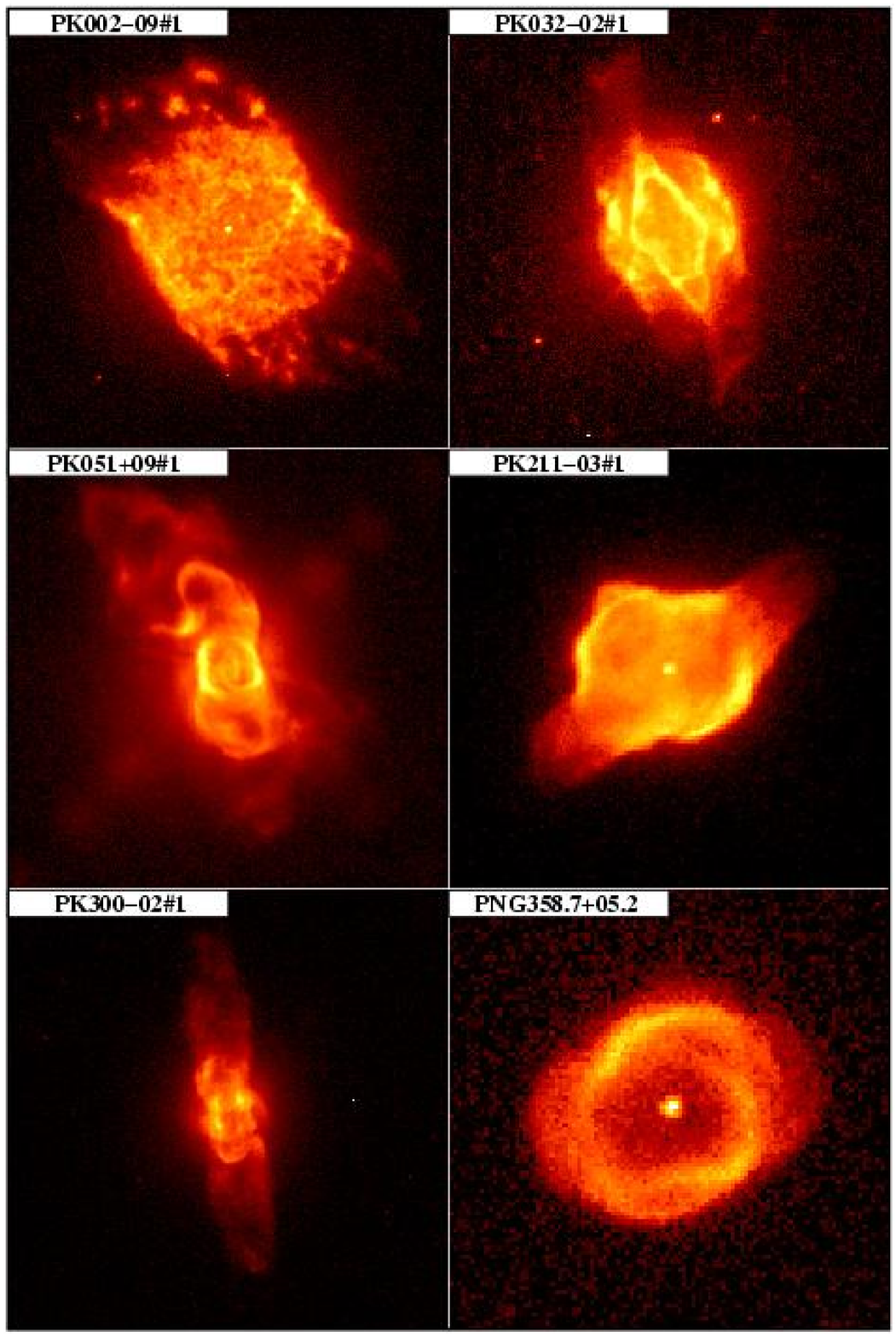}}
\caption{Young Planetary Nebulae with ``barrel-shaped central regions" which appear to be open at the barrel ends ($bcr(o)$)
}
\label{bcr-o}
\end{figure}

\begin{figure}[htb]
\resizebox{0.9\textwidth}{!}{\includegraphics{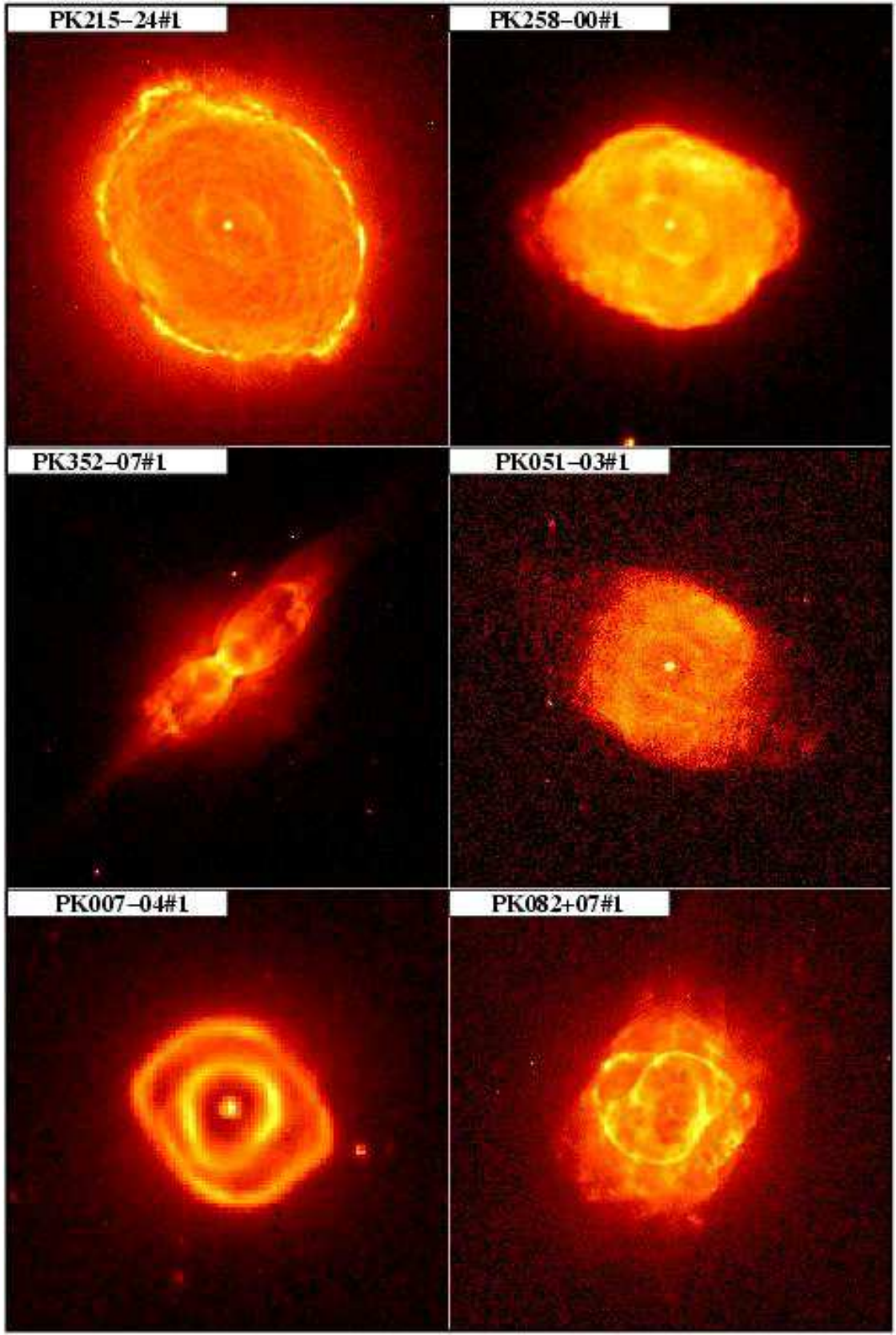}}
\caption{Young Planetary Nebulae with inner bubbles}
\label{ibmosaic}
\end{figure}

\begin{figure}[htb]
\resizebox{0.8\textwidth}{!}{\includegraphics{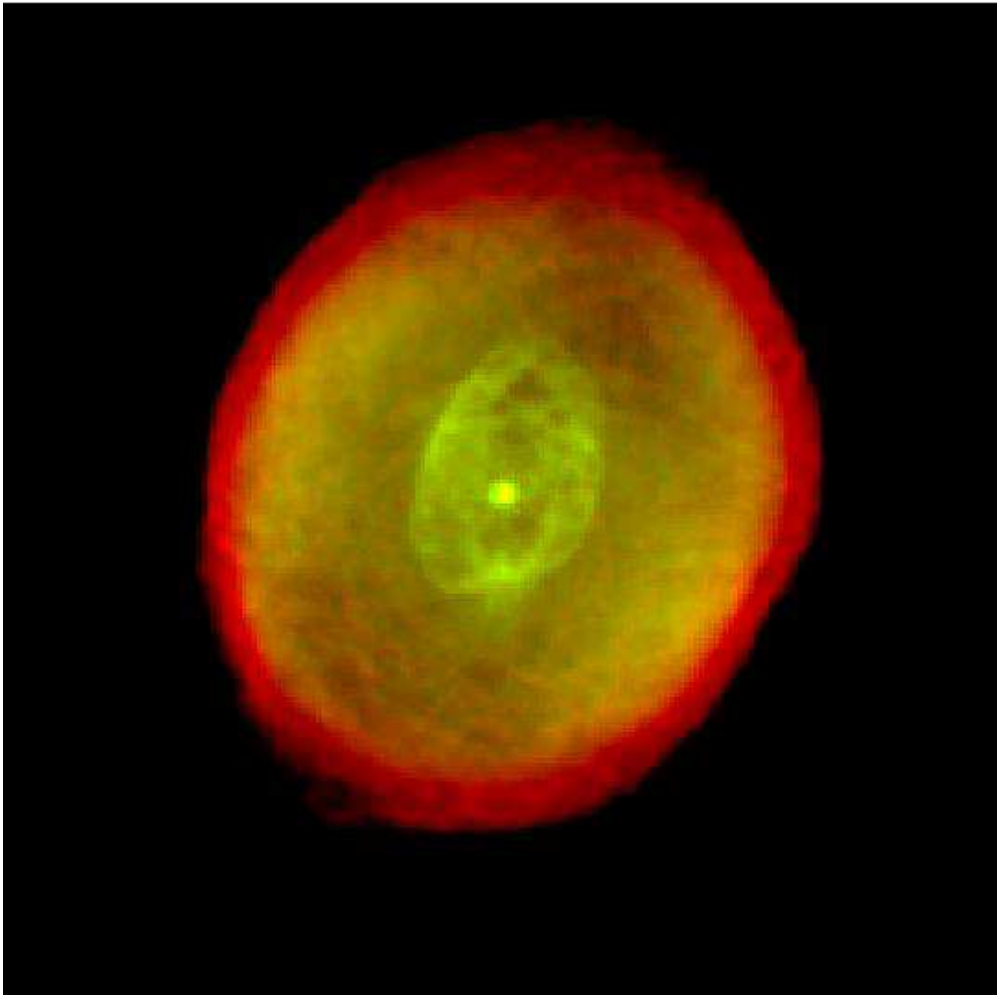}}
\caption{A composite image of PK215-24D1 (IC418) (H$\alpha$=red, [OIII]$\lambda$5007=green), showing the inner
bubble}
\label{oiii-ha-ic418}
\end{figure}

\begin{figure}[htb]
\resizebox{0.8\textwidth}{!}{\includegraphics{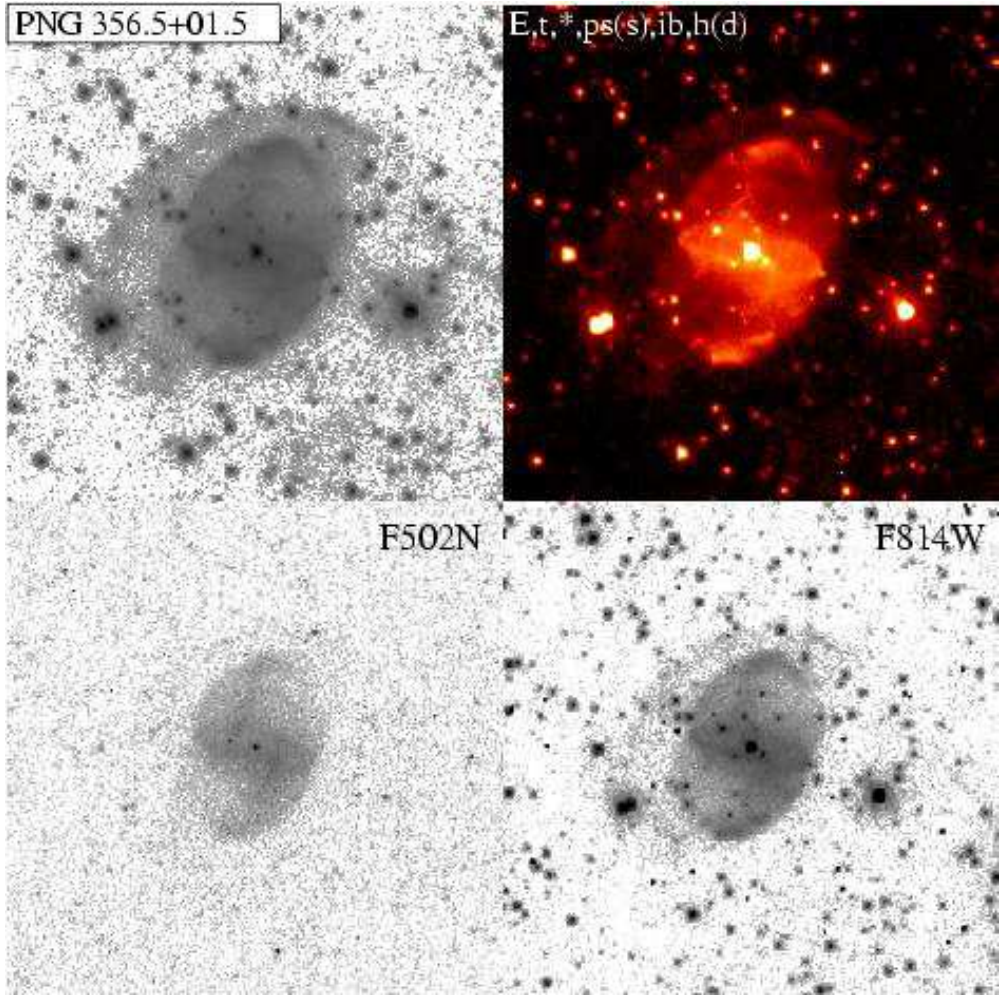}}
\caption{{\it top panel} As in Fig\,1., but for PNG356.5+01.5 and the broad-band filter F200LP. {\it
bottom panel} 
Log-stretch reverse-greyscale images of the same PN in the narrow [OIII]$\lambda$5007 (F502N) emission-line filter, and the
broad-band 0.8\micron~F814W filter}
\label{356.5+01.5}
\end{figure}

\clearpage
\begin{figure}[htb]
\vskip -0.6cm
\resizebox{0.77\textwidth}{!}{\includegraphics{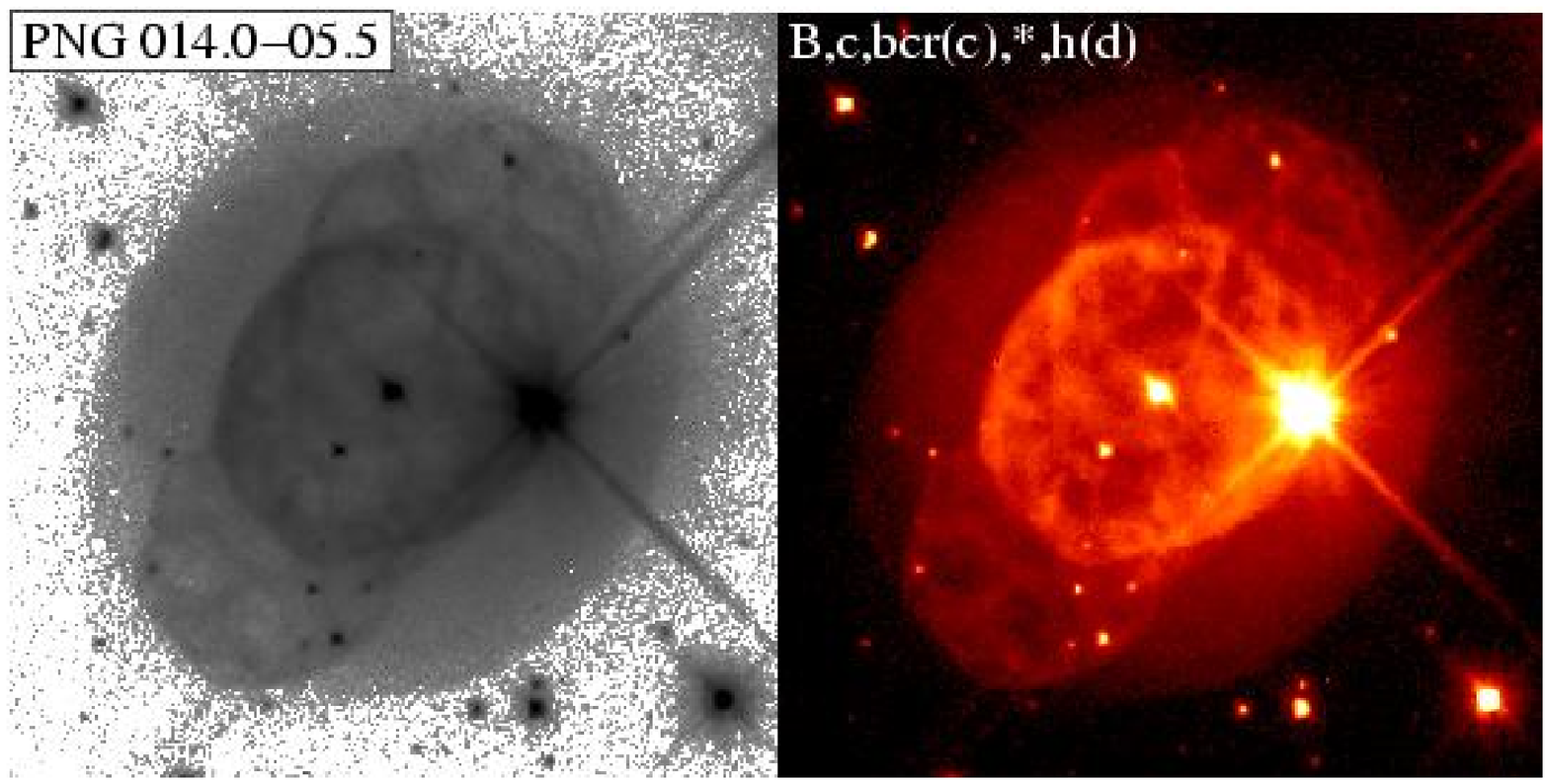}}
\vskip -0.4cm
\caption{As in Fig\,1., but for PNG014.0-05.5 and the broad-band filter F200LP
}
\label{014.0-05.5}
\end{figure}
5\vskip -0.2cm

\begin{figure}[htb]
\vskip -0.6cm
\resizebox{0.77\textwidth}{!}{\includegraphics{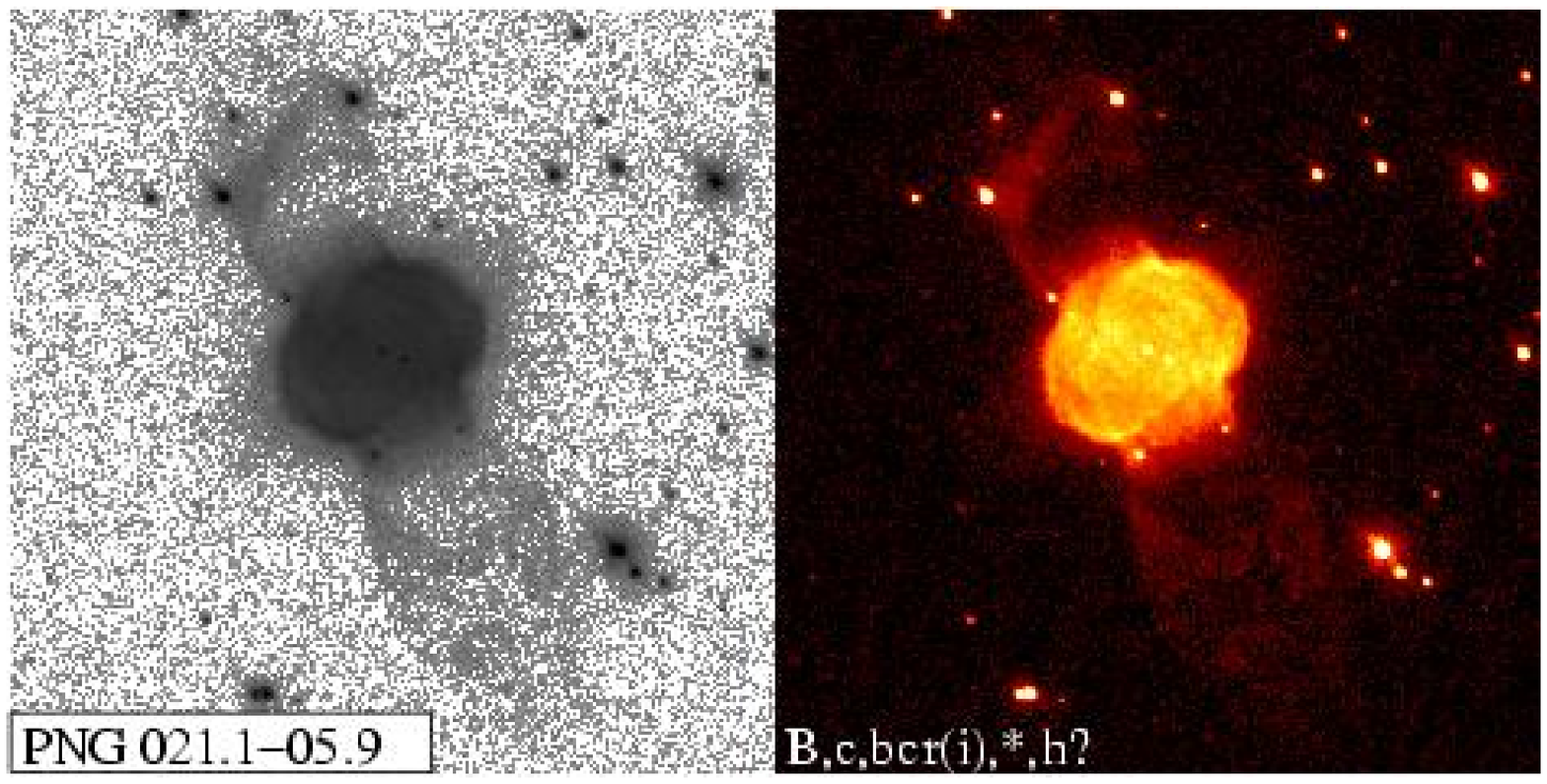}}
\vskip -0.4cm
\caption{As in Fig\,1., but for PNG021.1-05.9 and the broad-band filter F200LP
}
\label{021.1-05.9}
\end{figure}
\vskip -0.1cm

\begin{figure}[htb]
\vskip -0.8cm
\resizebox{0.77\textwidth}{!}{\includegraphics{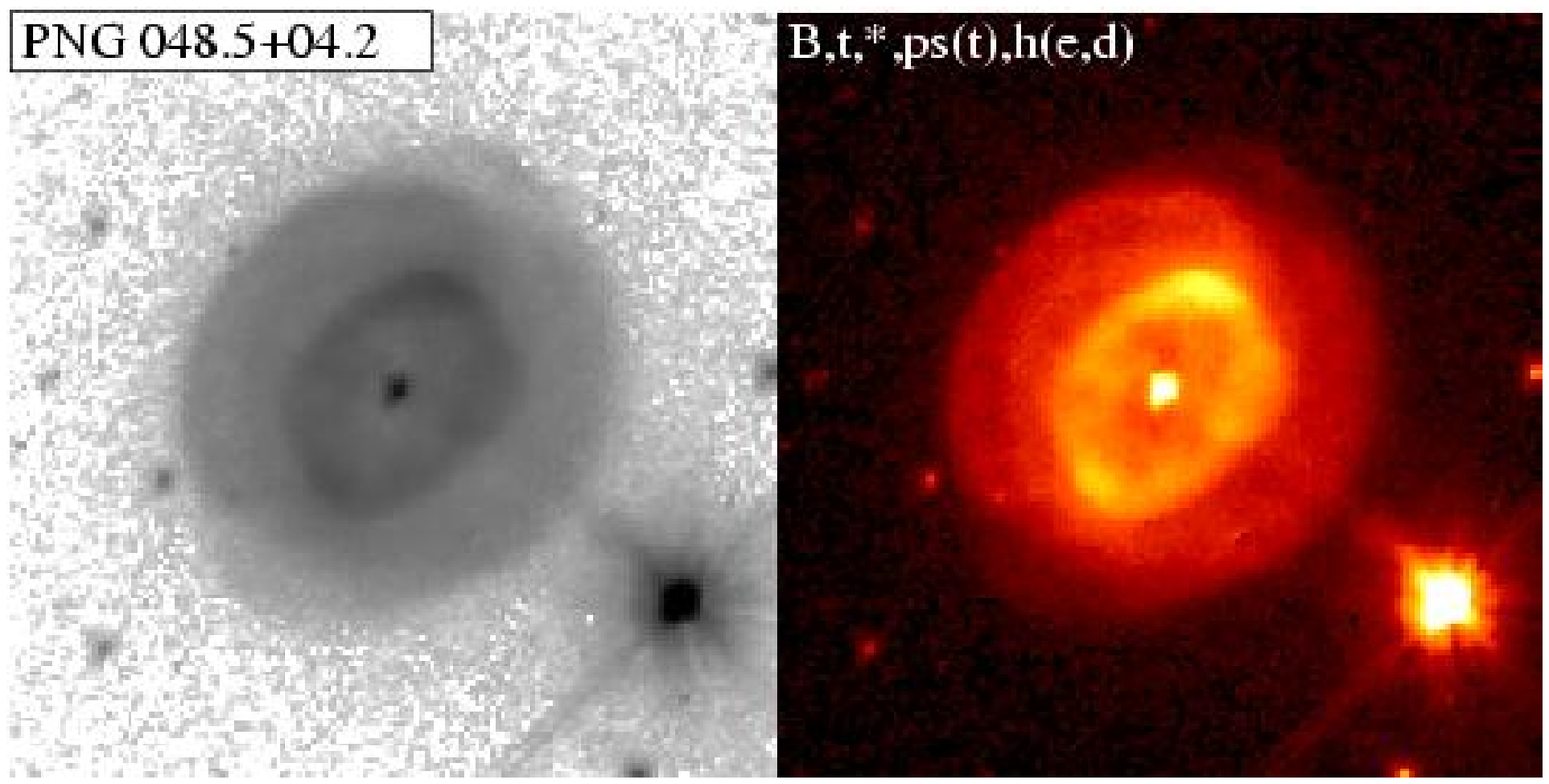}}
\caption{As in Fig\,1., but for PNG048.5+04.2 and the broad-band filter F200LP
}
\label{048.5+04.2}
\end{figure}

\begin{figure}[htb]
\vskip -0.8cm
\resizebox{0.77\textwidth}{!}{\includegraphics{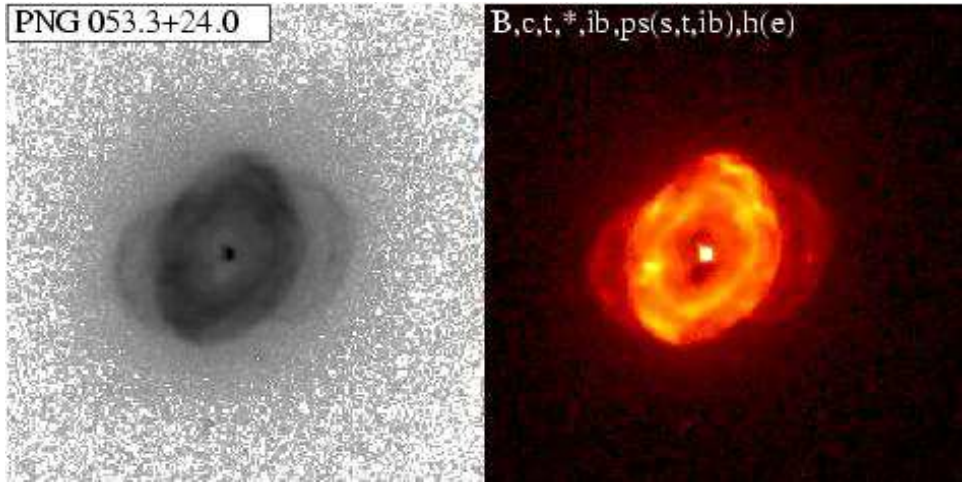}}
\caption{As in Fig\,1., but for PNG053.3+24.0 and the broad-band filter F200LP
}
\label{053.3+24.0}
\end{figure}

\begin{figure}[htb]
\vskip -0.6cm
\resizebox{0.77\textwidth}{!}{\includegraphics{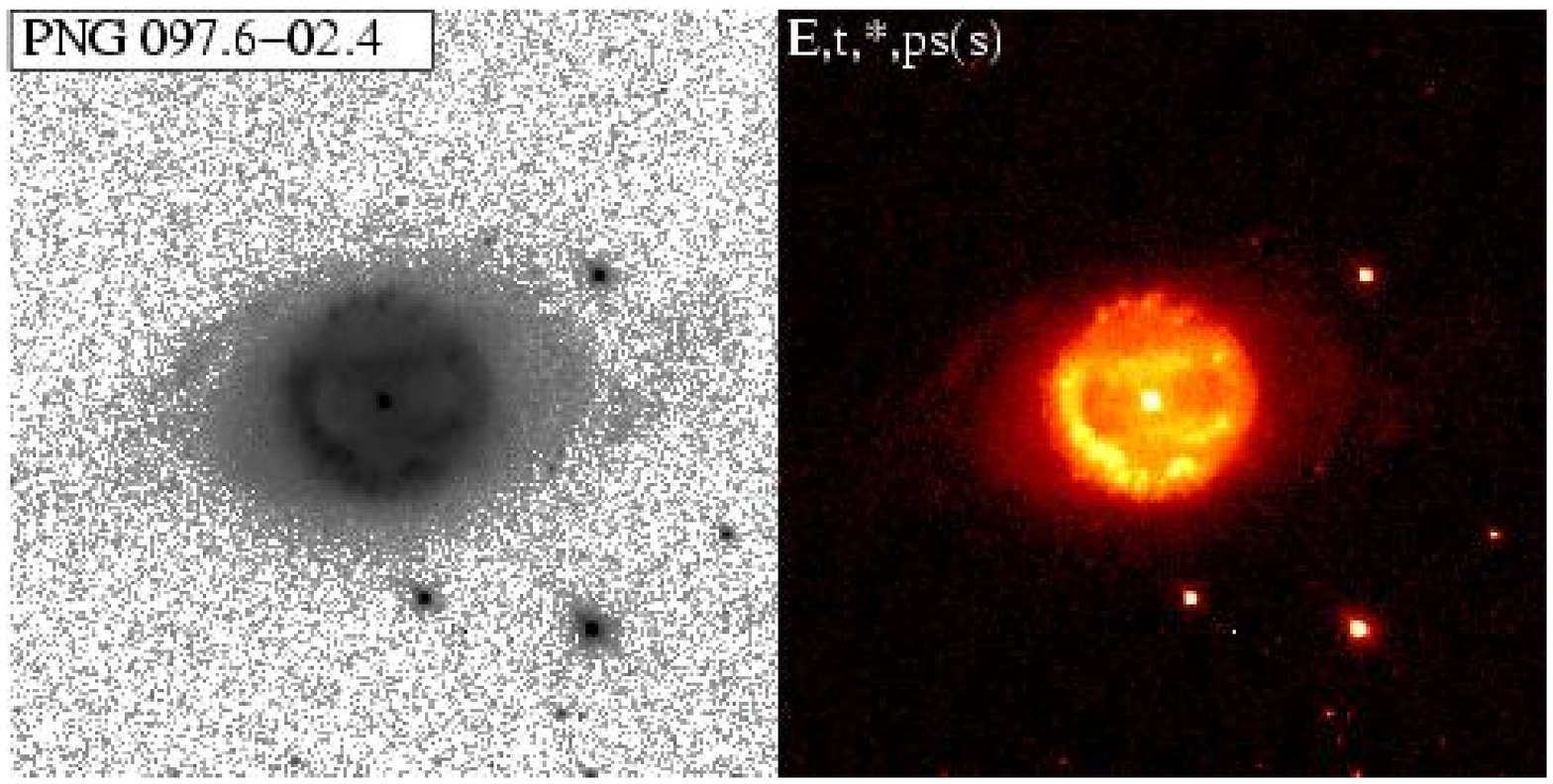}}
\caption{As in Fig\,1., but for PNG097.6-02.4 and the broad-band filter F200LP
}
\label{097.6-02.4}
\end{figure}

\begin{figure}[htb]
\vskip -0.6cm
\resizebox{0.77\textwidth}{!}{\includegraphics{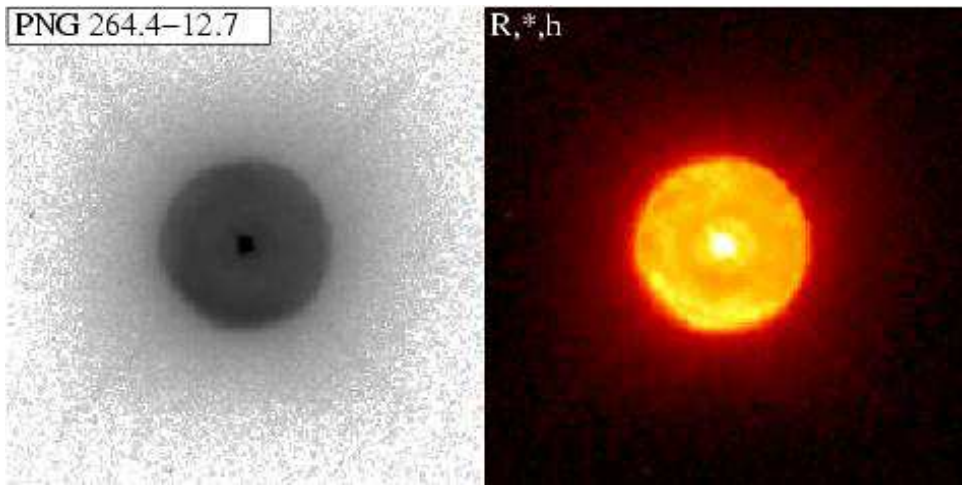}}
\caption{As in Fig\,1., but for PNG264.4-12.7 and the broad-band filter F200LP
}
\label{264.4-12.7}
\end{figure}

\begin{figure}[htb]
\vskip -0.6cm
\resizebox{0.77\textwidth}{!}{\includegraphics{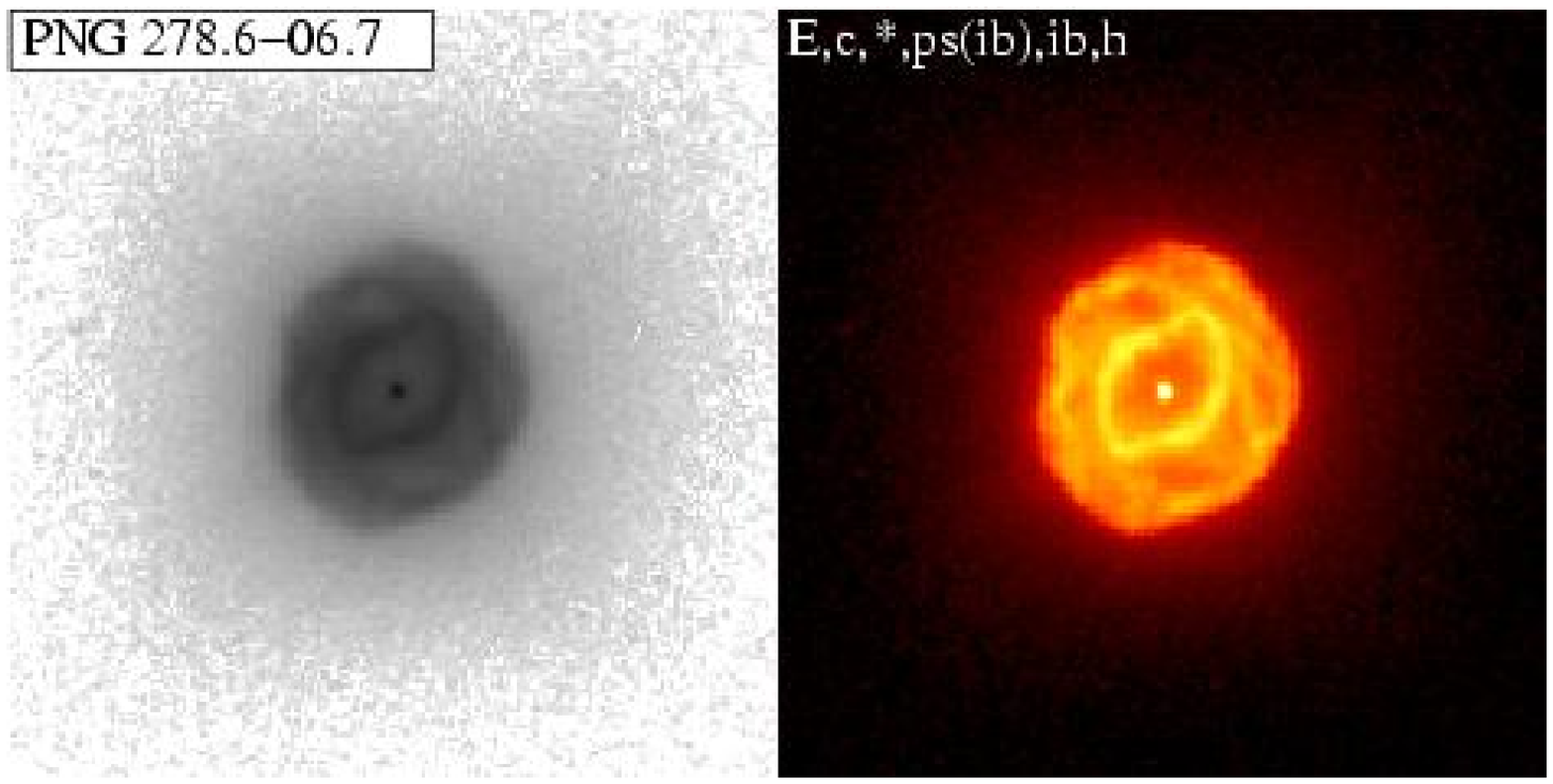}}
\caption{As in Fig\,1., but for PNG278.6-06.7 and the broad-band filter F200LP
}
\label{278.6-06.7}
\end{figure}

\begin{figure}[htb]
\vskip -0.6cm
\resizebox{0.77\textwidth}{!}{\includegraphics{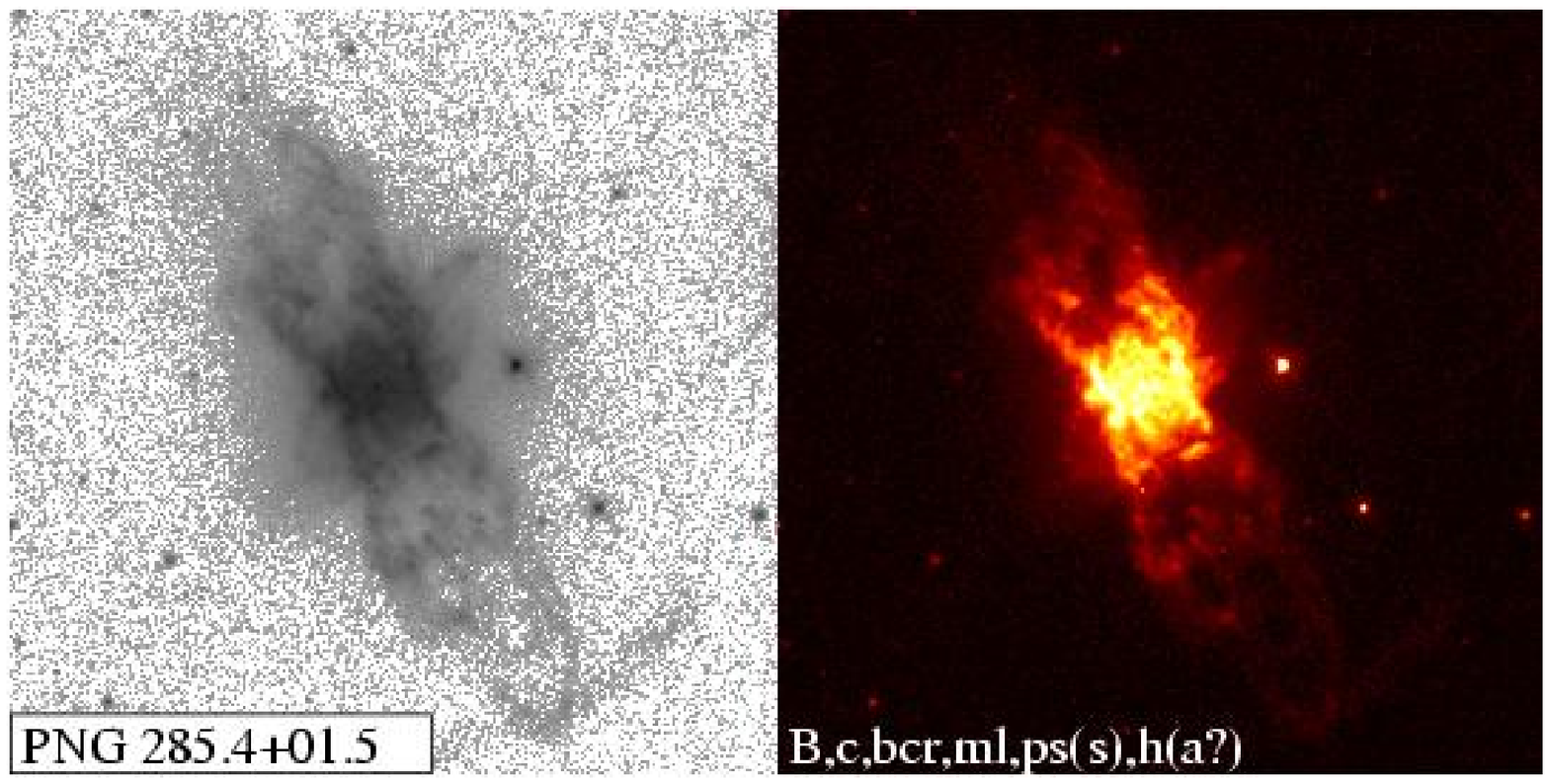}}
\caption{As in Fig\,1., but for PNG285.4+01.5 and the broad-band filter F200LP
}
\label{285.4+01.5}
\end{figure}

\begin{figure}[htb]
\vskip -0.6cm
\resizebox{0.77\textwidth}{!}{\includegraphics{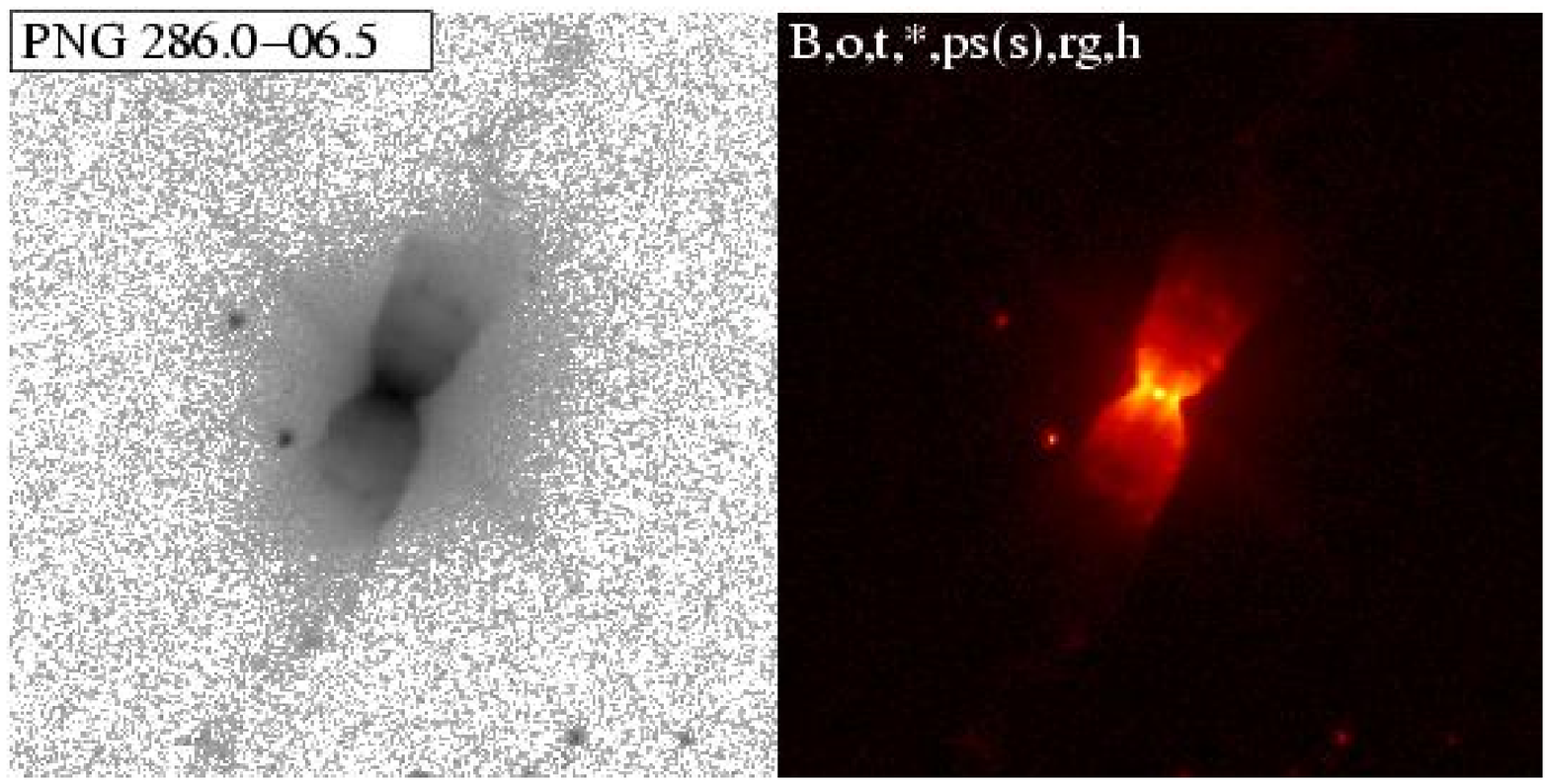}}
\caption{As in Fig\,1., but for PNG286.0-06.5 and the broad-band filter F200LP
}
\label{286.0-06.5}
\end{figure}

\begin{figure}[htb]
\vskip -0.6cm
\resizebox{0.77\textwidth}{!}{\includegraphics{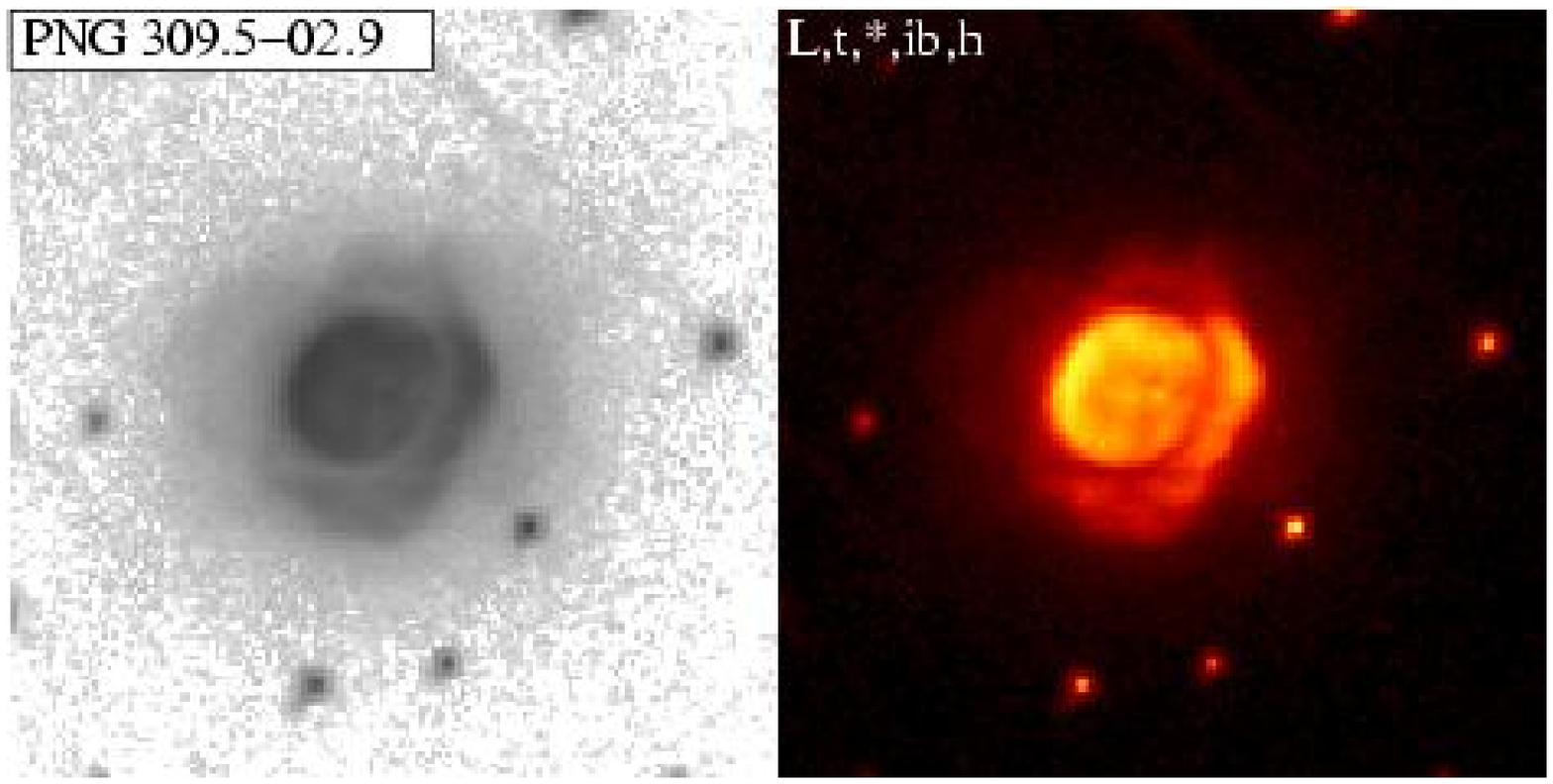}}
\caption{As in Fig\,1., but for PNG309.5-02.9 and the broad-band filter F200LP
}
\label{309.5-02.9}
\end{figure}

\begin{figure}[htb]
\vskip -0.6cm
\resizebox{0.77\textwidth}{!}{\includegraphics{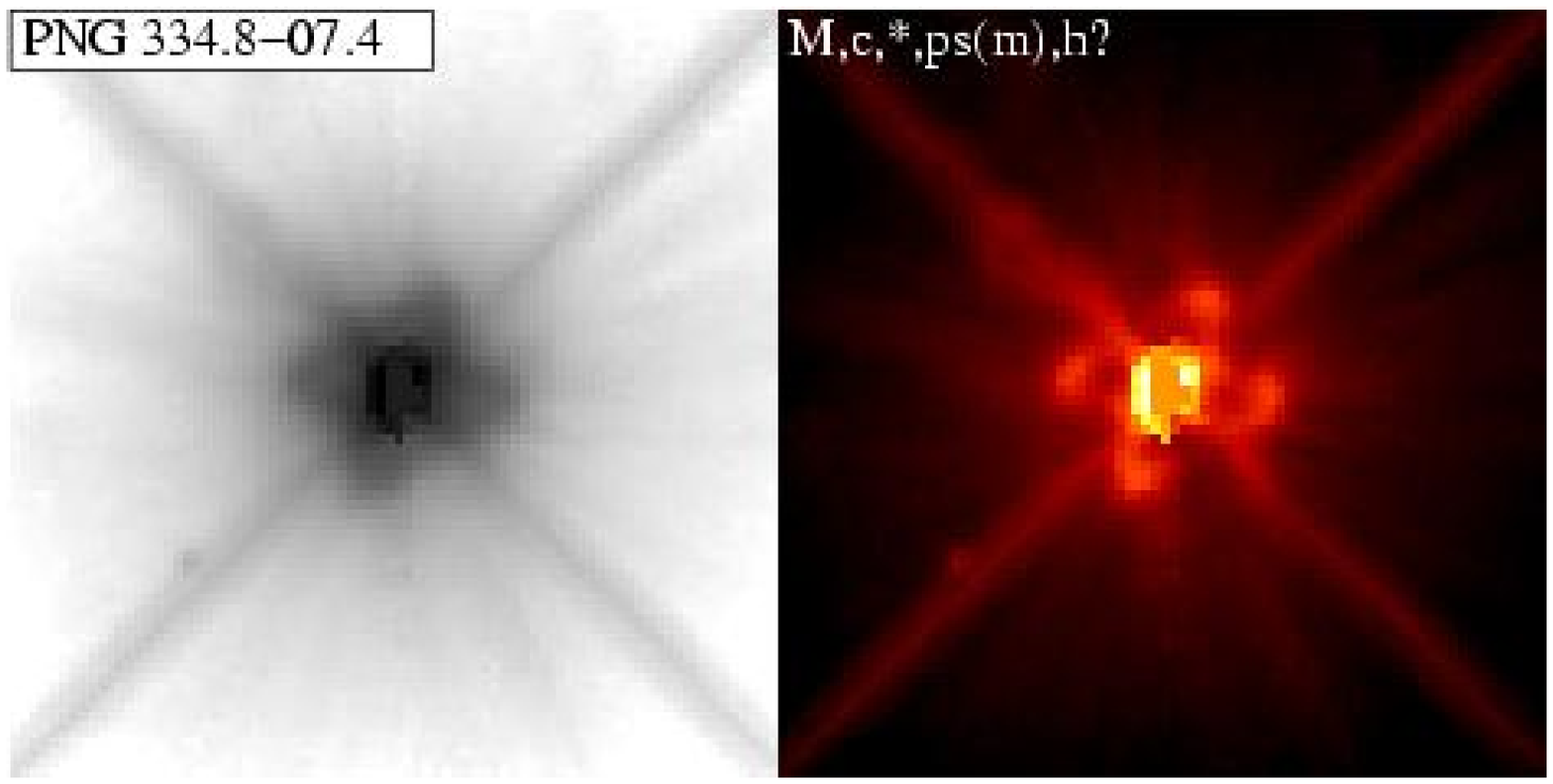}}
\caption{As in Fig\,1., but for PNG334.8-07.4 and the broad-band filter F200LP
}
\label{334.8-07.4}
\end{figure}

\begin{figure}[htb]
\vskip -0.7cm
\resizebox{0.77\textwidth}{!}{\includegraphics{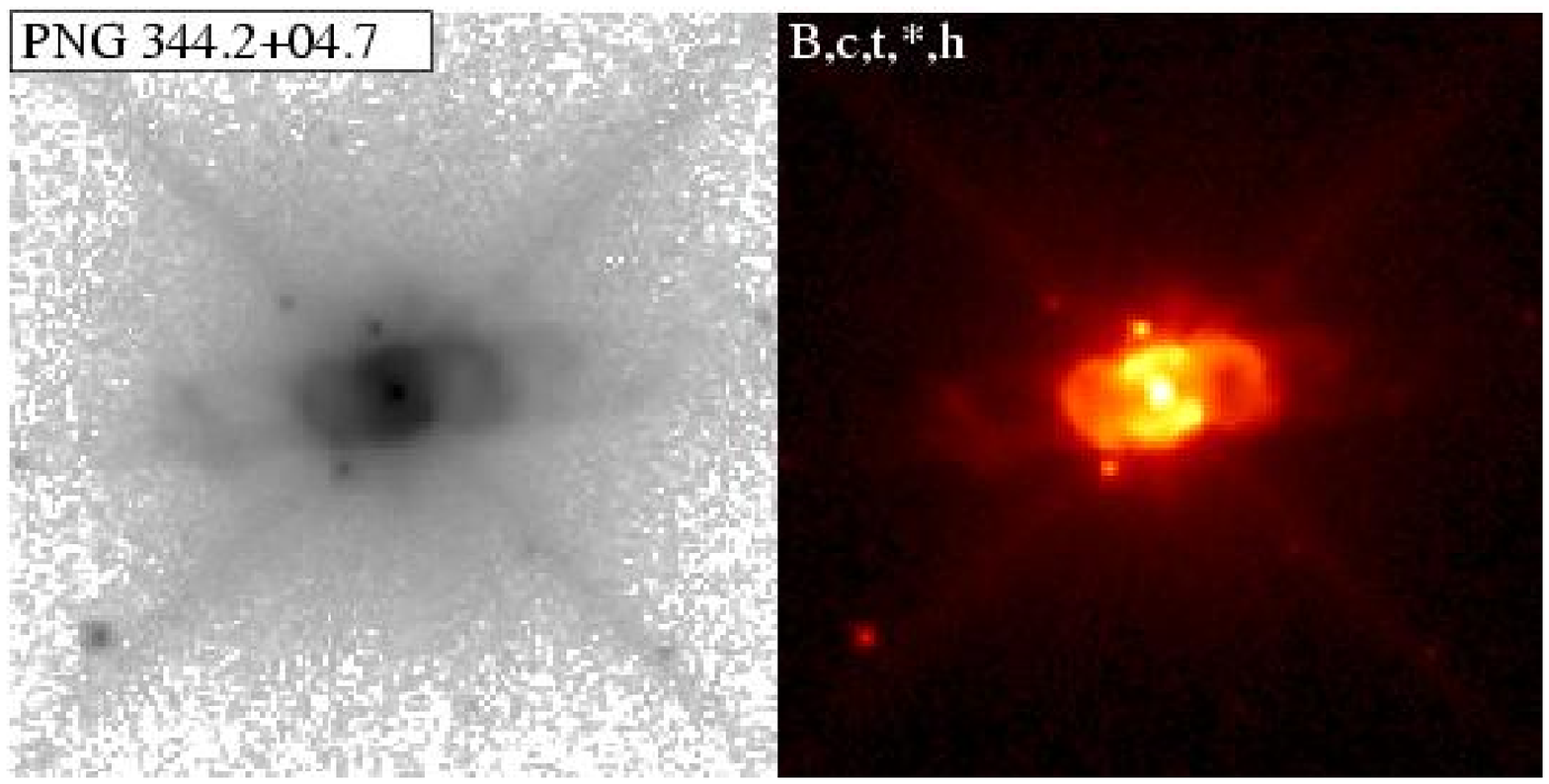}}
\caption{As in Fig\,1., but for PNG344.2+04.7 and the broad-band filter F200LP
}
\label{344.2+04.7}
\end{figure}

\begin{figure}[htb]
\vskip -0.7cm
\resizebox{0.77\textwidth}{!}{\includegraphics{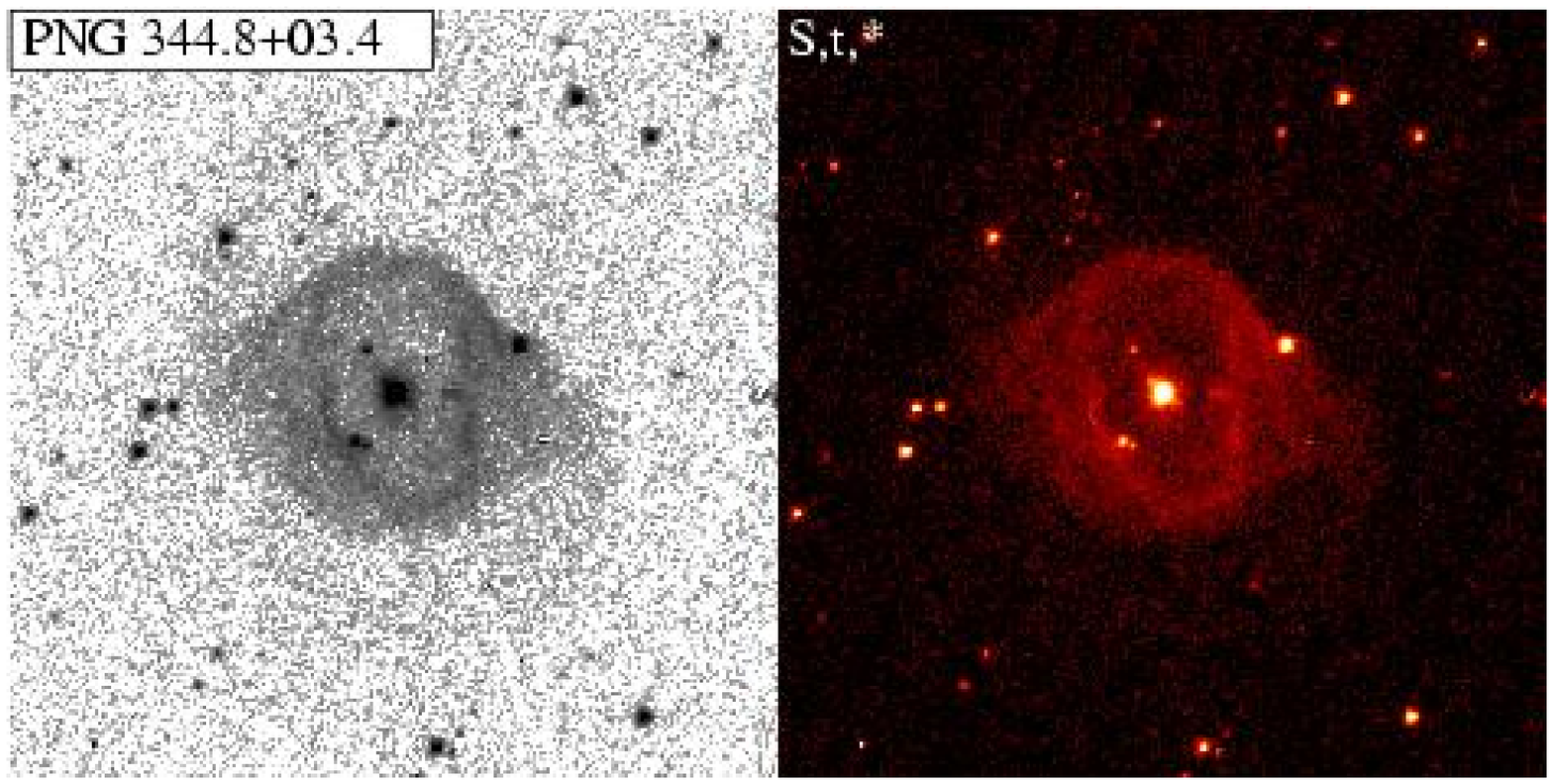}}
\caption{As in Fig\,1., but for PNG344.8+03.4 and the broad-band filter F200LP
}
\label{344.8+03.4}
\end{figure}

\begin{figure}[htb]
\vskip -0.7cm
\resizebox{0.77\textwidth}{!}{\includegraphics{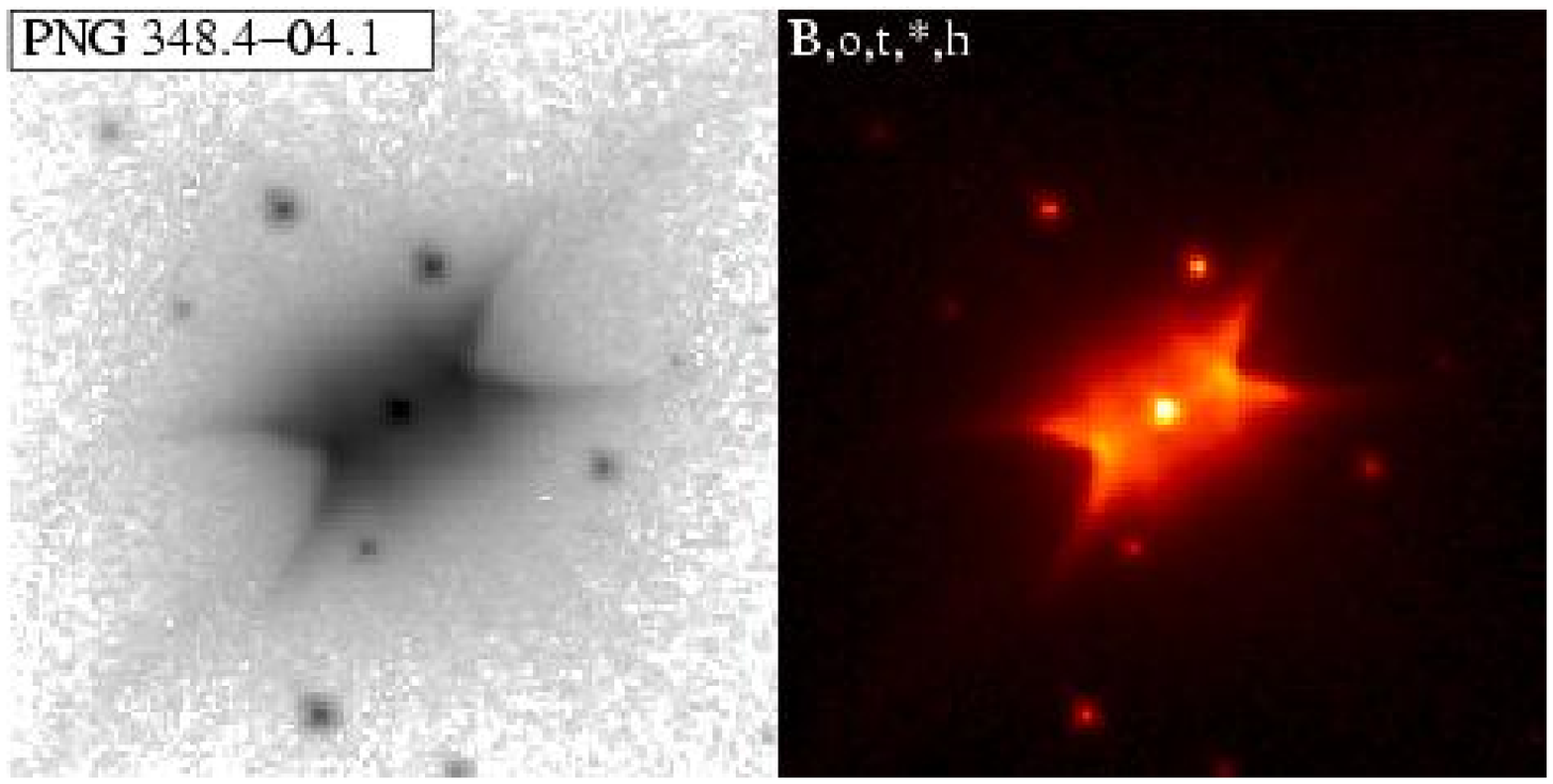}}
\caption{As in Fig\,1., but for PNG348.4-04.1 and the broad-band filter F200LP
}
\label{348.4-04.1}
\end{figure}

\begin{figure}[htb]
\vskip -0.7cm
\resizebox{0.77\textwidth}{!}{\includegraphics{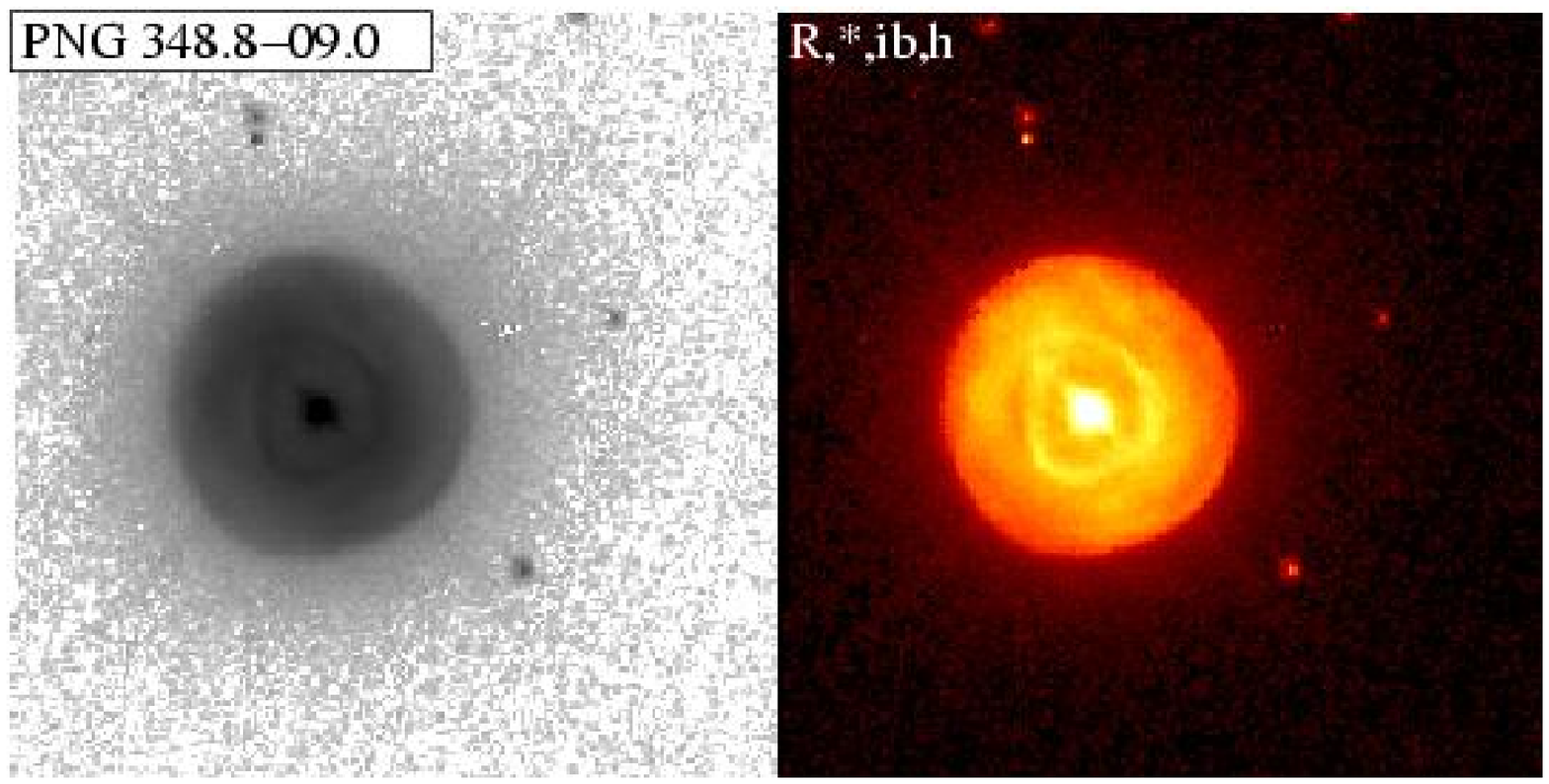}}
\caption{As in Fig\,1., but for PNG348.8-09.0 and the broad-band filter F200LP
}
\label{348.8-09.0}
\end{figure}
\clearpage
\scriptsize
\begin{center}
\begin{longtable}{llllll}
\caption[]{Log of Observations}\label{obslog}\\
\hline\hline \\[-2ex]
   \multicolumn{1}{l}{\textbf{Name\footnotemark[1]}} &
   \multicolumn{1}{l}{\textbf{Filter\footnotemark[2]}} &
   \multicolumn{1}{l}{\textbf{Exposure\footnotemark[3]}} &
   \multicolumn{1}{l}{\textbf{Date\footnotemark[4]}} &
   \multicolumn{1}{l}{\textbf{GO Prog\footnotemark[5]}} &
   \multicolumn{1}{l}{\textbf{dataset\footnotemark[6]}} \\[0.5ex] \hline
   \\[-1.8ex]
\endfirsthead

\multicolumn{2}{c}{{\tablename} \thetable{} -- Continued} \\[0.5ex]
  \hline \hline \\[-2ex]
   \multicolumn{1}{l}{\textbf{Name\footnotemark[1]}} &
   \multicolumn{1}{l}{\textbf{Filter\footnotemark[2]}} &
   \multicolumn{1}{l}{\textbf{Exposure\footnotemark[3]}} &
   \multicolumn{1}{l}{\textbf{Date\footnotemark[4]}} &
   \multicolumn{1}{l}{\textbf{GO Prog\footnotemark[5]}} &
   \multicolumn{1}{l}{\textbf{dataset\footnotemark[6]}} \\[0.5ex] \hline
  \\[-1.8ex]
\endhead

  \hline\\
  \multicolumn{3}{l}{{Continued on Next Page\ldots}} \\
\endfoot

  \\[-1.8ex] \hline \hline
\endlastfoot

\setcounter{footnote}{1}
Objects & with & R$_{exc}$ & $\le$\,\,1 \\
\hline 
 PK000+17D1 & F656N & 620 & 1999-08-12 & 8345 &  u5hh3102r \\  
 PNG001.2+02.1 & F656N & 280 & 2003-06-19 & 9356 &  u6mg1001m \\  
 PNG001.7-04.4 & F656N & 200 & 2002-08-12 & 9356 &  u6mg1301m \\ 
PNG002.8+01.7 & F656N & 280 & 2003-05-02 & 9356 &  u6mg0401m \\ 
    PK002-03D3 & F656N & 1240 & 1999-02-12 & 6353 &  u35t2105r \\  
    PK002-04D1 & F656N & 1120 & 1999-08-21 & 8345 &  u5hh4103r \\  
    PK002-09D1 & F656N & 480 & 2000-02-27 & 8345 &  u5hh5602r \\   
 PNG003.1+03.4 & F656N & 280 & 2002-07-02 & 9356 &  u6mg1401m \\  
 PNG003.6+03.1 & F656N & 280 & 2003-04-27 & 9356 &  u6mg1501m \\  
 PNG003.9-03.1 & F656N & 280 & 2003-04-26 & 9356 &  u6mg1701m \\  
 PNG004.0-03.0 & F656N & 200 & 2003-05-20 & 9356 &  u6mg5901m \\
 PNG004.8+02.0 & F656N & 400 & 2002-08-30 & 9356 &  u6mg1901m \\   
    PK004+04D1 & F656N & 620 & 2001-06-26 & 8345 &  u5hh4302r \\  
 PNG006.1+08.3 & F656N & 200 & 2002-07-06 & 9356 &  u6mg2201m \\  
 PNG006.8+04.1 & F656N & 200 & 2002-07-04 & 9356 &  u6mg2401m \\ 
PK006+02D5 & F656N & 620 & 1999-09-28 & 8345 &  u5hh6902r \\   
    PK007-04D1 & F656N & 620 & 1999-09-27 & 8345 &  u5hh0702m \\ 
 PNG008.2+06.8 & F656N & 200 & 2003-03-18 & 9356 &  u6mg2601m \\  
 PNG008.6-02.6 & F656N & 280 & 2002-09-15 & 9356 &  u6mg4701m \\  
PK010+00D1 (NGC6537) & F656N & 1240 & 1997-09-12 & 6502 & u42i0402b \\  
PK010+18D2 (M2-9) & F658N & 1300 & 2001-06-28 & 8773 &  u66b1405r \\
PK010-01D1 (NGC6578) & F658N & 800 & 1999-10-23 & 8390 &  u5hc0605r \\  
PNG012.2+04.9 (PM1-188) & F656N & 1120 & 2000-02-25 & 8345 &  u5hh3003r \\  
    PK015+03D1 & F656N & 1120 & 1999-07-23 & 8345 &  u5hh2803r \\  
    PK016-01D1 & F656N & 620 & 1999-07-17 & 8345 &  u5hh0902r \\  
    PK024+03D1 & F656N & 1120 & 2000-02-26 & 8345 &  u5hh2303r \\  
    PK027+04D1 & F656N & 520 & 2000-02-20 & 8307 &  u59b0604r \\  
    PK038+12D1 & F656N & 1020 & 1999-02-08 & 6353 &  u35t1905r \\  
    PK043+03D1 & F656N & 1120 & 2000-02-20 & 8345 &  u5hh1003r \\  
    PK045-02D1 & F656N & 1920 & 1996-11-29 & 6353 &  u35t3005t \\ 
    PK051+09D1 & F656N & 400 & 1996-10-17 & 6347 &  u39h0901t \\  
    PNG051.5+00.2 (IRAS19234+1627) & F606W & 676 & 2003-03-12 & 9463 & j8di68010 \\
    PK051-03D1 & F656N & 21 & 1999-07-14 & 8345 &  u5hh1101r \\  
    PK057-01D1 & F656N & 520 & 1999-11-09 & 8307 &  u59b0704r \\ 
PNG056.0+02.0 (IRAS19255+2123)  & F606W & 400 & 2002-02-17 & 9101 & u6fq0805r \\
    PK058-10D1 & F656N & 1120 & 1999-11-02 & 8345 &  u5hh1403r \\  
    PNG061.3+03.6 (IRAS19309+2646) & F606W & 600 & 2006-07-21 & 10536 & j9fj9010 \\
    PNG067.9-00.2 (IRAS20011+3024) & F606W & 1200 & 2006-03-29 & 10536 & j9fj97071 \\
    PK064+05D1 & F656N & 1000 & 1994-06-03 & 5403 &  u27q0208t \\
    PK068-02D1 & F656N & 1120 & 1999-11-06 & 8345 &  u5hh7303r \\  
    PK071-02D1 & F656N & 520 & 1999-11-04 & 8307 &  u59b0504r \\   
    PK082+07D1 & F656N & 1120 & 1999-08-11 & 8345 &  u5hh5203r \\  
    PK082+11D1 & F656N & 1240 & 1999-02-15 & 6353 &  u35t1005r \\  
PNG093.9-00.1 (IRAS21282+5050)\footnotemark & F656N & 1080 & 1999-08-23 & 8345 & u5hh7203r \\
                              & F606W & 500 & 2002-12-07 & 9463 &  j8di76011 \\
    PK100-08D1 & F656N & 2040 & 1996-08-05 & 6353 &  u35t1105t \\
PNG110.1+01.9 (IRAS22568+6141)  & F606W & 800  & 2002-12-18 & 9463 &  j8di58051 \\
    PK107-13D1 & F656N & 1120 & 1999-08-19 & 8345 &  u5hh1603r \\ 
PNG332.9-09.9 (IRAS17047-5650) & F435W & 450 & 2002-09-17 & 9463 & j8di80081 \\
PK111-02D1 (HB12-WFPC2) & F656N & 400 & 2001-09-24 & 9050 &  u6ci0401m \\  
    PK130-11D1 & F656N & 1120 & 1999-08-23 & 8345 &  u5hh5303r \\  
    PK146+07D1 & F656N & 700 & 1996-04-18 & 6353 &  u35t2401t \\  
    PK147-02D1 & F656N & 1120 & 1999-10-03 & 8345 &  u5hh5403r \\  
    PK165-06D1 & F656N & 1120 & 1999-08-24 & 8345 &  u5hh0103r \\  
    PK167-09D1 & F656N & 700 & 1996-04-08 & 6353 &  u35t2301t \\  
    PK211-03D1 & F656N & 620 & 1999-08-20 & 8345 &  u5hh1802r \\  
    PK215-24D1 & F656N & 888 & 1999-02-16 & 6353 &  u35t0905r \\  
    PK226-03D1 & F656N & 1120 & 1999-09-27 & 8345 &  u5hh1903r \\  
    PK232-04D1 & F656N & 620 & 1999-08-16 & 8345 &  u5hh3902r \\  
    PK235-01D1 & F656N & 1120 & 1999-09-26 & 8345 &  u5hh5503r \\  
    PK235-03D1 & F656N & 620 & 1999-10-08 & 8345 &  u5hh0502r \\  
    PK258-00D1 & F656N & 1120 & 1999-10-03 & 8345 &  u5hh2003r \\  
    PK285-02D1 & F656N & 720 & 1999-02-14 & 6353 &  u35t1407r \\  
    PK296-06D1 & F656N & 1240 & 1999-02-20 & 6353 &  u35t2605r \\  
    PK300-02D1 & F656N & 620 & 1999-09-24 & 8345 &  u5hh5802r \\  
PNG307.5-04.9 (MYCN18)  & F656N & 600  & 1995-07-30 & 6221 & u2rc0101t \\
    PK315+09D1 & F656N & 2040 & 1996-08-11 & 6353 &  u35t2805t \\  
    PK315-13D1 & F656N & 322 & 1996-10-04 & 6353 &  u35t0705t \\  
    PK320-09D1 & F656N & 700 & 1996-04-04 & 6353 &  u35t1501t \\  
    PK321+02D1 & F656N & 2040 & 1996-09-23 & 6353 &  u35t2905t \\  
    PK321+03D1 & F656N & 1120 & 1999-09-20 & 8345 &  u5hh7103r \\  
    PK325-12D1 & F656N & 1240 & 1999-02-20 & 6353 &  u35t0605r \\  
    PK326-06D1 & F656N & 1240 & 1999-02-13 & 6353 &  u35t2705r \\  
    PK327-02D1 & F656N & 2040 & 1996-08-12 & 6353 &  u35t2205t \\  
PK331-01D1 (MZ-3) & F656N & 900 & 1998-06-30 & 6856 & u47b0101b \\  
    PK331-02D2 & F656N & 1120 & 1999-06-06 & 8345 &  u5hh3303r \\  
    PK350+04D1 & F656N & 620 & 1999-08-29 & 8345 &  u5hh2502r \\  
 PNG351.1+04.8 & F656N & 160 & 2003-06-05 & 9356 &  u6mg2901m \\  
 PNG351.9-01.9 & F656N & 200 & 2003-05-26 & 9356 &  u6mg4801m \\ 
PNG352.6+03.0 & F656N & 200 & 2002-08-11 & 9356 &  u6mg3001m \\ 
    PK352-07D1 & F656N & 480 & 1999-08-23 & 8345 &  u5hh1302r \\  
 PNG354.5+03.3 & F656N & 280 & 2003-05-04 & 9356 &  u6mg5001m \\  
 PNG354.9+03.5 & F656N & 280 & 2002-10-15 & 9356 &  u6mg4901m \\  
 PNG355.4-02.4 & F656N & 200 & 2002-08-11 & 9356 &  u6mg3101m \\  
 PNG355.9+03.6 & F656N & 280 & 2003-06-20 & 9356 &  u6mg3301m \\ 
    PK355-03D3 & F656N & 1120 & 2001-06-26 & 8345 &  u5hh6503r \\  
    PK355-04D2 & F656N & 620 & 2000-02-23 & 8345 &  u5hh3702r \\  
  PNG356.5-03.6 & F656N & 360 & 2002-09-18 & 9356 &  u6mg4101m \\  
 PNG356.8+03.3 & F656N & 280 & 2002-07-16 & 9356 &  u6mg3501m \\ 
    PK356-03D3 & F656N & 1120 & 2001-06-26 & 8345 &  u5hh7503r \\   
 PNG357.1-04.7 & F656N & 200 & 2003-06-19 & 9356 &  u6mg3601m \\  
 PNG357.2+02.0 & F656N & 280 & 2002-08-14 & 9356 &  u6mg4201m \\  
PNG358.5-04.2 & F656N & 160 & 2003-06-19 & 9356 &  u6mg3801m \\
 PNG358.7+05.2 & F656N & 280 & 2002-09-18 & 9356 &  u6mg3901m \\  
 PNG358.9+03.4 & F656N & 200 & 2003-06-19 & 9356 &  u6mg5301m \\ 
    PK358-00D2 & F656N & 1720 & 1996-10-04 & 6353 &  u35t1807t \\  
 PNG359.2+04.7 & F656N & 280 & 2003-03-30 & 9356 &  u6mg5201m \\  
\hline
Objects & with & R$_{exc}$ & $>$\,\,1 \\
\hline 
PNG002.3-03.4 & F656N & 280 & 2003-04-26 & 9356 &  u6mg0501m \\  
PNG002.9-03.9 & F656N & 280 & 2003-04-27 & 9356 &  u6mg0601m \\  
PNG003.8+05.3 & F656N & 280 & 2002-09-12 & 9356 &  u6mg1601m \\
PK003+02D1 & F656N & 600 & 1996-08-17 & 6347 &  u39h0401t \\ 
PNG004.1-03.8 & F656N & 280 & 2002-09-11 & 9356 &  u6mg1801n \\
PNG004.8-22.7 (HE2-436) & F656N & 200 & 2003-05-03 & 9356 &  u6mg0201m \\  
PNG005.2-18.6 (STWR2-21) & F656N & 280 & 2003-03-16 & 9356 &  u6mg0301m \\  
PNG006.8-19.8 (WRAY16-423) & F656N & 200 & 2002-07-25 & 9356 &  u6mg0101m \\ 
PNG006.3+04.4 & F656N & 280 & 2003-04-02 & 9356 &  u6mg6001m \\
PK008-07D2 & F656N & 1120 & 2000-02-26 & 8345 &  u5hh4503r \\  
PK010-06D1 & F656N & 1120 & 1999-08-26 & 8345 &  u5hh4603r \\  
PK013+04D1 & F656N & 620 & 2001-06-25 & 8345 &  u5hh4702r \\ 
PK019-05D1 & F656N & 240 & 1999-11-08 & 8307 &  u59b0301r \\  
PK023-02D1 & F656N & 620 & 1999-09-24 & 8345 &  u5hh4802r \\
PK027-09D1 & F656N & 1120 & 2001-06-25 & 8345 &  u5hh4903r \\  
PK032+07D2 & F658N & 600 & 1999-06-29 & 6347 &  u39h2901r \\ 
PK032-02D1 & F656N & 600 & 1998-11-21 & 6347 &  u39h3201r \\  
PK037-06D1 & F656N & 160 & 1999-10-31 & 8307 &  u59b0101r \\ 
PK060-07D2 & F656N & 1120 & 1999-07-13 & 8345 &  u5hh5003r \\ 
PK074+02D1 & F656N & 280 & 1999-11-01 & 8307 &  u59b0401r \\  
PK089-05D1 & F656N & 240 & 1999-11-02 & 8307 &  u59b0201r \\  
PK304-04D1 & F656N & 1120 & 2000-01-29 & 8345 &  u5hh5903r \\
PNG356.9+04.4 & F656N & 280 & 2003-04-18 & 9356 &  u6mg5601m \\

\end{longtable}
\end{center}
\setcounter{footnote}{0}
\footnote{Col. 1: Object name as defined in GO program; if this name was not in standard PK or PNG format, then it is
provided
in parenthesis, and the PK or PNG name is given}
\footnote{Col. 2: The filter used for the images utilised for
morphological classification. This is generally the narrow-band F656N filter (which covers the H$\alpha$ line), but occasionally the
F658N filter (which covers the [NII] line) has been used. For a few objects, only broad-band filter images (F435W or F606W) were
available.}
\footnote{Col. 3: The total exposure time of the image.}
\footnote{Col. 4: The observation date (yyyy-mm-dd).}
\footnote{Col. 5: The GO program ID number.}
\footnote{Col. 6: the name for the image dataset as listed in the HST archive - when several exposures have been averaged, then it is
the name of the last dataset.}
\footnote{For this object, although both the F656N and F606W (broad-band) images are presented, only the latter has been used
for the morphological classification (explanation in text).}
\normalsize


\clearpage
\begin{footnotesize}
\begin{center}
\begin{longtable}{ll}
\caption[]{Morphological Classification Codes\label{codes}}\\
\hline\hline \\[-2ex]
\endfirsthead
\multicolumn{2}{c}{{\tablename} \thetable{} -- Continued} \\[0.5ex]
\endhead

\multicolumn{2}{c}{PRIMARY CLASSIFICATION: Nebular Shape} \\
B & Bipolar \\
M & Multipolar \\
E & Elongated \\
I & Irregular \\
{\it R} & {\it Round} \\
{\it L} & {\it Collimated Lobe Pair}\\
{\it S} & {\it Spiral Arm}\\
\tableline \\
\multicolumn{2}{c}{SECONDARY CHARACTERISTICS} \\
\multicolumn{2}{l}{{\bf Lobes}} \\
o & lobes open at ends\\
c & lobes closed at ends \\
\\
\multicolumn{2}{l}{{\bf Central Region}} \\
w & central region is (relatively) dark, and shows an obscuring waist \\
{\it t}   & {\it central region is bright and has a toroidal structure} \\
{\it bcr} & {\it central region is bright and barrel-shaped}\\
{\it bcr(c)} & {\it barrel has closed ends}\\
{\it bcr(o)} & {\it barrel has open ends}\\
{\it bcr(i)} & {\it irregular structure present in barrel interior}\\
\\
\multicolumn{2}{l}{{\bf Central Star}} \\
{\bf $\star$} & central star evident in optical images \\
{\bf $\star$}(nnn) & star is offset is offset from the center of symmetry of one or more nebular structures,\\ 
                   & {\it nnn} is maximum offset in milliarcsec\\
\\
\\
\\
\multicolumn{2}{l}{{\bf Other Nebular Characteristics}} \\
an & ansae \\
ml & minor lobes \\
sk & a skirt-like structure around the primary lobes \\
{\it ib} & {\it an inner bubble inside the primary nebular structure}\\
{\it wv} & {\it a patterned structure, such as a weave or a mottling}\\
{\it rg} & {\it rings projected on lobes}\\
{\it rr} & {\it radial rays}\\
{\it pr} & {\it one or more pairs of diametrically opposed protrusions on the primary geometrical shape }\\
{\it ir} & {\it additional unclassified nebular structure lacking symmetry, not covered by}\\ 
         & {\it the primary or secondary classifications}\\
\\
\multicolumn{2}{l}{{\bf Point Symmetry}} \\
ps(m)  & due to presence of two or more pairs of diametrically-opposed lobes \\
ps(an) & due to diametrically-opposed ansae\\
ps(s)  & overall geometric shape of lobes is point-symmetric \\
{\it ps(t)}  & {\it waist has point-symmetric structure}\\
{\it ps(bcr)}  & {\it barrel-shaped central region has point-symmetric structure}\\
{\it ps(ib)} & {\it inner bubble has point-symmetric structure}\\
\\

\multicolumn{2}{l}{{\bf Halo}} \\
h    & halo (relatively low-surface brightness diffuse region around primary nebular
structure) is present \\
h(e) & halo has elongated shape \\
h(i) & halo has indeterminate shape \\
h(a) & halo has centro-symmetric arc-like features \\
h(sb)  & halo shows searchlight-beams \\
{\it h(d)}  & {\it halo has a sharp outer edge, or shows a discontinuity in its interior} \\
\tableline
\end{longtable}
\end{center}
\end{footnotesize}
\clearpage
%
\scriptsize
\begin{landscape}
\begin{longtable}{p{0.55in}p{0.65in}p{0.1in}p{0.4in}p{0.5in}p{0.6in}p{0.2in}p{0.2in}p{0.15in}p{0.2in}p{0.2in}p{0.2in}p{0.2in}p{0.25in}}
\caption[]{Properties of Young Planetary Nebulae}\label{morphs}\\
\hline \\[-2ex]
   \multicolumn{1}{l}{\textbf{Name}} &
   \multicolumn{1}{l}{\textbf{Name}} &
   \multicolumn{4}{c}{\textbf{Morphology}} &
   \multicolumn{1}{l}{\textbf{R$_{exc}$}} &
   \multicolumn{1}{l}{\textbf{Size}} &
   \multicolumn{1}{l}{\textbf{V$_{1}$}} &
   \multicolumn{1}{l}{\textbf{V$_{2}$}} &
   \multicolumn{1}{l}{\textbf{D}} &
   \multicolumn{1}{l}{\textbf{Age}} &
   \multicolumn{1}{l}{\textbf{XBox}} &
   \multicolumn{1}{l}{\textbf{Fig\,\#}} \\[0.5ex]
   \multicolumn{1}{l}{(\footnotemark[1])} &
   \multicolumn{1}{l}{(\footnotemark[2])} &
   \multicolumn{4}{c}{(\footnotemark[3])} &
   \multicolumn{1}{l}{(\footnotemark[4])} &
   \multicolumn{1}{l}{(\footnotemark[5])} &
   \multicolumn{1}{l}{(\footnotemark[6])} &
   \multicolumn{1}{l}{(\footnotemark[7])} &
   \multicolumn{1}{l}{(\footnotemark[8])} &
   \multicolumn{1}{l}{(\footnotemark[9])} &
   \multicolumn{1}{l}{(\footnotemark[10])} &
  \multicolumn{1}{l}{(\footnotemark[11])}  \\[0.5ex]
   \multicolumn{1}{l}{\textbf{PK}} &
   \multicolumn{1}{l}{\textbf{PNG}} &
   \multicolumn{1}{l}{\textbf{P}} &
   \multicolumn{1}{l}{\textbf{Cen}} &
   \multicolumn{1}{l}{\textbf{PS}} &
   \multicolumn{1}{l}{\textbf{Other}} &
   \multicolumn{1}{l}{} &
   \multicolumn{1}{l}{\textbf{$''$}} &
   \multicolumn{2}{c}{\textbf{$\kms$}} &
   \multicolumn{1}{l}{\textbf{kpc}} &
   \multicolumn{1}{l}{\textbf{yr}} &
   \multicolumn{1}{l}{\textbf{$''$}} &
   \multicolumn{1}{l}{\textbf{}}  \\[0.5ex] \hline
   \\[-1.8ex]
\endfirsthead

\multicolumn{3}{c}{{\tablename} \thetable{} -- Continued} \\[0.5ex]
  \hline \\[-2ex]
   \multicolumn{1}{l}{\textbf{Name}} &
   \multicolumn{1}{l}{\textbf{Name}} &
   \multicolumn{4}{c}{\textbf{Morphology}} &
   \multicolumn{1}{l}{\textbf{R$_{exc}$}} &
   \multicolumn{1}{l}{\textbf{Size}} &
   \multicolumn{1}{l}{\textbf{V$_{1}$}} &
   \multicolumn{1}{l}{\textbf{V$_{2}$}} &
   \multicolumn{1}{l}{\textbf{D}} &
   \multicolumn{1}{l}{\textbf{Age}} &
   \multicolumn{1}{l}{\textbf{XBox}} &
   \multicolumn{1}{l}{\textbf{Fig\#}} \\[0.5ex]
   \multicolumn{1}{l}{\textbf{PK}} &
   \multicolumn{1}{l}{\textbf{PNG}} &
   \multicolumn{1}{l}{\textbf{P}} &
   \multicolumn{1}{l}{\textbf{Cen}} &
   \multicolumn{1}{l}{\textbf{PS}} &
   \multicolumn{1}{l}{\textbf{Other}} &
   \multicolumn{1}{l}{} &
   \multicolumn{1}{l}{\textbf{$''$}} &
   \multicolumn{2}{c}{\textbf{$\kms$}} &
   \multicolumn{1}{l}{\textbf{kpc}} &
   \multicolumn{1}{l}{\textbf{yr}} &
   \multicolumn{1}{l}{\textbf{$''$}} &
   \multicolumn{1}{l}{\textbf{}}  \\[0.5ex] \hline
   \\[-1.8ex]
\endhead

  \hline\\
  \multicolumn{3}{l}{{Continued on Next Page\ldots}} \\
\endfoot

  \\[-1.8ex] \hline \hline
\endlastfoot

\setcounter{footnote}{11}
Objects & with & R$_{exc}$ & $\le$\,\,1 \\
\hline 	 		 		 		 		 	 	 	 	 	 	 	 	 
000+17\#1 &	000.1+17.2 &	 B,c  	&	 t,*       	&	 ...                	&	h	&	0.31 &	6.02 &	22 &	... &	6.16 &	3991 &	8.203 &	\ref{000+17d1} \\
001+02\#1 &	001.2+02.1 &	 E    	&	 t,*       	&	 ...                	&	...	&	0.41 &	2.32 &	22 &	... &	5.42 &	1356 &	6.836 &	\ref{001.2+02.1} \\
001-04\#1 &	001.7-04.4\footnotemark &	 L    	&	 bcr(o),*  	&	 ...    &	...	&	0.02 &	3.29 &	22 &	... &	7.1 &	2514 &	5.013 &	\ref{001.7-04.4} \\
002+01\#1 &	002.8+01.7 &	 E,c  	&	 t,*       	&	 ps(s)              	&	...	&	0.01 &	3.89 &	22 &	... &	4.28 &	1796 &	5.468 &	\ref{002.8+01.7} \\
002-03\#3 &	002.6-03.4 &	 M,c  	&	 t,*       	&	 ps(m)              	&	h(a)	&	0 &	4.33 &	22 &	... &	4.2 &	1959 &	8.203 &	\ref{002-03d3} \\
002-04\#1 &	002.1-04.2 &	 M    	&	 bcr(o)   	&	 ps(m)              	&	h(a)	&	0.82 &	3.59 &	22 &	... &	4.9 &	1895 &	5.468 &	\ref{002-04d1} \\
002-09\#1 &	002.2-09.4 &	 B,o  	&	 bcr(o,i),*	&	 ps(s)              	&	h	&	$<1$ &	10.74 &	... &	18 &	4.93 &	6969 &	13.671 &	\ref{002-09d1} \\
003+03\#1 &	003.1+03.4 &	 E,c  	&	 *        	&	 ...                	&	...	&	0 &	3.9 &	22 &	... &	5.88 &	2471 &	6.836 &	\ref{003.1+03.4} \\
003+03\#2 &	003.6+03.1\footnotemark &	 B,c  	&	 t        	&	 ps(s)  &	ml,h	&	0.19 &	5.4 &	22 &	... &	8 &	4657 &	6.836 &	\ref{003.6+03.1} \\
... &	003.9-03.1 &	 I    	&	 ...      	&	 ...                	&	...	&	0.98 &	7.1 &	22 &	... &	8 &	6118 &	9.57 &	\ref{003.9-03.1} \\
004-03\#1 &	004.0-03.0\footnotemark &	 E    	&	 t,*       	&	 ...    &	h(d)	&	0.98 &	4.83 &	22 &	... &	4.04 &	2103 &	8.203 &	\ref{004.0-03.0} \\
004+02\#1 &	004.8+02.0 &	 R    	&	 *        	&	 ...    &	pr	&	             0.04    &	3.55 &	22 &	... &	5.23 &	2000 &	5.468 &	\ref{004.8+02.0} \\
004+04\#1 &	004.9+04.9\footnotemark &	 E,c  	&	 *        	&	 ...    &	h(e,d)	&	0.61 &	3.67 &	22 &	... &	4.18 &	1651 &	11.38 &	\ref{004+04d1} \\
006+08\#1 &	006.1+08.3 &	 E,c  	&	 ...      	&	 ...                	&	ib,h	&	0.52 &	2.56 &	22 &	... &	2.88 &	795 &	5.468 &	\ref{006.1+08.3} \\
006+04\#2 &	006.8+04.1 &	 L,c  	&	 bcr(c),*  	&	 ps(s)              	&	...	&	0.76 &	12.12 &	22 &	... &	3.19 &	4157 &	13.671 &	\ref{006.8+04.1} \\
006+02\#5 &	006.4+02.0 &	 M,c  	&	 t,*       	&	 ps(m)              	&	h	&	0.71 &	4.09 &	22 &	... &	4.22 &	1859 &	6.836 &	\ref{006+02d5} \\
007-04\#1 &	007.8-04.4 &	 E,c  	&	 *        	&	 ps(s)              	&	ib,h	&	0 &	2.04 &	22 &	... &	6.46 &	1416 &	4.557 &	\ref{007-04d1} \\
008+06\#1 &	008.2+06.8 &	 L    	&	 t,*       	&	 ps(s,t)            	&	h	&	0 &	1.93 &	22 &	... &	6.94 &	1442 &	4.557 &	\ref{008.2+06.8} \\
... &	008.6-02.6 &	 S    	&	 *        	&	 ...                	&	h	&	0.63 &	3.37 &	22 &	... &	8 &	2908 &	4.557 &	\ref{008.6-02.6} \\
010+00\#1 &	010.1+00.7 &	 B,o  	&	 ...      	&	 ps(s)              	&	...	&	0.78 &	46.18 &	... &	18 &	1.07 &	6507 &	68.28 &	\ref{010+00d1} \\
010+18\#2 &	010.8+18.0\footnotemark &	 B,o  	&	 *        	&	 ...    &	an,sk	&	0.11 &	61.3 &	... &	31 &	0.64 &	3000 &	69.671 &	\ref{010+18d2} \\
010-01\#1 &	010.8-01.8 &	 E,c  	&	 *        	&	 ...                	& an,wv,ir,h(ir) &	1.02 &	6.82 &	22 &	... &	2.31 &	1697 &	15.932 &	\ref{010-01d1} \\
... &	012.2+04.9\footnotemark &	 R    	&	 *        	&	 ...            &	ib	&	$<1$ &	14 &	22 &	... &	1.6 &	2413 &	18.228 &	\ref{012.2+04.9} \\
015+03\#1 &	015.9+03.3 &	 B,c  	&	 bcr(o),*  	&	 ...                	&	ib?	&	0.03 &	7.69 &	22 &	... &	2.69 &	2229 &	11.39 &	\ref{015+03d1} \\
016-01\#1 &	016.4-01.9 &	 R    	&	 t,*       	&	 ...                	&	ib,pr	&	0.09 &	11.09 &	... &	7 &	2.3 &	8634 &	19.139 &	\ref{016-01d1} \\
024+03\#1 &	024.1+03.8 &	 B,c  	&	 bcr(i),*  	&	 ...                	&	h	&	0.2 &	9.86 &	22 &	... &	4.82 &	5120 &	11.393 &	\ref{024+03d1} \\
027+04\#1 &	027.6+04.2 &	 L,c  	&	 bcr(c)   	&	 ...                	&	pr,h(e)	&	0.12 &	2.33 &	22 &	... &	1.43 &	359 &	6.828 &	\ref{027+04d1} \\
038+12\#1 &	038.2+12.0 &	 E,c  	&	 t,*       	&	 ...                	&	h	&	0.03 &	4.60 &	... &	10 &	3.58 &	3900 &	7.972 &	\ref{038+12d1} \\
043+03\#1 &	043.1+03.8\footnotemark &	 E,c  	&	 *        	&	 ...    &	wv,h(a)	&	0.02 &	3.68 &	11.5 &	... &	5.87 &	4453 &	13.671 &	\ref{043+03d1} \\
045-02\#1 &	045.4-02.7 &	 I    	&	 ...      	&	 ...                	&	...	&	0.59 &	3.24 &	25 &	17.5 &	2.03 &	624 &	8.203 &	\ref{045-02d1} \\
051+09\#1 &	051.4+09.6 &	 B,c  	&	 bcr(o),*  	&	 ps(s)              	&	ib,ir,h	&	$<1$ &	10.34 &	9.5 &	17 &	1 &	1441 &	10.025 &	\ref{051+09d1} \\
... & 051.5+00.2\footnotemark &  E,c    &        *              &        ps(s)                  &       pr      &       $<1$ &   9.35 & ...  &   ... &   ...  &   ... &   20    & \ref{067.9-00.2-f606w} \\
051-03\#1 &	051.9-03.8 &	 E    	&	 bcr,*     	&	 ps(s)              	&	ib	&	0.56 &	8.76 &	... &	11 &	3.88 &	7327 &	11.393 &	\ref{051-03d1} \\
056+02\#1 &	056.0+02.0 &	 B    	&	 w        	&	 ps(s)              	&	an,h(e)	&	0.62 &	6.72 &	22 &	... &	3.98 &	2880 &	7.791 &	\ref{056.0+02.0-ha-cont} \\
057-01\#1 &	057.9-01.5 &	 M,c  	&	 t,*?      	&	 ps(m,t)            	&	h	&	0.25 &	4.36 &	22 &	... &	2.87 &	1349 &	6.834 &	\ref{057-01d1} \\
058-10\#1 &	058.3-10.9\footnotemark &	 B,c  	&	 t?,*      	&	 ...    &	...	&	0.54 &	7.2 &	21.5 &	14.5 &	2.3 &	1825 &	9.114 &	\ref{058-10d1} \\
061+03\#1 &	061.3+03.6\footnotemark &	 B,c  	&	 *        	&	ps(s,ib)&	sk,ib	&	0.18 &	44.08 &	22 &	... &	1 &	4749 &	48.000 &	\ref{061.3+03.6-f606w} \\
064+05\#1 &	064.7+05.0\footnotemark &	 E,c  	&	 *        	&	 ...    &	rr,an?,h(e,d)	&	0 &	5.27 &	28 &	23 &	0.95 &	422 &	22.370 &	\ref{064+05d1} \\
067-00\#1 &	067.9-00.2\footnotemark &	 I    	&	 ...      	&	 ...    &	...	&	0.35 &	3.17 &	22 &	... &	2.46 &	841 &	7.500 &	\ref{067.9-00.2-f606w} \\
068-02\#1 &	068.3-02.7 &	 I    	&	 t ,*       	&	 ...                	&	h(i)	&	0 &	3.35 &	22 &	... &	3.35 &	1210 &	6.608 &	\ref{068-02d1} \\
071-02\#1 &	071.6-02.3\footnotemark &	 M,c  	&	 bcr,*     	&	 ps(m)  &	h	&	0.73 &	2.89 &	30 &	17 &	1.76 &	402 &	5.924 &	\ref{071-02d1} \\
082+07\#1 &	082.1+07.0 &	 M,o  	&	 t        	&	 ps(m,s,ib)         	&	ib,h	&	0 &	9.59 &	... &	23 &	1.7 &	1680 &	18.224 &	\ref{082+07d1} \\
082+11\#1 &	082.5+11.3 &	 E    	&	 ...      	&	 ...                	&	h	&	$<1$ &	0.82 &	... &	13 &	4.76 &	713 &	6.836 &	\ref{082+11d1} \\
... &	093.9-00.1\footnotemark &	 M,c  	&	 *        	&	 ps(m)          &	...	&	0 &	9.56 &	22 &	... &	3 &	3088 &	12.5 &	\ref{093.9-00.1} \\
100-08\#1 &	100.0-08.7 &	 M,c  	&	 bcr(o)   	&	 ps(m)              	&	h	&	$<1$ &	3.02 &	15.5 &	8 &	3.45 &	1595 &	8.203 &	\ref{100-08d1} \\
107-13\#1 &	107.6-13.3 &	 E,c  	&	 *        	&	 ...                	&	wv,h(d)	&	$<1$ &	9.39 &	... &	14 &	8.6 &	13676 &	13.668 &	\ref{107-13d1} \\
... &	110.1+01.9\footnotemark &	 B    	&	 w        	&	 ...            &	an?	&	0 &	5.48 &	... &	70 &	6 &	1112 &	9.000 &	\ref{cpd-22568} \\
111-02\#1 &	111.8-02.8 &	 B,o  	&	 ...      	&	 ...                	&	rg,h	&	$<1$ &	10.12 &	... &	14 &	2.05 &	3504 &	13.671 &	\ref{111-02d1} \\
130-11\#1 &	130.3-11.7\footnotemark &	 I    	&	 *?       	&	 ...    &	...	&	0.73 &	10.43 &	38 &	39 &	6.1 &	3968 &	13.671 &	\ref{130-11d1} \\
146+07\#1 &	146.7+07.6\footnotemark &	 B,c  	&	 t,*       	&	 ps(s)  &	h	&	0 &	2.77 &	17 &	7.5 &	7.15 &	2763 &	5.468 &	\ref{146+07d1} \\
147-02\#1 &	147.4-02.3\footnotemark &	 E,c  	&	 t,*       	&	 ...    &	ib	&	0.68 &	7.36 &	... &	13 &	2.3 &	3087 &	10.025 &	\ref{147-02d1} \\
165-06\#1 &	165.5-06.5 &	 B,c  	&	 t        	&	 ...                	&	ir,h(i)	&	0.7 &	6.82 &	22.5 &	20.5 &	3.45 &	2476 &	8.201 &	\ref{165-06d1} \\
167-09\#1 &	167.4-09.1 &	 B,o  	&	 t,*       	&	 ps(an,s)           	&	an,	&	0.19 &	5.49 &	30.5 &	16.5 &	6.68 &	2852 &	6.836 &	\ref{167-09d1} \\
211-03\#1 &	211.2-03.5 &	 L,c  	&	 bcr(o),*  	&	 ...                	&	h	&	0.1 &	5.76 &	22 &	... &	2.65 &	1644 &	6.836 &	\ref{211-03d1} \\
215-24\#1 &	215.2-24.2 &	 E,c  	&	 *        	&	 ps(s)              	&	ib,wv,pr,h	&	$<1$ &	16.87 &	12 &	$<6$ &	0.8 &	2666 &	22.785 &	\ref{215-24d1} \\
226-03\#1 &	226.4-03.7 &	 E,c  	&	 *        	&	 ps(s)              	&	wv,h(e,d)	&	$<1$ &	11.12 &	22 &	... &	... &	... &	15.946 &	\ref{226-03d1} \\
232-04\#1 &	232.8-04.7\footnotemark &	 E,c  	&	 t,*       	&	 ...    &	h(e,d)	&	0.01 &	2.07 &	22 &	... &	1.29 &	286 &	5.97 &	\ref{232-04d1} \\
235-01\#1 &	234.9-01.4 &	 E,c  	&	 *        	&	 ...                	&	ib,wv,h	&	0.67 &	5.72 &	22 &	... &	4.53 &	2791 &	10.025 &	\ref{235-01d1} \\
235-03\#1 &	235.3-03.9 &	 L,c  	&	 bcr,*     	&	 ps(s,bcr)          	&	h	&	0.03 &	4.35 &	22 &	... &	3.75 &	1758 &	7.291 &	\ref{235-03d1} \\
258-00\#1 &	258.1-00.3 &	 E,c  	&	 *        	&	 ps(s)              	& ib,wv,pr,h	&	0.55 &	6.37 &	22 &	... &	1.52 &	1043 &	9.568 &	\ref{258-00d1} \\
285-02\#1 &	285.6-02.7 &	 M,c  	&	 t,*       	&	 ps(t)              	&	h(i)	&	0.02 &	8.81 &	22 &	... &	3.58 &	3392 &	11.393 &	\ref{285-02d1} \\
296-06\#1 &	296.4-06.9\footnotemark &	 E,c  	&	 bcr?     	&	 ...    &	ir,h	&	0.17 &	1.6 &	22 &	... &	7.8 &	1342 &	6.836 &	\ref{296-06d1} \\
300-02\#1 &	300.7-02.0\footnotemark &	 M,c  	&	 bcr(o),*   	&	 ...    &	ml,h(a)	&	0.74 &	10.69 &	22 &	... &	2.31 &	2660 &	13.671 &	\ref{300-02d1} \\
307-04\#1 &     307.5-04.9\footnotemark &        B,o    &        t,*            &  ps(s,an)     & an,ib,wv,rg   &       0.26 &  17.6  & 41 &    10  &   2.4  &  2530 &  ...   & ... \\
315+09\#1 &	315.4+09.4\footnotemark &	 B,o  	&	 *        	&	 ps(an) &	an,ib	&	0.35 &	37.32 &	22 &	... &	0.8 &	3217 &	54.67 &	\ref{315+09d1} \\
315-13\#1 &	315.1-13.0 &	 E,c  	&	 t,*       	&	 ps(s)              	&	pr,h	&	$<1$ &	6.64 &	11 &	12 &	0.87 &	1248 &	13.671 &	\ref{315-13d1} \\
320-09\#1 &	320.1-09.6 &	 M,c  	&	 *        	&	 ps(s,m)            	&	pr,h	&	0 &	6.67 &	... &	11 &	5 &	8757 &	14.582 &	\ref{320-09d1} \\
321+02\#1 &	321.3+02.8 &	 B,c  	&	 t,*       	&	 ps(s)              	&	ib,h	&	0.41 &	5.87 &	9 &	... &	2 &	3093 &	8.203 &	\ref{321+02d1} \\
321+03\#1 &	321.0+03.9\footnotemark &	 M,c  	&	 w,*       	&	 ...                	&	h	&	0.02 &	4.03 &	22 &	... &	1.5 &	651 &	8.201 &	\ref{321+03d1} \\
325-12\#1 &	325.8-12.8 &	 E,c  	&	 ...      	&	 ...                	&	h	&	$<1$ &	1.58 &	... &	21.5 &	7.4 &	1287 &	8.203 &	\ref{325-12d1} \\
326-06\#1 &	326.0-06.5 &	 E    	&	 *        	&	 ps(s)              	&	h	&	0 &	1.12 &	... &	4 &	8 &	5303 &	6.836 &	\ref{326-06d1} \\
327-02\#1 &	327.1-02.2 &	 E?,c 	&	 *        	&	 ps(s)              	&	rr,h	&	0.01 &	3.59 &	22 &	... &	3.7 &	1430 &	9.114 &	\ref{327-02d1} \\
331-01\#1 &	331.7-01.0\footnotemark &	 B,o  	&	 w,*       	&	 ...                	&	ib,rr	&	0.02 &	48.87 &	22 &	... &	1.34 &	7029 &	66.004 &	\ref{331-01d1} \\
331-02\#2 &	331.5-02.7\footnotemark &	 E,o  	&	 t,*?      	&	 ...                	&	an	&	0.15 &	20.55 &	22 &	... &	3.35 &	7418 &	27.342 &	\ref{331-02d2} \\
... &	332.9-09.9\footnotemark &	 M    	&	 *        	&	 ps(m)?             	&	h	        &	0 &	12.24 &	22 &	... &	1.5 &	1978 &	13.375 &	\ref{cpd-22568} \\
350+04\#1 &	350.9+04.4\footnotemark &	 I,c  	&	 t,*       	&	 ...                	&	ir,h(i)	&	0.11 &	4.93 &	... &	13 &	4.6 &	4134 &	11.393 &	\ref{350+04d1} \\
351+04\#1 &	351.1+04.8 &	 E,c  	&	 *        	&	 ps(s)              	&	ib,pr,ir	&	0.78 &	3.24 &	22 &	... &	5.54 &	1935 &	5.924 &	\ref{351.1+04.8} \\
... &	351.9-01.9 &	 B,c? 	&	 t        	&	 ...                	&	ib	&	0.69 &	5.01 &	22 &	... &	8 &	4317 &	6.836 &	\ref{351.9-01.9} \\
352+03\#2 &	352.6+03.0 &	 M,c  	&	 bcr,*?    	&	 ps(m)              	&	...	&	0.39 &	3.77 &	22 &	... &	5.26 &	2136 &	5.013 &	\ref{352.6+03.0} \\
352-07\#1 &	352.9-07.5 &	 B,o  	&	 t        	&	 ...                	&	an,sk,ib,h(e)	&	$<1$ &	10.82 &	22 &	... &	2.13 &	2478 &	15.95 &	\ref{352-07d1} \\
354+03\#1 &	354.5+03.3 &	 B,c  	&	 t        	&	 ...                	&	...	&	0.86 &	2.4 &	22 &	... &	8 &	2068 &	3.646 &	\ref{354.5+03.3} \\
355+03\#3 &	354.9+03.5 &	 E,o  	&	 *?       	&	 ...                	&	...	&	0.13 &	5.87 &	22 &	... &	8 &	5061 &	8.203 &	\ref{354.9+03.5} \\
355-02\#1 &	355.4-02.4\footnotemark &	 B,o? 	&	 bcr(o)?  	&	 ...                	&	...	&	0.91 &	9.89 &	22 &	... &	3.48 &	3706 &	12.76 &	\ref{355.4-02.4} \\
355+03\#2 &	355.9+03.6 &	 E    	&	 ...      	&	 ps(s)              	&	h	&	0.29 &	0.66 &	... &	40 &	4.36 &	170 &	2.279 &	\ref{355.9+03.6} \\
355-03\#3 &	356.5-03.9 &	 B,c  	&	 bcr(o)   	&	 ...                	&	h	&	0.79 &	4.59 &	22 &	... &	2.61 &	1289 &	6.836 &	\ref{355-03d3} \\
355-04\#2 &	355.9-04.2\footnotemark &	 M,c  	&	 t,*       	&	 ps(m)              	&	h(a)	&	0.25 &	4.73 &	22 &	... &	3 &	1530 &	7.291 &	\ref{355-04d2} \\
356-03\#2 &	356.5-03.6\footnotemark &	 B    	&	 t?       	&	 ps(s)              	&	...	&	0.56 &	12.9 &	22 &	... &	8 &	11119 &	14.127 &	\ref{356.5-03.6} \\
356+03\#1 &	356.8+03.3 &	 S    	&	 t        	&	 ...                	&	h(i)	&	0.04 &	1.34 &	22 &	... &	8 &	1156 &	4.557 &	\ref{356.8+03.3} \\
356-03\#3 &	356.5-03.9 &	 L    	&	 bcr(c),*  	&	 ps(s,bcr)          	&	ib?,h	&	0.02 &	5.21 &	22 &	... &	7.53 &	4226 &	6.836 &	\ref{356-03d3} \\
357-04\#3 &	357.1-04.7 &	 B    	&	 t,*       	&	 ps(s)              	&	...	&	0 &	2.63 &	22 &	... &	13.2 &	3739 &	4.557 &	\ref{357.1-04.7} \\
357+02\#6 &	357.2+02.0 &	 R    	&	 ...      	&	 ...                	&	h	&	1.04 &	2.67 &	22 &	... &	8 &	2304 &	6.836 &	\ref{357.2+02.0} \\
358-04\#1 &	358.5-04.2 &	 M,c  	&	 t        	&	 ps(m,an,s)         	&	an,ml	&	0.6 &	8 &	22 &	... &	3.92 &	3375 &	9.114 &	\ref{358.5-04.2} \\
358+05\#2 &	358.7+05.2\footnotemark &	 E?    	&	 bcr(o),*  	&	 ps(s)              	&	...	&	0 &	3.22 &	22 &	... &	5.51 &	1910 &	4.557 &	\ref{358.7+05.2} \\
358+03\#4 &	358.9+03.4 &	 M,c  	&	 t        	&	 ps(m)              	&	...	&	0.22 &	3.21 &	22 &	... &	4.9 &	1693 &	4.557 &	\ref{358.9+03.4} \\
358-00\#2 &	358.9-00.7\footnotemark &	 M,c  	&	 bcr(c),*  	&	 ...                	&	ir,h(i)	&	0.07 &	5.13 &	5.5 &	13.5 &	1.51 &	3325 &	13.671 &	\ref{358-00d2} \\
359+04\#1 &	359.2+04.7\footnotemark &	 E,c  	&	 *        	&	 ...                	&	...	&	0 &	1.81 &	22 &	... &	5 &	974 &	3.646 &	\ref{359.2+04.7} \\
\hline 	 		 		 		 		 	 	 	 	 	 	 	 	 
Objects & with & R$_{exc}$ & $>$\,\,1 \\
\hline 	 		 		 		 		 	 	 	 	 	 	 	 	 
002-03\#2 &	002.3-03.4 &	 B    	&	 t        	&	 ...                	&	...	&	1.45 &	4.43 &	22 &	... &	3.63 &	1731 &	6.836 &	\ref{002.3-03.4} \\
002-03\#6 &	002.9-03.9\footnotemark &	 S    	&	 *        	&	 ...                	&	...	&	2 &	4.47 &	22 &	... &	8 &	3852 &	9.114 &	\ref{002.9-03.9} \\
003+05\#1 &	003.8+05.3\footnotemark &	 I    	&	 ...      	&	 ...                	&	...	&	1.16 &	3.91 &	22 &	... &	6.93 &	2920 &	6.38 &	\ref{003.8+05.3} \\
003+02\#1 &	003.1+02.9 &	 M,c  	&	 bcr(i),*  	&	 ps(m,an)           	&	an	&	1.14 &	30.05 &	... &	23 &	2.15 &	6658 &	31.899 &	\ref{003+02d1} \\
... &	004.1-03.8\footnotemark &	 E    	&	 t?       	&	 ...                	&	...	&	1.31 &	2.67 &	22 &	... &	8 &	2304 &	4.557 &	\ref{004.1-03.8} \\
004-22\#1 &	004.8-22.7 &	 E    	&	 ...      	&	 ...                	&	h	&	1.87 &	0.61 &	22 &	... &	4.76 &	314 &	2.279 &	\ref{004.8-22.7} \\
... &	005.2-18.6\footnotemark &	 E,o  	&	 t        	&	 ps(t)              	&	...	&	2.73 &	3.02 &	22 &	... &	24.8 &	8078 &	3.646 &	\ref{005.2-18.6} \\
... &	006.8-19.8 &	 E    	&	 ...      	&	 ps(s)              	&	...	&	3.64 &	1.62 &	22 &	... &	24.8 &	4332 &	4.101 &	\ref{006.8-19.8} \\
006+04\#1 &	006.3+04.4 &	 L,c  	&	 bcr,*      	&	 ...                	&	ir,h	&	1.91 &	5.13 &	22 &	... &	5.8 &	3201 &	6.836 &	\ref{006.3+04.4} \\
008-07\#2 &	008.3-07.3\footnotemark &	 M,c  	&	 bcr?,*    	&	 ...                	&	ib,h	&	4.0 &	7.72 &	22 &	... &	2.67 &	2217 &	11.39 &	\ref{008-07d2} \\
010-06\#1 &	010.7-06.4\footnotemark &	 I    	&	 *        	&	 ps(s)              	&	h	&	2.9 &	2.04 &	22 &	... &	4.9 &	1076 &	4.101 &	\ref{010-06d1} \\
013+04\#1 &	013.1+04.1\footnotemark &	 B?,c 	&	 bcr?,*    	&	 ps(s)              	&	h	&	1.13 &	5.29 &	... &	12.5 &	3.41 &	1944 &	8.194 &	\ref{013+04d1} \\
019-05\#1 &	019.4-05.3 &	 M,c  	&	 bcr,*     	&	 ps(s)              	&	ml,ib,h	&	1.41 &	5.72 &	22 &	... &	2.37 &	1460 &	8.203 &	\ref{019-05d1} \\
023-02\#1 &	023.9-02.3 &	 B,c  	&	 bcr,*     	&	 ...                	&	ml,h	&	1.34 &	17.89 &	... &	13 &	1.23 &	3996 &	21.874 &	\ref{023-02d1} \\
027-09\#1 &	027.6-09.6\footnotemark &	 M,c  	&	 bcr(i),*  	&	 ps(m,an)           	&	an,ib?,h(i)	&	2.8 &	9.36 &	... &	13.1 &	6.5 &	6558 &	11.393 &	\ref{027-09d1} \\
032+07\#2 &	032.1+07.0 &	 S    	&	 *        	&	 ...                	&	...	&	1.99 &	4.67 &	22 &	... &	5.71 &	2871 &	5.468 &	\ref{032+07d2} \\
032-02\#1 &	032.7-02.0 &	 L    	&	 bcr(o)   	&	 ps(ib)             	&	an,ib,h(i)	&	2.01 &	7.13 &	22 &	... &	3.44 &	2644 &	9.114 &	\ref{032-02d1} \\
037-06\#1 &	037.8-06.3 &	 L,c  	&	 bcr(c)   	&	 ...                	&	ib,h	&	2.34 &	4.98 &	... &	15 &	1.54 &	1213 &	7.291 &	\ref{037-06d1} \\
060-07\#2 &	060.1-07.7\footnotemark &	 B,c  	&	 bcr(o,i),* 	&	 ...    &	wv,h	&	3.9 &	11.12 &	25.5 &	20 &	1.7 &	1757 &	13.671 &	\ref{060-07d2} \\
074+02\#1 &	074.5+02.1 &	 B,o  	&	 t        	&	 ps(s,t)            	&	rg	&	1.88 &	17.09 &	... &	16.5 &	3.2 &	7856 &	30.988 &	\ref{074+02d1} \\
089-05\#1 &	089.8-05.1 &	 M,c  	&	 bcr(c)   	&	 ps(m)              	&	h(e,a)	&	3.8 &	4.16 &	21.5 &	16.5 &	2.1 &	964 &	9.114 &	\ref{089-05d1} \\
304-04\#1 &	304.5-04.8 &	 B,c  	&	 bcr(o,i)?	&	 ps(s)              	&	an,ml,ir	&	4.5 &	19.59 &	... &	12 &	2.1 &	8124 &	33.266 &	\ref{304-04d1} \\
356+04\#2 &	356.9+04.4 &	 B    	&	 ...      	&	 ps(an,s)           	&	an	&	1.56 &	5.65 &	22 &	... &	4.73 &	2878 &	6.836 &	\ref{356.9_04.4} \\

\end{longtable}
\end{landscape}
\setcounter{footnote}{0}
\footnote{Col. 1: Object name in PK format}
\footnote{Col. 2: Object name in PNG format}
\footnote{Cols. 3,4,5,6: morphological classification, divided into 4 parts: (i) the primary classification and the secondary
descriptor
related to whether the lobes (or the shell in the case of an E primary classification) are open or closed, (ii) the secondary
descriptors
for the central region, (iii) the secondary descriptors describing point-symmetry, and (iv) all remaining secondary descriptors}
\footnote{Col. 7: the [OIII]$\lambda$5007/H$\alpha$ line flux ratio}
\footnote{Col. 8: The angular size of the object in arcseconds}
\footnote{Col. 9: The expansion velocity measured from the
[NII]$\lambda$6583 as listed in the Acker et al. catalog}
\footnote{Col. 10: The expansion velocity measured from the
[OIII] line as listed in the Acker et al. catalog}
\footnote{Col. 11: The distance to the object}
\footnote{Col. 12: The derived expansion age}
\footnote{Col. 13: The size of the panel (along the horizontal axis) in the figure of the object}
\footnote{Col. 14: The number of the figure showing the object}
\footnote{PNG001.7$-$04.4: $bcr$ since major/minor axis ratio of central region not consistent with tilted torus}
\footnote{PNG003.6+03.1, PNG003.9$-$03.1, PNG008.6$-$02.6, PNG326.0$-$06.5, PNG351.9$-$01.9, PNG354.5+03.3, PNG354.9+03.5,
PNG356.5$-$03.6, PNG356.8+03.3, PNG002.9-03.9 and PNG004.1$-$03.8 are Galactic Bulge PNs, and their
distance is taken to be 8\,kpc}
\footnote{PNG004.0$-$03.0: note the jet emanating at $pa\sim30\arcdeg$, a structure not covered in our
morphological scheme. Making the assumption that the jet is launched roughly orthogonally to the waist of the
nebula, the bright elliptical ring is identified as a tilted, toroidal waist ($t$ descriptor) with its major axis
along $pa\sim-55\arcdeg$}
\footnote{PNG004.9+04.9: the halo has an inner irregularly-shaped brighter component, with a discernible
periphery, and an outer, more typical diffuse component with a surface-brightness limited size}
\footnote{PNG010.8+18.0: the distance has been taken from Schwarz et al. (1997)}
\footnote{PNG012.2+04.9: distance taken from Preite-Martinez (1988), and R$_{exc}$ estimated from spectrum in Su{\'a}rez et al. (2006);
object is listed as IRAS\,17514-1555 in these studies}
\footnote{PNG043.1+03.8: the $h(a)$ descriptor is used to denote the presence of the faint,
somewhat irregular, but complete ring in the halo, which surrounds the central bright nebula}
\footnote{PNG051.5+00.2: R$_{exc}$ estimated from spectrum in Kerber et al. (1996); no distance estimate available, hence no estimate
of age possible. Note that finding chart in Kerber et al. (1996) incorrectly points to a star 11.5{''} west of the PN, and the
coordinates given in Table 1 are incorrect. The J2000 SIMBAD coordinates of this source, RA=19h25m40.68s +16d33m05.6s, however, are
consistent with the coordinates derived for the central star, from the HST image, RA=19h25m40.48s +16d33m05.4s}
\footnote{PNG058.3$-$10.9: the bipolar lobes can be seen very faintly along $pa\sim25\arcdeg$; the $t$ denotes
the bright region with its major axis orthogonal to the lobe axis}
\footnote{PNG061.3+03.6: the distance has been taken from Dobrin{\v c}i{\'c} et al.(2008)}
\footnote{PNG064.7+05.0: the [OIII]/H$\alpha$ ratio is much less than unity; since the H$\alpha$ line is saturated in
Acker et al, we have set the ratio to 0}
\footnote{PNG067.9$-$00.2: the object may be bipolar (with lobes aligned along $pa\sim90\arcdeg$), with an obscuring
waist and irregular structure, i.e., B,w,ir; however we have conservatively chosen I as the primary classification for this
object}
\footnote{PNG071.6-02.3: the elongated structure at $pa\sim60\arcdeg$ is considered to be a second lobe pair,
leading to its primary classification of M}
\footnote{PNG093.9-00.1: the [OIII]/H$\alpha$ ratio is much smaller than unity, based on spectroscopic data
described in Sanchez Contreras et al. 2008; distance is taken from Meixner et al. (1997)}
\footnote{PNG110.1+01.9: the [OIII]/H$\alpha$ ratio has been set to zero based on the non-detection of
[OIII]$\lambda5007$ by Garcia Lario et al. (1991), and the distance has been taken from the same reference}
\footnote{PK130$-$11\#1: we use ``*?", since given the diffuse, irregular shape of this object, and the presence
of additional stars in the vicinity, we cannot be certain that the star near the center of this object is really
the CSPN}
\footnote{PNG146.7+07.6: because of the apparent large tilt of the nebular axis towards the line-of-sight, the
lobes may also be open structures}
\footnote{PNG147.4$-$02.3: the primary PN shape is assumed to be defined by the outer periphery in this image;
the inner bright structure is an inner bubble (i.e., $ib$)}
\footnote{PNG232.8$-$04.7: the halo has an inner irregularly-shaped brighter component, with a discernible
periphery, and an outer, more typical diffuse component with a surface-brightness limited size}
\footnote{PNG296.4$-$06.9: the central region is described as ``$bcr?$" because its extent and structure are not well
resolved}
\footnote{PNG300.7$-$02.0: this object shows a dusty structure which produces obscuration as it cuts across the lower
lobes, which is not captured in our primary and secondary descriptors}
\footnote{PNG307.5$-$04.9: we have adopted values of the size and distance for this object (also known as MyCNn\,18), from Sahai et al.
(1999). The radially-varying [NII] expansion velocity, V$_1$, is estimated from eqn. (4) of Dayal et al. (2000) using a radius R set
to half the size of the object. A ground-based image (Bryce et al. 1997) shows distant ansae on both sides of the center}
\footnote{PNG315.4+09.4: distance from Schwarz, Aspin \& Lutz (1989), note that the long-slit spectra in Corradi \&
Schwarz (1993) suggest much higher expansion velocities ($\gtrsim100$\,\kms), who derive a smaller expansion age, 
$\lesssim$920\,yr}
\footnote{PNG321.0+03.9: distance from  Sahai et al. (2000b)}
\footnote{PNG331.07$-$01.0: the $w$ descriptor is used to denote the belt of obscuration (oriented along the
minor axis of the object) that cuts across the inner regions of the bright bipolar lobes}
\footnote{PNG331.5$-$02.7: we use ``*?", since the ``central" star is noticeably offset from the geometric center
of the toroidal waist, and since there are many additional stars visible in its vicinity, we cannot be sure that
this star is really the CSPN}
\footnote{PNG332.9$-$09.9: the [OIII]/H$\alpha$ ratio has been set to zero based on the non-detection of
[OIII]$\lambda5007$ by De Marco \& Crowther (1998), and the distance has been taken from the same reference. The object most
likely possesses point-symmetry by virtue of having at least two pairs of diametrically-opposed lobes, but we assign it
``$ps(ml)?$" because the $pa=180\arcdeg$ counterpart of the $pa=0\arcdeg$,\,$180\arcdeg$ is only partially visible}
\footnote{PNG350.9+04.4: the $t$ descriptor is used to describe bright inner ring structure, which is most likely a
torus seen nearly face-on}
\footnote{PNG355.4$-$02.4: we assign ``$bcr(o)?$" to the central region, because although its extensions along the polar
direction cannot be separated unambiguously from the low-latitude regions of the bipolar lobes}
\footnote{PK355$-$04\#2: the lobe at $pa\sim0\arcdeg$ has a complex, ``layered" periphery, with 3 layers}
\footnote{PNG356.5$-$03.6: we desribe the central region as ``$t?$" because it appears to be somewhat less extended along
the polar axis than along the equatorial plane; however its structure makes this assessment a bit uncertain}
\footnote{PNG358.7+05.2: We have assigned this ``E?", but, like PNG013.1+04.1, it may be a bipolar object with its major axis
oriented at a small angle to the line-of-sight, since a slight ``pinching-in" of the primary shape can be seen along $pa\sim55\deg$}
\footnote{PNG358.9$-$00.7: the central region is described as $bcr(c)$, even though its shape is almost spherical, not
barrel-like; the plethora of protrusions emanating from the main nebula body represent structures
not covered in our morphological scheme}
\footnote{PNG359.2+04.7: distance from Gesicki and Zijlstra (2007)}
\footnote{PNG002.9$-$03.9: since there are two stars near the center, we speculate that these may represent a 
binary CSPN responsible for the spiral structure}
\footnote{PNG003.8+05.3: although classified as I, faint nebulosity can be seen extending along
$pa\sim170\arcdeg$ and possibly $pa\sim-15\arcdeg$, suggesting that this may be really a B or L object with very
faint lobes}
\footnote{PNG004.1$-$03.8: the curved bright structure seen near the waist may represent a partial torus
structure}
\footnote{PNG005.2$-$18.6 \& PNG006.8-19.8: based on their radial velocities, these PNs are believed to belong to
the Sagittarius dwarf spheroidal galaxy (Zijlstra et al 2006) at a distance of 24.8\,kpc (Kunder \& Chaboyer 2009)}
\footnote{PNG008.3$-$07.3: this object is clasified as multipolar (M) because it has two pairs of
lobes at slightly different orients (in projection), i.e., at $pa\sim5\arcdeg,\,17\arcdeg$. We assign ``$bcr?$" to the central
region because the structure seen there could result from a pair of lobes projected almost directly towards the
line-of-sight}
\footnote{PNG010.7-06.4: inspite of its primary classification as I, there are two parallel, almost linear
features on the left and right sides of the central star which motivate including the ps(s) descriptor}
\footnote{PNG013.1+04.1: we have assigned this ``B?", because it is quite possibly a bipolar object with its major axis
oriented at a small angle to the line-of-sight. For the same reason, although the central region appears extended along the
major axis, it is difficult to assess this extent, hence we use ``$bcr?$" to describe this region}
\footnote{PNG027.6$-$09.6: we assign the secondary descriptor ``$ib?$" to this PN because although there appears to be an
elliptical shaped structure in the center, the additional structure within the central region makes its identification
somewhat ambiguous}
\footnote{PNG060.1$-$07.7: the central star and round halo are seen clearly in the F555W image, and are therefore
included in the morphological classification}
\normalsize
\clearpage
\begin{footnotesize}
\begin{center}
\begin{longtable}{p{1.0in}p{0.6in}p{0.6in}p{0.6in}p{0.6in}p{0.6in}}
\caption{Statistics}\label{t5-stats}\\
\hline\hline \\[-2ex]
   \multicolumn{1}{l}{\textbf{Classification}} &
   \multicolumn{1}{l}{\textbf{Number$^1$}} &
   \multicolumn{1}{l}{\textbf{Fraction$^1$}} &
   \multicolumn{1}{l}{} &
   \multicolumn{1}{l}{\textbf{Number$^2$}} &
   \multicolumn{1}{l}{\textbf{Fraction$^2$}} \\
  \multicolumn{1}{c}{\textbf{~}} &  
  \multicolumn{2}{c}{\textbf{R$_{exc}\le1$}} & 
  \multicolumn{1}{l}{} &
  \multicolumn{2}{c}{\textbf{All Objects}}  \\[0.5ex] \hline
  \\[-1.8ex]
\endfirsthead

\multicolumn{2}{c}{{\tablename} \thetable{} -- Continued} \\[0.5ex]
  \hline \hline \\[-2ex]
   \multicolumn{1}{l}{\textbf{Classification}} &
   \multicolumn{1}{l}{\textbf{Number$^1$}} &
   \multicolumn{1}{l}{\textbf{Fraction$^1$}} &
   \multicolumn{1}{l}{} &
   \multicolumn{1}{l}{\textbf{Number$^2$}} &
   \multicolumn{1}{l}{\textbf{Fraction$^2$}} \\
  \multicolumn{1}{c}{\textbf{~}} &  
  \multicolumn{2}{c}{\textbf{R$_{exc}\le1$}} &  
  \multicolumn{1}{l}{} &
  \multicolumn{2}{c}{\textbf{All Objects}}   \\[0.5ex] \hline
  \\[-1.8ex]
\endhead

  \hline\\
  \multicolumn{3}{l}{{Continued on Next Page\ldots}} \\
\endfoot

  \\[-1.8ex] \hline \hline
\endlastfoot

B         &   27  &   0.28   &   & 33    & 0.28 \\
M         &   18  &   0.19   &   & 23    & 0.20 \\
E         &   32  &   0.34   &   & 37    & 0.31 \\
I         &    6  &   0.063  &   & 8     & 0.068\\
R         &    4  &   0.042  &   & 4     & 0.034\\
L         &    7  &   0.074  &   & 10    & 0.085\\
S         &    2  &   0.021  &   & 4    & 0.034\\
\hline
\multicolumn{6}{c} {Point Symmetry}\\
\hline
B, ps$^3$ &   12  &   0.44   &   & 14    & 0.45 \\
M, ps$^4$ &   15  &   0.83   &   & 19    & 0.83 \\
E, ps$^5$ &   13  &   0.41   &   & 15    & 0.42 \\
ps$^4$    &   42  &   0.44   &   & 53    & 0.45 \\

\end{longtable}
\end{center}
\setcounter{footnote}{0}
\footnote{Number of objects in given class, and as a fraction of the total (96) for which the [OIII]$\lambda$5007/H$\alpha$ flux ratio,
R$_{exc}\le1$,  }
\footnote{Number of all objects in given class, and as a fraction of the total sample (119)}
\footnote{Number of point-symmetric objects in class B, and as a fraction of the total in class B}
\footnote{Number of point-symmetric objects in class M, and as a fraction of the total in class M}
\footnote{Number of point-symmetric objects in class E, and as a fraction of the total in class E}
\footnote{Total number and fraction of point-symmetric objects, including objects of the S primary class, which is point-symmetric by
definition}
\end{footnotesize}
\clearpage
\begin{scriptsize}
\begin{center}
\begin{longtable}{p{0.65in}p{0.1in}p{0.4in}p{0.5in}p{0.6in}p{0.3in}p{0.3in}}
\caption{Morphological Classification for PNs from HST program GO 11657}\label{t5}\\
\hline\hline \\[-2ex]
   \multicolumn{1}{l}{\textbf{Name}} &
   \multicolumn{4}{c}{\textbf{Morphology}} &
   \multicolumn{1}{l}{\textbf{R$_{exc}$}} &
   \multicolumn{1}{l}{\textbf{Fig\#}} \\[0.5ex] \hline
   \multicolumn{1}{l}{\textbf{PNG}} &
   \multicolumn{1}{l}{\textbf{Prim}} &
   \multicolumn{1}{l}{\textbf{Cen}} &
   \multicolumn{1}{l}{\textbf{PS}} &
   \multicolumn{1}{l}{\textbf{Other}} &
   \multicolumn{2}{l}{\textbf{}} \\[0.5ex] \hline
   \\[-1.8ex]
\endfirsthead

\multicolumn{2}{c}{{\tablename} \thetable{} -- Continued} \\[0.5ex]
  \hline \hline \\[-2ex]
   \multicolumn{1}{l}{\textbf{Name}} &
   \multicolumn{4}{c}{\textbf{Morphology}} &
   \multicolumn{1}{l}{\textbf{R$_{exc}$}} &
   \multicolumn{1}{l}{\textbf{Fig\#}} \\[0.5ex] \hline
   \multicolumn{1}{l}{\textbf{PNG}} &
   \multicolumn{1}{l}{\textbf{Prim}} &
   \multicolumn{1}{l}{\textbf{Cen}} &
   \multicolumn{1}{l}{\textbf{PS}} &
   \multicolumn{1}{l}{\textbf{Other}} & 
   \multicolumn{2}{l}{\textbf{}} \\[0.5ex] \hline
  \\[-1.8ex]
\endhead

  \hline\\
  \multicolumn{3}{l}{{Continued on Next Page\ldots}} \\
\endfoot

  \\[-1.8ex] \hline \hline
\endlastfoot
\setcounter{footnote}{0}
Objects & with & R$_{exc}$ & $\le$\,\,1 	 \\
\hline	
014.3-05.5  	&	 E,c    	&	 bcr(c),* 	&	 ...                	&	h	&	 0   	& ... \\
059.9+02.0  	&	 B,c    	&	 t,*       	&	 ...                	&	h	&	 0   	& ... \\
063.8-03.3  	&	 E,c    	&	 t?,*      	&	 ...                	&	h	&	 0   	& ... \\
079.9+06.4  	&	 R    	&	*       	&	 ...                	&	h	&	 0.68 		& ... \\
098.2+04.9  	&	 E,c    	&	*       	&	 ...                	&	h	&	 0.88 	& ... \\
104.1+01.0  	&	 E,c    	&	*       	&	 ...                	&	h	&	 0.57 	& ... \\
107.4-02.6  	&	 E,c    	&	*       	&	 ...                	&	h	&	 0.72 	& ... \\
294.9-04.3  	&	 E,c    	&	*       	&	 ...                	&	h	&	 0.16 	& ... \\
309.0+00.8  	&	 B    	&	 bcr(i),* 	&	 ...                	&	h(e)	&	 0.86 		& ... \\
324.8-01.1  	&	 E,c    	&	*       	&	 ps(s)?             	&	ib,ml,h	&	 0.37 	& ... \\
327.1-01.8  	&	 E,c    	&	*       	&	 ...                	&	pr,h	&	 0.036 	& ... \\
334.8-07.4  	&	 M,c    	&	*       	&	 ps(m)                	&	h?	&	 0.36  	& \ref{334.8-07.4} \\
344.2+04.7\footnotemark  	&	 B,c  	&	 t,*       	&	 ...            &	h	&	 0.59  	& \ref{344.2+04.7} \\
344.8+03.4  	&	 S    	&	 t,*       	&	 ...                	&	...	&	 0.22  		& \ref{344.8+03.4} \\
356.5+01.5  	&	 E    	&	 t,*       	&	 ps(s)              	&	ib,h(d)	&	 0.18  		& \ref{356.5+01.5} \\
184.0-02.1  	&	 E,c    	&	*       	&	 ...          	&	pr,h	&	 $<1$  		& ... \\
\hline	
Objects & with & R$_{exc}$ & $>$\,\,1 	\\
\hline	
000.8-07.6  	&	 B,o    	&	 t,*       	&	 ...                	&	h(i)	&	 2.6 	& ... \\
014.0-05.5  	&	 B,c  	&	 bcr(c),* 	&	 ...                	&	h(d)	&	 1.5 		& \ref{014.0-05.5} \\
021.1-05.9  	&	 B,c  	&	 bcr(i),* 	&	 ...                	&	h?	&	 3.1 		& \ref{021.1-05.9} \\
025.3-04.6  	&	 E    	&	 bcr(o),* 	&	 ps(an,bcr)         	&	an,h	&	 3.2 		& ... \\
026.5-03.0  	&	 B,c  	&	*       	&	 ps(s)              	&	h(d)	&	 1.4 		& ... \\
042.9-06.9  	&	 L    	&	 t,*       	&	 ...                	&	ir,h	&	 3.0 		& ... \\
048.5+04.2\footnotemark  	&	 B    	&	 t,*       	&	 ps(t)  &	h(e,d)?	&	 1.3 		& \ref{048.5+04.2} \\
052.9-02.7  	&	 E,c    	&	*       	&	 ps(s)              	&	h	&	 2.1 	& ... \\
053.3+24.0\footnotemark  	&	 B,c  	&	 t,*        	&	 ps(s,t,ib)    	&	ib,h(e)	&	 3.4 	& \ref{053.3+24.0} \\
068.7+14.8  	&	 R    	&	*       	&	 ...                	&	h	&	 1.2 		& ... \\
068.7+01.9\footnotemark  	&	 E    	&	 t,*       	&	 ...    &	ib,h(a?)	&	 1.2 	& ... \\
097.6-02.4  	&	 E    	&	 t,*        	&	 ps(s)              	&	...	&	 2.1 		& \ref{097.6-02.4} \\
205.8-26.7  	&	 R    	&	*       	&	 ...                	&	ib,h	&	 4.0 		& ... \\
263.0-05.5  	&	 E,c    	&	*       	&	 ...                	&	ib,h	&	 2.3 	& ... \\
264.4-12.7  	&	 R    	&	*       	&	 ...                	&	h	&	 1.7 		& \ref{264.4-12.7} \\
275.3-04.7  	&	 E,c    	&	*       	&	 ...                	&	h(d)	&	 2.5  	& ... \\
278.6-06.7  	&	 E,c    	&	*       	&	 ps(ib)             	&	ib,h	&	 1.1  	& \ref{278.6-06.7} \\
285.4+02.2  	&	 I    	&	*       	&	 ...                	&	h(i)	&	 2.4  		& ... \\
285.4+01.5\footnotemark  	&	 B,c  	&	 bcr      	&	 ps(s)  &	ml,h(a?) &	 1.1  		& \ref{285.4+01.5} \\
286.0-06.5  	&	 B,o  	&	 t,*       	&	 ps(s)              	&	rg,h	&	 2.1  		& \ref{286.0-06.5} \\
289.8+07.7  	&	 R    	&	*       	&	 ...                	&	h	&	 3.1  		& ... \\
295.3-09.3  	&	 B    	&	 bcr,*     	&	 ...                	&	h	&	 2.4  		& ... \\
296.3-03.0  	&	 B,c  	&	 bcr(o)   	&	 ps(s,ib)           	&	ib,h	&	 2.3  		& ... \\
309.5-02.9\footnotemark  	&	 L    	&	 t,*       	&	 ...    &	ib,h	&	 1.2  		& \ref{309.5-02.9} \\
336.9+08.3  	&	 E,c    	&	*       	&	 ...            &	h(e,d)	&	 1.2  		& ... \\
340.9-04.6  	&	 E    	&	*       	&	 ...                	&	ib,h	&	 1.5   		& ... \\
343.4+11.9\footnotemark  	&	 E    	&	*       	&	 ...    &	h	&	 3.7   		& ... \\
348.4-04.1      &        B,o    &       t,*             &        ...    &       h       &        1.5     	& \ref{348.4-04.1} \\
348.8-09.0  	&	 R    	&	*       	&	 ...    &	ib,h	&	 1.2   		& \ref{348.8-09.0} \\
351.3+07.6  	&	 E    	&	*       	&	 ps(s)  &	h	&	 1.3   		& ... \\
358.6+07.8  	&	 E    	&	*       	&	 ...    &	h(d)	&	 1.4 		& ... \\

\end{longtable}
\end{center}
\setcounter{footnote}{0}
\footnote{PNG344.2+04.7: Although we assign B as the primary class, this object may be intrinsically multipolar (M), with the inner
bright lobe pair component having its axis
projected along the axis of the larger, faint primary lobe pair}
\footnote{PNG048.5+04.2: The axis of the bipolar lobes, which are apparently inclined at a relatively small inclination to the
line-of-sight, is oriented at $pa\sim50\arcdeg$. An elongated diffuse nebulosos structure can be seen with its major axis at
$pa\sim140\arcdeg$, but gives its rather limited radial extent, we tentatively classify it as a halo, i.e., as h(e,d)?}
\footnote{PNG053.3+24.0: The ``t" descriptor refers to the bright structure oriented with its long axis along $pa\sim-20\arcdeg$;
however,
it is possible that this structure is actually a lobe structure, similar to that seen in the multipolar PPN, IRAS19475+3119 (Sahai et
al.
2007b), suggesting the alternative classification M,c,*,ib,ps(m,ib),h(e)}
\footnote{PNG068.7+01.9: We use ``$a?$" for the qualifier in the halo descriptor, since only a single, partial arc-like feature is
seen}
\footnote{PNG285.4+01.5: We assign the feature at $pa\sim-50\arcdeg$ the minor lobe descriptor $ml$, but it is possible that this
is a planar structure whose radial density distribution has a sharp outer edge, in which case it would be better described by the
$w(b)$ descriptor used for PPNs (SMSC07). We use ``$a?$" for the qualifier in the halo descriptor, since only a single, partial
arc-like feature is seen}
\footnote{PNG309.5-02.9: We assign this a primary class L; the collimated lobes are seen weakly at $pa\sim70\arcdeg$. The bright
structure surrounding the central star is an inner bubble, and the torus is the ring-like structure which is brightest on the
right side of the central star in the image}
\footnote{PNG343.4+11.9: A well-defined nebular shell is not seen in this object; we assume that the inner bright region oriented
at $pa\sim-15\arcdeg$ represents the primary nebula, and the surrouding diffuse structure is the halo}

\end{scriptsize}
\end{document}